\documentclass[11pt,letterpaper,oneside,openany]{book}

\title{Initial conditions of bulk matter in ultrarelativistic nuclear collisions}
\author{J.\ Scott Moreland}

\usepackage[utf8]{inputenc}
\usepackage[T1]{fontenc}
\usepackage{lmodern}
\usepackage[light]{roboto}

\usepackage{mathtools}
\usepackage{amssymb}
\usepackage{amsmath}
\usepackage{amsthm}
\theoremstyle{remark}
\newtheorem*{remark}{Remark}

\usepackage{graphicx}
\graphicspath{{figures/}}

\usepackage{wrapfig}
\usepackage{tikz}
\usetikzlibrary{arrows,backgrounds,calc,external,matrix,shapes}

\definecolor{deepred}{RGB}{146,11,36}
\definecolor{offblack}{RGB}{35,31,32}
\usepackage[pdfusetitle,colorlinks=true,allcolors=deepred,bookmarksdepth=2,linktoc=section,bookmarksopen=false]{hyperref}

\usepackage{titlesec}
\titleformat{\chapter}[display]{\Huge\sffamily\color{offblack}\filcenter
}{\thechapter}{1ex}{}
\titleformat*{\section}{\Large\bfseries\color{deepred}}
\titleformat*{\subsection}{\large\bfseries\color{deepred}}
\titleformat*{\subsubsection}{\bfseries\color{deepred}}
\titleformat*{\paragraph}{\bfseries\color{offblack}}
\titlespacing{\paragraph}{0pt}{*1.4}{*2.5}

\usepackage[font=small,labelfont={bf,color=deepred},labelsep=quad]{caption}

\usepackage{lettrine}

\usepackage[inline]{enumitem}

\usepackage{booktabs}
\usepackage{multirow}

\usepackage[style=phys,biblabel=brackets,eprint=true,url=true,
maxcitenames=3]{biblatex}
\usepackage{listings}
\newcommand{\fmc}{\ensuremath{\text{fm}/c}}
\newcommand{\sqrts}{\sqrt{s_\mathrm{NN}}}
\newcommand{\sigmann}{\sigma^\text{inel}_\mathrm{NN}}
\newcommand{\sigmaf}{\sigma_\text{fluct}}
\newcommand{\nch}{N_\text{ch}}
\newcommand{\ntrk}{N_\text{trk}^\text{offline}}
\newcommand{\Np}{N_\text{part}}
\newcommand{\Nc}{N_\text{coll}}
\newcommand{\np}{n_\text{part}}
\newcommand{\nc}{n_\text{coll}}
\newcommand{\Qs}[1]{Q_{s,\text{#1}}}
\newcommand{\avg}[1]{\langle #1 \rangle}
\newcommand{\vnk}[2]{v_{#1}\{#2\}}
\newcommand{\order}[1]{\ensuremath{\mathcal{O}(#1)}}

\newcommand{\trento}{T\raisebox{-0.5ex}{R}ENTo}
\newcommand{\T}{\tilde{T}}
\newcommand{\dmin}{d_\text{min}}
\newcommand{\tfs}{\tau_\text{fs}}
\newcommand{\Tsw}{\ensuremath{T_\text{switch}}}

\newcommand{\vv}{\mathbf v}
\newcommand{\xv}{\mathbf x}
\newcommand{\yv}{\mathbf y}
\newcommand{\zv}{\mathbf z}

\newcommand{\tran}{^\mathrm{T}}
\newcommand{\etas}{\eta/s}
\newcommand{\etashrg}{(\eta/s)_\text{hrg}}
\newcommand{\etasmin}{(\eta/s)_\text{min}}
\newcommand{\etasslope}{(\eta/s)_\text{slope}}
\newcommand{\etascrv}{(\eta/s)_\text{crv}}
\newcommand{\zetas}{\zeta/s}
\newcommand{\zetasnorm}{(\zeta/s)_\text{norm}}
\newcommand{\zetasmax}{(\zeta/s)_\text{max}}
\newcommand{\zetaswidth}{(\zeta/s)_\text{width}}
\newcommand{\zetasloc}{(\zeta/s)_{T_0}}

\newcommand{\sigf}{\sigma_\text{fluct}}
\newcommand{\eccratio}{\sqrt{\langle \varepsilon_2^2 \rangle}/\sqrt{\langle \varepsilon_3^2 \rangle}^{\,0.6}}

\DeclareMathOperator{\expi}{Ei}
\DeclareMathOperator{\acosh}{acosh}
\DeclareMathOperator{\diag}{diag}
\DeclareMathOperator{\cov}{cov}
\DeclareMathOperator{\corr}{corr}
\DeclareMathOperator{\SC}{SC}
\DeclareMathOperator{\NSC}{NSC}

\DeclareMathOperator{\trace}{tr}

\begin{document}

\frontmatter

\makeatletter
\begin{titlepage}
  \centering
  \sffamily
  \vspace*{.25\textheight}
  \Huge\@title \\
  \vspace{.05\textheight}
  \LARGE\@author \\
  \vspace{.05\textheight}
  \Large Ph.D.\ dissertation \\[.25ex]
  Advisor: Steffen A.\ Bass \\[.25ex]
  Department of Physics, Duke University \\
  \vspace{.05\textheight}
  \today
\end{titlepage}
\makeatother

\chapter{Abstract}
\label{ch:abstract}

\enlargethispage{1\baselineskip}

Dynamical models based on relativistic fluid dynamics provide a powerful tool to extract the properties of the strongly-coupled quark-gluon plasma (QGP) produced in the first ${\sim}10^{-23}$ seconds of an ultrarelativistic nuclear collision.
The largest source of uncertainty in these model-to-data extractions is the choice of theoretical initial conditions used to model the distribution of energy or entropy at the hydrodynamic starting time.

Descriptions of the QGP initial conditions are generally improved through iterative cycles of testing and refinement.
Individual models are compared to experimental data; the worst models are discarded and best models retained.
Consequently, successful traits (assumptions) are passed on to subsequent generations of the theoretical landscape.
This so-called bottom-up approach correspondingly describes a form of theoretical trial and error, where each trial proposes an \emph{ab initio} solution to the problem at hand.

A natural complement to this strategy, is to employ a top-down or data-driven approach which is able to reverse engineer properties of the initial conditions from the constraints imposed by the experimental data.
In this dissertation, I motivate and develop a parametric model for initial energy and entropy deposition in ultrarelativistic nuclear collisions which is based on a family of functions known as the generalized means.
The ansatz closely mimics the variability of \emph{ab initio} calculations and serves as a reasonable parametric form for exploring QGP energy and entropy deposition assuming imperfect knowledge of the complex physical processes which lead to its creation.

With the parametric model in hand, I explore broad implications of the proposed ansatz using recently adapted Bayesian methods to simultaneously constrain properties of the initial conditions and QGP medium using experimental data from the Large Hadron Collider.
These analyses show that the QGP initial conditions are highly constrained by available measurements and provide evidence of a unified hydrodynamic description of small and large nuclear collision systems.

\pagebreak

\thispagestyle{empty}
\begin{center}
  \topskip0pt
  \vspace*{\fill}
  \noindent
  I dedicate this dissertation to my mother and father and to Erin, for their unyielding love, encouragement, and unwavering support.
  \vspace*{\fill}
\end{center}

\pagebreak

% reduce spacing around toc header so that it fits on one page
\titlespacing{\chapter}{0pt}{0pt}{30pt}
\setcounter{tocdepth}{1}
\tableofcontents
% reset to default
\titlespacing{\chapter}{0pt}{50pt}{40pt}

\mainmatter

\chapter{Introduction}
\label{ch:introduction}

\lettrine{A}{} central scientific endeavor is to investigate the reducible nature of matter---to classify its elementary quanta and understand its fundamental interactions.
This search is aided by particle accelerators, fantastic machines that collide together nature's smallest particles in search of hidden substructure and unifying symmetries.
The crowning jewel of this effort is the so-called Standard Model of particle physics which describes the strong, weak, and electromagnetic forces observed in nature.

Quantum chromodynamics (QCD), the theory of the strong nuclear force, explains the zoo of strongly interacting particles produced by high-energy nuclear collisions as combinations of two or more fundamental particles known as quarks.
Each quark carries color charge---analogous to the more familiar electric charge of classical electromagnetism---and interacts by exchanging particles known as gluons which mediate the strong force.

A property of QCD known as color confinement stipulates that free quarks can never be observed in nature; quarks may only combine to form color-neutral bound states known as hadrons, of which the proton and neutron are just two examples.
Although the existence of quark and gluon degrees of freedom cannot be observed directly, their presence has been inferred by examining the properties of final-state hadrons produced by energetic nuclear collisions.

One of the primary goals of the high-energy nuclear physics community is to understand the emergent behavior which arises from fundamental quark and gluon interactions over different time and distance scales.
This encompasses both the complex dynamics which occur inside a relativistic nuclear collision, as well as other more exotic nuclear phenomena such as the primordial interactions of quarks and gluons shortly after the big bang.
This particular endeavor is distinguished from the more general effort to specify the fundamental forces and elementary particles of the Standard Model, in that it seeks to understand \emph{bulk} properties of quark-gluon matter, i.e.\ attributes of the aggregate substance and not just its individual components.

Relativistic nuclear collisions are a powerful experimental tool to study quark and gluon interactions experimentally, because unlike the one-of-a-kind event which produced the big bang, high-energy particle physics experiments are repeatable and configurable.
They therefore provide an experimental sand box to develop and test theoretical ideas.
This general concept of using high-energy collisions to study the bulk properties of nuclear matter dates back to the early 1950's, when Landau proposed a hydrodynamic description of hadronic collisions \cite{Landau:1953gs}.
His general argument followed a simple line of reasoning.
When two nucleons collide at relativistic energies, they release a large amount of energy into a very small volume which may be viewed from the center of mass frame of the colliding nucleon pair.
If the collision energy is sufficiently high, the resulting density of secondary particles will be large, and their mean free path will be short relative to the system size.
The resulting interparticle interactions will thus be governed by statistical laws, and the produced fireball will expand hydrodynamically until the mean free path of the particles becomes comparable to that of the system size.
The system will then ultimately break up and disintegrate into a shower of separate particles \cite{Belenkji1956}.

Landau's original hydrodynamic model never described the quanta of a nuclear collision in terms of quarks and gluons; in fact, the existence of these particles was not even postulated until nearly a decade later \cite{Gellmann:1961, Zweig:570209}.
His hydrodynamic model would, however, ultimately lay the groundwork for a new way of thinking about fundamental interactions between quarks and gluons in the context of a thermalized fluid.
This modern hydrodynamic picture of relativistic nuclear collisions began to emerge when Gross, Wilczek, and Politzer discovered asymptotic freedom in 1973: a phenomenon that predicts a weakening of the strong interaction between quarks as the quarks get closer together \cite{PhysRevLett.30.1343, PhysRevLett.30.1346}.
Their finding had broad phenomenological implications, and it lead to the realization that quarks and gluons would become liberated in high energy nuclear collisions to produce a new state of deconfined matter subsequently referred to as \emph{quark-gluon plasma} or QGP for short \cite{CABIBBO197567, PhysRevLett.34.1353}.

About a decade later, Bjorken famously synthesized these ideas and developed a revolutionary model of relativistic nuclear collisions which remains largely accurate to this day \cite{PhysRevD.27.140}.
The ideas were based on Landau's model of ideal hydrodynamics.
Bjorken's insight was to apply additional symmetries to the problem in order to derive simple solutions for the hydrodynamic equations of motion.
These equations allowed Bjorken to elucidate the space-time evolution of the collision and provide estimates for its initial energy density and temperature.
From these estimates he reasoned that it was likely the produced system would be in the deconfined QGP phase.

In the years that followed, hydrodynamic modeling of nuclear collisions grew from a nascent qualitative science into a quantitative one.
Viscosity was added to the simulations \cite{Israel:1976tn, Teaney:2004qa, Muronga:2001zk, Muronga:2003ta, Muronga:2004sf, Chaudhuri:2005ea, Heinz:2005bw}.
Crude estimates for the QGP energy density and pressure were replaced with realistic calculations derived from first principles \cite{Borsanyi:2013bia, Bazavov:2014pvz}.
Models were updated to include event-by-event fluctuations in the density of initial nuclear matter \cite{Alver:2010gr}, and descriptions of dilute regions of the collision were also greatly improved \cite{Bass:2000ib, Teaney:2001av, Nonaka:2006yn, Petersen:2008dd, Song:2010aq}.
The refined simulations began to accurately reproduce and even predict a large number of seemingly unrelated experimental observables, substantiating the veracity of the hydrodynamic framework.

Modern hydrodynamic computer models allow researchers to simulate the full time history of the QGP produced in relativistic nuclear collisions in all its gory detail.
The models recreate events exactly as they are believed to occur inside the detector and output simulated observables that can be directly compared to experimental data.
Free parameters of the framework such as its dissipative transport coefficients are then calibrated to optimally reproduce experimental measurements in order to infer intrinsic properties of the produced matter.

In this manner, data-driven methods are used to extract fundamental properties of hot and dense nuclear matter which are not directly accessible to first principle calculations due to the complexity of the system's microscopic dynamics.
The accuracy of these model-based QGP parameter extractions is of course limited by the fidelity of the simulations.
If any aspect of the simulation is incorrectly modeled, it will generally affect the inferred values of the model parameters.
Estimating these QGP parameters with quantitative uncertainty thus involves a careful accounting of all sources of potential error in the assumed framework.

The hydrodynamic initial conditions---which describe the energy density and flow velocity of the QGP medium at the hydrodynamic starting time ${\sim}1~\fmc$ after the nuclei first collide---are the single largest source of uncertainty impeding the extraction of QGP medium properties by comparing simulation predictions to data.
They are simulated using a variety of different computer models, and there is no unified consensus regarding their correct theoretical treatment.
Different initial condition models generally predict different descriptions of the QGP space-time evolution and hence prefer different values for the QGP medium parameters.
Their understanding is thus a limiting factor when using models to reverse engineer properties of the produced matter.

The QGP initial conditions are therefore important for two separate reasons.
First, they are interesting in their own right.
They evolve out of a highly chaotic dynamical process which tests our current understanding of nuclear matter under extreme conditions.
Second, they provide a necessary ingredient for dynamical simulations of the collision.
If the initial conditions are incorrectly modeled, the simulation predictions will be misleading, and all derivative conclusions will be tenuous at best.
In this latter sense, the initial conditions act as a nuisance parameter.

Ultimately, one seeks a correct first principles description of the QGP initial conditions as it would appropriately address both of these objectives.
Deriving the QGP initial conditions from first principles, however, is exceptionally challenging.
QCD is so difficult to solve in practice, that \emph{ab initio} initial condition calculations only exist for approximations of QCD and related quantum field theories.
These calculations generally involve different starting assumptions and hence result in descriptions of the QGP initial conditions which are always in some degree of mutual tension.

Such \emph{ab initio} calculations are commonly refined through iterative cycles of trial and error.
Individual theoretical assumptions are tested by comparing model predictions to experimental data.
Successful assumptions are then passed on to subsequent iterations of the theoretical landscape and problematic assumptions discarded.
Each step of the validation process is slow and typically involves significant computational effort.
Model-to-data comparison has thus emerged as a rich field of research in and of itself.

Generally speaking, these efforts describe a so-called bottom-up approach that searches for a solution to the problem derived from deeper fundamental laws.
In this dissertation, I apply an alternative, albeit complementary, approach to study the QGP initial conditions which addresses the problem from the opposite direction.
I start with the observations of the experimental data and work backwards to infer the requisite starting point of hydrodynamic simulations.
This data-driven or top-down approach is commonly known as solving the inverse problem.

Data-driven methods naturally require a new way of thinking about the QGP initial conditions, one that embraces theoretical uncertainty instead of fighting it.
For this purpose, I develop an extremely simple parametric model of the QGP initial conditions which is flexible enough to span a wide range of reasonable theoretical descriptions.
In this sense, I create a meta-model for the landscape of mutually incompatible theory calculations.
With the parametric model in hand, I then proceed to rigorously constrain its free parameters with experimental data, using Bayesian methods recently developed for heavy-ion collisions.
I find that the functional form of the QGP initial conditions is highly constrained by existing measurements, regardless of the theoretical uncertainty surrounding the details of its derivation.
This eliminates, to a large degree, the confounding uncertainty introduced by different microscopic models of the initial conditions, enabling quantitative QGP parameter estimates with meaningful uncertainty and unprecedented precision.

\chapter{Ultrarelativistic nuclear collisions}
\label{ch:overview}

\enlargethispage{1\baselineskip}

\lettrine{S}{hortly} after the quark-gluon nature of nuclear matter was discovered in the 1960s and 1970s, physicists began to seriously consider the idea of using high-energy nuclear collisions to study the properties of nuclear matter at extreme temperatures and densities \cite{Baym:2016wox}.
It was believed that heavy-ion collisions, e.g.\ two gold nuclei, would maximize the produced matter's lifetime and system size, thereby enhancing the QGP's effect on final state observables.
Simple estimates based on the energy released per unit rapidity in nucleon-nucleon collisions indicated that relativistic heavy-ion collisions could reasonably attain energy densities in excess of ${\sim}3$~GeV/fm$^3$, conditions which were generally expected to be sufficient to produce thermalized matter in the deconfined QGP phase \cite{PhysRevD.27.140}.

\section{Characterizing hot and dense nuclear matter}

Motivated in part by these general ideas, the US and international nuclear theory communities invested significant resources over the next few decades developing ultrarelativistic heavy-ion programs at the Relativistic Heavy-ion Collider (RHIC) located in Brookhaven, New York and the Large Hadron Collider (LHC) situated on the border of France and Switzerland.
I'll discuss these experiments in more detail shortly.
First, I want to explain some of the big picture questions which these programs sought to address.

Broadly, the goal of these investments is to quantify the bulk properties of hot and dense nuclear matter.
These properties can be subdivided into two general categories: equilibrium properties which characterize the matter's steady-state behavior and dynamical properties which describe its response to deviations from equilibrium.

\subsection{Equilibrium properties}

The equilibrium properties of a substance depend on the conditions of its static environment.
For example, water is a liquid at room temperature and atmospheric pressure, while it exists as a solid and gas at other temperature and pressure combinations.
This information is typically plotted as a \emph{phase diagram} which illustrates the pressure and temperature combinations needed to reproduce each phase of matter.

\begin{figure}[t]
  \centering
  \includegraphics{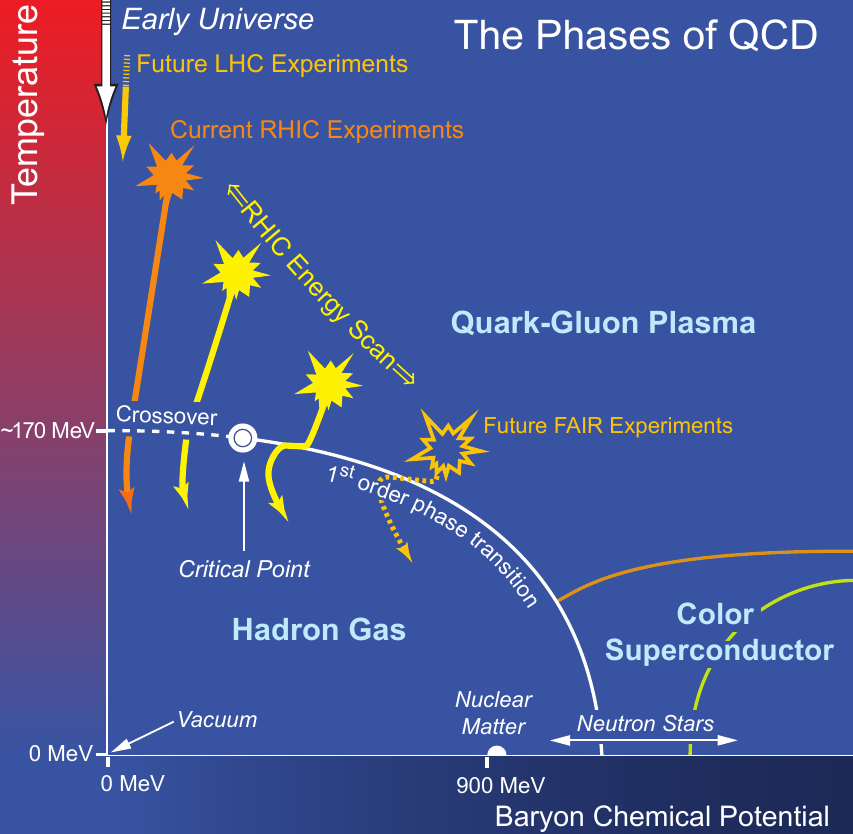}
  \caption{
    \label{fig:qcd_phase_diagram}
    Schematic of the QCD phase diagram in the temperature and baryon chemical potential plane \cite{Frontiers:2008}.
    Trajectories show regions of the phase diagram probed by various nuclear collider experiments.
  }
\end{figure}

The same general picture is also used to classify different phases of nuclear matter.
Figure \ref{fig:qcd_phase_diagram} shows the current picture of the QCD phase diagram as a function of temperature $T$ and baryon chemical potential $\mu_B$, a quantity related to the net baryon density (net imbalance of matter and antimatter).
At low temperatures and baryon densities, nuclear matter exists as hadrons, color-neutral combinations of two or three bound-quark states, while at higher temperatures and/or baryon densities, these hadrons ``melt'' to form the deconfined QGP phase.
Lattice QCD calculations (see below) have established that the transition from the hadronic phase to the QGP phase is a smooth crossover at $\mu_B=0$.
Meanwhile, at larger $\mu_B$, the transition is expected to become first-order \cite{Frontiers:2008}, although it is not yet clear from first principles where this should occur.
I'll briefly summarize now the basis of our current theoretical understanding at small baryon chemical potential.
RHIC and LHC collisions produce almost equal parts matter and antimatter at $\sqrts > 100$~GeV, so zero net baryon density is a very good approximation for the collisions studied in this dissertation.

\subsubsection{Equation of state}

At each point in the QCD phase diagram, the equilibrium properties of nuclear matter are quantified by an equation of state (EoS), specifying the energy density, entropy density, and pressure (among other quantities) at fixed temperature and baryochemical potential.
At zero baryochemical potential (left edge of figure \ref{fig:qcd_phase_diagram}), the QCD EoS is rigorously calculable using non-perturbative methods based on the Feynman path integral approach.
The key to this method is the realization that the density operator ${\hat{\rho} = e^{- \beta \hat{H}}}$ resembles a time-evolution operator $e^{i \hat{H} t}$ if one replaces $\beta$ (inverse temperature) with imaginary time $\tau = -i t$.
Therefore, by substituting $t \to i \tau$, the path integral formulation of the field theory can be made to resemble a partition function $Z = \trace \hat{\rho}$, thereby specifying the system's statistical properties in thermodynamic equilibrium.

The partition function can be evaluated using \emph{lattice QCD}, an algorithm to discretize the path integral onto a hypercubic lattice of $N_\sigma^3 N_\tau$ space-time points, where $N_\sigma$ and $N_\tau$ are the number of steps used to discretize the spatial and temporal dimensions respectively.
These lattice sites are separated by lattice spacing $a$ which relates the number of grid steps $N_\sigma, N_\tau$ to the simulation's effective equilibrium temperature and volume
\begin{align}
  T &= 1/(a N_\tau),\\
  V &= 1/(a N_\sigma)^3.
\end{align}
The calculation is repeated for different grid dimensions to vary the system's equilibrium temperature and grid resolution.
Then, the results of successively finer grids are extrapolated to the continuum limit to remove finite lattice effects.

Lattice calculations are typically presented in terms of the trace of the stress energy tensor $\Theta^{\mu\mu}$, equal to the difference of the energy density and three times the pressure.
This quantity is commonly referred to as the trace anomaly or interaction measure because it measures the deviation of the fluid from the conformal EoS.
Defined on the lattice, the trace anomaly is related to the total derivative of $\log Z$ with respect to the lattice spacing $a$:
\begin{equation}
  \Theta^{\mu\mu} = -\frac{T}{V} \frac{d \log Z}{d \log a}.
\end{equation}
Scaled by powers of the temperature $T$, the trace anomaly $\Theta^{\mu\mu}$ forms a dimensionless interaction measure
\begin{equation}
  I \equiv \frac{\Theta^{\mu\mu}(T)}{T^4} = \frac{e - 3 P}{T^4}.
\end{equation}
The thermodynamic pressure is then calculated from the interaction measure using the relation
\begin{equation}
  \frac{P(T)}{T^4} = \frac{P_0}{T_0^4} + \int_{T_0}^T dT' \frac{I(T')}{T'},
\end{equation}
where $P_0$ and $T_0$ are a reference pressure and temperature, typically calculated from the hadron resonance gas model.
The energy density $e$ and entropy density $s$ are then easily obtained from the thermodynamic relations
\begin{align}
  \frac{e(T)}{T^4} &= I(T) + 3 \frac{P(T)}{T^4}, \\[1ex]
  \frac{s(T)}{T^3} &= \frac{e(T) + P(T)}{T^4}.
\end{align}

\begin{figure}
  \centering
  \includegraphics[scale=.8]{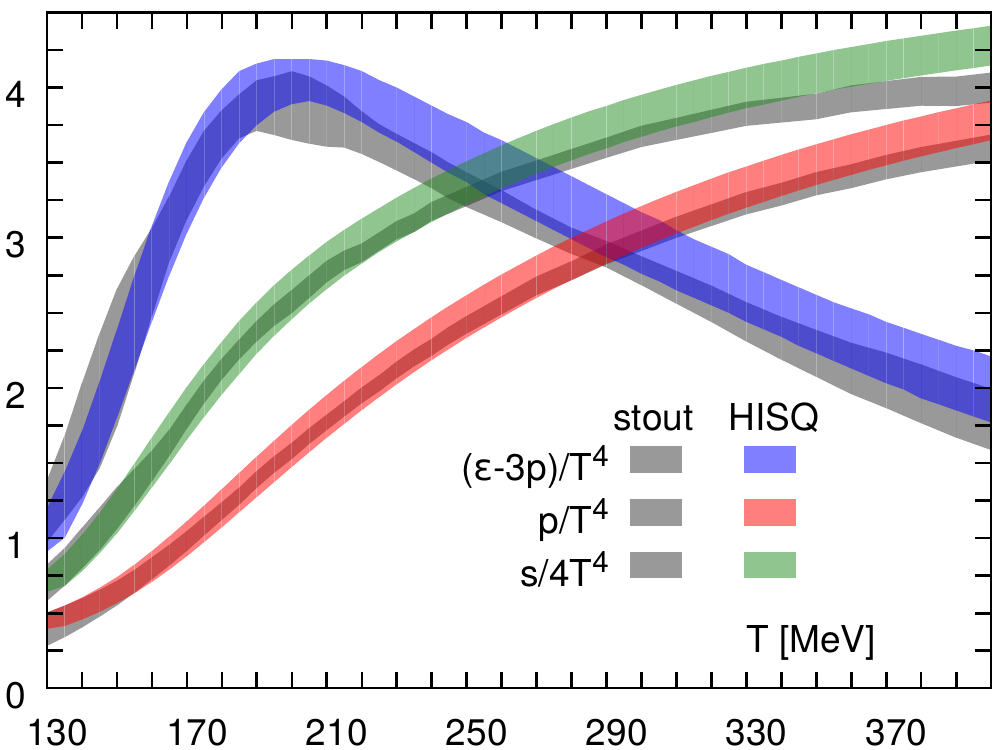}
  \caption{
    \label{fig:hotqcd_eos}
    Lattice equation of state for (2+1)-flavor QCD ($u$, $d$, and $s$ quarks) at zero net baryon density calculated by the HotQCD collaboration (colored) and Wuppertal-Budapest collaboration (gray) \cite{Bazavov:2014pvz, Borsanyi:2013bia}.
  }
\end{figure}

Figure \ref{fig:hotqcd_eos} shows the trace anomaly, pressure, and entropy density divided by powers of the temperature for (2+1)-flavor QCD ($u$, $d$, and $s$ quarks) at zero net baryon density obtained from lattice calculations performed by two independent collaborations.
The gray bands are calculations by the Wuppertal-Budapest collaboration using the stout fermion action \cite{Borsanyi:2013bia}, and the colored bands are calculations by the HotQCD collaboration using the HISQ/tree action \cite{Bazavov:2014pvz}.
Both collaborations observe a smooth crossover phase transition to the QGP phase located at the pseudocritical temperature $T_c \sim 150$--155~MeV at $\mu_B=0$.
Considering the complexity of each calculation, the agreement between the two groups is a remarkable accomplishment.

Recent developments in lattice QCD include more precise estimates for the QGP pseudocritical temperature $T_c = 156.5 \pm 1.5$~MeV \cite{Bazavov:2018mes} and new calculations in (2+1+1)-flavors, i.e.\ with thermalized charm quarks \cite{Borsanyi:2016ksw}.
The addition of charm quarks modifies the trace anomaly at very high temperatures, but the corrections are modest for $T \lesssim 400$ MeV.
Therefore, the QCD EoS is generally considered to be well constrained by first-principles theory at vanishing net baryon density.

These calculations, however, describe just one edge of the QCD phase diagram at zero baryon chemical potential $\mu_B$.
As I mentioned previously, an outstanding question facing the nuclear physics community is whether QCD switches from a smooth crossover at $\mu_B=0$ to a first-order transition at some $\mu_B > 0$, as predicted by multiple theories \cite{Stephanov:2004wx}.
This feature in the phase diagram is known as the critical point (see figure \ref{fig:qcd_phase_diagram}).

Strictly speaking, it is not yet possible to calculate the QCD EoS at significant baryon density on the lattice due to the existence of the fermion sign problem \cite{deForcrand:2010ys}.
Nevertheless, several lattice-based methods exist to calculate the QCD EoS in the presence of a small quark potential.
For example, one can take derivatives of quark and gluonic observables with respect to $\mu_B$ to calculate the leading-order Taylor expansion of the theory at the edge of the phase diagram $\mu_B=0$ \cite{Allton:2002zi}.
The truncated Taylor expansion can then be used to extrapolate to small baryon densities $\mu_B > 0$.
Alternatively, the QCD EoS can be solved for imaginary quark potentials, thereby circumventing the sign problem, and analytically continued to real $\mu_B$ \cite{deForcrand:2002hgr, DElia:2002tig, Gunther:2016vcp}.
The QCD EoS at nonzero baryon density is therefore an evolving picture, and an ongoing area of theoretical research.

\subsection{Dynamical properties}

To this point, I've only discussed the steady-state properties of bulk nuclear matter at fixed temperature and chemical potential.
Such ideal systems, however, seldom exist in nature.
The physical processes that produce QGP matter are typically violent and far from equilibrium.
Ultrarelativistic heavy-ion collisions, for example, produce small ($10^{-14}$ m), short-lived ($10^{-23}$ s) QGP fireballs that rapidly expand and cool, tracing complex trajectories through the QCD phase diagram.

The collision's dynamical evolution contains additional information---specific to the form of matter---which is not specified by the QCD EoS.
Therefore, it is important to supplement thermodynamic measures with additional numbers to characterize these properties.

\subsubsection{Dissipative hydrodynamics}

Hydrodynamics is a mathematical framework that describes the response of a system to small perturbations from local thermal equilibrium, constructed by applying basic conservation laws to gradient expansions of the stress-energy tensor $T^{\mu\nu}$.
It relates the system's extended non-equilibrium dynamics to the properties of its locally equilibrated matter.
Expanded to first-order in gradients of the fluid flow velocity, the stress-energy tensor can be written as
\begin{equation}
  \label{eq:hydro_eqn1}
  T^{\mu\nu} = (e + P) u^\mu u^\nu - P g^{\mu\nu} + \pi^{\mu\nu} - \Delta^{\mu\nu} \Pi,
\end{equation}
where $e$ and $P$ are the energy density and pressure in the local fluid rest frame, $u^\mu$ is the local fluid velocity, and $\Delta^{\mu\nu} = g^{\mu\nu} - u^\mu u^\nu$ is the projector onto the space orthogonal to $u^\mu$.
The terms $\pi^{\mu\nu}$ and $\Pi$, meanwhile, are the first-order shear and bulk viscous corrections to the zeroth-order theory which I'll describe shortly.

The hydrodynamic equations of motion are obtained from equation \eqref{eq:hydro_eqn1} by applying energy-momentum and charge conservation,
\begin{equation}
  \partial_\mu T^{\mu\nu} = 0 \quad\text{and}\quad \partial_\mu j^\mu=0,
\end{equation}
to the energy-momentum tensor $T^{\mu\nu}$ and charge-current $j^\mu = n u^\mu$ in combination with an EoS $P = P(e)$ and initial conditions for $e$, $u^\mu$, $\Pi$, and $\pi^{\mu\nu}$.
Typically for heavy-ion collisions, the charge-current is associated with the system's baryon density $n$.
Throughout this dissertation, I study ultrarelativistic nuclear collisions with vanishing net baryon density, so I'll neglect discussing this latter conserved current.

The shear viscous pressure tensor $\pi^{\mu\nu}$ and bulk pressure $\Pi$ apply dissipative corrections to the stress-energy tensor $T^{\mu\nu}$.
In relativistic Navier-Stokes theory, these viscous terms can be further decomposed in the form \cite{Landau:1987bk}
\begin{equation}
  \pi^{\mu\nu} = 2\eta \Delta^{\mu\nu\alpha\beta} \partial_\alpha u_\beta \quad\text{and}\quad \Pi = -\zeta \partial_\mu u^\mu,
\end{equation}
where
\begin{equation}
  \Delta^{\mu\nu\alpha\beta} = \frac{1}{2}(\Delta^{\mu\alpha} \Delta^{\nu\beta} + \Delta^{\nu\alpha} \Delta^{\mu\beta}) - \frac{1}{3} \Delta^{\mu\nu} \Delta^{\alpha\beta}
\end{equation}
is a symmetric direct product of projection operators orthogonal to $u^\mu$ \cite{deSouza:2015ena}.
The quantities $\eta$ and $\zeta$ multiplying each term are hydrodynamic \emph{transport coefficients}.
They are free parameters of the theory describing fundamental dynamical properties of the fluid.

\begin{remark}
  When discretized on a grid, the first-order Navier-Stokes equations generate superluminal hydrodynamic modes which render the numerical scheme unstable.
  Therefore, in practice, the gradient expansion is implemented at second-order to maintain stability.
  I'll introduce the second-order equations of motion later in subsection \ref{subsec:boostinv_hydro}.
  The second-order equations introduce additional transport coefficients, but it is reasonable to expect their effect on the system dynamics to be much smaller than the first-order coefficients.
  Indeed, it has been shown, for example, that the system is relatively agnostic to the value of the second-order relaxation time transport coefficient $\tau_\pi$ \cite{Song:2008si, Bernhard:2015hxa}.
\end{remark}

\begin{figure}[t]
  \centering
  \begin{tikzpicture}
    \node[draw,thick,trapezium,trapezium left angle=70,trapezium right angle=-70,minimum height=2cm] {\Large$\eta$};

    % shear viscosity
    \foreach \xshift/\yshift in {-1.3/-.7, -1.17/-.35, -.92/.35, -.79/.7}{
      \draw[thick, ->] (\xshift,\yshift) -- ++(2,0);
    }
    \draw[thick, -] (-1.05, 0) to ++(.5, 0);
    \draw[thick, ->] (.45, 0) to ++(.5, 0);

    % bulk viscosity
    \draw[thick,xshift=5cm] circle (1.0cm) node{\Large$\zeta$};
    \foreach \theta in {60, 120, 180, 240, 300, 360}{
      \draw[->,thick,xshift=5cm,rotate around={\theta:(0, 0)}] (0.5, 0) -- ++(1, 0);
    }
  \end{tikzpicture}
  \caption{
    \label{fig:shear_bulk_schematic}
    The shear viscosity $\eta$ applies a force that opposes shearing flows (left), while the bulk viscosity $\zeta$ applies a force that opposes radial expansion and compression (right).
  }
\end{figure}
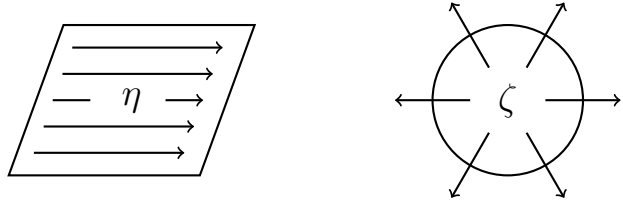

\subsubsection{First-order transport coefficients}

The shear viscosity $\eta$ and bulk viscosity $\zeta$ (natural units fm$^{-3}$) describe the fluid's dissipative corrections at leading order.
Determining these transport coefficients for QCD matter is therefore a primary goal of fundamental importance.
Both coefficients are generally expected to depend on the temperature $T$ and baryon chemical potential $\mu_B$.

In the hydrodynamic equations, the viscosities appear as dimensionless ratios, $\etas$ and $\zetas$, where $s$ is the fluid entropy density.
These so-called \emph{specific viscosities} are generally more interesting and meaningful than the unscaled $\eta$ and $\zeta$ values, because they describe the magnitude of stresses inside the medium relative to its natural scale.

\paragraph{Shear viscosity}

The shear viscosity $\eta$ applies a force that opposes shearing flows in the fluid (see figure \ref{fig:shear_bulk_schematic}), converting the damped motion to heat.
Microscopically, it describes how well the fluid transmits momentum across adjacent layers of fluid flow.
In weakly-coupled kinetic theory, the specific shear viscosity $\etas$ relates to the inter-particle mean free path $\lambda_\mathrm{mfp}$ \cite{CasalderreySolana:2011us, York:2008rr, Romatschke:2017ejr}:
\begin{equation}
  \eta/s \propto T \lambda_\text{mfp}.
\end{equation}
Therefore a larger mean free path (weaker coupling) corresponds to a larger value of $\etas$.
Conversely, in the strongly-coupled limit, $\eta/s$ vanishes and the matter behaves like a ``perfect fluid'' with minimal resistance to shearing flow.
It's important to note that quasi-particle descriptions of the fluid only make sense up to some maximum coupling strength.
Beyond this point, the particles' mean free path $\lambda_\mathrm{mfp}$ becomes smaller than their de Broglie wavelength $1/T$, at which point the notion of quasi-particles is ill defined \cite{deSouza:2015ena}.
Such strongly-coupled systems are thus fundamentally field-like.

\paragraph{Bulk viscosity}

The bulk viscosity $\zeta$ introduces an effective pressure $\Pi$ that modifies the ideal pressure $P$.
When the local fluid velocity divergence $\partial_\mu u^\mu$ is positive, this effective pressure is negative and \emph{vice versa}.
The bulk viscosity therefore opposes radial expansion and compression (see figure \ref{fig:shear_bulk_schematic}).
Microscopically, the mechanisms that explain bulk viscosity are complicated.
However, they generally relate to a certain reconfiguration energy needed for the fluid to expand or contract.
The bulk viscosity of a diatomic gas, for example, is nonzero due to the exchange of molecular energy between translational and rotational degrees of freedom \cite{Chapman:1990ceb}.
For scale invariant\footnote{A scale invariant system is one that appears self-similar at all scales. For example, an equilateral triangle is scale invariant.} theories, the bulk viscosity of the system must vanish.
However, QCD is known to break scale invariance, particularly near the QGP phase transition, so the bulk viscosity of QCD matter could be large near $T_c$.

\subsubsection{Jet \& hard-probe interactions}

Hydrodynamic transport coefficients describe the medium's bulk interactions among its constituents.
It's also interesting to study interactions between the hydrodynamic medium and highly energetic probes that are initially far from equilibrium.
For example, suppose I shoot an energetic quark through an infinite brick of equilibrated QGP matter.
There are many interesting questions that I might ask, for example:
\begin{itemize}
  \item
    How does the quark scatter inside the medium and lose energy?
  \item
    How does the quark deflect perpendicular to its direction of motion and diffuse inside the medium?
  \item
    What is the path-length dependence of its energy loss?
  \item
    How does the medium absorb the energy that is lost by the quark?
\end{itemize}

These types of questions broadly pertain to a subfield of the QGP research effort dedicated to studying \emph{jets} and \emph{hard-probes}.
The term jet refers to a highly energetic cone of hadrons and other material ejected by an initial hard-scattering process, while the term hard-probe usually refers to a single energetic particle (possibly inside a jet), e.g.\ a high-momentum charm quark which traverses the medium.
This subject matter is beyond the scope of the present work, so I will not delve into it here.
For an overview, see \cite{dEnterria:2009xfs, Bedjidian:2004gd, Baier:1996sk}.
Nevertheless, for the sake of completeness, I'll describe a few of the primary quantities that jet and hard-probe studies seek to measure.
These coefficients are similar in importance to the specific shear viscosity $\etas$ and bulk viscosity $\zetas$ used to quantify the properties of bulk matter interactions.

The majority of interactions between the probe and the medium are soft small-angle scatterings which each transfer a small amount of momentum from the fluid to the probe such that the fractional change of the probe's momentum is small.
The probe's response to these soft kicks is summarized by the Fokker-Planck equation \cite{Svetitsky:1987gq, Moore:2004tg}.
Much like hydrodynamics, the Fokker-Planck equation introduces several transport coefficients which specify important properties of the probe-medium interaction.

\paragraph{Drag coefficient}

One fundamental measure of the probe-medium interactions is the longitudinal drag coefficient \cite{Ghiglieri:2015ala}
\begin{equation}
  \eta_D = -\frac{1}{p_L} \frac{dp_L}{dt},
\end{equation}
where $p_L$ is the longitudinal component of the probe momentum $\mathbf{p}$.
This quantity measures the percentage longitudinal momentum loss per unit time.
It is sensitive to the stopping power of the medium, and hence the coupling strength between the probe and the locally equilibrated QGP.

\paragraph{Longitudinal momentum broadening}

The second-moment of the longitudinal momentum transfer distribution is quantified by the longitudinal broadening coefficient \cite{Ghiglieri:2015ala}
\begin{equation}
  \hat{q}_L = \frac{d}{dt} \langle (\Delta p_L)^2 \rangle,
\end{equation}
defined as the typical longitudinal momentum kick squared per unit time incurred by the probe as it traverses the medium.

\paragraph{Transverse momentum broadening}

Perhaps the most studied transport parameter is the transverse momentum broadening coefficient $\hat{q}$, defined as the typical transverse momentum kick squared per unit time incurred by a jet or hard-probe as it traverses the QGP medium \cite{Baier:1996sk, Baier:2000mf, Ghiglieri:2015ala}:
\begin{equation}
  \hat{q} = \frac{d}{dt} \langle (\Delta p_\perp)^2 \rangle.
\end{equation}
It is expected to measure important properties of hot and dense QCD matter, such as its coupling strength (strong vs weak) and its constituent nature (quasi-particles vs non-localized fields) \cite{Bass:2012hkg}.
As such, it is considered a fundamental QCD quantity of primary interest.

\section{Hadron collider experiments}

To this point, I've described the QGP largely theoretically, as something believed to exist based on our current knowledge of QCD.
How do we know that it \emph{actually} exists?
The primary experimental evidence for the QGP's existence is provided by ultrarelativistic nuclear collisions conducted at RHIC and the LHC which I mentioned briefly.
These facilities are massive, each involving thousands of scientists and numerous nuclear collision experiments.

\paragraph{Relativistic Heavy-ion Collider (RHIC)}

This circular accelerator collides primarily heavy-ions, but also protons and light-ions, at center-of-mass energies per nucleon pair\footnote{Beam energies are commonly measured using $\sqrts$, equal to the total energy of each colliding nucleon pair in its center-of-mass frame.} ranging from $\sqrts=7.7$ to 200 GeV \cite{Kumar:2012fb}.
It is a lower energy collider than the LHC, but it has several unique advantages which make it an excellent probe of the QGP.
For instance, it supports longer heavy-ion operation times, and its beam is highly configurable, enabling researchers to study numerous collision partners and beam energies.

\paragraph{Large Hadron Collider (LHC)}

Like RHIC, the LHC is a large circular hadron collider.
A distinctive feature of the LHC is its unprecedented beam energy.
To date, it has run proton-proton collisions up to $\sqrts=13$ TeV \cite{Aad:2016mok, Adam:2015pza, Khachatryan:2015jna, Aaij:2015bpa} and heavy-ion collisions up to $\sqrts=5.02$ TeV \cite{Adam:2015ptt, Aaboud:2017tql, Khachatryan:2016odn}.
This is over an order of magnitude larger than the highest energies achieved at RHIC.
Heavy-ion collisions, however, are a smaller fraction of the overall physics program at the LHC compared to RHIC, so fewer collision systems and beam energies have been studied.

The experimentalists running these colliders are able to directly control two quantities: the species of the colliding nuclei and the energy of the collision.
They fix these quantities, accelerate two counter-rotating circular beams of nuclei, and perform measurements on the random collisions that occur between the accelerated ions.
Isolated events are then selected from the stream of detector activity using an experimental \emph{trigger} to identify the existence of individual inelastic nuclear-nuclear collisions.
These raw unfiltered events form a \emph{minimum-bias} sample, i.e.\ an unbiased subsample drawn from the population of all equal probability inelastic collision events.

\begin{figure}[t]
  \centering
  \includegraphics[scale=.25]{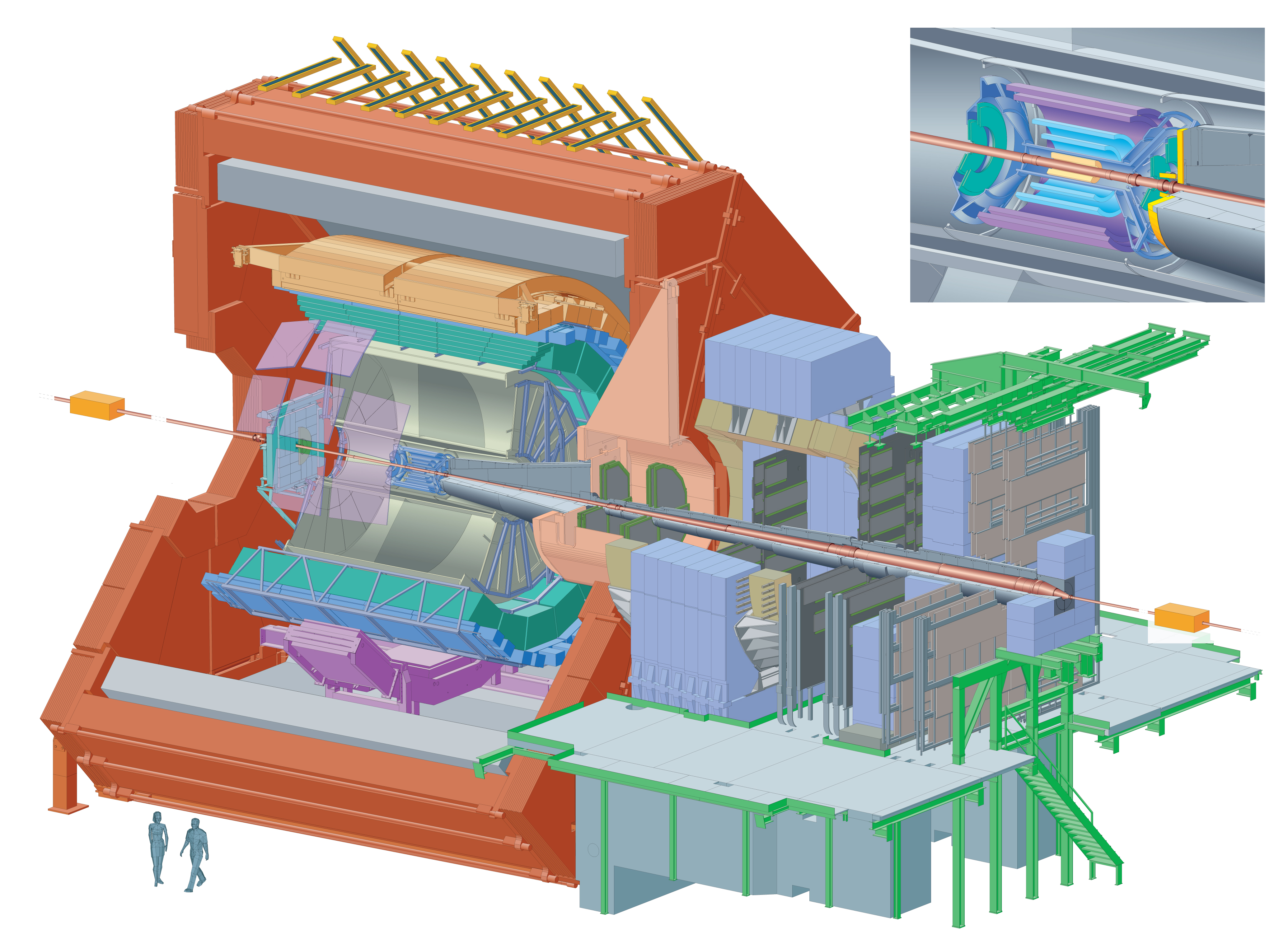}
  \caption{
    \label{fig:alice_detector}
    Computer rendering of the ALICE detector experiment \cite{ALICE:2008cad}.
  }
\end{figure}

The nuclear collision events are measured by one of several detectors situated on each beam.
Each detector is a large apparatus wrapped around the symmetry axis of the beam pipe which captures the flux of particles generated by each collision event.
Both facilities have multiple detectors, each managed by an independent experimental collaboration sharing the name of the detector.
RHIC has the BRAHMS, PHENIX, PHOBOS, and STAR detectors, while the LHC has ALICE, ATLAS, CMS, and LHCb.

These detectors vary in their design.
Each is specifically optimized to perform a certain task.
For example, ALICE (A Large Ion Collider Experiment) is optimized to detect the tens of thousands of particles produced by a lead-lead collision, specifically those particles emitted with low momentum.
Figure \ref{fig:alice_detector} shows a computer rendering of the ALICE experiment.
Notice the two people in the lower left corner to appreciate the sense of scale.

\subsection{Event properties}

The properties of the particles produced by each collision are determined using an ensemble of particle trackers and energy calorimeters layered around the nominal interaction vertex.
These detector components allow the experiments to measure properties of each particle (when possible) such as its momentum, charge, mass, and particle type.
These raw particle properties are then post-processed into experimental observables which describe features of the event sample.

\subsubsection{Kinematic variables}

Consider a collision in which the nuclei move through a beam aligned with the $\hat{z}$ direction.
In high-energy particle physics, it is common to specify each particle's four-momentum in a transformed coordinate system:
\begin{align}
  (E, p_x, p_y, p_z) &= (m_T \cosh y,\,p_T \cos \phi,\, p_T \sin \phi,\, m_T \sinh y),\\
  m_T &= \sqrt{E^2 - p_z^2},\\
  y &= \frac{1}{2} \log \frac{E + p_z}{E - p_z},\\
  p_T &= \sqrt{p_x^2 + p_y^2},\\
\phi &= \mathrm{atan2}(p_y, p_x),
\end{align}
where $m_T$ is the particle's transverse mass, $y$ is its rapidity, $p_T$ is its average transverse momentum, and $\phi$ is its azimuthal angle in the plane orthogonal to the beam axis.

Note, the transverse mass $m_T$ and the rapidity $y$ both require knowledge of the particle's total energy $E$ which depends on its mass $m$.
This information is often inaccessible for technical reasons, so typically the experiments replace the rapidity $y$ with a similar quantity known as the pseudorapidity $\eta$.
It is defined as
\begin{equation}
  \eta = - \log [\tan (\theta/2)] = \frac{1}{2} \log \frac{|\mathbf{p}| + p_z}{|\mathbf{p}| - p_z},
\end{equation}
where $\theta$ is the momentum vector's polar angle with respect to the beam axis, i.e.\ $\cos \theta = p_z / |\mathbf{p}|$.
For massless particles, the rapidity $y$ and the pseudorapidity $\eta$ are equivalent.
They are also equivalent at midrapidity, i.e.\ for $y = \eta = 0$.
This quantity is convenient because the particle's polar angle $\theta$ is easily measured inside the detector.

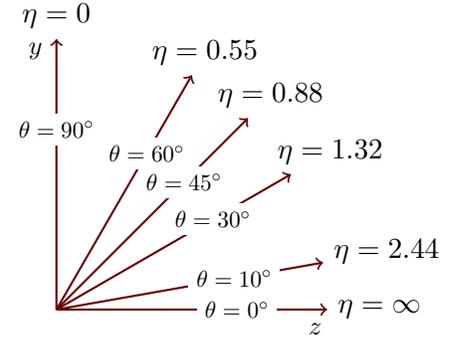
\begin{wrapfigure}[12]{r}[.2\textwidth]{.5\textwidth}
  % Author: Izaak Neutelings (July 2017)
  % CC Attribution-Share Alike license
  \centering
  \vspace{-20pt}
  \begin{tikzpicture}[scale=3]
   % limits
   \def\R{1.2}

   % axis labels
   \node[scale=0.9,below=5pt,left=2pt] at (0,\R) {$y$};
   \node[scale=0.9,left=5pt,below=2pt] at (\R,0) {$z$};

   % lines
   \foreach \t/\e in {90/0,60/0.55,45/0.88,30/1.32,10/2.44,0/\infty}{
     \draw[->,black!60!red,thick]
       (0,0) -- ({\R*cos(\t)},{\R*sin(\t)})
       node[anchor=180+\t,black] {$\eta=\e$};
     \node[fill=white,scale=0.8] at ({0.8*cos(\t)},{0.8*sin(\t)}) {$\theta=\t^\circ$};
   }
  \end{tikzpicture}
  \caption{
    \label{fig:pseudorapidity}
    Diagram illustrating the relationship between the pseudorapidity $\eta$ and the polar angle $\theta$ \cite{Neutelings:2017web}.
  }
\end{wrapfigure}

Figure \ref{fig:pseudorapidity} visualizes the relationship between the particle's pseudorapidity $\eta$ and its polar angle $\theta$.
For $y=\eta=0$, the particle emerges orthogonal to the beam axis.
This two-dimensional $(x, y)$ plane is thus commonly referred to as the \emph{transverse plane}.
Meanwhile, for $\eta=\infty$, the particle remains inside the beam pipe.
Thus, due to detector limitations, it is only possible for the experiments to measure particles out to some maximum rapidity.

\subsubsection{Collision centrality}

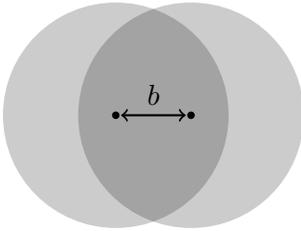
\begin{wrapfigure}[9]{l}[.2\textwidth]{.5\textwidth}
  \centering
  \begin{tikzpicture}
    \fill[black, fill opacity=0.2] (-0.5, 0) circle (1.5);
    \fill[black, fill opacity=0.2] (0.5, 0) circle (1.5);
    \fill[black] (-0.5, 0) circle (.05);
    \fill[black] (0.5, 0) circle (.05);
    \draw[<->, black, thick] (-.43, 0) to (.43, 0) node[above, midway] {$b$};
  \end{tikzpicture}
  \caption{
    \label{fig:impact_param}
    Two nuclei move in and out of the plane of the page.
    Arrow is the collision impact parameter $b$.
  }
\end{wrapfigure}

Once the beam is running, there is no way to control the orientation of the collisions.
Each pair of nuclei collides randomly, separated by an impact parameter $b$ in the transverse plane, defined as the distance between the two nuclei's centers of mass at the moment of closest approach; see figure \ref{fig:impact_param}.

In principle, it would be useful to measure the collision's properties as a function of the impact parameter $b$.
However, this quantity cannot be directly measured, so it is typically replaced with a related observable known as \emph{centrality}.

The collision centrality is defined by sorting all events in a minimum-bias event sample according to some measure of the underlying event activity (see below).
Once the events are sorted, they are partitioned into equal sized bins, where each bin is associated with some percentage of the overall event sample.
For example, if the events are partitioned into $n=10$ equal sized bins by their event activity, then the bin with the highest event activity is the 0--10\% centrality class.

\begin{figure}[t]
  \centering
  \begin{tikzpicture}
    \node {\includegraphics[scale=.5]{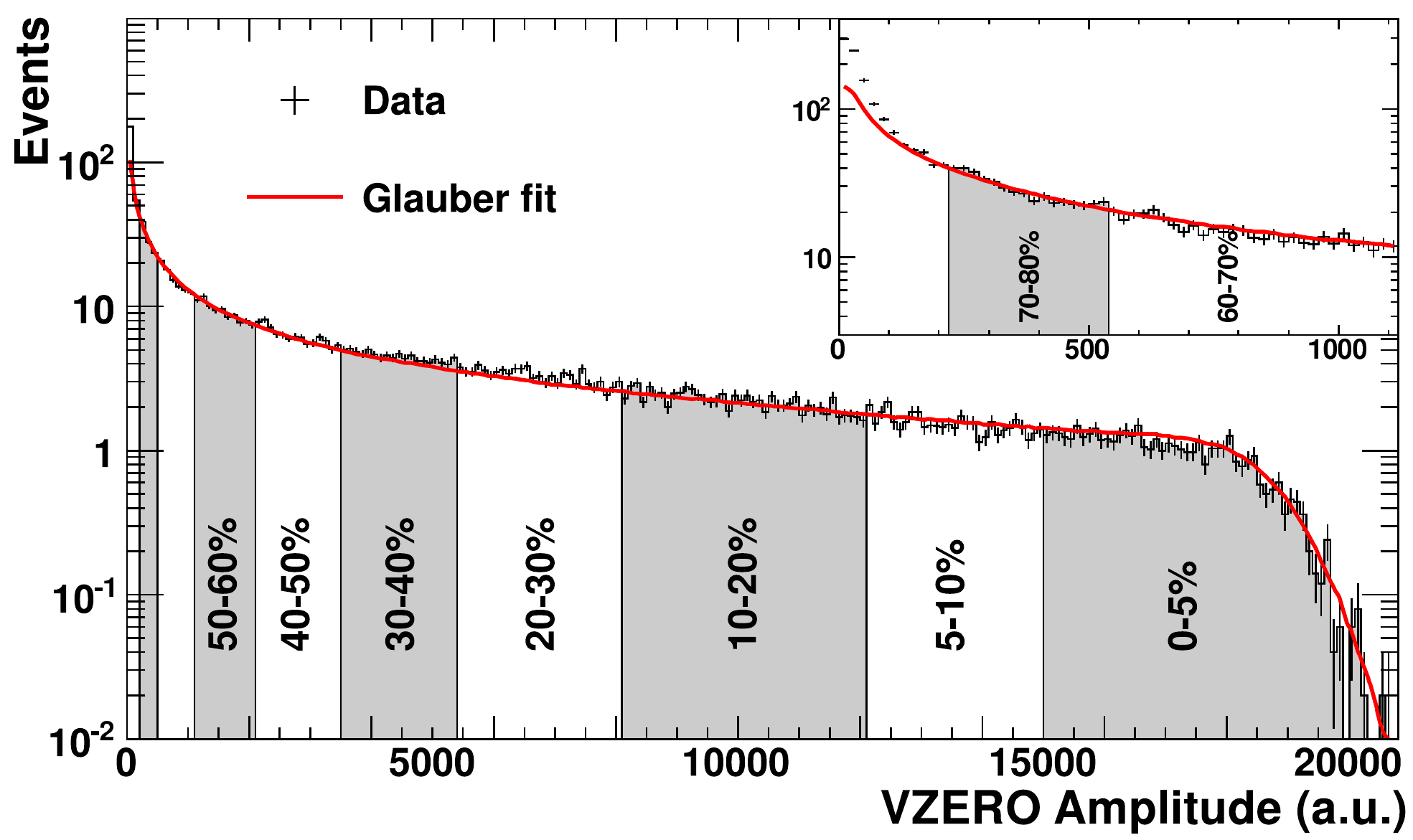}};
    \fill[black, fill opacity=0.2] (-3.5, -3.4) circle (.3);
    \fill[black, fill opacity=0.2] (-3.0, -3.4) circle (.3);
    \fill[black, fill opacity=0.2] (0.0, -3.4) circle (.3);
    \fill[black, fill opacity=0.2] (0.26, -3.4) circle (.3);
    \fill[black, fill opacity=0.2] (3.55, -3.4) circle (.3);
    \fill[black, fill opacity=0.2] (3.66, -3.4) circle (.3);
  \end{tikzpicture}
  \caption{
    \label{fig:alice_centrality}
    Centrality classes defined by the ALICE experiment using the sum of amplitudes in the detector's VZERO scintillators \cite{Aamodt:2010cz}.
    This quantity is monotonically related to the charged-particle yield in the same rapidity region.
    The shaded circles below the figure show the approximate average impact parameter of three centrality classes.
  }
\end{figure}
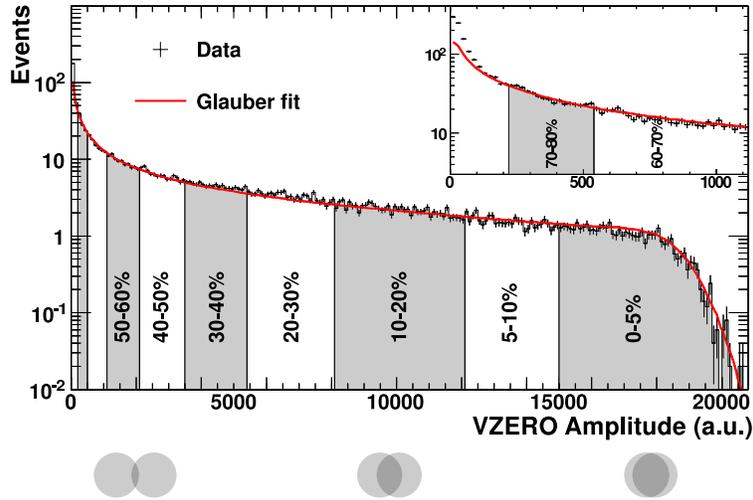

The definition of the underlying event activity used to sort the events varies from experiment to experiment.
A common choice is to measure some proxy for the event's charged-particle yield in a given rapidity window.
For example, the ALICE experiment commonly defines the collision centrality according to the sum of amplitudes in the detector's VZERO scintillators, covering $2.8 < \eta < 5.1$ (VZERO-A) and $-3.7 < \eta < -1.7$ (VZERO-C), signals which are monotonically related to the charged-particle yield \cite{Abelev:2013qoq}.
Figure \ref{fig:alice_centrality} shows an example of this centrality binning procedure applied to Pb-Pb collision data measured by the ALICE experiment \cite{Aamodt:2010cz}.

\subsection{Signatures of the quark-gluon plasma}
\label{subsec:qgp_signatures}

Now that I've broadly motivated and described heavy-ion collision experiments at RHIC and the LHC, I want to summarize some of their key results, particularly those results which evidence the production of the QGP.
This subsection is not meant to be an exhaustive list; doing so would require far more than a few pages.
Rather, these are several experimental observations that are commonly cited when discussing QGP formation.
Ultimately, I will explain at the end of the chapter that these features are collectively explained by a standard hydrodynamic model of relativistic heavy-ion collisions.
Once I've motivated and explained this model, I'll be able to frame the central problem addressed by this dissertation.

\subsubsection{Thermal particle yields}

One intriguing indicator that ultrarelativistic heavy-ion collisions produce QGP is provided by statistical hadronization models.
These models calculate hadron yields in nuclear collisions by sampling particles from a common chemical freeze-out surface at fixed temperature $T$ and baryon chemical potential $\mu_B$, i.e.\ by sampling from an emitter in thermal equilibrium.
The observed particle yields are consequently assumed to arise from the decay of fully equilibrated hadronic matter comprising all known hadron states.
The model is then calibrated to optimally fit the data by adjusting the temperature $T$ and chemical potential $\mu_B$ of the emitter, together with its freeze-out volume.
A detailed description of this approach is presented in \cite{BraunMunzinger:2003zd}.

\begin{figure}[!b]
  \centering
  \makebox[\textwidth]{
    \includegraphics[height=.525\textwidth]{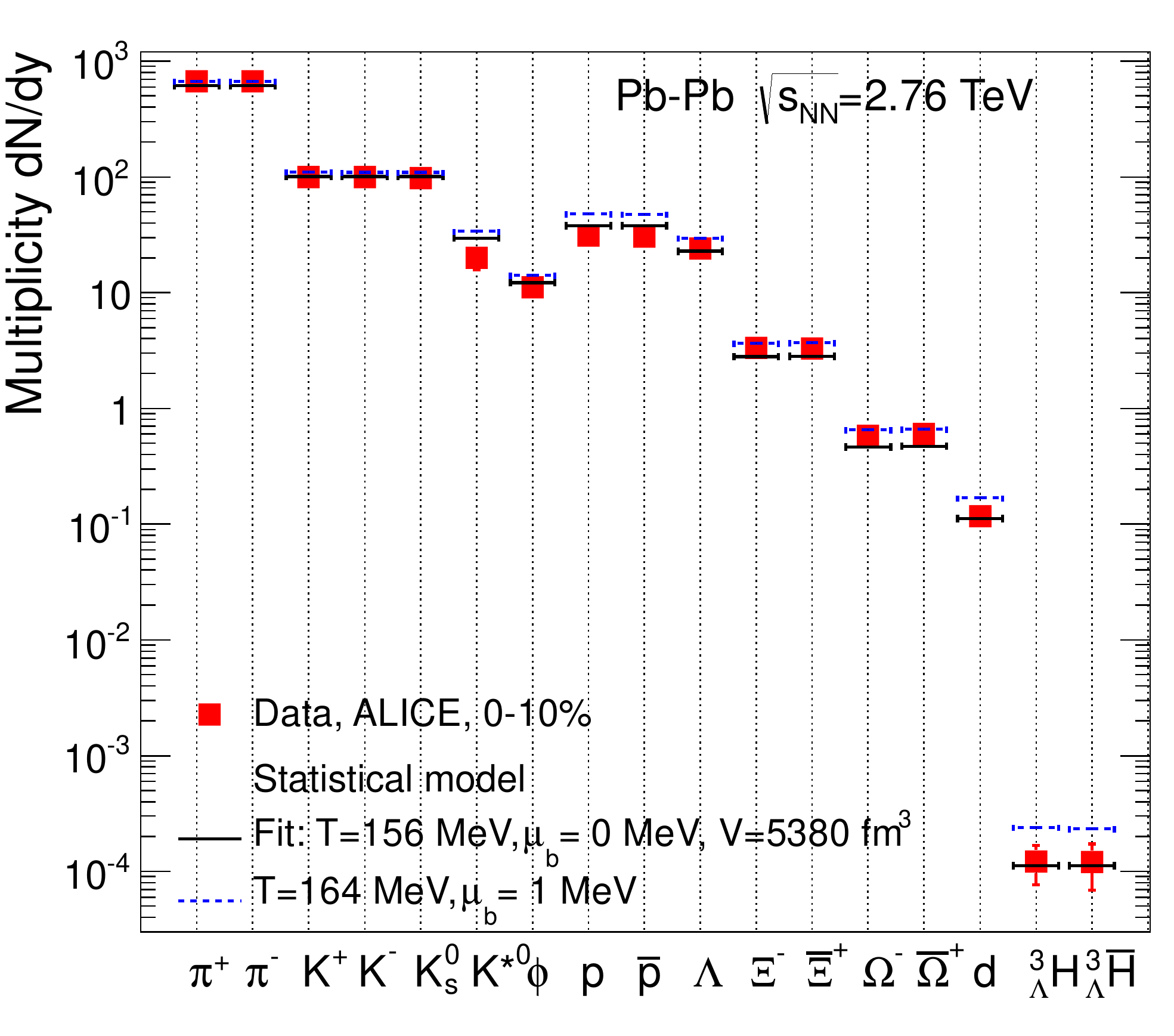}
    \includegraphics[height=.5\textwidth]{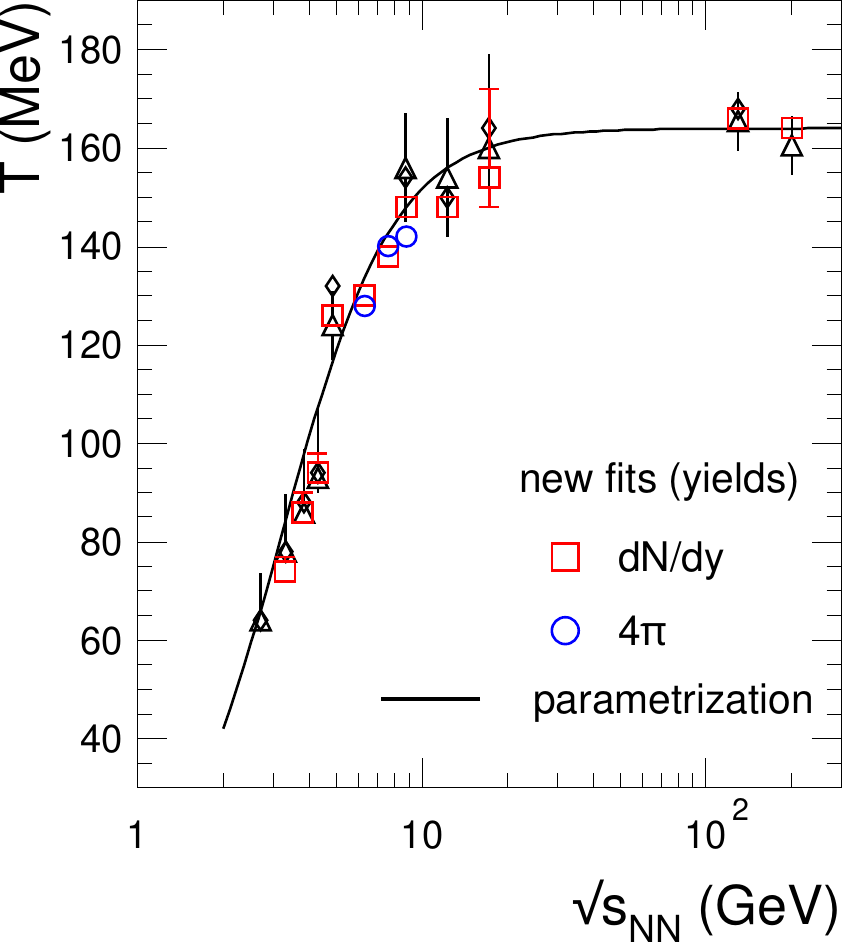}
  }
  \caption{
    \label{fig:stat_hadron_yields}
    Left: Particle yields for various hadrons predicted by the statistical hadronization model calibrated to fit Pb-Pb collisions at $\sqrts=2.76$ TeV \cite{Stachel:2013zma} using experimental data from ALICE \cite{Abelev:2013vea, Abelev:2013xaa, Abelev:2013zaa, Knospe:2013tda, Abelev:2014uua, Adam:2015vda}.
    Right: Best fit statistical hadronization parameter $T$ as a function of beam energy $\sqrts$ \cite{BraunMunzinger:2011ze}.
  }
\end{figure}

The left side of figure \ref{fig:stat_hadron_yields} shows the particle yields predicted by such a model \cite{Stachel:2013zma}, calibrated on and compared to Pb-Pb collision data at $\sqrts=2.76$~TeV measured by ALICE \cite{Abelev:2013vea, Abelev:2013xaa, Abelev:2013zaa, Knospe:2013tda, Abelev:2014uua, Adam:2015vda}.
The fit obtains a hadronization temperature $T = 156$~MeV and baryon chemical potential $\mu_B = 0$~MeV, which is in perfect agreement with the location of the pseudocritical transition temperature $T_c = 156.5 \pm1.5$~MeV at $\mu_B=0$ predicted by lattice QCD \cite{Bazavov:2018mes}.
Meanwhile, a study of the energy-dependence of the fit parameter $T$ presented on the right-side of figure \ref{fig:stat_hadron_yields} shows that the chemical freeze-out temperature increases as a function of beam energy before flat-lining at $\sqrts \sim 10$~GeV \cite{BraunMunzinger:2011ze}.
This suggests that the hadron resonance gas cannot be heated above some maximum temperature, presumably the temperature of the QGP phase transition.

\subsubsection{Collective flow}

Perhaps the most famous observation associated with QGP formation is the existence of collective flow.
Prior to the first ultrarelativistic heavy-ion collisions at RHIC, many believed that QGP would behave like a weakly-coupled gas characterized by a large mean free path.
Assuming particle production occurs independently at different points inside the heavy-ion collision, this conjecture would imply final hadron yields that are weakly correlated with respect to the azimuthal angle $\phi$.
The only significant azimuthal correlations would arise from jets and other hard scatterings which produce back-to-back showers of particles close to midrapidity.

However, the first measurements at RHIC revealed a very different picture of the collision.
The particles produced by each collision were found to be strongly correlated with respect to the azimuthal angle $\phi$, and these correlations persisted far from midrapidity \cite{Ackermann:2000tr} at odds with weakly-coupled predictions \cite{Shuryak:2004cy, Molnar:2001ux}.
The signal was consistent with a strongly-coupled picture of the collision, in which the QGP flows like a nearly inviscid liquid.

\begin{figure}[t]
  \centering
  \makebox[\textwidth]{
    \begin{tikzpicture}
      \node{\includegraphics[scale=.6]{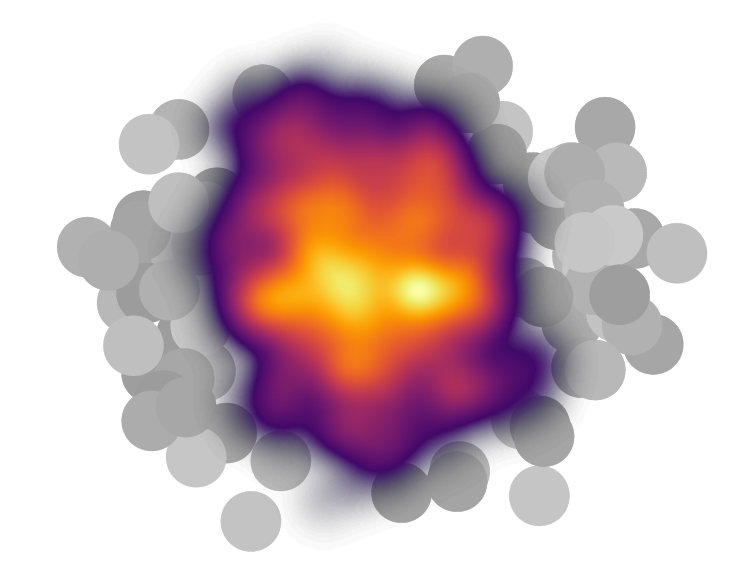}};
      \node[xshift=4.75cm]{\includegraphics[scale=.35]{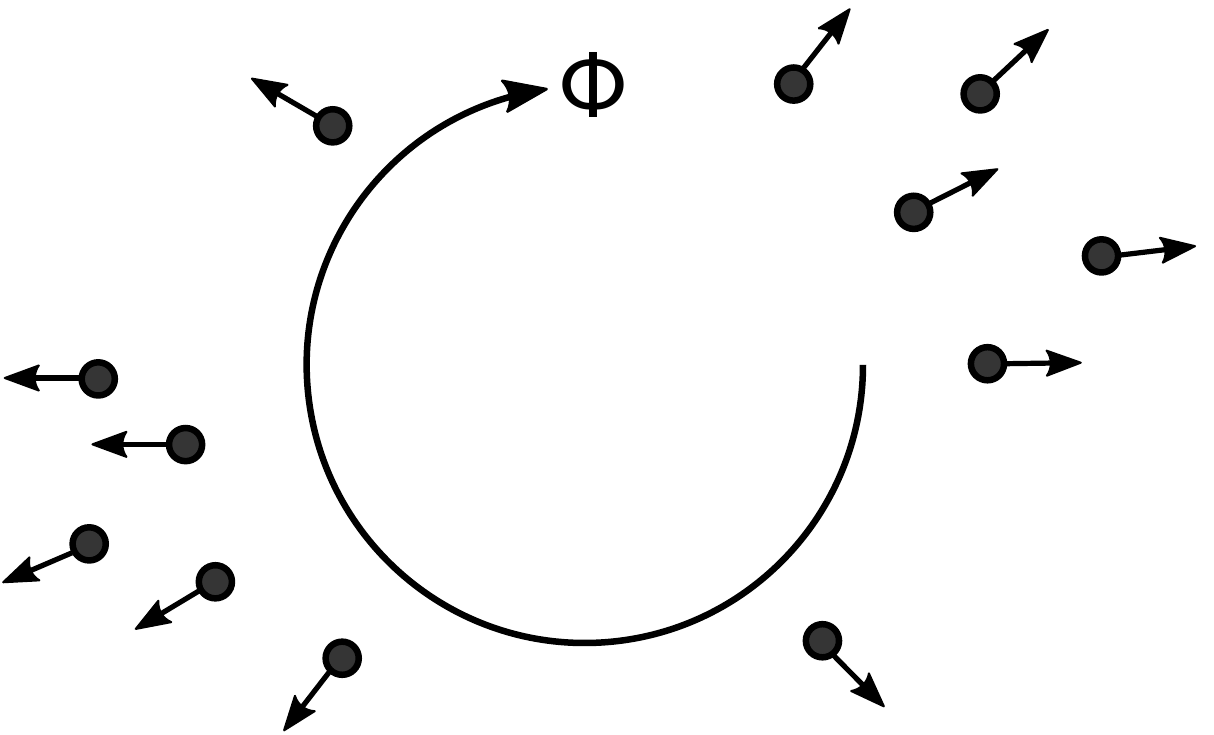}};
      \node[xshift=10cm, yshift=-0.5cm]{\includegraphics[scale=1]{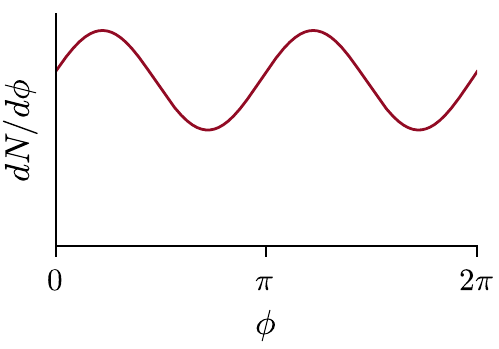}};
    \end{tikzpicture}
  }
  \caption{
    \label{fig:anisotropic_flow}
    Left: Spatially deformed initial energy density profile in the transverse plane (heatmap) and non-interacting nucleons (gray circles).
    Middle: Hydrodynamics converts the spatial anisotropy into momentum anisotropy, resulting in an azimuthally deformed flow field and anisotropic particle emission.
    Right: The measured particle distribution $dN/d\phi$ depends on the azimuthal angle $\phi$.
  }
\end{figure}

To understand how hydrodynamic flow gives rise to these correlations, consider a generic collision between two highly relativistic nuclei as shown in figure~\ref{fig:anisotropic_flow}.
When these nuclei collide, they generate an initial transverse energy density profile (left) which is spatially deformed due to the ``almond'' shape of the overlap region at nonzero impact parameter (see figure \ref{fig:impact_param}) and the fluctuations of nucleon positions inside each nucleus.
These spatial inhomogeneities create pressure gradients along the radial direction which vary as a function of the azimuthal angle $\phi$, producing stronger radial expansion along some directions and less along others.
This generates an azimuthally anisotropic flow field which preferentially emits particles in the direction of strongest fluid flow (middle), imparting this signal on the final azimuthal hadron distribution $dN/d\phi$ (right).
In essence, hydrodynamics converts spatial anisotropy into momentum anisotropy, which also shows up in the detector as a particle yield anisotropy.

Experimentally, this yield anisotropy is quantified by expanding the azimuthal particle distribution as a Fourier series \cite{Ollitrault:1993ba, Voloshin:1994mz, Poskanzer:1998yz}
\begin{equation}
  \frac{dN}{d\phi} \propto 1 + 2\sum\limits_{n=1}^\infty v_n \cos[n (\phi - \Psi_n)],
\end{equation}
where $\Psi_n$ is the phase or ``event plane'' angle, equal to the direction of maximum final-state particle density.
Here the number $n$ indexes the order of the harmonic.
The first harmonic $v_1$ is called directed flow, the second harmonic $v_2$ elliptic flow, the third harmonic $v_3$ triangular flow, and so on.

These coefficients are calculated using the relation
\begin{equation}
  v_n = \langle \langle \cos[n(\phi - \Psi_n)] \rangle \rangle,
\end{equation}
where the double angular brackets mean averaging over all particles in a given event, then averaging over all events in a given event class selected to satisfy certain centrality, rapidity, and transverse momentum requirements.
When the flow is calculated as a function of transverse momentum $p_T$ using narrow $p_T$ bins it is called \emph{differential flow}, and when it is calculated using all particles irrespective of their $p_T$ over a wide kinematic range, it is referred to as \emph{integrated flow}.

\begin{figure}[t]
  \centering
  \makebox[\textwidth]{
    \includegraphics[height=.45\textwidth]{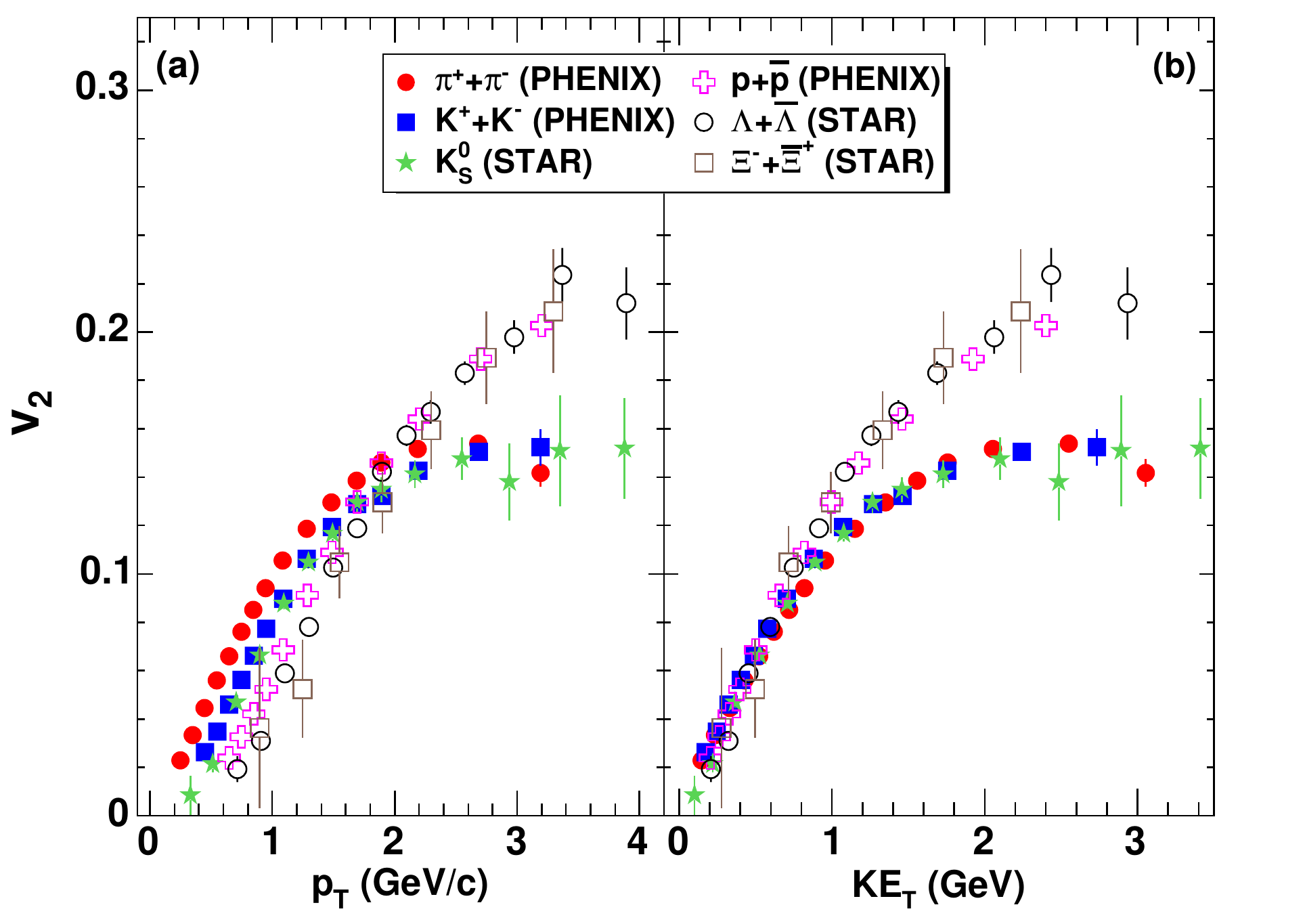}
    \includegraphics[height=.45\textwidth]{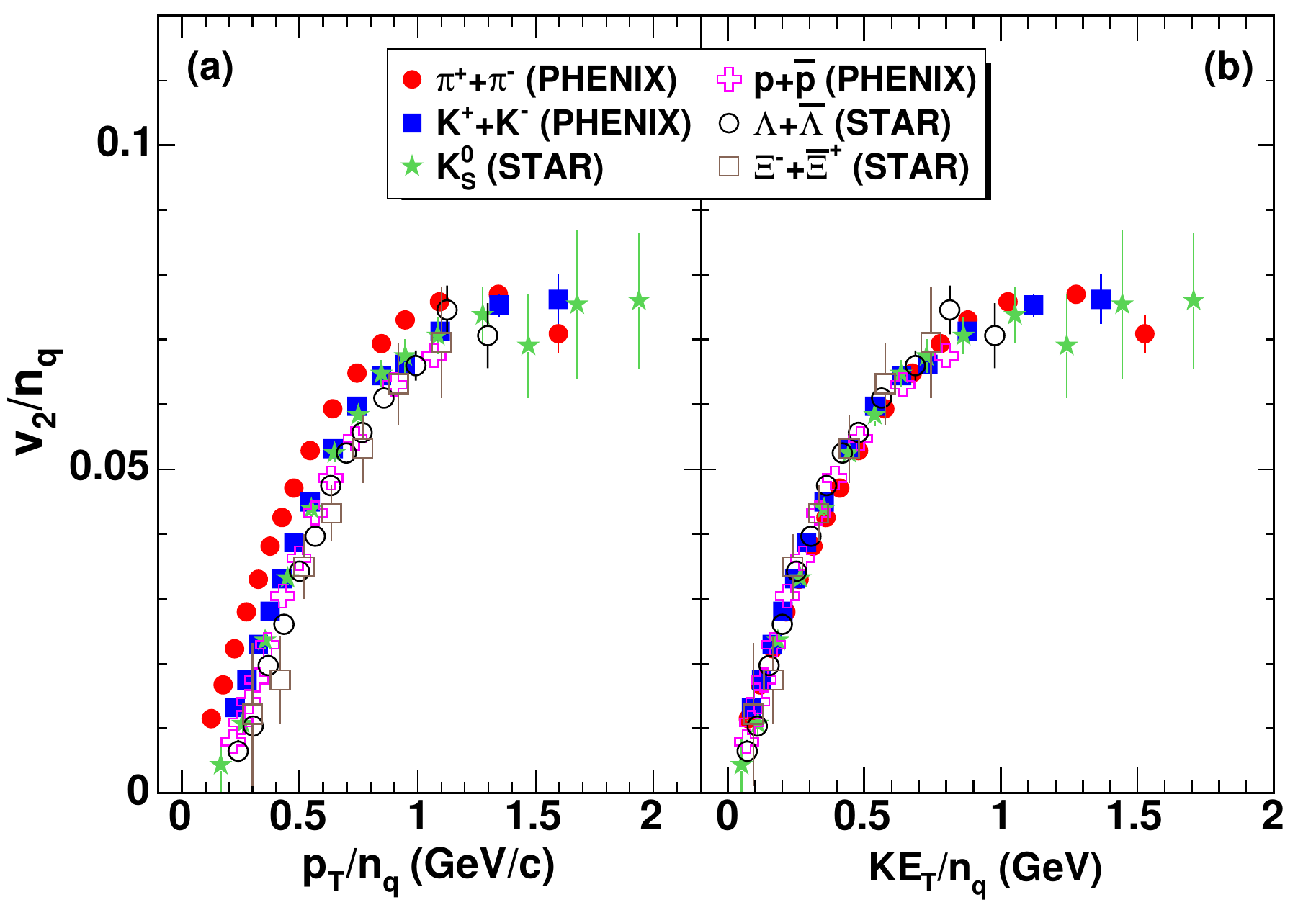}
  }
  \caption{
    \label{fig:quark_flow_scaling}
    Left two figures: Differential elliptic flow $v_2$ for various hadron species plotted as a function of $p_T$ and the transverse kinetic energy $K\!E_T = m_T - m$, where $m_T$ is the transverse particle mass.
    Right two figures: Same as on the left, but with both axes scaled by the quark number $n_q$.
    Figure from the PHENIX collaboration \cite{Adare:2006ti}.
  }
\end{figure}

Remarkably, early RHIC experiments showed that the produced collectivity is best understood if it is assumed to develop from a flowing liquid of deconfined quarks, rather than a super hot gas of hadrons \cite{Adare:2006ti, Adams:2003am, Adams:2005zg}.
This preference for quark degrees of freedom is illustrated in figure \ref{fig:quark_flow_scaling}.

First, look at the far left plot which shows the $p_T$-differential elliptic flow $v_2(p_T)$ for various hadron species in minimum-bias Au-Au collisions at RHIC.
The mass splitting visible among the different species is a characteristic signature of hydrodynamic flow.
If the mass-ordering of $v_2$ is driven by hydrodynamic pressure gradients, then the differential $v_2$ of each particle should scale with the transverse kinetic energy ${K\!E}_T = m_T - m$, where $m_T$ is the particle's transverse mass.

The second figure from the left shows the differential elliptic flow $v_2$ plotted against the transverse kinetic energy $K\!E_T$.
Notice how the elliptic flow curves split into two branches.
The upper branch contains all the baryons (three-quark states) while the lower branch contains all the mesons (two-quark states).
Presumably, the baryons carry more elliptic flow because they carry one extra quark than the mesons.

Finally, look at the figure on the far right which shows both these quantities divided by the number of valence quarks in each particle.
Suddenly, all of the differential flow measurements collapse to a single curve.
This signifies that the elliptic flow is carried by individual quarks, and that the elliptic flow is transmitted from the quarks to the hadrons by the hadronization process \emph{after} the flow has already developed.
This strongly evidences the creation of a fluid comprised of free flowing quarks.

\subsubsection{Jet quenching}

When two nuclei collide at ultrarelativistic energies, the quarks and gluons inside the nuclei occasionally scatter at large angles, producing two or more energetic partons carrying very large transverse momenta, anywhere from one to several orders of magnitude larger than the typical transverse particle momentum inside the event.
These energetic partons penetrate the produced QGP medium and fragment into softer particles, emerging from the interaction region as columnated sprays of nuclear matter known as jets.

\begin{wrapfigure}[17]{r}[.2\textwidth]{.5\textwidth}
  \centering
  \begin{tikzpicture}
    \node{\includegraphics[width=.5\textwidth]{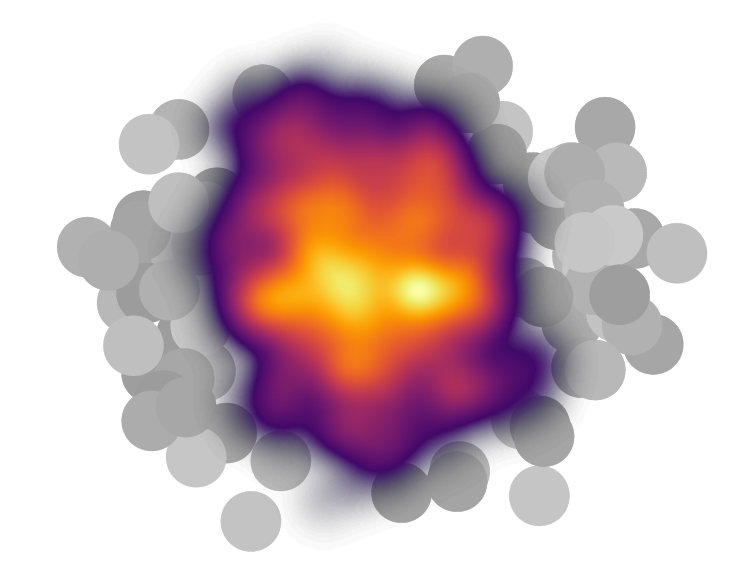}};
    \draw[very thick, draw=teal, dashed, rotate around={45:(1.15,-1.425)}] (1.15, -1.425) arc (170:10:.4cm and 0.2cm)coordinate[pos=0] (a);
    \draw[very thick, draw=teal, rotate around={45:(1.15,-1.425)}] (1.15,-1.425) arc (-170:-10:.4cm and 0.2cm)coordinate (b);
    \draw[very thick, draw=teal] (a) -- ([xshift=-1cm, yshift=1cm]$(a)!0.5!(b)$) -- (b);

    \draw[very thick, draw=teal, dashed, rotate around={225:(-.7,1.95)}] (-.7,1.95) arc (170:10:.7cm and 0.2cm)coordinate[pos=0] (a);
    \draw[very thick, draw=teal, rotate around={225:(-.7,1.95)}] (-.7,1.95) arc (-170:-10:.7cm and 0.2cm)coordinate (b);
    \draw[very thick, draw=teal] (a) -- ([xshift=1.5cm, yshift=-1.5cm]$(a)!0.5!(b)$) -- (b);
    \node[align=center] at (2.2, -1.7) {near\\side};
    \node[align=center] at (-1.7, 2.0) {away\\side};
  \end{tikzpicture}
  \caption{
    \label{fig:near_away_side}
    Di-jet event (teal cones) superimposed on the initial transverse energy density of an ultrarelativistic heavy-ion collision (heatmap).
  }
\end{wrapfigure}

As each penetrating jet moves through the QGP medium, it images the properties of the produced matter analogous to an x-ray radiograph.
If the QGP is strongly-coupled, each jet is expected to lose significant energy to the medium via induced gluon radiation such that the final jet is strongly modified or ``quenched''.
The existence of jet quenching is therefore a key prediction of a strongly-coupled QGP.
Presumably, this effect should depend on fundamental properties of interest such the color-charge density of the QGP and its short-distance structure \cite{Gyulassy:2003mc}.

Naturally, if a hard-scattering process produces back-to-back jets near the periphery of the fireball, with one jet moving into the medium and the other moving out of it, then the jet moving into the medium (away-side jet) should be more strongly modified than the jet moving out of it (near-side jet).
This setup is depicted in figure \ref{fig:near_away_side}.
One way to test this hypothesis, is to measure two-particle azimuthal correlations, using a high-$p_T$ trigger particle to orient the correlation function relative to the dominant jet.

\begin{figure}[t]
  \centering
  \includegraphics[scale=.44]{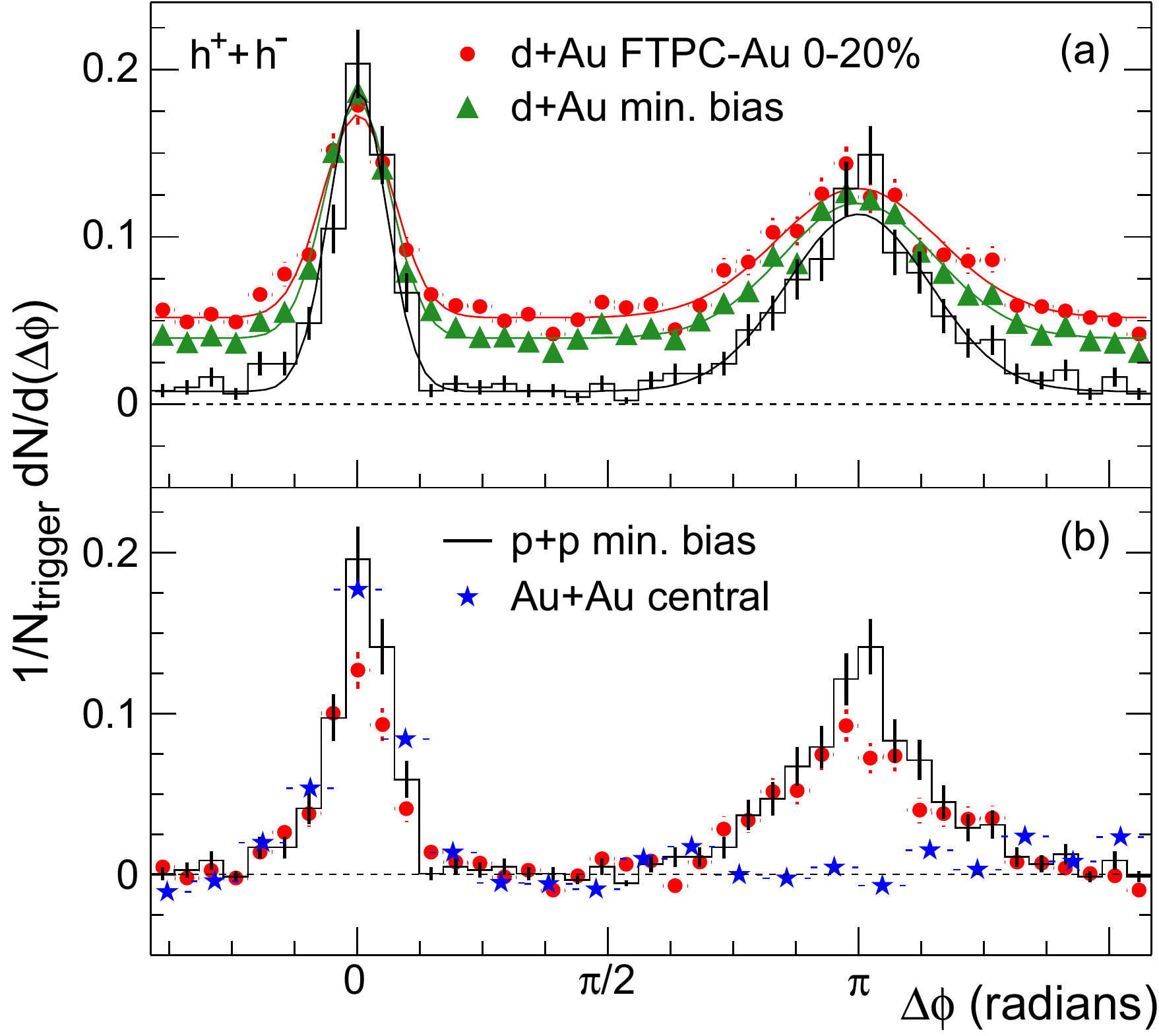}
  \caption{
    \label{fig:jet_energy_loss}
    Top: Two-particle azimuthal distributions for minimum-bias and central $d$-Au collisions, and for $p$-$p$ collision.
    Bottom: Comparison to the distribution for central Au-Au collisions \cite{Adams:2003im}.
  }
\end{figure}

Figure \ref{fig:jet_energy_loss} shows such a test applied to $p$-$p$, $d$-Au, and Au-Au collisions at $\sqrts=200$~GeV by the STAR collaboration \cite{Adams:2003im}.
The quantity plotted is the two-particle azimuthal distribution
\begin{equation}
  D(\Delta \phi) = \frac{1}{N_\text{trigger}} \frac{dN}{d (\Delta \phi)},
\end{equation}
constructed by correlating a high-$p_T$ trigger particle with angle $\phi_a$ and transverse momentum $4 < (p_T)_a < 6$~GeV with all partner particles in the same event having angle $\phi_b$ and transverse momentum $(p_T)_a > (p_T)_b > 2$~GeV.
The constant $N_\text{trigger}$ is the number of selected trigger particles, and the quantity $\Delta \phi = \phi_a - \phi_b$ is the azimuthal angle between each particle pair.

First, look at the top panel of the figure which shows this two-particle azimuthal distribution for minimum-bias and central $d$-Au collisions, and for minimum-bias $p$-$p$ collisions.
All three distributions show a sharp peak at $\Delta \phi = 0$, corresponding to particles that are emitted at small angles with respect to the high-$p_T$ trigger particle.
There is also a second peak centered on $\Delta \phi = \pi$, which is somewhat smaller in stature and smeared out.
Now look at the bottom figure, which shows the $p$-$p$ and $d$-Au two-particle azimuthal distributions compared to the same distribution for central Au-Au collisions.
In the Au-Au system, the $\Delta \phi = 0$ peak is clearly visible, but the $\Delta \phi = \pi$ peak is absent.

This result is naturally explained by the existence of a strongly-coupled QGP.
The peak at $\Delta\phi = 0$ is produced by particles emitted from the near-side jet, while the peak at $\Delta \phi = \pi$ is produced by the away-side jet.
In a di-jet event, the initial partons are produced back-to-back so the jets are separated by 180$^\circ$.
The near-side jet is produced closer to the surface of the QGP fireball, so it escapes with little modification, while the away-side jet plows into the medium where it is strongly quenched.
Given that the jet loses several GeV of energy as it traverses a relatively short distance, this observation corroborates that the matter is strongly-coupled.
Meanwhile, independent studies show that electromagnetic probes, e.g.\ direct photons and $Z$ bosons, show no evidence of jet quenching \cite{Adler:2006bv, Gardner:2013wya}.
Hence, the opaqueness of the matter appears specific to particles which interact via the strong force, consistent with the picture of QGP formation.

\section{Hydrodynamic computer simulations}
\label{sec:initial_condition_problem}

Hydrodynamic computer simulations are a powerful tool to refine our current understanding of hot and dense nuclear matter.
These simulations recreate entire nuclear collision events, exactly as they are believed to occur inside the detector, and output virtual particles that can be post-processed and analyzed using the same methods applied to the experimental data.
Important QGP medium parameters, e.g.\ the QGP specific shear viscosity $\etas$ and specific bulk viscosity $\zetas$, are then extracted by tuning their values to maximize the agreement of the simulation with experiment.

Hydrodynamic computer simulations vary in their exact implementation, but they generally follow a canonical framework which is constantly being updated and refined.
This section briefly summarizes the current picture of the hydrodynamic framework and explains how it can be used to extract QGP transport coefficients.
Finally, I conclude by discussing the largest obstacle limiting the precision of these simulation-based extractions, the so-called QGP ``initial condition problem'', which is the subject of this dissertation research.

\subsection{Space-time picture of a single event}
\label{subsec:spacetime_picture}

\begin{figure}[h]
  \centering
  \includegraphics{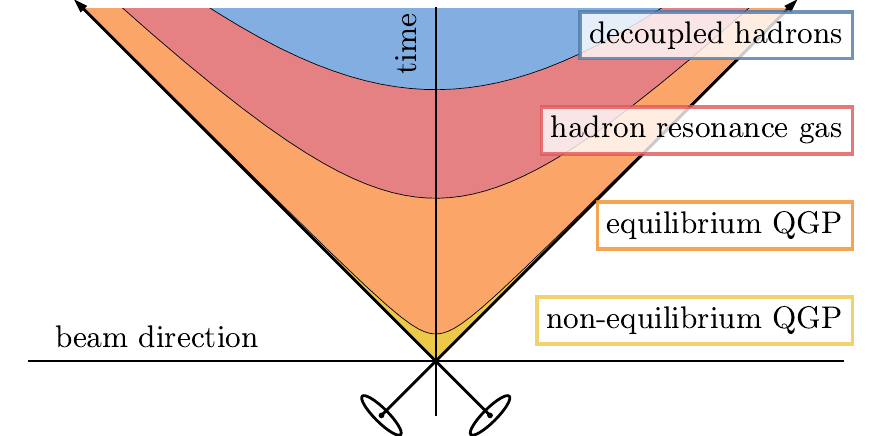}
  \caption{
    \label{fig:spacetime_diagram}
    Qualitative space-time diagram of a relativistic heavy-ion collision.
  }
\end{figure}

Consider two nuclei barreling toward each other at nearly the speed of light inside the beam pipe, as visualized by the space-time diagram figure \ref{fig:spacetime_diagram}.

\subsubsection{Preparing the nuclei}

Hydrodynamic computer simulations begin by preparing the two ions for a simulated collision.
The ions are constructed by sampling their three-dimensional nucleon densities, mimicking the spatial fluctuations seeded by the ultimate collapse of each nuclear wave function.
The nuclei are given a random rotation and impact parameter offset and boosted to their respective beam velocities, causing each ion to appear as a Lorentz contracted disk in the stationary lab frame.
The Lorentz factor is about half the value of the center-of-mass energy per nucleon pair when expressed in units of GeV.
Thus for a collision at $\sqrts=200$~GeV, each nucleus is contracted by $\gamma \sim 100$ along its direction of motion.

\subsubsection{Initial state}

If the sampled impact parameter offset is sufficiently small, the two nuclei interpenetrate and briefly overlap.
This convolves the three-dimensional density of each nucleus, depositing tremendous energy in the process.
The produced secondary matter fills the space between the receding ion fragments and forms an extended tube of deconfined quarks and gluons characterized by very small baryon density.
This initial overlap process is so brief, $\Delta \tau_\text{overlap} \lesssim 0.1~\fmc$, that computer models commonly assume it to happen instantaneously.
Simulations therefore start by calculating the matter's energy or entropy density at some early time $\tau_0 \lesssim 1~\fmc$ shortly after the nuclei interpenetrate.
Alternatively, more advanced simulations calculate all components of the initial stress-energy tensor $T^{\mu\nu}$ \cite{Schenke:2012wb, Gale:2012rq}.

\subsubsection{Pre-equilibrium evolution}

The stress-energy tensor of the initially produced matter is locally anisotropic and far from equilibrium.
These conditions preclude the direct application of hydrodynamics at very early times $\tau_\text{hydro} \lesssim 1~\fmc$.
Pre-equilibrium transport models based on strongly and weakly-coupled effective field theories are therefore used to evolve the system forward in time until the local stress-energy tensor more closely resembles the form predicted by second-order hydrodynamics \cite{Schenke:2012wb, Kurkela:2018vqr, Romatschke:2015gxa, vanderSchee:2013pia}.
Computer simulations that properly model the pre-equilibrium stage of the collision are a relatively recent development, so often hydrodynamic models evolve the system to the hydrodynamic starting time using simple free-streaming approximations \cite{Liu:2015nwa, Broniowski:2008qk} or they opt to skip the pre-equilibrium stage entirely.

\subsubsection{Hydrodynamic evolution}

The pre-equilibrium phase is then matched to viscous hydrodynamics to simulate the space-time evolution of the QGP liquid.
The hydrodynamic simulation is provided initial conditions for the energy density $e$, fluid velocity $u^\mu$, and shear and bulk viscous corrections $\pi^{\mu\nu}$ and $\Pi$, an EoS from lattice QCD, and values for the temperature-dependent QGP transport coefficients $\etas$ and $\zetas$.\footnote{If the initial fluid velocity $u^\mu$ and viscous corrections $\pi^{\mu\nu}$ and $\Pi$ are not provided by the initial condition model, they are typically set to zero. This approximation is known as static initialization.}
The hydrodynamic equations of motion are then solved numerically on a discretized grid.

Hydrodynamic simulations vary in their approximations and numerical schemes.
One common variant of the framework applies a simplifying symmetry known as \emph{boost-invariance} which asserts Lorentz invariance to boosts along the beam direction.
In his seminal paper on relativistic heavy-ion collisions, Bjorken argued that boost-invariance should hold for ultrarelativistic heavy-ion collisions, since the nuclei are already so highly boosted ($\gamma \gtrsim 100$), that the collision will appear essentially identical to any observer in a moderately boosted reference frame \cite{PhysRevD.27.140}.

This assumed symmetry reduces (3+1) space-time dimensions to (2+1) dimensions and dramatically simplifies the hydrodynamic equations of motion.
Boost-invariant hydrodynamic codes therefore run an order of magnitude faster than their three-dimensional counterparts.
In this dissertation, I perform calculations using both boost-invariant and three-dimensional hydro codes.
Boost-invariance generally works well near midrapidity $|\eta| \lesssim 1$ \cite{Shen:2016zpp, Vredevoogd:2012ui}, but it is a poor approximation if used to analyze particles detected at moderate to large rapidities.

\subsubsection{Particlization and hadronic evolution}

After $\order{10}~\fmc$ of hydrodynamic evolution, the medium cools past the QGP transition temperature and freezes into individual hadrons.
These emitted particles continue to scatter and decay, then eventually decouple and free stream into the detector.
Hydrodynamic mean-field approximations begin to break down as the system disintegrates, so the hydrodynamic evolution is commonly spliced onto microscopic kinetic theory which is better suited to handle the system's non-equilibrium break-up.

\begin{figure}[t]
  \centering
  \makebox[\textwidth]{
    \begin{tikzpicture}
      \node{\includegraphics[width=\textwidth]{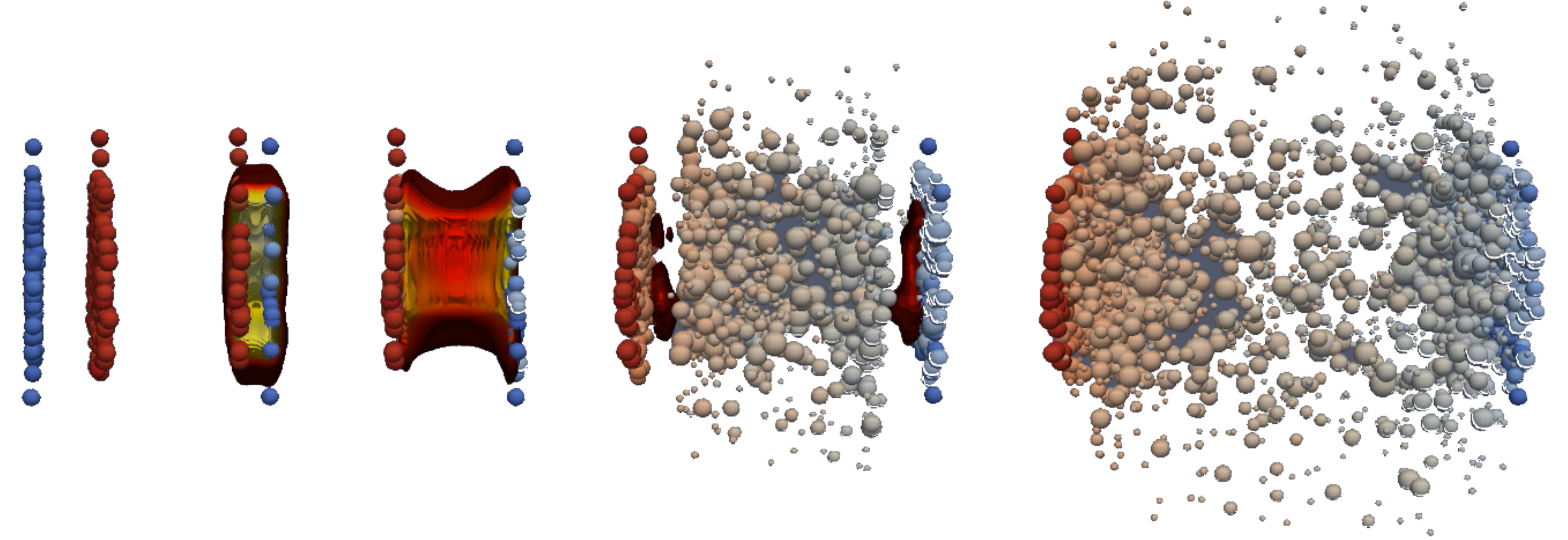}};
      \draw[->, thick] (-6, -2.5) node[above right]{$0~\fmc$} -- node[above] {Time} (6, -2.5) node[above left]{${\sim}20~\fmc$};
    \end{tikzpicture}
  }
  \caption{
    \label{fig:zhi_hydro}
    Hybrid model computer simulation of a typical heavy-ion collision event.
    Figure is adapted from a visualization originally constructed by H. Petersen for the MADAI collaboration.
  }
\end{figure}

This \emph{hybrid model} prescription \cite{Bass:2000ib, Nonaka:2006yn, Petersen:2008dd} converts the fluid to hadrons assuming thermal particle emission from a pre-specified switching isotherm $\Tsw$, typically required to lie near the pseudocritical temperature $T_c$ in order to fit the observed particle yields.
Once the fluid is ``particlized'', its subsequent interactions are modeled by the Monte Carlo implementation of the Boltzmann equation which follows each hadron microscopically until the last interactions cease and the system freezes out, yielding a list of final particle data for each event.
For a visualization, see figure \ref{fig:zhi_hydro} which shows several snapshots of a typical event simulated using the hybrid model framework.

\subsection{Extracting QGP transport coefficients}

The QGP transport coefficients can be inferred from hydrodynamic simulations, by analyzing their effect on bulk particle properties.
Typically, this is accomplished by identifying key observables which are particularly sensitive to a given parameter of interest.
The parameter's true value is then inferred by adjusting its assumed value until the simulation optimally agrees with experiment.

One notable example of this procedure, is the use of the flow harmonics $v_n$ to constrain the QGP specific shear viscosity $\etas$.
Recall that these harmonics $v_n$ measure the final particle distribution's anisotropy with respect to the azimuthal angle $\phi$.
Elliptic flow $v_2$ measures its ellipticity, triangular flow $v_3$ measures its triangularity, and so on.

\begin{wrapfigure}[11]{r}[.1\textwidth]{.5\textwidth}
  \centering
  \vspace{-15pt}
  \begin{tikzpicture}
    \draw[very thick] (0,0) ellipse (10pt and 20pt) node[above, yshift=1cm]{Initial};
    \draw[->,very thick] (-1,-1.5)-- ++(0,.5) node[above]{$\hat{y}$};
    \draw[->,very thick] (-1,-1.5)-- ++(.5,0) node[right]{$\hat{x}$};

    \draw[very thick] (3,0) ellipse (20pt and 10pt) node[above, yshift=1cm]{Final};
    \draw[->,very thick] (2,-1.5)-- ++(0, .5) node[above]{$\hat{p}_y$};
    \draw[->,very thick] (2,-1.5)-- ++(.5, 0) node[right]{$\hat{p}_x$};
  \end{tikzpicture}
  \caption{
    \label{fig:hydro_anisotropy}
    Elliptically deformed spatial profile (left) generates elliptically deformed momentum profile (right).
  }
\end{wrapfigure}
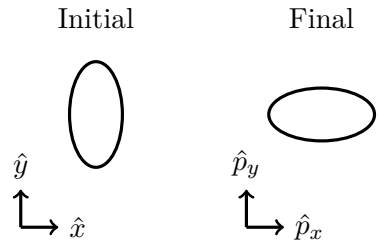

These final-state \emph{momentum} anisotropies originate as initial-state \emph{spatial} anisotropies.
Crudely speaking, hydrodynamics converts spatial anisotropy into momentum anisotropy.
For example, if the initial state is elliptically deformed, its hydrodynamic evolution will generate elliptic flow (see figure \ref{fig:hydro_anisotropy}).
Similarly, triangular profiles generate triangular flow, quadrangular profiles generate quadrangular flow, etc.

Much like the final momentum anisotropy, the initial spatial anisotropy can be quantified by its azimuthal harmonics
\begin{equation}
  \varepsilon_n e^{i n \Phi} = - \frac{\int dx\, dy\, r^n e^{i n \phi} \rho(x, y)}{\int dx\, dy\, r^n \rho(x, y)},
\end{equation}
where $\varepsilon_n$ is the \emph{eccentricity harmonic} of order $n$, $\Phi$ is its phase angle, and $\rho$ is the density profile of interest, typically assumed to be the event's transverse energy or entropy density.
Generally speaking, initial profiles with large $\varepsilon_n$ generate large $v_n$.
Linear scaling $v_n \propto \varepsilon_n$ is observed for $n=2$, 3 in heavy-ion collisions to good approximation \cite{Gardim:2011xv, Niemi:2012aj, Gardim:2014tya}, but the scaling breaks down for $n > 3$ due to non-linear mode mixing \cite{Gardim:2011xv}.

The QGP specific shear viscosity $\etas$ governs the efficiency with which the hydrodynamic evolution converts spatial anisotropy into momentum anisotropy.
Hence, it is directly related to the ratio $v_n / \varepsilon_n$ which quantifies the flow that's produced per unit eccentricity.
Small values of $v_n / \varepsilon_n$ correspond to large shear viscosities and large values of $v_n / \varepsilon_n$ correspond to small shear viscosities.
Note, the flow anisotropies $v_n$ are directly measurable, while the initial state eccentricities $\varepsilon_n$ are not; they can only be estimated theoretically.
Therefore, $\etas$ extractions are directly limited by one's ability to calculate the QGP initial conditions precisely.

\begin{figure}[t]
  \centering
  \makebox[\textwidth]{
    \includegraphics[scale=.6]{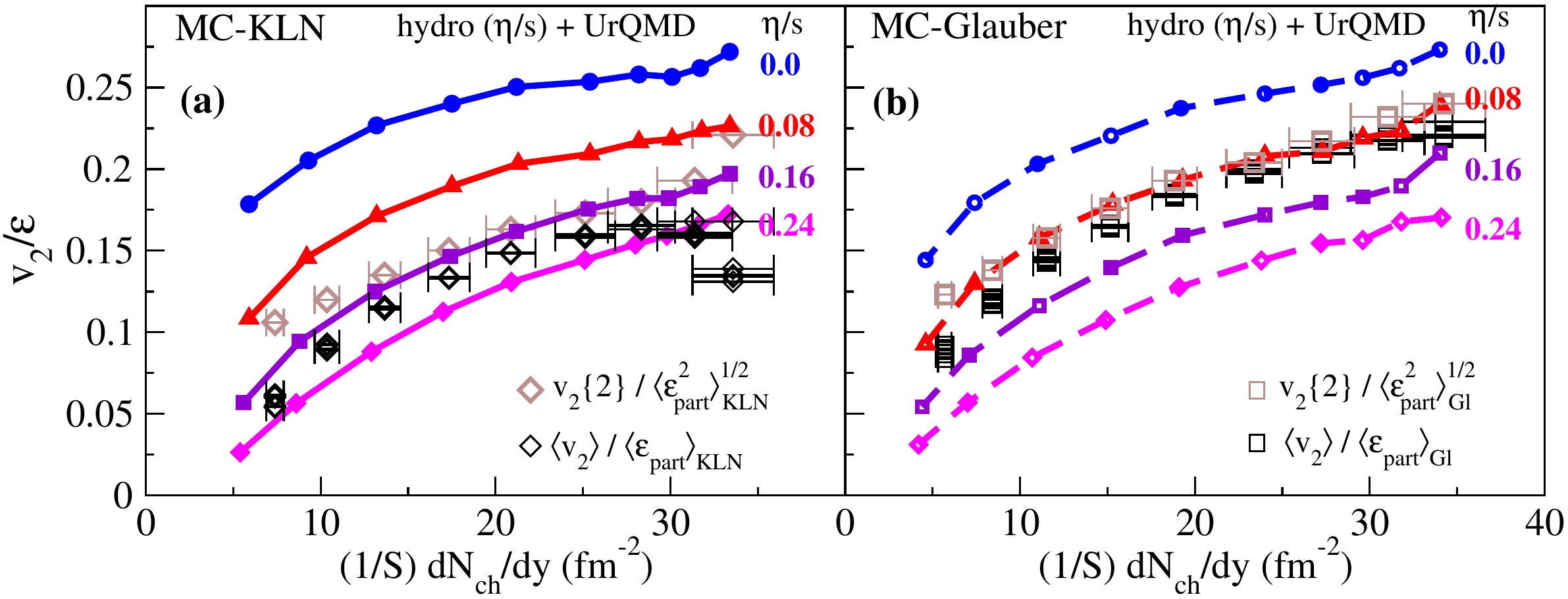}
  }
  \caption{
    \label{fig:etas_glb_kln}
    Eccentricity-scaled elliptic flow $v_2/\varepsilon_2$ plotted versus the charged-particle density per unit overlap area $(1/S)\, d\nch/dy$ \cite{Song:2010mg}.
    Symbols are constructed using the experimentally measured elliptic flow mean $\langle v_2 \rangle$ \cite{Ollitrault:2009ie} and two-particle cumulant $\vnk{2}{2}$ \cite{Adams:2004bi}, along with the charged-particle density $d\nch/dy$ \cite{Abelev:2008ab}, while colored lines are constructed using simulated values for these quantities, calculated for several values of the specific shear viscosity $\etas=0.0$, 0.08, 0.16, and 0.24.
    The eccentricity $\varepsilon_2$ and overlap area $S = \sqrt{\langle x^2 \rangle \langle y^2 \rangle}$ are obtained from the initial condition model.
    Results are shown for the MC-KLN initial conditions \cite{Drescher:2006ca, Drescher:2007rap} (left) and MC-Glauber initial conditions \cite{Miller:2007ri} (right).
  }
\end{figure}

Figure \ref{fig:etas_glb_kln} shows the eccentricity-scaled elliptic flow $v_2/\varepsilon_2$ (second harmonic) plotted versus the charged-particle density per unit overlap area $(1/S)\, d\nch/dy$ for \mbox{Au-Au} collisions at $\sqrts=200$~GeV using two different models for the QGP initial conditions (left and right plots).
The symbols are calculations using the experimentally measured elliptic flow $v_2$ and charged-particle density $d\nch/dy$, while the colored lines are constructed using simulated values for these quantities, calculated for several different values of the QGP specific shear viscosity $\etas$.
The eccentricity $\varepsilon_2$ and root-mean-square overlap area $S = \sqrt{\langle x^2 \rangle \langle y^2 \rangle}$, meanwhile, are provided by the respective initial condition model.
The panel on the left shows an extraction using MC-KLN initial conditions \cite{Drescher:2006ca, Drescher:2007rap}, and the panel on the right shows an extraction using MC-Glauber initial conditions \cite{Miller:2007ri}.
It's not important that I describe these models in detail at the moment---suffice to say, each initial condition model predicts different eccentricities $\varepsilon_n$.

The ratio $v_2/\varepsilon_2$ can be thought of as a ``ruler'' which measures the fluid's specific shear viscosity $\etas$.
The experimentally extracted viscosity is read from the plot by matching the symbols with the colored lines, each corresponding to a specific value of $\etas$.
Hence, the extraction based on MC-KLN initial conditions obtains $\etas \sim 0.2$, while the extraction based on MC-Glauber initial conditions obtains $\etas \sim 0.08$ (each with large errors).
The authors of the study were therefore able to conclude that the QGP specific shear viscosity for $T_c  < T \lesssim 2 T_c$ lies within the range $0.08 < (\etas)_\text{QGP} < 0.20$, with the remaining uncertainty arising from insufficient theoretical control over the initial source eccentricity $\varepsilon_2$.
Consequently, the primary means to improve this estimate is to reduce the model's systematic initial condition uncertainty.
This is one example of what I refer to as the \emph{initial condition problem}.

The initial condition problem is, of course, far more general than the relationship between the elliptic flow, shear viscosity, and eccentricity.
The initial conditions strongly affect essentially every model output, so their uncertainty is strongly correlated with the uncertainty of the inferred medium properties.
For example, if a given initial condition model predicts QGP energy densities which are too compact, the simulation will expand more explosively than it should and require an artificially large bulk viscosity to compensate.

\section{The initial condition problem}

To date, there exist \emph{numerous} theoretical models for the QGP initial conditions, of which the MC-Glauber and MC-KLN models are two examples.
Different initial condition models generally predict different energy density and flow velocity profiles, so their hydrodynamic evolutions consequently prefer different values of the QGP transport coefficients.
Studies of the initial condition and QGP medium properties are thus inextricably linked.

The most straightforward procedure to reduce the list of mutually incompatible theory calculations is to validate candidate models using sensitive experimental observables.
Each initial condition model typically includes several free parameters which can be tuned to selectively fit one or two observables at a time, so it is important to test models self-consistently using a large cross section of the available experimental data.
Presumably, the correct model will reproduce all observables within the realm of its applicability, assuming the subsequent hydrodynamic evolution is well understood.

\subsection{\emph{Ab initio} theory calculations}

Over the last decade, tremendous progress has been made in understanding the initial stages of ultrarelativistic nuclear collisions.
The discovery process has been accelerated by several important theoretical developments, resulting in a handful of credible bottom-up initial condition approaches based on approximations of QCD and related field theories.
This subsection summarizes two such models which have demonstrated broad agreement with the experimental data, far surpassing the MC-Glauber and MC-KLN models mentioned previously.
I should emphasize that this is not meant to be an exhaustive list of \emph{all} credible initial condition models, and I apologize to the authors whose work is not discussed.

\subsubsection{IP-Glasma model}

One \emph{ab initio} model which successfully describes a large number of experimentally measured bulk observables is IP-Glasma \cite{Schenke:2012wb, Schenke:2012fw}.
This model obtains the QGP initial conditions from Color Glass Condensate effective field theory, by combining the impact-parameter dependent saturation model (IP-Sat) \cite{Bartels:2002cj, Kowalski:2003hm} with the classical Yang-Mills description of initial gluon fields.
Color Glass Condensate (CGC) effective field theory is a general theoretical framework which describes the small-$x$\footnote{Bjorken $x = Q^2/(2\, p \cdot q)$ is a common variable in deep-inelastic scattering related to the fraction of the proton momentum carried by a certain parton. Here $p$ is the incoming proton momentum, $q$ is its momentum transfer with the probe, and $Q^2 = -q^2$.} behavior of the hadronic wave function in QCD \cite{Iancu:2003xm}.
In this approach, the system's large-$x$ color-charge degrees of freedom act as static sources for small-$x$ gauge fields $A^\mu$.
At high energies, the density of produced partons at small-$x$ becomes large, leading to a saturation of the parton distribution function which occurs at the characteristic saturation momentum $Q_s$.

\begin{figure}[t]
  \centering
  \makebox[\textwidth]{
    \includegraphics[height=.4\textwidth]{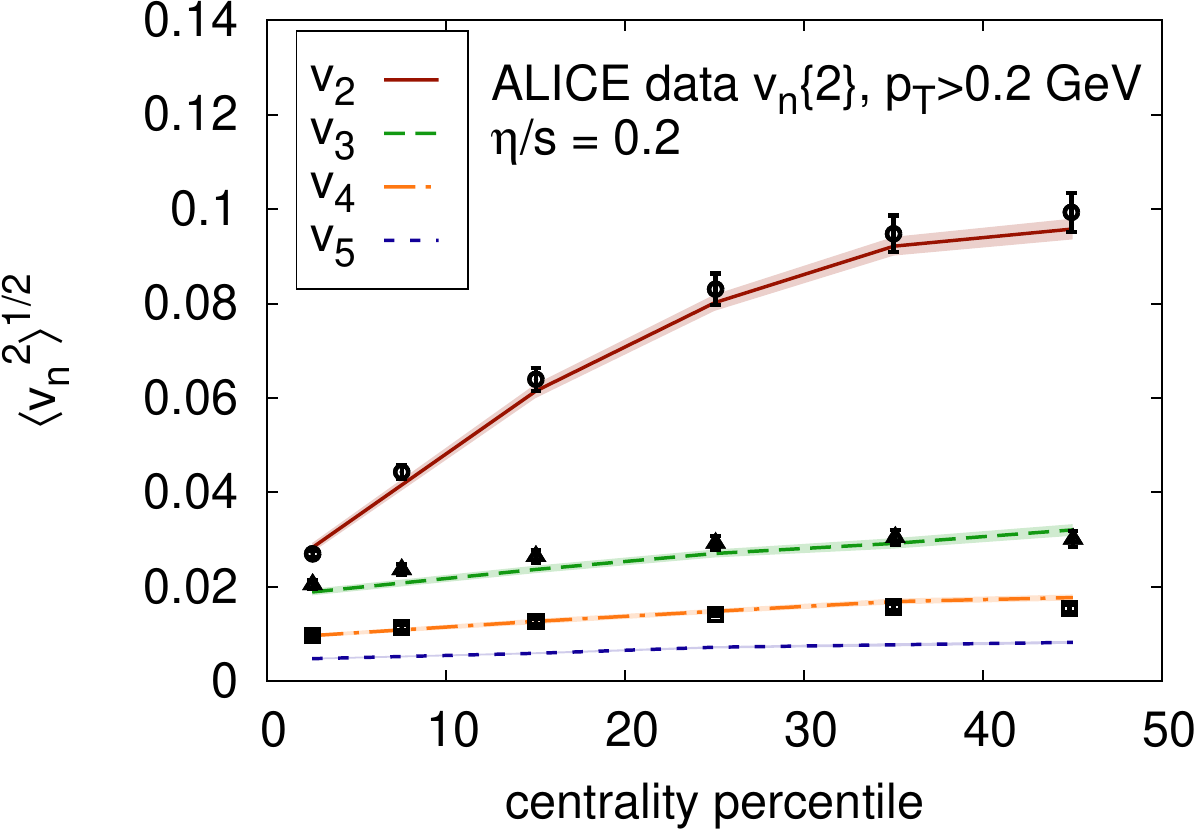}
    \includegraphics[height=.4\textwidth]{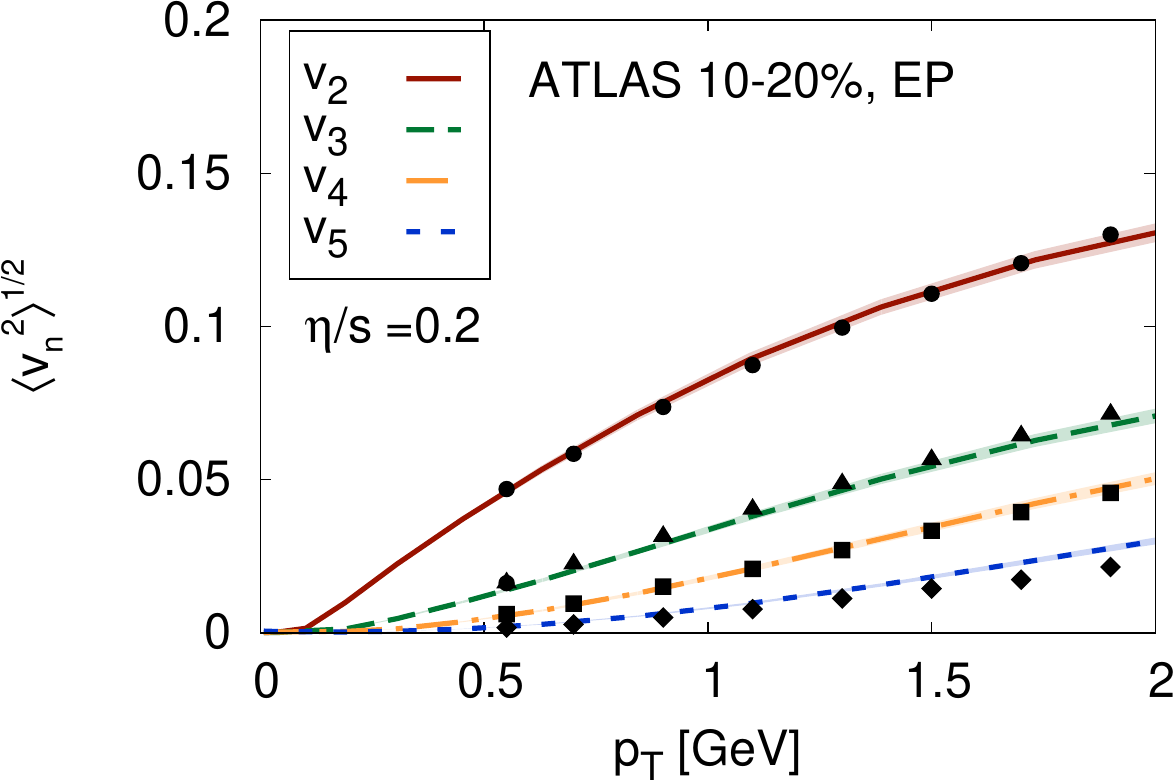}
  }
  \caption{
    \label{fig:ipglasma_flows}
    Left: Root-mean-square anisotropic flow coefficients $\sqrt{\langle v_n \rangle^2}$ as a function of centrality calculated from IP-Glasma initialized hydrodynamic simulations (lines) \cite{Gale:2012rq} compared to the experimentally measured two-particle flow cumulants $\vnk{n}{2}$ measured by ALICE (symbols) \cite{ALICE:2011ab}.
    Right: Root-mean-square anisotropic flow coefficients $\sqrt{\langle v_n^2 \rangle}$ as a function of transverse momentum $p_T$ (lines) compared to experimental data from ATLAS (symbols) using the event-plane method \cite{ATLAS:2012at}.
    All calculations are for $\etas = 0.2$ and $\zeta/s=0$.
  }
\end{figure}

The IP-Glasma model starts by sampling the positions of nucleons within each nucleus from a Fermi distribution (more on this later).
Once the nucleon positions are known, the IP-Sat model provides the saturation scale $Q^2_s(x, \mathbf{b}_\perp)$ as a function of Bjorken $x$ and the transverse impact parameter $\mathbf{b}_\perp$ relative to each nucleon's center.
The color-charge density squared per unit transverse area $g^2 \mu^2$ is then assumed to be proportional to the saturation scale $Q_s^2$.

For a nucleus with $A$ nucleons, the quantity $g^2 \mu_{A(B)}^2(x, \xv_\perp)$ is obtained for each nucleus by adding the color-charge contributed by each nucleon.
Provided this mean square color-charge density, random color charges $\rho^a$ are sampled from the Gaussian distribution
\begin{equation}
  \langle \rho_{A(B)}^a(\xv_\perp) \rho_{A(B)}^b(\yv_\perp) \rangle = \delta^{ab} \delta^{(2)}(\xv_\perp - \yv_\perp) g^2 \mu_{A,B}^2(x, \xv_\perp),
\end{equation}
for nucleus $A$ and $B$.

\begin{wrapfigure}[24]{r}[.2\textwidth]{.67\textwidth}
  \centering
  \vspace{-10pt}
  \includegraphics[width=.6\textwidth]{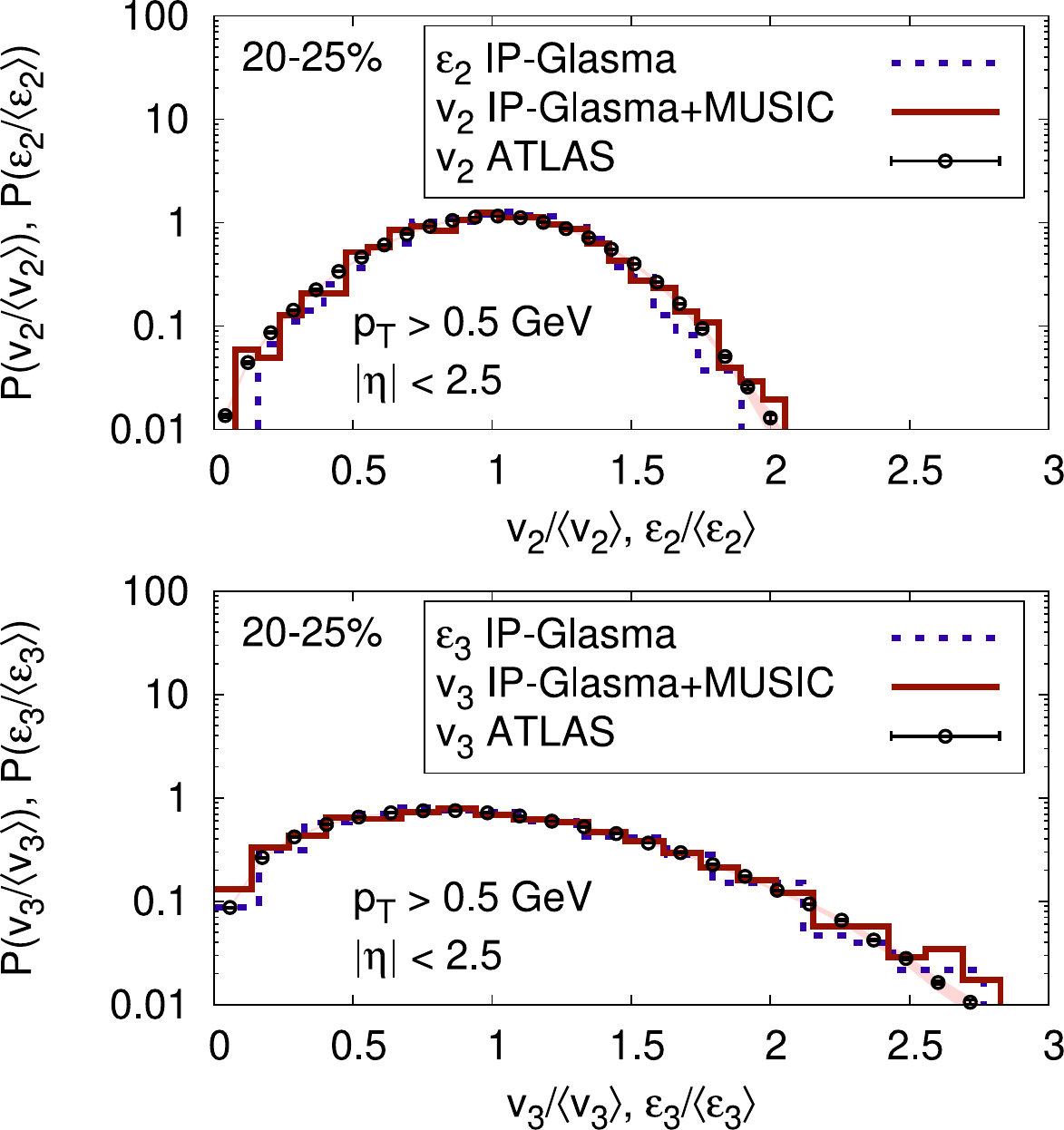}
  \caption{
    \label{fig:ipglasma_ebe_flow}
    Probability distributions for the scaled anisotropic flows $v_n/\langle v_n \rangle$ and scaled eccentricities $\varepsilon_n / \langle \varepsilon_n \rangle$ predicted by IP-Glasma initialized hydrodynamic calculations \cite{Gale:2012rq} compared to ATLAS data \cite{Jia:2012ve}.
  }
\end{wrapfigure}

After this sampling, the random color-charge distribution of each nucleus is used to calculate the electric and magnetic color fields by solving the classical Yang-Mills equations
\begin{equation}
  [D_\mu, F^{\mu\nu}] = J^\nu,
\end{equation}
where $F^{\mu\nu}$ is the field strength tensor and $J^\nu$ is the color current density, calculated from each Lorentz contracted sheet of boosted color-charge density.
Finally, the QGP's initial energy density $e$ and flow velocity $u^\mu$ are calculated from the produced gluon fields evolved to a pre-specified hydrodynamic starting time shortly after the collision.

IP-Glasma is perhaps best known as the first initial condition model to correctly reproduce the first few harmonics of the azimuthal flow anisotropy $v_n$ generated by heavy-ion collisions \cite{Gale:2012rq}.
Figure \ref{fig:ipglasma_flows} shows the root-mean-square anisotropic flow coefficient $\sqrt{\langle v_n^2 \rangle}$ for $n=2,3,4,5$ plotted as a function of collision centrality (left) and as a function of transverse momentum $p_T$ (right) for Pb-Pb collisions at $\sqrts=2.76$~TeV compared to experimental data from ALICE \cite{ALICE:2011ab} and ATLAS \cite{ATLAS:2012at}.
The model provides a superb description of these observables, suggesting that a proper modeling of the eccentricity harmonics $\varepsilon_n$ is achieved.
At the time, this level of agreement with the data was truly unprecedented.

Even more impressive, the IP-Glasma model correctly describes the \emph{full} probability distribution $P(v_n)$ of each flow harmonic as a function of collision centrality \cite{Gale:2012rq}.
In other words, the model doesn't just describe one moment of the flow distribution, it correctly describes its non-trivial shape as well.
Figure \ref{fig:ipglasma_ebe_flow} shows IP-Glasma initialized hydrodynamic calculations for the mean-scaled eccentricity distribution $P(\varepsilon_n / \langle \varepsilon_n \rangle)$ and mean-scaled flow distribution $P(v_n / \langle v_n \rangle)$ \cite{Gale:2012rq} compared to the corresponding flow distributions measured by ATLAS \cite{Jia:2012ve}.
The model calculations nicely track the experimental data, validating the assumptions of the framework.

The IP-Glasma initial condition model is generally well tested, and has been compared to numerous other experimental observables at RHIC and LHC energies as well, including e.g.\ the centrality-dependence of the charged-particle yield and mean transverse momenta \cite{McDonald:2016vlt, Schenke:2019ruo}.
Extractions of $(\etas)(T)$ and $(\zetas)(T)$ obtained using IP-Glasma initial conditions vary somewhat in the literature due to the specifics of each analysis; however, recent estimates \cite{McDonald:2016vlt, Schenke:2019ruo} find good agreement with the data using an effective specific shear viscosity $\etas \sim 0.095$--0.12 and a temperature-dependent specific bulk viscosity $(\zetas)(T)$ which peaks near $T \sim 165$--180~MeV and obtains a maximum value $\zetasmax \sim 0.24$--0.3.

\subsubsection{EKRT model}

The recently updated NLO EKRT model, which combines next-to-leading-order (NLO) collinearly factorized pQCD minijet production with a conjecture for low-$p_T$ gluon saturation, is another highly successful initial condition model named for its original authors Eskola, Kajantie, Ruuskanen, and Tuominen \cite{Eskola:1999fc, Niemi:2015qia}.
In this approach, the collision deposits energy in the form of low-$p_T$ partons (predominantly gluons) and high-$p_T$ minijets which are separated by a transverse momentum scale $p_0 \gg \lambda_\text{QCD}$.

Consider two nuclei, labeled $A$ and $B$, with three-dimensional nuclear densities $\rho_A$ and $\rho_B$ respectively.
Assume the nuclei collide with impact parameter vector $\mathbf{b}$ in the transverse plane $\xv_\perp=(x, y)$.
Let $T_A = \int dz\, \rho_A(\xv_\perp, z)$ define the transverse density of nucleus $A$, and assume $T_B$ follows accordingly.
For a given beam energy $\sqrts$, the initial transverse-area density of minijet transverse energy, $dE_T/d^2\xv_\perp$, produced perturbatively into a rapidity window $\Delta y$ above the transverse momentum cut-off $p_0$ is given by
\begin{equation}
  \label{eq:ekrt1}
  \frac{dE_T}{d^2\xv_\perp} = T_A(\xv_\perp)\, T_B(\xv_\perp -\, \mathbf{b})\, \sigma \langle E_T \rangle_{p_0, \Delta y, \beta},
\end{equation}
where $\sigma \langle E_T \rangle_{p_0, \Delta y, \beta}$ is the $E_T$-weighted minijet cross section computed from NLO pQCD.
This quantity depends on the transverse momentum cut-off $p_0$, the width of the rapidity interval $\Delta y$, and a phenomenological parameter $\beta$ which controls the minimum transverse energy $E_T$ allowed in $\Delta y$.
For a detailed formulation of $\sigma \langle E_T \rangle$, see \cite{Paatelainen:2012at, Niemi:2015qia}.

Here it is assumed that only minijets with transverse momenta ${p_T > p_0}$ contribute significantly to $dE_T/d^2\xv_\perp$.
Below the transverse momentum cut-off $p_0$, contributions from \mbox{$(3 \to 2)$} and higher-order partonic processes begin to dominate conventional \mbox{$(2 \to 2)$} processes causing the parton density to saturate.
This condition leads to the saturation criteria \cite{Paatelainen:2013eea}
\begin{equation}
  \label{eq:ekrt2}
  \frac{dE_T}{d^2\xv_\perp} = \frac{K_\text{sat}}{\pi} p_0^3 \Delta y,
\end{equation}
where $K_\text{sat}$ is an unknown normalization constant determined by the fitting the experimentally measured charged-particle density $d\nch/d\eta$ using a single narrow centrality interval.

Equations \eqref{eq:ekrt1} and \eqref{eq:ekrt2} are finally equated and solved numerically to determine the transverse momentum cut-off $p_0$ where the soft-gluon production saturates.
Provided this saturation momentum $p_\text{sat} \equiv p_0$, the local energy density $e$ at the local formation time $\tau_s(\xv_\perp) = 1/p_\text{sat}(\xv_\perp)$ at midrapidity follows from equation \eqref{eq:ekrt2}:
\begin{equation}
  e(\xv_\perp, \tau_s) = \frac{dE_T(p_\text{sat})}{d^2 \xv_\perp} \frac{1}{\tau_s \Delta y} = \frac{K_\text{sat}}{\pi} \left [ p_\text{sat}(\xv_\perp) \right]^4.
\end{equation}
This energy density is then evolved to a universal proper time $\tau_0 = 0.2~\fmc$ using one-dimensional Bjorken hydrodynamics.
The EKRT model does not provide the initial flow velocity $u^\mu$ or shear corrections $\pi^{\mu\nu}$ and $\Pi$, so these additional components are typically set to zero.

\begin{figure}[t]
  \centering
  \makebox[\textwidth]{
    \includegraphics[width=1.4\textwidth]{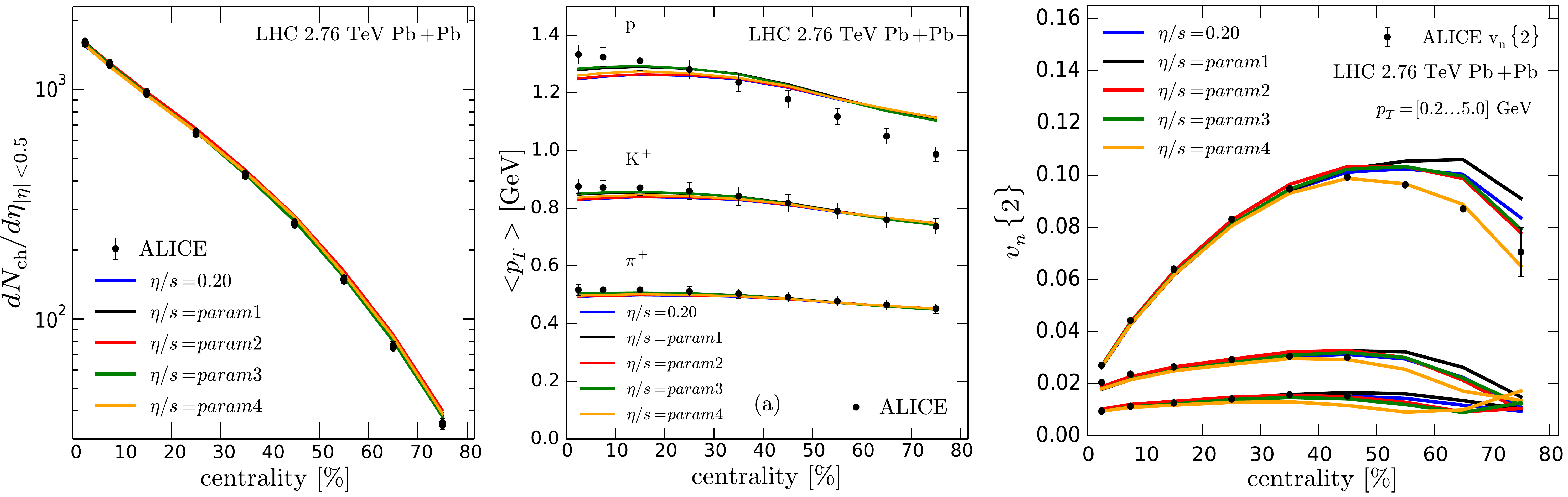}
  }
  \caption{
    \label{fig:ekrt_megaplot}
    Hydrodynamic calculations using EKRT initial conditions \cite{Niemi:2015qia} for the centrality-dependence of the charged-particle density $d\nch/d\eta$ (left), identified-particle mean $p_T$ (middle), and two-particle anisotropic flow cumulants $\vnk{n}{2}$ for $n=2$, 3, and 4 (right) compared to experimental data from ALICE \cite{ALICE:2011ab, Aamodt:2010cz, Abelev:2013vea}.
  }
\end{figure}

Figure \ref{fig:ekrt_megaplot} shows EKRT initialized hydrodynamic calculations for the centrality-dependence of the midrapidity charged-particle density $d\nch/d\eta$ (left), identified-particle mean $p_T$ (middle), and two-particle flow cumulants $\vnk{n}{2}$ for $n=2$, 3, and 4 (right) using several different specific shear viscosity parametrizations (lines) compared to experimental data from ALICE (symbols) \cite{ALICE:2011ab, Aamodt:2010cz, Abelev:2013vea}.
The model provides an excellent description of these observables, and also explains several other observables not pictured including the experimentally measured anisotropic flow probability distributions \cite{Aad:2013xma} and event-plane correlations \cite{Aad:2014fla}.
See reference \cite{Niemi:2015qia} for a comprehensive overview.

To extract the QGP specific shear viscosity, the authors of reference \cite{Niemi:2015qia} ran EKRT initialized hydrodynamic simulations with several different piecewise-linear $(\etas)(T)$ parametrizations (assuming zero bulk viscosity) and calculated numerous RHIC and LHC flow observables.
Of the parametrizations that they tested, the two that provided the best overall description of the data were a constant (flat) parametrization $\etas = 0.2$, and a sloped parametrization with a small hadronic viscosity and a minimum specific shear viscosity ${\etasmin = 0.12}$ located at ${T=150}$~MeV.

Presumably, these preferred specific shear viscosities would also change in the presence of non-zero bulk viscosity, which has been shown to affect extracted shear viscosity estimates \cite{Ryu:2015vwa, Ryu:2017qzn}.
Therefore, it is difficult at the present time to directly compare the viscosities extracted by IP-Glasma and EKRT initial conditions.
The estimates are obviously different, but it is not yet clear how much should be attributed to the initial conditions versus other components of the hydrodynamic simulation framework.

\subsection{Case for a new approach}

The IP-Glasma and EKRT models have significantly improved our current theoretical understanding of the initial stages of the collision.
However, neither model describes the experimental data perfectly within the realm of its applicability, so it stands to reason that neither model is complete.
This residual modeling error is a form of systematic uncertainty which biases current estimates of the QGP transport coefficients.

In the next chapter, I motivate and develop a complementary top-down approach for studying the QGP initial conditions using the constraints provided by the experimental data.
This will allow me to investigate the correlated effect of initial condition uncertainties on QGP parameter estimates, and it will allow me to independently validate the effective scaling predicted by the IP-Glasma and EKRT initial condition frameworks.
I start by deconstructing the initial condition problem into its simplest form.

\chapter[Initial conditions of bulk matter]{Initial conditions of bulk matter}
\label{ch:initialization}

\lettrine{E}{very} simulation needs a starting point.
For hydrodynamic simulations of relativistic nuclear collisions, the starting point is
the energy density $e$, fluid velocity $u^\mu$, and initial values of the bulk correction $\Pi$ and shear correction $\pi^{\mu\nu}$ at the hydrodynamic starting time.
Generally speaking, models of the QGP initial conditions strive to be parameter free, predictive, and established on a firm theoretical footing.
The holy grail would be an initial condition model that is elegantly derived from first principles, void of free parameters, and in perfect agreement with experimental measurements, barring the existence of confounding model errors.
This idealized description would effectively eliminate the uncertainty in the QGP initial conditions and enable simulation-based extractions of fundamental QGP properties with unprecedented precision.

Over the past decade, theoretical progress has brought the field closer to this ultimate goal.
In section \ref{sec:initial_condition_problem}, I discussed two of the more successful \emph{ab initio} theoretical calculations, the so-called EKRT \cite{Eskola:1999fc} and IP-Glasma \cite{Schenke:2012wb} initial condition models which are based on general concepts of gluon saturation physics.
There are of course many other theoretical models which have been proposed in the literature, but these two models in particular have arguably reproduced the largest swath of experimental data using a rather small (albeit non-zero) number of free parameters.

These models, of course, do not provide all of the answers.
It remains unclear, for example, to what extent the IP-Glasma and EKRT frameworks are mutually compatible.
While both theoretical models are based on similar ideas, their theoretical and computational implementations diverge in subtle ways which are difficult to quantify.
Moreover, the experimental data can only validate the \emph{result} of each model calculation.
It thus becomes difficult to assess the veracity of competing initialization frameworks when the candidate models provide comparable descriptions of global experimental measurements.
Additionally, it is not fully understood why these models reproduce certain experimental measurements which other models fail to describe.
In order to address this question, it is important to identify the essential and non-essential features of each initial condition model which are needed to describe the data.
Hydrodynamic simulations, however, often blur cause-and-effect relationships which makes it difficult to enumerate evidence for (or against) individual theoretical assumptions.

While the IP-Glasma and EKRT models provide global descriptions of soft-sector observables in relativistic nuclear collisions which are---all things considered---quite good, their descriptions of the data are of course imperfect.
Often imperfections reflect missing features, e.g.\ nuclear structure modifications, which are easily added to the models without modifying their essential substance.
It is of course also likely that at least \emph{some} of the observed tension is attributable to errors in the adopted frameworks themselves.
This is only natural; theoretical models are rarely perfect, and modeling errors are unavoidable.

Parameters of the EKRT and IP-Glasma models are generally fixed by their respective theoretical frameworks.
In this sense, they are rigid models.
When such models fail to describe the experimental data, there is little one can do to resolve the observed tension short of reworking each calculation.
Initial condition errors are often reabsorbed by hydrodynamic model parameters when calibrating simulations to describe experimental data.
For instance, if an initial condition model generates too little radial flow, the simulation may prefer a smaller QGP bulk viscosity than it should to compensate as I mentioned before.
In this manner, initial condition errors propagate through the entire simulation framework.
Hydrodynamic parameter estimates are thus often (and rightly) criticized for being highly dependent on the choice of initial conditions.

In this chapter, I propose an alternative approach to \emph{ab initio} theory calculations, which seeks to reverse engineer the properties of the QGP initial conditions using systematic model-to-data comparison.
I develop for this purpose a new parametric model of the QGP initial conditions which is designed to be flexible.
This flexibility allows the model to mimic specific theory calculations as well as interpolate between them.
It describes, in this sense, a sort of meta-model which spans a semi-exhaustive space of reasonable theoretical descriptions.
I then constrain free parameters of the model using top-down data-driven methods that rigorously account for different sources of uncertainty in the hydrodynamic framework.
The method hence claims to know very little about the QGP initial conditions \emph{a priori} in order to see what can be learned from the data and hydrodynamic framework alone.
Such conclusions are thus less model dependent, and more robust to theoretical uncertainties.

This chapter is intended for the pragmatist.
My goal is to explain the QGP initial conditions simply, using notation that is readily expressed as computer code.
I also make a concerted effort to describe \emph{all} relevant components of the initial conditions, including those components which are often neglected in the literature because they are deemed theoretically uninteresting, or because they are relatively generic.

\section{Approximations in the high-energy limit}
\label{sec:high_energy_approx}

Throughout this dissertation, I apply approximations which are only valid in the so-called ultrarelativistic limit, i.e.\ collisions where the nuclei are Lorentz contracted by $\gamma \gtrsim 100$ along their direction of motion in the lab frame.
This definition is somewhat arbitrary, but I will explain why it is necessary in a moment, and it will become clear why this choice is a reasonable cutoff.

Consider, for example, two identical spherical nuclei, each with radius $R$, that move with velocities $\pm \beta_z$ along the $\hat{z}$ direction.
Each nucleus is Lorentz contracted by a factor
\begin{equation}
  \gamma = \frac{1}{\sqrt{1 - \beta_z^2}}
\end{equation}
along its direction of motion and thus has a diameter $D = 2 R / \gamma$ along the $\hat{z}$ direction when viewed from the lab frame.
Assuming the two nuclei collide head on, they will pass through each other after an overlap time
\begin{equation}
  \label{eq:overlap_time}
  \tau_\text{overlap} = \frac{2 R}{\gamma \beta_z} = \frac{2 R}{\sinh(y_\text{beam})},
\end{equation}
in the lab frame, where $y_\text{beam} =  \acosh(\sqrts / (2 m_p))$ is the rapidity of the beam \cite{Shen:2017bsr}.
Here $\sqrts$ is the center of mass energy per nucleon pair of the accelerated ions, and $m_p=0.938$~GeV is the proton mass.
\begin{figure}
  \centering
  \includegraphics{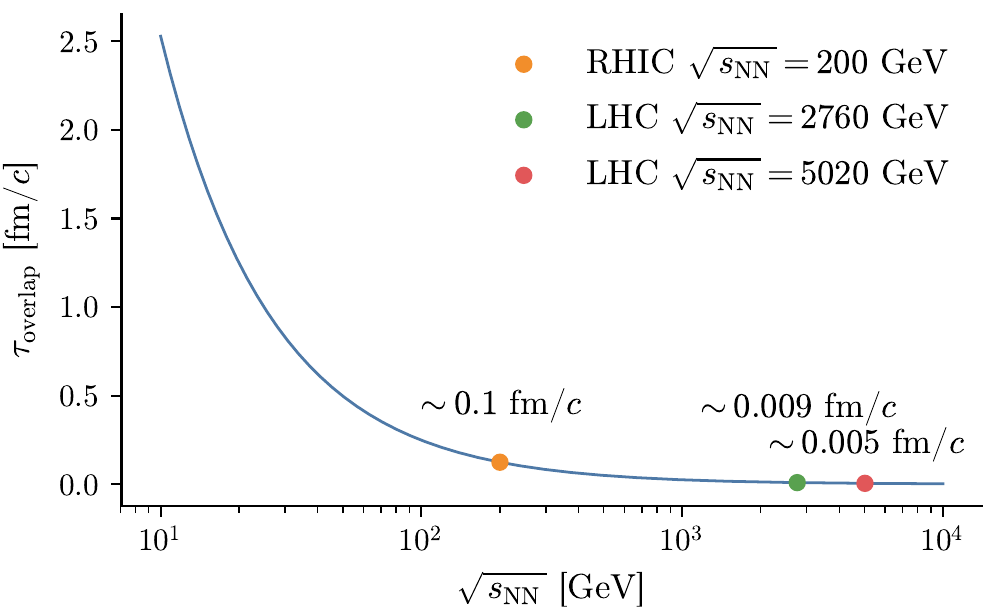}
  \caption{
    \label{fig:overlap_time}
    Nuclear overlap times $\tau_\text{overlap}$ measured in the lab frame for Pb-Pb collisions at several RHIC and LHC beam energies.
    Based on figure from \cite{Shen:2017bsr}.
  }
\end{figure}
Several example overlap times are shown in figure~\ref{fig:overlap_time} for Pb-Pb collisions at $\sqrts=200$, 2760, and 5020 GeV, beam energies which are used at RHIC and the LHC.
For all three of these beam energies, $\gamma \ge 100$ and $\tau_\text{overlap} \lesssim 0.1~\fmc$.

I now argue that these overlap times are sufficiently short to neglect transverse dynamics which occur while the nuclei pass through each other.
Let's consider a single vertex for an interaction between two partons, each located on the leading edge of the colliding nuclei.
When these primary partons scatter, they produce secondary partons which emerge from their interaction vertex with some velocity $\beta \le 1$.
If $\tau_\text{overlap} \lesssim 0.1 ~\fmc$, all partons involved in the interaction---secondary or otherwise---may propagate for an equivalent amount of time as the nuclei continue to interpenetrate.

The absolute farthest each parton can move from its original interaction vertex in this time is $\Delta x_\text{max} = 0.1$~fm, and hence the same is true for its displacement in the transverse plane.
Given that hydrodynamics is an effective theory which (in its kinetic formulation) averages thermal quantities over length scales of the interparticle mean free path, density fluctuations over distances of $\ell \lesssim 0.1$~fm should not significantly affect the bulk dynamics of the system at much larger scales \cite{Noronha-Hostler:2015coa}.
I therefore assume that the interacting matter moves along straight-line trajectories parallel to the beam axis as it pierces each nucleus.
This is a central assumption of my work, and it underlies nearly all of the approximations that follow.

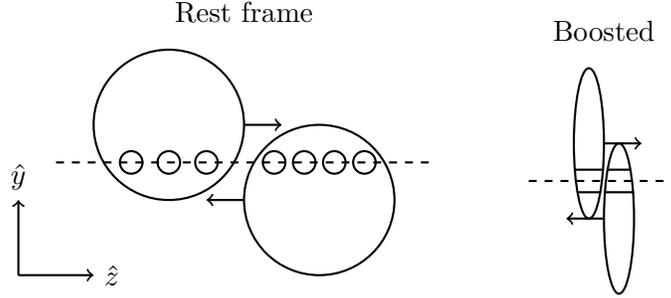
\begin{figure}[t]
  \centering
  \begin{tikzpicture}
    \node at (0, 2) {Rest frame};
    \draw[thick] (-1, .5) circle (1);
    \draw[thick, ->] (0, .5) to (.5, .5);
    \draw[thick] (-.5, 0) ellipse (.15 and .15);
    \draw[thick] (-1, 0) ellipse (.15 and .15);
    \draw[thick] (-1.5, 0) ellipse (.15 and .15);
    \draw[thick] (1, -.5) circle (1);
    \draw[thick, ->] (0, -.5) to (-.5, -.5);
    \draw[thick] (.4, 0) ellipse (.15 and .15);
    \draw[thick] (.8, 0) ellipse (.15 and .15);
    \draw[thick] (1.2, 0) ellipse (.15 and .15);
    \draw[thick] (1.6, 0) ellipse (.15 and .15);
    \draw[thick, ->] (-3, -1.5) to (-2, -1.5) node[right]{$\hat{z}$};
    \draw[thick, ->] (-3, -1.5) to (-3, -.5) node[above]{$\hat{y}$};
    \draw[thick, dashed] (-2.5, 0) to (2.5, 0);
  \end{tikzpicture} \hspace{1cm}
  \begin{tikzpicture}
    \node at (0, 2) {Boosted};
    \draw[thick] (-.4, .15) to (0, .15);
    \draw[thick] (-.35, -.15) to (-.05, -.15);
    \draw[thick] (0, -.15) to (.4, -.15);
    \draw[thick] (.05, .15) to (.35, .15);
    \draw[thick] (-.2, .5) ellipse (.2 and 1);
    \draw[thick, ->] (0, .5) to (.5, .5);
    \draw[thick] (.2, -.5)  ellipse (.2 and 1);
    \draw[thick, ->] (0, -.5) to (-.5, -.5);
    \draw[thick, dashed] (-1, 0) to (1, 0);
  \end{tikzpicture}
  \caption{
    \label{fig:side_view}
    Cartoon of the local nuclear density overlap in an ultrarelativistic nuclear collision.
    Left: Side view of the collision when both nuclei are at rest.
    Right: The same picture for a highly boosted system where each nucleus is Lorentz contracted along its direction of motion.
    The dashed line is parallel to the beam axis and pierces a single point in the transverse plane.
  }
\end{figure}

Consider now a single straight-line trajectory defined by $(x, y)=(x',y')$, which is parallel to the beam axis and pierces the interaction region of the collision, as depicted by the dashed line in figure~\ref{fig:side_view}.
Moreover, let $e_0$ define the three-dimensional energy density deposited by the collision immediately after the nuclei pass through each other.
The energy density $e_0$ deposited along this line at very early times is causally disconnected from all positions $(x, y, z)$ in each colliding nucleus where
\begin{equation}
  (x - x')^2 + (y - y')^2 > \tau^2_\text{overlap}.
\end{equation}
In other words, the collision dynamics can only traverse distances less than or equal to the collision's age.
For ultrarelativistic collisions with $\gamma \ge 100$ and $\tau_\text{overlap} \lesssim 0.1 ~\fmc$, the energy density $e_0(x, y, z)$ is essentially insensitive to all transverse coordinates $(x', y') \neq (x, y)$.

The aforementioned ultrarelativistic limit consequently factorizes the dynamics which occur at different points in the transverse plane immediately after first impact.
Each point in the transverse plane thus describes an \emph{independent} one-dimensional initialization problem.
The situation is conceptually analogous at each transverse grid location to a head-on collision between two oncoming trains of variable length, where each train represents a stack of Lorentz contracted nucleons.

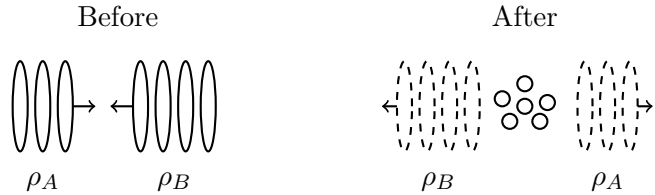
\begin{figure}[t]
  \centering
  \begin{tikzpicture}
    \draw[thick, ->] (-.5, 0) to (-.1, 0);
    \draw[thick, ->] (.5, 0) to (.1, 0);
    \draw[thick, black, fill=white] (-1.1, 0) ellipse (.1 and .6);
    \draw[thick, black, fill=white] (-.8, 0) ellipse (.1 and .6);
    \draw[thick, black, fill=white] (-.5, 0) ellipse (.1 and .6);
    \draw[thick, black, fill=white] (.5, 0) ellipse (.1 and .6);
    \draw[thick, black, fill=white] (.8, 0) ellipse (.1 and .6);
    \draw[thick, black, fill=white] (1.1, 0) ellipse (.1 and .6);
    \draw[thick, black, fill=white] (1.4, 0) ellipse (.1 and .6);
    \node[thick] at (-.8, -1) {$\rho_A$};
    \node[thick] at (.95, -1) {$\rho_B$};
    \node[thick] at (.2, 1.2) {Before};
  \end{tikzpicture}
  \hspace{5em}
  \begin{tikzpicture}
    \draw[thick, ->] (-1.4, 0) to (-1.7, 0);
    \draw[thick, ->] (1.6, 0) to (1.9, 0);
    \draw[thick, dashed, black, fill=white] (1.6, 0) ellipse (.1 and .6);
    \draw[thick, dashed, black, fill=white] (1.3, 0) ellipse (.1 and .6);
    \draw[thick, dashed, black, fill=white] (1, 0) ellipse (.1 and .6);
    \draw[thick, dashed, black, fill=white] (-.5, 0) ellipse (.1 and .6);
    \draw[thick, dashed, black, fill=white] (-.8, 0) ellipse (.1 and .6);
    \draw[thick, dashed, black, fill=white] (-1.1, 0) ellipse (.1 and .6);
    \draw[thick, dashed, black, fill=white] (-1.4, 0) ellipse (.1 and .6);
    \draw[thick] (-.1, .1) circle (.1);
    \draw[thick] (.4, -.2) circle (.1);
    \draw[thick] (.5, .05) circle (.1);
    \draw[thick] (.2, 0) circle (.1);
    \draw[thick] (.2, .3) circle (.1);
    \draw[thick] (0, -.2) circle (.1);
    \node[thick] at (1.3, -1) {$\rho_A$};
    \node[thick] at (-.95, -1) {$\rho_B$};
    \node[thick] at (.2, 1.2) {After};
  \end{tikzpicture}
  \caption{
    \label{fig:traincar}
    In the ultrarelativistic limit, causality reduces the collision dynamics to a one-dimensional problem at each transverse grid location.
    The resulting picture is analogous to that of two colliding trains, where each train is a stack of Lorentz contracted nucleons.
    The secondary matter shown on the right is produced locally and is some function of the projectile and target densities $\rho_A$ and $\rho_B$ in each nucleus.
  }
\end{figure}

This analogue is depicted in figure~\ref{fig:traincar} which shows a stack of three nucleons barreling down the beam axis to collide head-on with a stack of four nucleons moving in the opposite direction.
Let's assume that these stacks fully interpenetrate after an overlap time $\tau_\text{overlap} < 0.1 ~\fmc$ as is generally the case for nuclear collisions at top RHIC and LHC energies.
Our goal is to determine the energy (or entropy) which is deposited by the collision shortly after this overlap time into a small volume element centered on the collision epicenter.

This highly simplified one-dimensional picture of a relativistic nuclear collision may be further subdivided into three distinct modular components which each describe different aspects of the collision problem:
\begin{enumerate}[label=\roman*)]
  \item
    Prior to the collision, the state of the system is described by the density of nuclear matter $\rho_A$ and $\rho_B$ in each nucleus which passes through the transverse coordinate of interest.
    Following the freight train analogy, these densities represent the number of boxcars in each train and hence the total energy and momentum carried toward the collision epicenter.
    The first component of the problem therefore describes \emph{nuclear structure}.
  \item
    Nucleons are quantum objects which interact probabilistically.
    Even for the seemingly head-on collision in figure~\ref{fig:traincar}, there is a small chance the nucleons interpenetrate without interacting.
    The existence of an inelastic nucleus-nucleus collision hence presupposes the existence of one or more inelastic nucleon-nucleon collisions.
    The second component of the problem thus describes the collision's \emph{inelastic nuclear cross sections}.
  \item
    Given the density of participant matter in each nucleus which passes through a certain transverse grid location, one must ultimately determine the energy or entropy deposited by the collision into a small volume element centered at that point.
    The last component of the problem hence describes \emph{local energy and entropy deposition}.
  \end{enumerate}

I now proceed to describe each of these modular components in detail.
Collectively, they form a framework for modeling the energy and entropy deposited by the collision immediately after the two nuclei interpenetrate.

\section{Nuclear structure}
\label{sec:nuclear_structure}

The starting point of every initial condition model is the three-dimensional density $\rho(\xv)$ of nuclear matter in the rest frame of each colliding nucleus.
Strictly speaking, this density is governed by quantum mechanics, and thus it is characterized by a multi-body nuclear wave function $\Psi_N$ subject to the normalization condition
\begin{equation}
  \int d^3x_1\dots d^3x_N \, |\Psi_N(\xv_1, \dots, \xv_N)|^2 = N,
\end{equation}
where $N$ is the total number of protons and neutrons inside the nucleus.
The act of the collision collapses the wave function and samples inside each nucleus a set of discrete nucleon positions
\begin{equation}
  \label{eq:nucleon_positions}
  |\Psi_N(\xv_1, \dots, \xv_N)|^2 \mapsto \{\xv_1, \dots, \xv_N\},
\end{equation}
where each nucleon is itself an extended object described by its own multi-body wave function consisting of smaller quark and gluon degrees of freedom.
The density of nuclear matter probed by the collision is thus given by
\begin{equation}
  \label{eq:mc_nucleus}
\rho(\xv) = \sum\limits_{i=1}^N \rho_n(\xv - \xv_i),
\end{equation}
where $\xv_i$ is the position of each nucleon, $\rho_n$ is its sampled nucleon density, and $N$ is the number of nucleons in the nucleus.
For the moment, I choose to ignore nucleon density fluctuations and blithely model each nucleon density $\rho_n$ as a generic blob described by a three-dimensional Gaussian distribution
\begin{equation}
  \label{eq:nucleon_density}
  \rho_n(\xv) = \frac{1}{(2 \pi w^2)^{3/2}} \exp \biggl ({-}\frac{|\xv|^2}{2 w^2} \biggr ),
\end{equation}
with a free parameter $w$ which varies the nucleon's effective size.
Here I've chosen a Gaussian for convenience; it is a crude simplifying assumption commonly used in the literature \cite{Niemi:2015qia, Holopainen:2010gz}, but it is by no means realistic.
I will revisit this approximation later in subsection \ref{subsec:nucleon_substructure}.

\subsection{Sampling heavy-ions}
\label{subsec:sample_heavy_ions}

The distribution used to sample the nucleon positions is typically inferred from electron scattering experiments which measure the average charge-density of the nucleus \cite{Devries:1987vjv}.
This charge-density is carried by the distribution of protons inside the nucleus, which is generally somewhat different than the distribution of neutrons \cite{Tanihata:1992wf, Tarbert:2013jze}.
However, for reasons which I will explain shortly, the difference between the two distributions is rather small, and thus the measured charge-density distribution is commonly used to sample the positions of \emph{all} nucleons inside the nucleus.

Electron scattering experiments use the Born approximation to relate the charge-density distribution of the nucleus $\rho(r)$ to the differential cross section $d\sigma / d\Omega$ of electrons scattered off it \cite{Sakurai:1167961}.
The radial charge-density of a spherical heavy nucleus is commonly modeled by a so-called two-parameter Fermi (2PF) distribution
\begin{equation}
  \label{eq:2pf}
  \rho(r) = \frac{\rho_0}{1 + \exp \left(\frac{r - R}{a} \right)},
\end{equation}
where $\rho_0$ is the local charge-density inside the nucleus, $R$ is the half-height radius of the nucleus, and $a$ is a diffuseness parameter which gives the nucleus a soft edge.

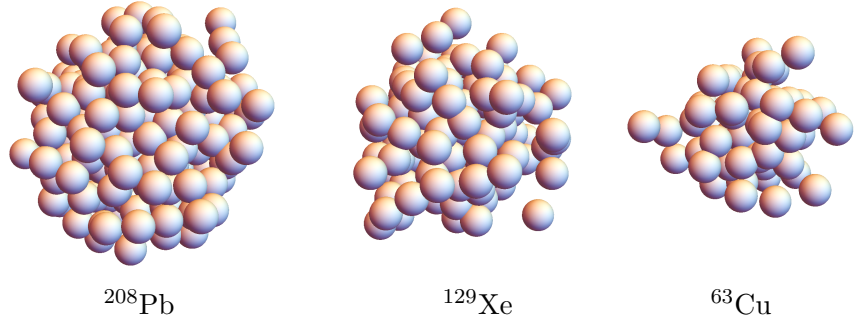
\begin{figure}
  \centering
  \makebox[\textwidth]{
    \begin{tikzpicture}
      \node at (0, 0) {\includegraphics[trim={2cm 2cm 1.5cm 2cm}, clip, scale=.5]{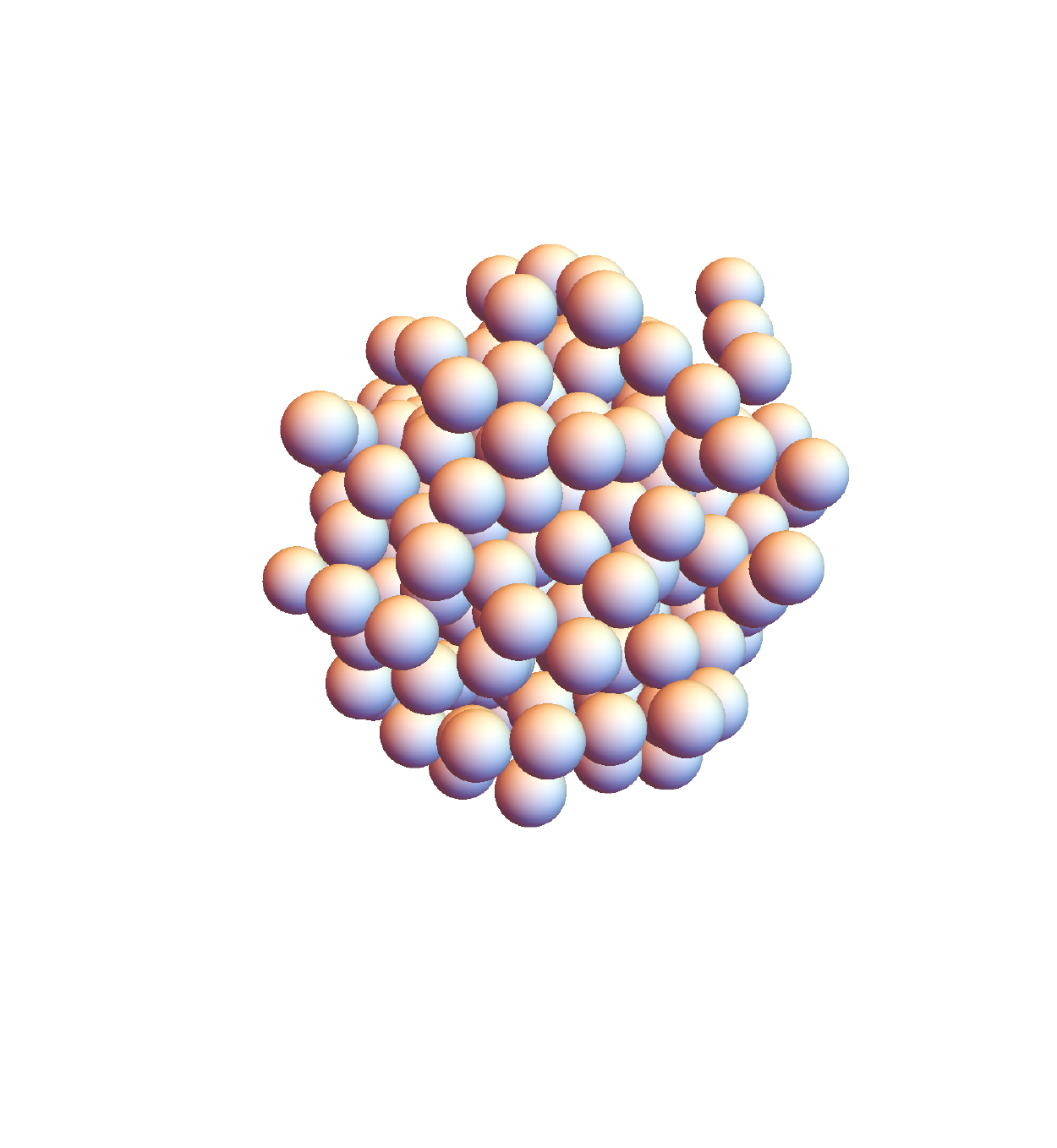}};
      \node at (4.5, 0) {\includegraphics[trim={3cm 2cm 1cm 2cm}, clip, scale=.5]{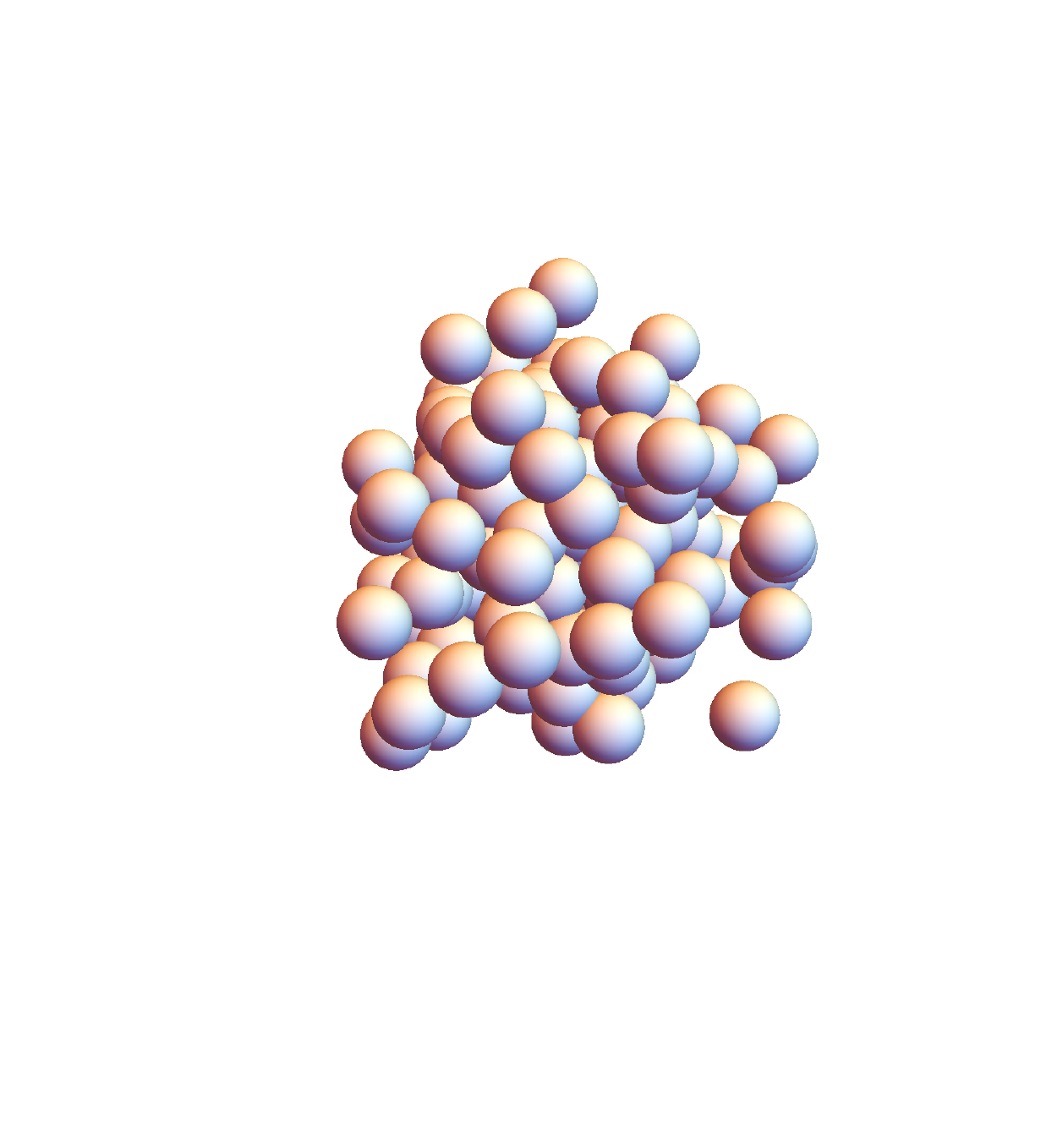}};
      \node at (8, 0) {\includegraphics[trim={3cm 2cm 3cm 2cm}, clip, scale=.5]{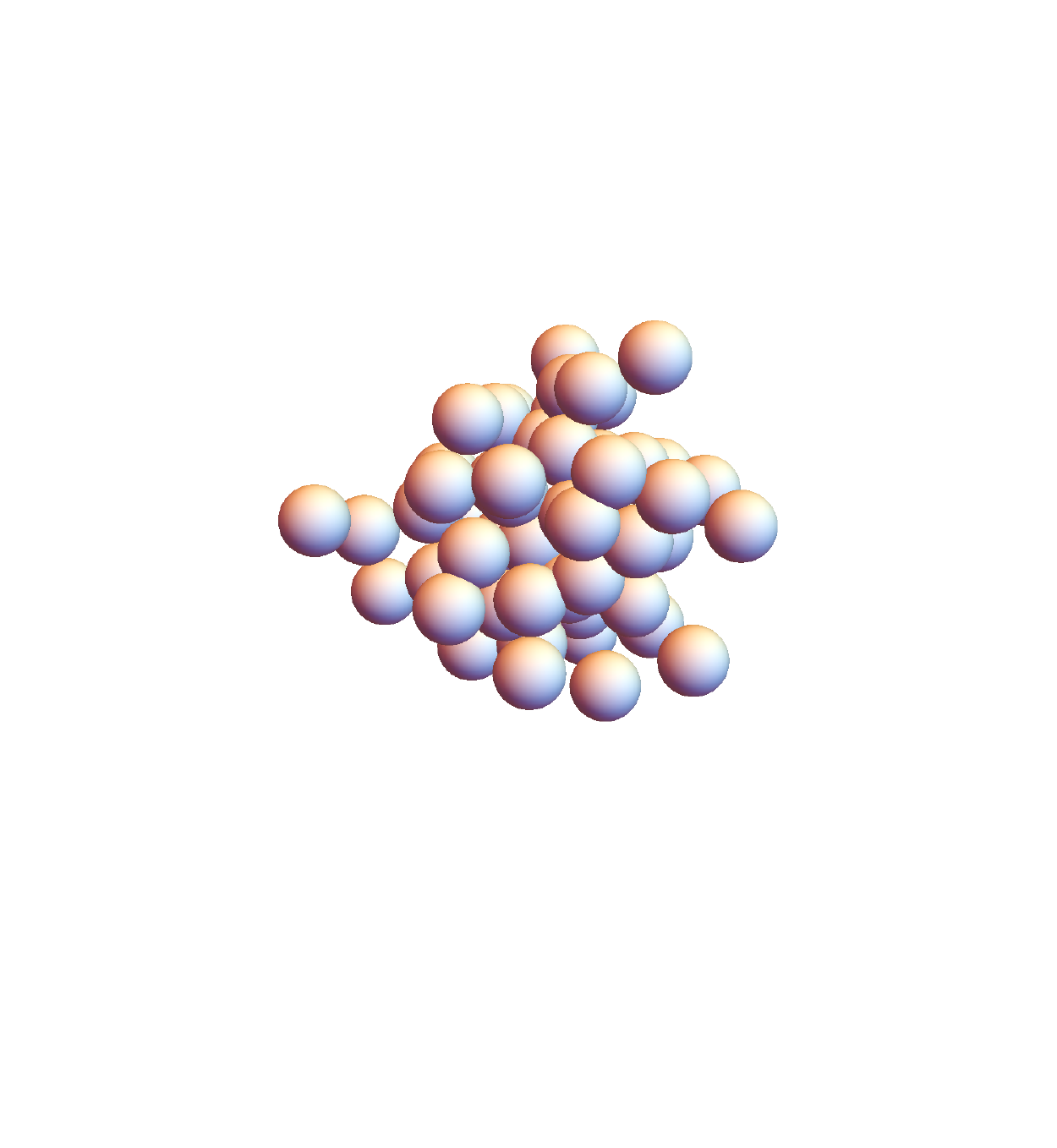}};
      \node at (0, -2) {$^{208}$Pb};
      \node at (4.5, -2) {$^{129}$Xe};
      \node at (8, -2) {$^{63}$Cu};
    \end{tikzpicture}
  }
  \caption{
    \label{fig:example_nuclei}
    Examples of lead, xenon, and copper isotopes generated by the Woods-Saxon sampling procedure.
  }
\end{figure}

Deviations from spherical symmetry are then added using spherical harmonics $Y_\ell^m(\theta, \phi)$ to deform the half-height radius of the distribution such that $R \rightarrow R'(\theta)$.
For example, $Y_2^0$ and $Y_4^0$ harmonics are commonly used to deform $R$ along the polar angle $\theta$ according to
\begin{equation}
  R'(\theta) = R\, [1 + \beta_2 Y_2^0(\theta) + \beta_4 Y_4^0(\theta)],
  \label{eq:deformation}
\end{equation}
where $\beta_2$ and $\beta_4$ are dimensionless coefficients that control the degree of deformation.
For instance, $^{238}U$ is roughly shaped like a rugby ball and is described by deformation parameters $\beta_2=0.28$ and $\beta_4=0.093$ \cite{Adare:2015bua, Masui:2009qk}.
I list the 2PF and deformation parameters for several common heavy nuclei at RHIC and the LHC in table~\ref{tab:nuclear_param}.

Each heavy-ion is then modeled---to first approximation---by sampling independent nucleon positions from the 2PF distribution equation~\eqref{eq:2pf}, using the experimentally measured half-height radius $R$, diffuseness parameter $a$, and (when necessary) deformation parameters $\beta_2$ and $\beta_4$ using supplementary equation~\eqref{eq:deformation}.
Figure~\ref{fig:example_nuclei} shows, for illustration purposes, nucleon configurations for lead, xenon, and copper isotopes generated using this method.

\begin{table}[t]
  \centering
  \captionsetup{width=.75\textwidth}
  \caption{
    \label{tab:nuclear_param}
    Estimates of the parameters $R$ and $a$ for the two-parameter Fermi distribution \eqref{eq:2pf} used to model the charge-density of heavy nuclei.
    Values are shown for several common isotopes used at RHIC and the LHC along with their corresponding deformation parameters $\beta_2$ and $\beta_4$.
    The $^{238}$U parameters are somewhat contentious so several sets are listed.
  }
  \small
  \begin{tabular}{lcccccc}
    \toprule
    Name   & $A$ & $R$ [fm] & $a$ [fm] & $\beta_2$ & $\beta_4$ & Ref. \\
    \midrule
    Copper                   & 63                   & 4.2  & 0.596 & 0.162  & -0.006 & \cite{Devries:1987vjv, Moller:1995mnm} \\[1ex]
    Xenon                    & 129                  & 5.36 & 0.59  & 0.162  & -0.003 & \cite{Loizides:2014vua, Tsukada:2017llu, Moller:2015fba} \\[1ex]
    Gold                     & 197                  & 6.38 & 0.535 & -0.131 & -0.031 & \cite{Devries:1987vjv, Moller:1995mnm} \\[1ex]
    Lead                     & 208                  & 6.62 & 0.546 & ---    & ---    & \cite{Loizides:2014vua, Devries:1987vjv} \\[1ex]
    \multirow{3}{*}{Uranium} & \multirow{3}{*}{238} & 6.81 & 0.550 & 0.280  & 0.093  & \cite{Adare:2015bua, Masui:2009qk} \\
                             &                      & 6.86 & 0.420 & 0.265  & 0.000  & \cite{Adare:2015bua, Shou:2014eya} \\
                             &                      & 6.67 & 0.440 & 0.280  & 0.093  & \cite{Loizides:2014vua, Heinz:2004ir} \\
    \bottomrule
  \end{tabular}
\end{table}

There are several subtleties which are worth mentioning that pertain to sampling the nucleon positions.
First, as I mentioned before, the charge-density distribution is \emph{not} the same as the nucleon density.
Studies have shown that in spherical, neutron rich nuclei, the radial distribution of neutrons is generally somewhat larger than the radial distribution of protons \cite{Tarbert:2013jze, Abrahamyan:2012gp}.
This difference between the neutron and proton distributions is often expressed as a neutron skin thickness $\Delta r_{np}$, defined as the difference between the root-mean-square radii of the neutron and proton distributions.
Recent measurements of $^{208}$Pb nuclei found $\Delta r_{np} = 0.15 \pm 0.03 (\text{stat.})_{-0.03}^{+0.01}(\text{sys.})$~fm \cite{Tarbert:2013jze}, while previous estimates report an even larger effect $\Delta r_{np} = 0.33_{-0.18}^{+0.16}$~fm \cite{Abrahamyan:2012gp}.

Second, due to the finite size of nucleons inside the nucleus, the measured radial density is somewhat larger than the radial density of nucleon centers \cite{Hirano:2009ah}.
As a crude analogy, imagine a pepperoni pizza.
If you distribute the pepperoni centers all the way out to the edge of the pizza, the pepperoni will extend past the edge of the crust.
These nuclear densities are related by the convolution equation
\begin{equation}
  \label{eq:convolution}
  \rho(\xv) = \int d^3 x_0\, \hat{\rho}(\xv_0)\, \rho_n(\xv - \xv_0),
\end{equation}
where $\rho(\xv)$ is the observed radial density, $\hat{\rho}(\xv_0)$ is the density of nucleon positions, and $\rho_n(\xv)$ is the assumed density profile of the nucleon.
Consequently, if one naively samples the positions of nucleons according to a given target density distribution, the resulting ensemble-averaged nucleon density will be larger than desired.
It turns out, however, that correcting this artificial swelling tends to negate the effect of accounting for the neutron skin.
I therefore choose to ignore both effects when modeling various nuclei.

\subsection{Nucleon correlations in the nucleus}
\label{subsec:nucleon_corr}

When nucleon positions are sampled independently, it's possible that two or more nucleons land on top of one another.
This is of course unrealistic.
Repulsive forces between the nucleons introduce short-range correlations \cite{Frankfurt:2008zv} which discourage mutual overlap.
Nucleon correlations are commonly added to equation~\eqref{eq:2pf} by imposing a minimum distance criteria \cite{Abelev:2013qoq, Loizides:2014vua, Rybczynski:2013yba}
\begin{equation}
  \label{eq:min_dist}
  |\xv_i - \xv_j| > d_\text{min},
\end{equation}
between all pairs of nucleons $i,j$ in the sampled nucleus.

This minimum distance constraint is regularly implemented using basic rejection sampling; nucleon positions are sampled one-by-one, and candidate positions are rejected if they place a nucleon too close to any of its previously sampled neighbors.
Rejecting samples in this manner, however, modifies the target radial distribution and leads to an artificial swelling of the nucleus, similar to the effect caused by equation~\eqref{eq:convolution}.

This swelling is commonly remedied by readjusting the parameters of the 2PF distribution for every value of $d_\text{min}$ \cite{Rybczynski:2013yba}.
Unfortunately, such adjustments are cumbersome, and they often fail to recover the target radial distribution when $d_\text{min}$ is large.
Here, I describe a simple algorithm, developed by fellow graduate student Jonah Bernhard \cite{Bernhard:2018hnz}, which implements the minimum distance criteria in equation~\eqref{eq:min_dist} without modifying the target radial distribution.

First, consider a spherically symmetric heavy-ion with $A$ nucleons described by the radial density $\rho(r)$.
The algorithm starts by sampling the radii $(r_1, r_2, \dots, r_A)$ for all $A$ nucleon positions.
These radii are then sorted in ascending order, and a pair of spherical angles $(\theta_i, \phi_i)$ is sampled from a distribution of uniform solid angle $d\Omega = \sin\theta \, d\theta\, d\phi$ one-by-one for each nucleon position radius $r_i$.
If a sampled pair of spherical angles places a nucleon too close to any of its previously placed neighbors, $(\theta_i, \phi_i)$ is resampled (but not $r_i$) until the minimum distance criteria \eqref{eq:min_dist} is satisfied.
The algorithm will attempt to relocate each nucleon 1000 times by resampling $(\theta_i, \phi_i)$ until it gives up and leaves it in its last sampled position.
Note, the radii are sampled once and are never resampled; only the angles of each nucleon are resampled.
The target radial distribution is thus perfectly preserved.

The algorithm also works for deformed nuclei such as $^{238}$U, although, for deformed nuclei, only the azimuthal angle $\phi_i$ may be resampled since $\rho$ is no longer constant as a function of polar angle $\theta$.
This restricts the available phase space to readjust each nucleon position, and the algorithm breaks down for smaller values of $d_\text{min}$.
Despite this limitation, the algorithm works well (encounters limited failures) up to $d_\text{min} = 1.7$~fm for spherically symmetric nuclei and $d_\text{min} = 1.5$~fm for deformed nuclei.

\subsection{Sampling light-ions}
\label{subsec:sample_light_ions}

Experiments also commonly study collisions of light-ions including protons, deuterons, and helium-3 nuclei.
These ions are too small to be modeled by a two-parameter Fermi distribution, so their nucleon position distributions are modeled on a case-by-case basis.

The deuteron is a loosely bound system consisting of one proton and one neutron.
Its structure is commonly modeled using the Hulth\'en wave function
\begin{equation}
  \phi_d(r_{pn}) = \left( \frac{\alpha \beta (\alpha + \beta)}{2 \pi (\alpha - \beta)^2} \right)^{1/2} \left(\frac{e^{-\alpha\, r_{pn}} - e^{-\beta\, r_{pn}}}{r_{pn}} \right),
\end{equation}
where $r_{pn}$ is the distance between the proton and neutron, $\alpha = 0.228$~fm$^{-1}$, and $\beta = 1.18$~fm$^{-1}$ \cite{Adler:2006xd, Hulthen:1957hs}.
The corresponding density $\rho(r_{pn})$ is simply proportional to the square of this wave function, and thus
\begin{equation}
  \rho(r_{pn}) \propto \left(\frac{e^{-\alpha\, r_{pn}} - e^{-\beta\, r_{pn}}}{r_{pn}} \right )^2.
\end{equation}
In order to sample each nucleon position, we make the substitution $r_{pn} \rightarrow 2 r$, where $r$ is half the distance between the proton and neutron.
Then we sample the position of the first nucleon in the center of mass frame according to the radial distribution
\begin{equation}
  \label{eq:hulthen}
  \rho(r) \propto \biggl ( \frac{e^{-2\, \alpha\, r} - e^{-2\, \beta\, r}}{r} \biggr)^2.
\end{equation}
Once the position of the first nucleon is known, the second nucleon is placed directly across from it at a distance $r_{pn} = 2 r$ \cite{Loizides:2014vua}.

Unfortunately, the three-body system of the helium-3 nucleus is far too complicated to model using a simple analytic form.
When simulating collisions of $^3$He nuclei, I sample and randomly rotate pre-tabulated nucleon positions determined from computer simulations based on Green's-function Monte Carlo \cite{Carlson:1997qn}.
I do not show any results for collisions of deuteron or $^3$He nuclei in this dissertation, but the nuclei are nevertheless implemented in the computer code developed later in this chapter to facilitate future comparisons to RHIC data.

\section{Inelastic nuclear cross sections}
\label{sec:nuclear_cross_sections}

Consider now two particles which are simultaneously shot into opposite ends of a cylindrical pipe.
Assume that each particle moves through the pipe along a straight-line trajectory, parallel to its axis of symmetry.
Moreover, assume that each particle enters the pipe at a random entry point.
At some moment, the particles either collide inside the pipe or pass by each other unscathed.

Naturally, this collision probability depends on the properties of the particles and the properties of the pipe.
It's easy to reason the dependence on the pipe geometry, namely $P_\text{coll} \propto \mathcal{A}^{-1}$, where $\mathcal{A}$ is the cross sectional area of the cylinder.
The proportionality constant $\sigma = P_\text{coll}\, \mathcal{A}$ thus depends strictly on the properties of each particle.
This proportionality constant has units of area, is called a \emph{cross section}, and is often used as a proxy for particle collision probabilities.

Experimentally, the cross section $\sigma$ is determined by measuring the average collision rate between beams of particles.
It is defined by the relation
\begin{equation}
  \label{eq:expt_cross_section}
  \frac{dN}{dt} = \sigma \times \mathcal{L},
\end{equation}
where $dN/dt$ is the number of collisions per unit time, and $\mathcal{L}$ is the beam luminosity, defined as as the number of particles passing through the beam's cross sectional area $\mathcal{A}$ per unit time \cite{Abelev:2012sea}.

\begin{figure}
  \centering
  \begin{tikzpicture}[scale=.8]
    \draw[thick] (-1, 0) circle (1);
    \fill (-1, 0) circle(.05);
    \draw[thick] (1, 0) circle (1);
    \fill (1, 0) circle(.05);
    \draw[thick, dashed] (-1, 0) circle (2);
    \draw[thick, <->] (-1, 0) to (1, 0);
  \end{tikzpicture}
  \caption{
    \label{fig:sigma_geometric}
    Geometric interpretation of the cross section for a collision of two billiard balls of radius $R$.
    One billiard ball moves into the page, and the other moves out of the page (solid circles).
    The line connecting their centers is the impact parameter $b_\text{max} = 2 R$ corresponding to their maximal point of contact.
    Their geometric cross section is the area enclosed by the dashed circle $\sigma = \pi (2R)^2$.
  }
\end{figure}
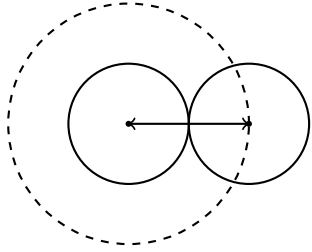

Theoretically, the cross section is defined a number of different ways.
In this dissertation, I focus exclusively on the \emph{geometric} cross section, defined as the area
\begin{equation}
  \label{eq:geometric_cross_section}
  \sigma = \int d^2b\, P_\text{coll}(\mathbf{b}).
\end{equation}
Here $\mathbf{b}$ is the impact parameter of the two-body system, defined as the vector between each particle's center of mass at the moment of closest approach, and $P_\text{coll}(\mathbf{b})$ is the probability of a collision at a given impact parameter.
One can easily verify that this geometric definition of the cross section also agrees with the experimental definition using Monte Carlo methods to sample random collisions between the particles.
Equation~\eqref{eq:geometric_cross_section} thus connects the experimentally measured cross section to the impact parameter dependent collision probability.

This geometric definition of the cross section is best explained by a simple example.
Consider for this purpose a classical collision between two billiard balls, each of radius $R$.
Their hard-sphere collision probability is given by
\begin{equation}
  P_\text{coll}(b) =
  \begin{cases*}
    1 & if $b < 2 R$, \\
    0 & otherwise.
  \end{cases*}
\end{equation}
The resulting geometric cross section is thus
\begin{equation}
  \sigma = \int_0^\infty 2 \pi\, b\, db\, P_\text{coll}(b) = \int_0^{2 R} 2 \pi\, b\, db = 4 \pi R^2.
\end{equation}
Equivalently, this area may be written in terms of the maximum impact parameter $b_\text{max} = 2 R$ between the two spheres which produces a collision,
\begin{equation}
  \sigma = \pi\, b_\text{max}^2.
\end{equation}
One sees that the classical billiard ball cross section is simply the area defined by the maximum point of contact between the two spheres as depicted in figure~\ref{fig:sigma_geometric}.

Conversely, given the experimentally measured value for the cross section $\sigma$, one can easily calculate the radius of the hard-sphere and its impact parameter dependent collision probability:
\begin{equation}
  P_\text{coll}(b) =
  \begin{cases*}
    1 & if $b < \sqrt{\sigma / \pi}$, \\
    0 & otherwise.
  \end{cases*}
  \label{eq:black_disk}
\end{equation}
This hard-sphere model of the cross section, while admittedly crude, is thus commonly used in computer models of high-energy particle collisions.
It is also referred to as the black-disk approximation, since it is functionally equivalent to a classical collision of two solid disks.

Up to this point, I've been rather vague about the definition of a collision event and similarly nondescript about the types of particles involved.
I now direct my attention collisions of ultrarelativistic nucleons, specifically \emph{inelastic} nucleon-nucleon collisions, i.e.\ collisions where at least one of the nucleons becomes excited or breaks up.
The aforementioned black-disk model is commonly used to sample inelastic nucleon-nucleon interactions, but what about the inelastic multi-body interaction of two heavy-nuclei?

\subsection{Glauber model of nuclear cross sections}
\label{subsec:glauber_cross_sections}

Inelastic nuclear cross sections are commonly described using a model of nucleus-nucleus collisions developed by Glauber \cite{Glauber:1959lom, Glauber:1955zkn, Glauber:1966aio}.
His model provides a theoretical foundation which relates pairwise nucleon-nucleon cross sections to the overall cross section of a larger nucleus.
I cover in this section practical aspects of the Glauber model which are relevant to nuclear collision simulations.
For a more detailed overview of the Glauber model including its historical origins, see reference~\cite{Miller:2007ri}.

Consider a collision of two heavy-ions, one with $A$ nucleons and the other with $B$ nucleons (labeled $A$ and $B$ respectively), which collide with impact parameter $\mathbf{b}$ in the transverse plane.
Moreover, assume that each nucleus contains an average density of nucleons $\rho_A$ and $\rho_B$ which are normalized so that $\int d^3x\, \rho_N(\xv) = N$, where $N$ is the number of nucleons in that nucleus.
The nuclear thickness function
\begin{equation}
  \label{eq:thickness}
  T_A(\xv_\perp) = \int dz\, \rho_A(\xv_\perp, z)
\end{equation}
describes the nucleon density in nucleus $A$ which penetrates each transverse coordinate $\xv_\perp$ during the collision.
Thickness function $T_B(\xv_\perp)$ follows in a similar fashion.

Now, let $i$ index a single nucleon in nucleus $A$, and let $j$ index a single nucleon in nucleus $B$.
I assert that each pair of nucleons $i, j$ collide inelastically with probability
\begin{equation}
  \label{eq:pairwise_prob}
  P_{i,j}^\text{coll}(\mathbf{b}) = \frac{\sigmann}{AB} \int d^2x\, T_A(\xv_\perp)\, T_B(\xv_\perp - \mathbf{b}).
\end{equation}
To see why this is the case, let's revisit the simple example discussed at the beginning of the section where each density $\rho_{A,B}$ describes a single proton distributed uniformly inside a beam pipe of cross sectional area $\mathcal{A}$.
The corresponding nuclear thickness functions are determined by equation~\eqref{eq:thickness}:
\begin{equation}
  T_{A,B} =
  \begin{cases}
  1/\mathcal{A} & \text{if $r < \sqrt{\mathcal{A} / \pi}$}, \\
    0             & \text{otherwise},
  \end{cases}
\end{equation}
where $r$ is the transverse distance from the symmetry axis of the pipe.
Assume that both thickness functions are centered in the middle of the beam pipe, i.e.\ $b=0$~fm, so their impact parameter offset may be ignored.
The collision probability of the two-nucleon system, defined by equation~\eqref{eq:pairwise_prob}, equals
\begin{equation}
  P_{i,j}^\text{coll} = \sigmann \int_0^{\sqrt{\mathcal{A}/ \pi}} 2 \pi\, r\, dr\, \mathcal{A}^{-2} = \sigmann / \mathcal{A}.
\end{equation}
Equation~\eqref{eq:pairwise_prob} thus recovers the original definition of the cross section $\sigma = \mathcal{A}\, P_\text{coll}$ as desired.
This nucleon-nucleon collision probability may be further simplified by defining a new quantity
\begin{equation}
  \label{eq:nuclear_overlap}
  T_{AB}(\mathbf{b}) \equiv \int d^2x\, T_A(\xv_\perp)\, T_B(\xv_\perp - \mathbf{b}),
\end{equation}
known as the nuclear overlap function.
The inelastic nucleon-nucleon collision probability may then be expressed as
\begin{equation}
  P_{i,j}^\text{coll}(\mathbf{b}) = \frac{\sigmann \,T_{AB}(\mathbf{b})}{A\,B}.
\end{equation}

We can now proceed to calculate the total inelastic nucleus-nucleus cross section $\sigma_{AB}^\text{inel}$.
It is related to the probability of observing at least one pairwise nucleon-nucleon collision between nucleus $A$ and $B$, equal to
\begin{align}
  P_{AB}^\text{coll}(\mathbf{b}) &= 1 - [1 - P_{ij}^\text{coll}(\mathbf{b})]^{A\,B} \\
                                 &= 1 - \left[1 - \frac{\sigmann \,T_{AB}(\mathbf{b})}{AB}\right]^{A\,B}.
  \label{eq:pcoll_AB}
\end{align}
The inelastic nucleus-nucleus cross section subsequently follows from the definition of the geometric cross section \eqref{eq:geometric_cross_section}.
This yields
\begin{equation}
  \label{eq:nuclear_cross_section}
  \sigma_{AB}^\text{inel}(\mathbf{b}) = \int d^2b\,
  \left\{ 1 - \left[1 - \frac{\sigmann \,T_{AB}(\mathbf{b})}{AB}\right]^{A\,B} \right \}.
\end{equation}
Noting that $\sigmann\, T_{AB} \ll AB$ in a typical heavy-ion collision, we can use the fact that $\lim\limits_{n \to \infty} (1 + x/n)^n = \exp(x)$ for small $x$ to further simplify equation~\eqref{eq:nuclear_cross_section}:
\begin{equation}
  \label{eq:approx_nuclear_cross_section}
  \sigma_{AB}^\text{inel}(\mathbf{b}) \approx \int d^2b\,
  \left\{ 1 - \exp[-\sigmann\, T_{AB}(\mathbf{b})] \right\}.
\end{equation}
Equations~\eqref{eq:nuclear_cross_section} and \eqref{eq:approx_nuclear_cross_section} thus reexpress, as desired, the inelastic nucleus-nucleus cross section in terms of the inelastic nucleon-nucleon cross section.

\subsection{Counting participants and binary collisions}
\label{subsec:npart_ncoll}

The Glauber model describes nucleus-nucleus collisions as a sum of all possible nucleon-nucleon collisions.
Previously, it was shown that a single nucleon pair $i, j$ collides with probability
\begin{equation}
  \label{eq:pcoll}
  P_{i,j}^\text{coll}(\mathbf{b}) = \frac{\sigmann \,T_{AB}(\mathbf{b})}{A\,B}.
\end{equation}
The probability of observing $n$ such pairwise collisions is described by a binomial distribution
\begin{equation}
  P(n, \mathbf{b}) = {AB \choose n} \left[ \frac{\sigmann \,T_{AB}(\mathbf{b})}{A\,B} \right]^n \, \left[1 - \frac{\sigmann \, T_{AB}(\mathbf{b})}{A\,B} \right]^{AB - n},
\end{equation}
where the prefactor is the number of combinations forming $n$ pairwise collisions, the first term is the probability of observing $n$ pairwise collisions, and the second term is the probability of observing $AB - n$ pairwise misses.

The average number of collisions is then given by
\begin{equation}
  \label{eq:bcoll_dens}
  N_\text{coll}(\mathbf{b}) = \sum\limits_{n=1}^{AB} n\,P(n, \mathbf{b}) = \sigmann\, T_{AB}(\mathbf{b}),
\end{equation}
which follows from the mean of the binomial distribution.
These pairwise inelastic nucleon collisions are commonly called \emph{binary collisions}.

In a similar fashion, one can estimate the average number of nucleons which participate in one of more pairwise collisions.
The fraction of nuclear thickness $T_A(\xv_\perp)$ which pierces the target thickness $T_B(\xv_\perp - \mathbf{b})$ without interacting is given by
\begin{equation}
  F_{A}^\text{miss}(\xv_\perp, \mathbf{b}) = \left[1 - \frac{\sigmann\, T_B(\xv_\perp - \mathbf{b})}{B}\right]^B.
\end{equation}
Correspondingly, the complementary fraction that is struck is
\begin{equation}
  F_A^\text{part}(\xv_\perp, \mathbf{b}) = 1 - \left[1 - \frac{\sigmann\, T_B(\xv_\perp - \mathbf{b})}{B}\right]^B.
\end{equation}
The average number of participant nucleons in nucleus $A$ and $B$ is therefore
\begin{equation}
  N_\text{part}(\mathbf{b}) = \int d^2x\, T_A(\xv_\perp)\, F_A^\text{part}(\xv_\perp, \mathbf{b}) + \int d^2x\, T_B(\xv_\perp - \mathbf{b})\, F_B^\text{part}(\xv_\perp, \mathbf{b}),
\end{equation}
which when expanded yields
\begin{align}
  \label{eq:participant_dens}
  N_\text{part}(\mathbf{b}) = & \int T_A(\xv_\perp) \left\{1 - \left[1 - \frac{\sigmann\, T_B(\xv_\perp - \mathbf{b})}{B} \right]^B \right\} \, d^2x~+ \nonumber \\
                              & \int T_B(\xv_\perp - \mathbf{b}) \left\{1 - \left[1 - \frac{\sigmann\, T_A(\xv_\perp)}{A} \right]^A \right\} \, d^2x.
\end{align}
These participant nucleons are commonly called \emph{wounded nucleons}.
All remaining nucleons are called \emph{spectators} since they rapidly leave the interaction region of the collision without meaningfully contributing to the dynamics of the produced fireball.

\subsection{Monte Carlo Glauber model}
\label{subsec:mc_glauber}

The Glauber model just described employed several approximations to simplify the multi-body nature of the problem.
The nuclei were modeled as smooth, cloud-like densities of nuclear matter $\rho_A$ and $\rho_B$.
Nucleon positions were assumed to be uncorrelated inside the nucleus and unspecified.
Each thickness function $T_A$ and $T_B$ was hence averaged over all possible nucleon positions sampled by the collision.
The implementation consequently neglected all effects which arise from event-by-event fluctuations in the positions of nucleons within each nucleus.

While it is difficult to account for these effects analytically, it is easy to incorporate them in numerical computer simulations.
This section describes an extension of the original Glauber model, known as the Monte Carlo Glauber model or MC-Glauber model for short.
It is more realistic than the analytic Glauber model, and hence it is commonly used when calculating quantities like $\sigma_{AB}^\text{inel}$, $N_\text{part}$, and $N_\text{coll}$.
It is also easier to explain and more intuitive.

Consider as before a collision between two ions, labeled $A$ and $B$, which are shot down a beam pipe aligned with the $\hat{z}$ direction.
Let $\{\xv_\perp^i\}$ denote the transverse positions of nucleons in nucleus $A$ and $\{\xv_\perp^j\}$ the transverse positions of nucleons in nucleus $B$ determined by the collapse of each wave function.
Assume for the moment that each pair of nucleons $i, j$ collides inelastically according to the black-disk collision profile, equation~\eqref{eq:black_disk}, separated by impact parameter $b_{ij} = |\xv_\perp^i - \xv_\perp^j|$ in the transverse plane.

The Monte Carlo Glauber model loops over all pairs of nucleons and samples their inelastic collision probability, determined by the measured inelastic nucleon-nucleon cross section $\sigmann$.
This asserts that every collision is independent of the previous collisions.
If a pair of nucleons collide, they are both labeled participants and the number of binary collisions is incremented.
Pseudo-code for the algorithm is shown below.
\begin{lstlisting}
  binary_collisions = 0
  nucleon_participants = 0

  for nucleon_i in nucleus_A:
    for nucleon_j in nucleus_B:
      xi = nucleon_i.position
      xj = nucleon_j.position

      bij = distance(xi, xj)

      if bij < sqrt(sigma_nn / pi):
        nucleon_i.is_participant = true
        nucleon_j.is_participant = true

        binary_collisions += 1

  for nucleon_i in nucleus_A:
    if nucleon_i.is_participant:
      nucleon_participants += 1

  for nucleon_j in nucleus_B:
    if nucleon_j.is_participant:
      nucleon_participants += 1
\end{lstlisting}

\begin{figure}
  \centering
  \includegraphics{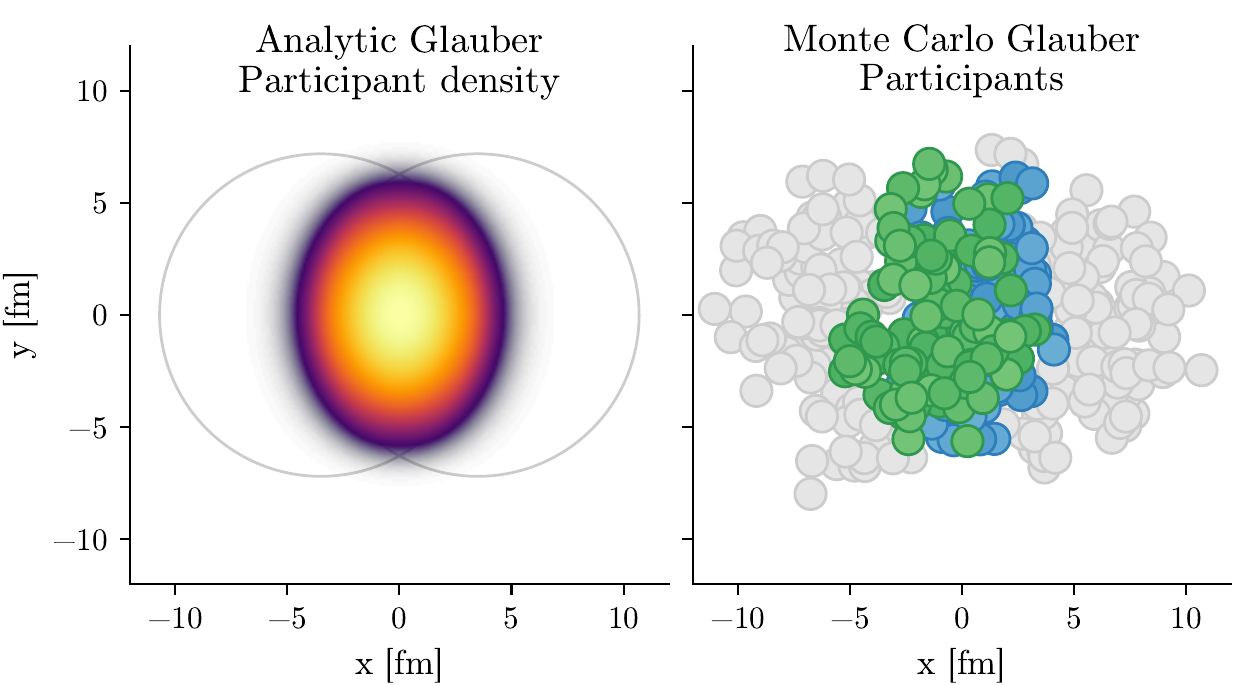}
  \caption{
    \label{fig:glauber_model_comparison}
    Left: Participant nucleon density (heat map) from the analytic Glauber model.
    Right: Participant nucleons (colored circles) from the Monte Carlo Glauber model.
    Both simulations show a Pb-Pb collision with $b=7$~fm impact parameter and $\sigmann=6.4$~fm$^2$.
  }
\end{figure}

Expressed as code, the Monte Carlo Glauber model is exceptionally simple.
It merely defines an impact-parameter dependent collision probability for each nucleon pair and then applies it to all pairs of nucleons independently.
Figure~\ref{fig:glauber_model_comparison} shows the average density and positions of participant nucleons predicted by the analytic and Monte Carlo Glauber models respectively.
Both panels show a single Pb-Pb collision with impact parameter $b=7$~fm using an inelastic nucleon-nucleon cross section $\sigmann=6.4$~fm$^2$.
The difference between the two models is obvious and striking.
Evidently, fluctuations in the positions of the nucleons within each nucleus are large, and thus one expects the local density of participant matter to vary significantly from event to event.

\subsection{Modeling the nucleon-nucleon interaction profile}
\label{subsec:nn_interaction}

Unlike the analytic Glauber model, the Monte Carlo Glauber model is sensitive to the impact-parameter dependence of the inelastic nucleon-nucleon collision probability.
I previously assumed for this purpose a black-disk interaction profile \eqref{eq:black_disk} which is commonly used in the literature.
This is, however, a crude assumption which only satisfies the most basic requirement of the interaction, namely $\int d^2b\, P_\text{coll}(\mathbf{b}) = \sigmann$.
Nearly everything that is known about protons and neutrons indicates that they do \emph{not} collide like hard spheres.

A far more realistic interaction profile may be obtained by modeling each nucleon as a composite object consisting of smaller constituents.
The analytic Glauber model may then be used to calculate the multi-body collision probability as a function of the nucleon-nucleon impact parameter \cite{Albacete:2011fw, Rybczynski:2011wv}.
Recall that for a heavy-ion collision, this probability \eqref{eq:pcoll_AB} was
\begin{equation}
  P_{AB}^\text{coll}(\mathbf{b}) = 1 - \left[1 - \frac{\sigmann \, T_{AB}(\mathbf{b})}{A\,B}\right]^{AB},
\end{equation}
where $A$ and $B$ were the number of nucleons in each nucleus, $T_{AB}$ was the nuclear overlap function \eqref{eq:nuclear_overlap}, and $\sigmann$ was the inelastic nucleon-nucleon cross section.
The analogous collision probability for two \emph{nucleons} is thus
\begin{equation}
  P_{nn}^\text{coll}(\mathbf{b}) = 1 - \left[1 - \frac{\sigma_\text{eff}^\text{inel} \, T_{nn}(\mathbf{b})}{N^2}\right]^{N^2},
\end{equation}
where $N$ is the average number of constituents inside each nucleon, $T_{nn}$ is the nucleon-nucleon overlap function, and $\sigma_\text{eff}^\text{inel}$ is the effective cross section between the constituents.

The content of each nucleon includes valence quarks, sea quarks and gluons.
At high energy, the gluon density grows and the total number of visible constituents $N \gg 1$ is large.
In this many-body limit, $\sigma_\text{eff}^\text{inel} \, T_{nn} \ll N^2$, and so
\begin{equation}
  \label{eq:pcoll_pp}
  P_{nn}^\text{coll}(\mathbf{b}) = 1 - \exp[-\sigma_\text{eff}^\text{inel} \, T_{nn}(\mathbf{b})].
\end{equation}
Each nucleon is, generally speaking, some blob of fluctuating constituent density.
To first approximation, this density can be modeled as a three-dimensional Gaussian \eqref{eq:nucleon_density} of width $w$.
The nucleon thickness is then
\begin{equation}
  T_n(\xv_\perp) = \frac{1}{2\pi w^2} \exp \left({-}\frac{|\xv_\perp|^2}{2w^2} \right),
\end{equation}
and the nucleon-nucleon overlap function is correspondingly
\begin{align}
  T_{nn}(b) &= \int d^2x\, T_{n}(\xv_\perp)\, T_{n}(\xv_\perp - \mathbf{b})\\
            &= \frac{1}{4 \pi w^2} \exp \left( - \frac{b^2}{4 w^2} \right).
\end{align}
Plugging this expression into equation~\eqref{eq:pcoll_pp} and using the definition of the geometric cross section $\int d^2b\, P_\text{coll}(\mathbf{b}) = \sigmann$, one finds that
\begin{equation}
  \sigmann = \int_0^\infty 2\pi b\, db\, \left\{ 1 - \exp \left[ -\frac{\sigma_\text{eff}^\text{inel}}{4\pi w^2} \exp \left(- \frac{b^2}{4 w^2} \right) \right] \right\}.
\end{equation}
Given a certain nucleon width $w$ and inelastic cross section $\sigmann$, this equation may be solved numerically to calculate the effective cross section parameter $\sigma_\text{eff}^\text{inel}$ which, together with equation~\eqref{eq:pcoll_pp}, specifies the inelastic nucleon-nucleon collision probability $P_{nn}^\text{coll}(b)$ at each impact parameter.

In practice, the relation may be solved by truncating the integral at a maximum impact parameter $b_\text{max} = C w$, expressed as some number of nucleon widths $C$.
After appropriate change of variables, it may be written
\begin{equation}
  \label{eq:cross_section_param}
  \frac{\sigmann}{4\pi w^2} = \frac{C^2}{4} + \expi \left( -e^{-A^2/4} \frac{\sigma_\text{eff}^\text{inel}}{4 \pi w^2} \right) - \expi \left( - \frac{\sigma_\text{eff}^\text{inel}}{4 \pi w^2} \right),
\end{equation}
where $\expi$ is the exponential integral.
This transcendental equation is then easily solved using a standard root finding algorithm.
Throughout this work $C=6$ is used, i.e.\ the maximum nucleon-nucleon impact parameter is sampled out to six nucleon widths.

\begin{figure}
  \centering
  \includegraphics{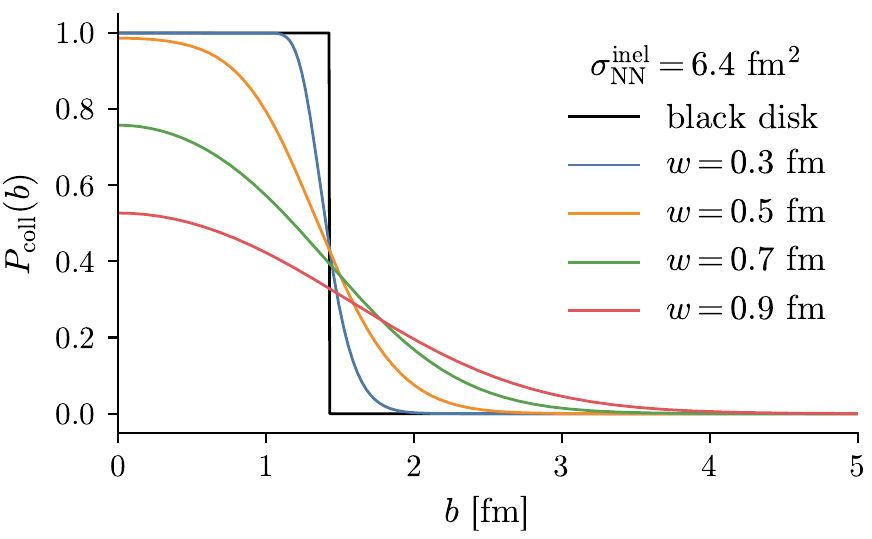}
  \caption{
    \label{fig:collision_profile}
    Inelastic nucleon-nucleon collision probability $P_\text{coll}$ as a function of the nucleon-nucleon impact parameter $b$.
    The black line shows the black-disk (hard sphere) interaction model, and the colored lines show a Glauber-like interaction model for different values of the nucleon width $w$.
  }
\end{figure}

Figure~\ref{fig:collision_profile} compares the black-disk nucleon interaction profile to the analytic Glauber nucleon profile for several values of the nucleon width $w$.
When the nucleon width $w$ is small, its density is compact, and the effective cross section parameter $\sigma_\text{eff}^\text{inel}$ determined by equation~\eqref{eq:cross_section_param} is large.
The nucleon thus becomes opaque and asymptotically approaches the black-disk limit.
Equation~\eqref{eq:pcoll_pp} is thus a more general case of the black-disk limit that naturally accommodates nucleons of different widths.

\begin{figure}[t]
  \medskip
  \makebox[\textwidth]{\includegraphics{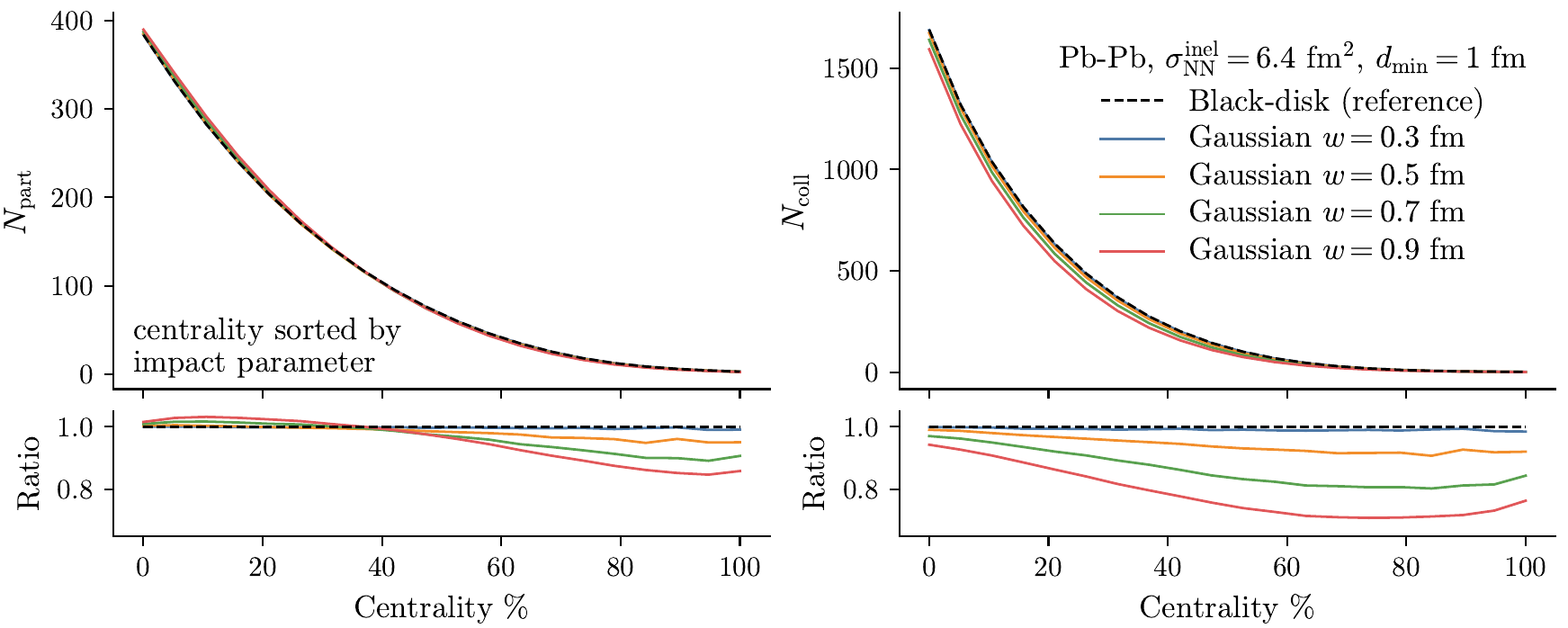}}
  \caption{
    \label{fig:glauber_attr}
    Monte Carlo Glauber model predictions for the number of nucleon participants (left) and binary collisions (right) as a function of collision centrality for Pb-Pb collisions assuming $\sigmann=6.4$~fm$^2$ and $d_\text{min}=1$~fm.
    Each line shows the prediction of a different nucleon-nucleon collision profile; black dashed-lines are the black-disk nucleon collision profile \eqref{eq:black_disk}, and colored lines are the Glauber-like collision profile \eqref{eq:pcoll_pp} using Gaussian nucleons of variable width $w$.
    Sub-panels show the ratio of each calculation to the black-disk model.
  }
\end{figure}

Figure~\ref{fig:glauber_attr} shows the average number of nucleon participants and binary collisions predicted by the model as a function of collision centrality, calculated for different models of the nucleon-nucleon collision profile.
The black-disk interaction profile (black dashed line) is compared to a Glauber interaction profile (colored lines) using Gaussian nucleons of various width $w$.
While the different collision profiles predict roughly the same number of participants, they predict dramatically different numbers of binary collisions as illustrated by the ratio plot in the bottom-right panel.
The difference is ${\sim}10\%$ for $w = 0.5$~fm, and it grows as large as ${\sim}30\%$ for $w = 0.9$~fm.

The nucleon-nucleon collision profile is thus an important source of uncertainty which should be accounted for when estimating the number of nucleon participants and binary collisions.
Throughout this work, I use the analytic Glauber interaction profile \eqref{eq:pcoll_pp} for sampling nucleon-nucleon collisions because it is unequivocally more realistic than the black-disk approximation, and because it contains the black-disk approximation as a specific limiting case.
Moreover, it allows me to vary the nucleon width $w$ over a large range of values while maintaining a sensible nucleon collision profile.

\subsection{Constructing the participant thickness functions}
\label{subsec:participant_thickness}

The Monte Carlo Glauber model is a useful tool for calculating the QGP initial conditions because it samples inelastic collisions between each pair of nucleons.
Each simulated nuclear collision event can then be categorized as elastic or inelastic based on the existence (or non-existence) of at least one inelastic nucleon-nucleon collision, roughly mimicking the inelastic event selection used by experiment.

Even when the inelastic event trigger is satisfied, usually only a small fraction of nucleons in each nucleus collide inelastically.
The remaining nucleons disassociate from their parent nuclei and are ejected at large rapidities.
These spectators have negligible effect on the energy and entropy deposited at midrapidity so their contribution can be safely ignored.
Therefore, it is only necessary to model the participant density of each nucleus.

This participant density is easily calculated in the local rest frame of nucleus $A,B$ by summing the density of each participant nucleon
\begin{equation}
  \label{eq:participant_density}
  \rho_{A,B}^\text{part}(\xv) = \sum\limits_{i=1}^{N_{A,B}} \rho_n(\xv - \xv_i \pm \mathbf{b}/2),
\end{equation}
over $N_{A,B}$, the total number of participants in nucleus $A,B$ respectively.
The function $\rho_n$ under the summation is the nucleon density specified by equation~\eqref{eq:nucleon_density}, $\xv_i$ is the location of each nucleon in the nuclear center of mass frame, and $\mathbf{b}$ is the impact parameter vector separating each nuclear center of mass.
Hence, for a simple collision involving two protons, each participant density consists of a single Gaussian shifted by the appropriate impact parameter offset.
In a larger nucleus-nucleus collision, the number and distribution of participants fluctuates event-by-event, leading to fluctuations in the collision's overall energy and entropy deposition.

This fluctuating nuclear density, however, does not account for all the observed variance in the resulting event activity.
Proton-proton collisions, for example, evidence large multiplicity fluctuations \cite{Alner:1987wb, Khachatryan:2010nk, Aamodt:2010ft} which cannot be explained by differences in the geometric overlap density alone.
It is therefore common to multiply each nucleon density using a randomly sampled weight factor \cite{Shen:2014gfa, Dumitru:2012yr, Bozek:2013uha}.
The resulting fluctuated participant density may then be written as
\begin{equation}
  \label{eq:fluct_participant_density}
  \tilde{\rho}_{A,B}^\text{part}(\xv) = \sum\limits_{i=1}^{N_{A,B}} \gamma_i\, \rho_n(\xv - \xv_i \pm \mathbf{b}/2),
\end{equation}
where $\gamma_i$ is the random weight used to fluctuate each nucleon (note, I've added a tilde to $\tilde{\rho}_{A,B}^\text{part}$ to indicate the presence of the additional fluctuations).
Throughout this work, I sample the weights $\gamma_i$ from a gamma distribution with unit mean and variance $1/k$, where $k$ is the so-called gamma shape parameter.
I've chosen this distribution because particlization fluctuations are Poissonian, and the folding of the gamma distribution with a Poisson distribution yields a negative binomial distribution which is commonly used to fit the multiplicity distribution of high-energy proton-proton and nucleus-nucleus collisions \cite{Alner:1987wb, Adare:2008ns}.

Equation \eqref{eq:fluct_participant_density} describes the three-dimensional density of participant matter in the rest frame of each nucleus.
Following the convention used in Glauber model, I project this density onto the transverse plane $\xv_\perp$ orthogonal to the beam axis to calculate the density of participant matter visible to a probe piercing the transverse coordinate $\xv_\perp$:
\begin{equation}
  \label{eq:participant_thickness}
  \T_{A,B}(\xv_\perp) = \int dz\, \tilde{\rho}_{A,B}^\text{part}(\xv_\perp, z).
\end{equation}
I call this quantity the \emph{participant thickness function}.
It equals the canonical thickness function defined by equation~\eqref{eq:thickness} with a few modifications: it excludes spectator matter, it re-weights each nucleon density by a gamma random variable, and it has the impact parameter offset $\pm \mathbf{b}/2$ already baked in.
The participant thickness functions $\T_A, \T_B$ provide \emph{almost} all the information that is needed to model the initial conditions at midrapidity.
All that remains is to specify a scalar mapping which translates this information to an energy or entropy density profile.

\section{Energy and entropy deposition}
\label{sec:energy_entropy_deposition}

\begin{figure}
  \centering
  \begin{tikzpicture}
    \draw[thick, ->] (-.5, 0) to (-.1, 0);
    \draw[thick, ->] (.5, 0) to (.1, 0);
    \draw[thick, black, fill=white] (-1.1, 0) ellipse (.1 and .6);
    \draw[thick, black, fill=white] (-.8, 0) ellipse (.1 and .6);
    \draw[thick, black, fill=white] (-.5, 0) ellipse (.1 and .6);
    \draw[thick, black, fill=white] (.5, 0) ellipse (.1 and .6);
    \draw[thick, black, fill=white] (.8, 0) ellipse (.1 and .6);
    \draw[thick, black, fill=white] (1.1, 0) ellipse (.1 and .6);
    \draw[thick, black, fill=white] (1.4, 0) ellipse (.1 and .6);
    \draw[thick, |<->|] (-1.1, -.8) to (-.5, -.8);
    \draw[thick, |<->|] (0.5, -.8) to (1.4, -.8);
    \node[thick] at (-.8, -1.2) {$\T_A$};
    \node[thick] at (.95, -1.2) {$\T_B$};
    \node[thick] at (.2, 1.2) {Before};
  \end{tikzpicture}\hspace{5em}
  \begin{tikzpicture}
    \draw[thick, ->] (-2.1, 0) to (-2.4, 0);
    \draw[thick, ->] (2.2, 0) to (2.5, 0);
    \draw[thick, dashed, black, fill=white] (2.2, 0) ellipse (.1 and .6);
    \draw[thick, dashed, black, fill=white] (1.9, 0) ellipse (.1 and .6);
    \draw[thick, dashed, black, fill=white] (1.6, 0) ellipse (.1 and .6);
    \draw[thick, dashed, black, fill=white] (-1.2, 0) ellipse (.1 and .6);
    \draw[thick, dashed, black, fill=white] (-1.5, 0) ellipse (.1 and .6);
    \draw[thick, dashed, black, fill=white] (-1.8, 0) ellipse (.1 and .6);
    \draw[thick, dashed, black, fill=white] (-2.1, 0) ellipse (.1 and .6);
    \draw[thick, ->] (.2, -.3) -- ++(.15,-.3);
    \draw[thick, black, fill=white] (.2, -.3) circle (.1);
    \draw[thick, ->] (-.1, -.2) -- ++(-.15,-.3);
    \draw[thick, black, fill=white] (-.1, -.2) circle (.1);
    \draw[thick, ->] (.2, .2) -- ++(.15,.3);
    \draw[thick, black, fill=white] (.2, .2) circle (.1);
    \draw[thick, ->] (-.1, .25) -- ++(-.2, .3);
    \draw[thick, black, fill=white] (-.1, .25) circle (.1);
    \draw[thick, ->] (-.2, 0) -- ++(-.3,.1);
    \draw[thick, black, fill=white] (-.2, 0) circle (.1);
    \draw[thick, ->] (.25, -.05) -- ++(.3,-.1);
    \draw[thick, black, fill=white] (.25, -.05) circle (.1);
    \draw[thick] (-.5, -.6) rectangle (.5, .4);
    \draw[thick] (-.3, -.4) rectangle (.7, .6);
    \node at (0.05, -.9) {$\tau_0\, d\eta_s$};
    \node at (0.9, -.6) {$dx$};
    \node at (1.0, 0.0) {$dy$};
    \draw[thick] (-.5, -.6) to (-.3, -.4);
    \draw[thick] (.5, -.6) to (.7, -.4);
    \draw[thick] (.5, .4) to (.7, .6);
    \draw[thick] (-.5, .4) to (-.3, .6);
    \node[thick, white] at (-.8, -1.2) {$\T_A$};
    \node[thick, white] at (.95, -1.2) {$\T_B$};
    \node[thick] at (0, 1.2) {After};
  \end{tikzpicture}
  \caption{
    \label{fig:density_to_particles}
    Left: Local participant thickness functions $\T_A$ and $\T_B$.
    Right: Average energy (or entropy) contained in the volume element $dV=dx\, dy\, \tau_0\, d\eta_s$ centered at midrapidity $\eta_s=0$ at proper time $\tau=\tau_0$.
  }
\end{figure}
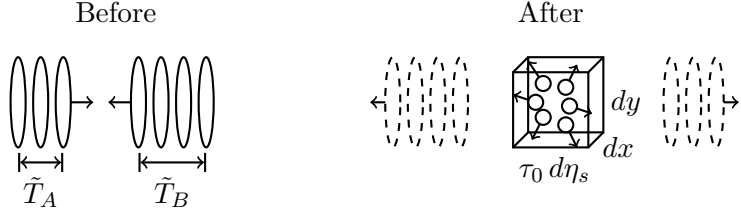

Consider again the situation depicted in figure~\ref{fig:density_to_particles} which was discussed at the beginning of the chapter.
Three nucleons barrel down a beam pipe to collide head-on with four nucleons moving in the opposite direction.
Suppose these nucleons collide inelastically and convert some of their energy into secondary matter, e.g.\ liberated quarks and gluons, which subsequently interact and undergo hydrodynamic expansion.

In the previous section, I characterized the initial state of the collision just before impact by two fields $\T_A, \T_B$, which describe the density of participant matter in each nucleus projected onto the plane orthogonal to the beam axis.
The collision depicted in figure~\ref{fig:density_to_particles} is similarly characterized by two participant densities $\T_A, \T_B$, each describing the transverse density of three and four stacked nucleons respectively.
Since I am interested in modeling the collision hydrodynamically, suppose that these participant densities rapidly interact to produce a fluid which is in local thermal equilibrium shortly after the nucleons interpenetrate.
Moreover, assume that this fluid is approximately boost-invariant near midrapidity as discussed in subsection~\ref{subsec:spacetime_picture}.

Generally speaking, the three-dimensional fluid produced by the collision is \emph{not} boost-invariant.
The longitudinal density of produced matter fluctuates both locally point-to-point in the transverse plane as well as globally event-by-event due to asymmetries in the sampled density of participant matter \cite{Ke:2016jrd, Bozek:2010vz}.
Nevertheless, boost-invariance has been shown to be a good approximation when analyzing particles detected close to midrapidity \cite{Shen:2016zpp}.
I therefore assume approximate boost-invariance and work in Milne coordinates $(\xv_\perp, \eta_s, \tau)$, where $\mathbf{x}_\perp$ specifies a Cartesian coordinate $(x, y)$ in the transverse plane, $\eta_s = \tfrac{1}{2} \ln [(t+z)/(t-z)]$ is the system's space-time rapidity, and $\tau=\sqrt{t^2 - z^2}$ is its proper time.

Consider now the infinitesimal volume element $dV = d^2x_\perp \tau_0\, d\eta_s$ centered on the space-time coordinate $(\xv_\perp, \eta_s=0, \tau_0)$, where $\tau_0$ is some time shortly after the nucleons interpenetrate (right-side of figure~\ref{fig:density_to_particles}).
Additionally, let $e_0 = e(\xv_\perp, \eta_s=0, \tau_0)$ and $s_0 = s(\xv_\perp, \eta_s=0, \tau_0)$ denote the energy and entropy density inside the cell at this time, averaged over an ensemble of collisions with the same values of $\T_A$ and $\T_B$.
The collision correspondingly maps the initial state of the system just before impact, characterized by $\T_A$ and $\T_B$, to some new state immediately after impact, characterized by its energy density $e_0$ or entropy density $s_0$.
These mappings describe scalar functions
\begin{equation}
  f_e: \T_A, \T_B \mapsto e_0 \quad\text{and}\quad f_s: \T_A, \T_B \mapsto s_0,
\end{equation}
which provide an effective description of early-time dynamics.
The goal of this section is to parametrize the functions $f_e$ and $f_s$.
Each parametrization need not arise from a first-principles calculation, but it must obey basic physical constraints.
Specifically, it should be:
\begin{itemize}
  \item
    Symmetric under interchange of $\T_A$ and $\T_B$.
    The average system at midrapidity is invariant to reflections through the transverse plane.
  \item
    Conserving of total energy.
    Let $E_0$ be the total energy deposited into the volume $dV = d^2x_\perp \tau_0\, d\eta_s$ at time $\tau_0$.
    This energy is bounded above by $E_\text{tot} = (\sqrts/2) \, (\T_A + \T_B)\, d^2x_\perp$, the total energy carried by all nucleons interacting with that volume element.
  \item
    Monotonically increasing as a function of $\T_A$ and $\T_B$.
    Denser, more energetic collisions should deposit more energy and entropy on average.
\end{itemize}

Subject to these stipulations, we can conjecture a reasonable ansatz for the initial energy density $e_0$ or entropy density $s_0$ as a function of the participant thickness functions $\T_A$ and $\T_B$.
Perhaps the simplest such function is a sum:
\begin{equation}
  \label{eq:participant_scaling}
  \left \{
  \begin{aligned}
    e_0\\
    s_0
  \end{aligned}
  \right \}
  = \text{Norm} \times (\T_A + \T_B),
\end{equation}
where Norm is an energy-dependent normalization constant carrying the appropriate units.
The term in parentheses is simply the local participant or ``wounded'' nucleon density
\begin{equation}
  \np = \T_A + \T_B,
\end{equation}
so this ansatz is commonly referred to as the \emph{wounded nucleon model}.
It effectively assigns each nucleon a finite amount of energy or entropy which is fully deposited by its first nucleon-nucleon collision.

Indeed, this simple ansatz was actually one of the first models used in the literature to describe the initial state of nucleus-nucleus collisions.
The idea dates back to a seminal paper by Bialas \emph{et al.}\ \cite{Bialas:1976ed} which conjectured a linear relation between the number of produced particles and the number of inelastic nucleon participants.
This conjecture was supported by measurements of Pb-Pb collisions at the Super Proton Synchrotron (SPS) which showed that the transverse energy density $dE_T/d\eta$ and charged-particle multiplicity $d\nch/d\eta$ both scale linearly with the number of participant nucleons \cite{Aggarwal:2000bc}.
Ideal boost-invariant hydrodynamics conserves the system's energy and entropy per unit rapidity, so naturally this seemed to suggest that $e_0$ and $s_0$ scaled with the wounded nucleon density as well.

\begin{figure}[t]
  \centering
  \begin{tikzpicture}
    \node {\includegraphics{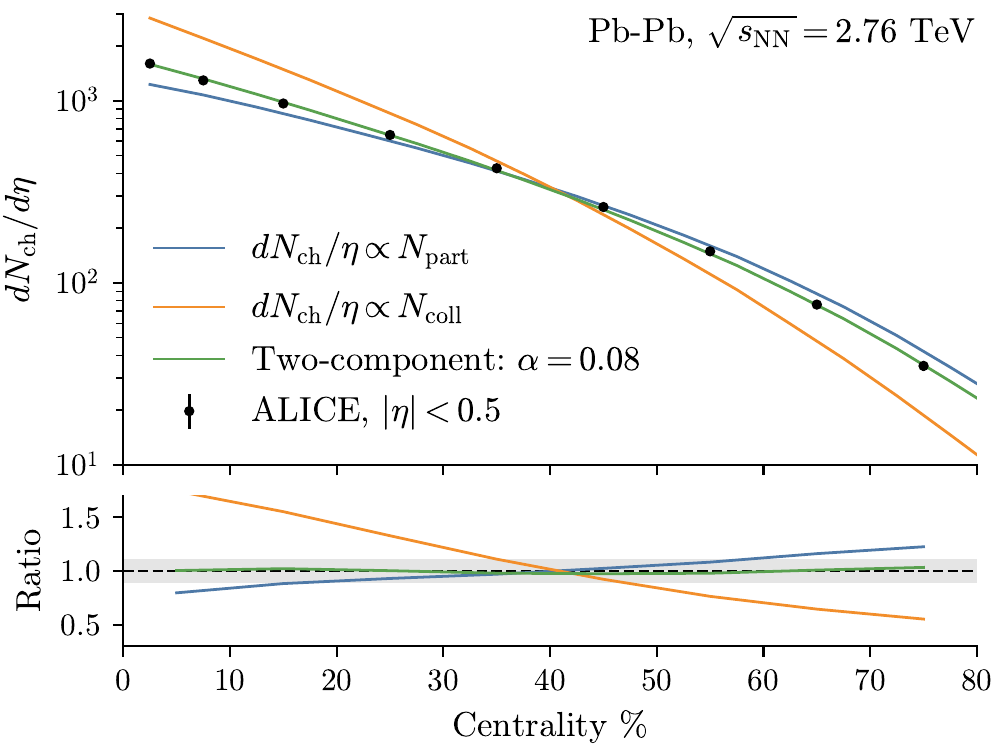}};
    \fill[black, fill opacity=0.2] (-2.85, -3.6) circle (.3);
    \fill[black, fill opacity=0.2] (-2.65, -3.6) circle (.3);
    \fill[black, fill opacity=0.2] (3.55, -3.6) circle (.3);
    \fill[black, fill opacity=0.2] (4, -3.6) circle (.3);
  \end{tikzpicture}
  \caption{
    \label{fig:npart_ncoll}
    Charged-particle density $d\nch/d\eta$ as a function of collision centrality for Pb-Pb collisions at $\sqrts=2.76$~TeV.
    Colored lines are model predictions based on wounded nucleon scaling $d\nch/d\eta \propto \Np$, binary collision scaling $d\nch/d\eta \propto \Nc$, and the two-component ansatz \eqref{eq:two_comp_ansatz} with $\alpha=0.08$.
    Black symbols show experimental data from the ALICE experiment \cite{Aamodt:2010cz}.
    The bottom plot shows the ratio of the model over the data (gray band $\pm 10\%$), and circles below the x-axis show the approximate nuclear overlap at two centralities.
  }
\end{figure}

It was later realized, however, that the wounded nucleon model fails to reproduce the centrality dependence of particle production observed by experiment.
For example, consider the so-called entropy variant of the wounded nucleon model, $s_0 \propto \np$.
In ideal boost-invariant hydrodynamics, the midrapidity charged-particle yield $d\nch/d\eta$ scales linearly with the initial entropy density $dS/d\eta_s$ at fixed freeze-out temperature and chemical potential \cite{Kolb:2001qz}:
\begin{equation}
  d\nch/d\eta \propto dS/d\eta_s.
\end{equation}
Hence, the conjectured scaling $s_0 \propto \np$ is easily verified by comparing the predicted centrality dependence of $\Np$ against the measured centrality dependence of $d\nch/d\eta$.
Figure~\ref{fig:npart_ncoll} shows this comparison for Pb-Pb collisions at $\sqrts=2.76$~TeV.
The blue line is the aforementioned scaling $d\nch/d\eta \propto \Np$ scaled up to fit the data, and the black symbols with errors are data from the ALICE collaboration \cite{Aamodt:2010cz}.
The wounded nucleon model significantly underpredicts the steep rise in particle production observed in central collisions.
A similar line of reasoning can also be used to invalidate the alternative wounded nucleon variant $e_0 \propto \np$; see reference \cite{Kolb:2001qz}.
The wounded nucleon model is therefore excluded by the data.

\subsection{Two-component ansatz}
\label{subsec:two_comp_ansatz}

Motivated by this finding, it was suggested that particle production may receive an additional contribution from hard processes which scale like the number of binary collisions $\Nc$ \cite{Eskola:198937, PhysRevLett.59.2527}.
Unlike the participant number $\Np$ which grows approximately linearly with increasing nuclear thickness, the binary collision number grows quadratically,
\begin{equation}
  \Nc \propto \int d^2x_\perp \,T_A(\xv_\perp) T_B(\xv_\perp).
\end{equation}
The orange line in figure~\ref{fig:npart_ncoll} shows the curve $d\nch/d\eta \propto \Nc$ predicted by binary collision scaling.
It suffers from the opposite problem of participant scaling.
The number of binary collisions rises much faster than the density of produced particles.
Such observations led to the creation of the so-called \emph{mixed} or \emph{two-component} ansatz \cite{Kharzeev:2000ph, Wang:2000bf}
\begin{equation}
  \label{eq:two_comp_ansatz}
  \left \{
  \begin{aligned}
    e_0 \\
    s_0
  \end{aligned}
  \right \}
  = \text{Norm} \times \left( \frac{1 - \alpha}{2}\, \np + \alpha\, \nc \right),
\end{equation}
which linearly interpolates between the local wounded nucleon density $\np$ and the local binary collision density $\nc$ using a dimensionless parameter $\alpha$.
The parameter $\alpha$ is tuned once at each collision energy by fitting the centrality dependence of the charged-particle yield \cite{Back:2002uc} after which it remains fixed.
The model was particularly appealing because it could exactly fit the centrality dependence of the charged-particle density $d\nch/d\eta$ at midrapidity as illustrated by the green line in figure~\ref{fig:npart_ncoll}.

I argue now that this fit is a red herring and that the two-component ansatz is \emph{wrong}.
There are in fact many problems with the ansatz; I'll explain the simplest one.
Imagine a stack of $N$ projectile nucleons colliding head-on with a stack of $N$ target nucleons.
Moreover, assume for simplicity's sake that all of the nucleons collide with each other.
Let $S$ denote the total entropy deposited into the rapidity interval $|\eta| < 1$ immediately after the collision.
According to the two-component ansatz \eqref{eq:two_comp_ansatz}
\begin{equation}
  S/N \propto (1 - \alpha) + \alpha N.
\end{equation}
This implies that I can make the entropy deposited per nucleon $S/N$ arbitrarily large by increasing $N$, the number of columnated nucleons in each stack, at \emph{fixed} beam energy $\sqrts$.
Similarly, since the energy density grows faster than the entropy density $e \sim s^{4/3}$, I can make $E/N$, the energy deposited per nucleon, arbitrarily large as well.
This is clearly absurd as it violates energy conservation in the dense-dense limit at fixed beam energy.
Note, this violation will occur for \emph{any value of $\alpha$}, for sufficiently large $N$.

The two-component model fails to conserve energy because it treats collision $N$ the same as collision $N+1$.
Presumably, the incremental energy and entropy deposited by each collision is somewhat smaller each time on average \cite{Ghosh:2016npo}.
The average energy (or entropy) deposited by the collision should therefore be boundable by participant scaling.
Functionally, this means there exists some constant $C$ such that
\begin{equation}
  \label{eq:stipulation1}
  f(\T_A, \T_B) \le C (\T_A + \T_B), \quad\forall\, \T_A, \T_B.
\end{equation}
This stipulation, however, is somewhat puzzling.
Figure~\ref{fig:npart_ncoll} seems to require a term which rises \emph{faster} than $\Np$ with increasing density.
How can this also be true?

\subsection{Generalized mean ansatz}
\label{subsec:gen_mean_ansatz}

There is, in fact, a superior ansatz for QGP energy and entropy deposition which alleviates this problem.
Let's back up for a moment and return to the wounded nucleon ansatz in equation~\eqref{eq:participant_scaling}.
This function is equivalent to the \emph{arithmetic mean} of participant nucleon density
\begin{equation}
  \left \{
  \begin{aligned}
    e_0\\
    s_0
  \end{aligned}
  \right \}
  = \text{Norm} \times \frac{\T_A + \T_B}{2},
\end{equation}
up to a meaningless factor of two in the denominator, which may be extracted from the normalization coefficient.

Evidently, this scaling fails to describe the centrality dependence of charged-particle production, but what about \emph{other} types of means?
With this in mind, I replace the arithmetic mean of the wounded nucleon model with a more flexible parametrization \cite{Moreland:2014oya}
\begin{equation}
  \label{eq:gmean_ansatz}
  \left \{
  \begin{aligned}
    e_0\\
    s_0
  \end{aligned}
  \right \}
  = \text{Norm} \times M_p(\T_A, \T_B),
\end{equation}
where $M_p$ is a family of functions known as the generalized means
\begin{equation}
  \label{eq:gmean}
  M_p(x, y) = \left( \frac{x^p + y^p}{2} \right)^{1/p}.
\end{equation}
This parametrization introduces a dimensionless parameter $p$ which varies the scaling behavior of initial energy (or entropy) deposition at midrapidity.
For certain discrete values, it reduces to well known functional forms such as the arithmetic, geometric, and harmonics means:
\begin{equation}
  \newlength{\extraspace}
  \setlength{\extraspace}{0.5ex}
  M_p(x, y) =
  \begin{cases}
    \max(x, y) & p \rightarrow +\infty, \\[\extraspace]
    (x + y)/2 & p = +1, \hfill \text{ (arithmetic)} \\[\extraspace]
    \sqrt{x\, y} & p = 0, \hfill \text{ (geometric)} \\[\extraspace]
    2\, x y/(x + y) & p = -1, \hfill \text{ (harmonic)} \\[\extraspace]
    \min(x, y) & p \rightarrow -\infty,
  \end{cases}
  \label{eq:trento_p}
\end{equation}
although it is not limited to these values.
The parameter $p$ is a continuous variable which can take any value $-\infty < p < \infty$.

The generalized mean has a number of interesting properties which make it suitable for the present context:
\begin{itemize}
  \item
    It is symmetric: $M_p(x, y) = M_p(y, x)$. This is required; the average energy or entropy produced at midrapidity should naturally be invariant under reflections $\T_A \leftrightarrow \T_B$.
  \item
    It is bounded: $M_p(x, y) \leq x + y$ for all positive $x$ and $y$.
    Therefore I can always choose a normalization that satisfies equation~\eqref{eq:stipulation1}, ensuring that the available energy is not violated in the dense-dense limit (the same is not true for the two-component ansatz).
  \item
    It is scale-invariant: $M_p(c \T_A, c \T_B) = c M_p(\T_A, \T_B)$. I'll explain the significance of this property later in the chapter.
\end{itemize}

The transverse field $M_p(\T_A, \T_B)$ carries units of $\text{fm}^{-2}$, the same as $\T_A, \T_B$, so I refer to this quantity as the collision's \emph{reduced thickness function}
\begin{equation}
  \label{eq:reduced_thickness}
  T_R \equiv M_p(\T_A, \T_B),
\end{equation}
so named because it takes two thickness functions $\T_A$ and $\T_B$ and reduces them to a third thickness function, similar to a reduced mass.
Indeed, for generalized mean parameter $p=-1$, the reduced thickness function and the reduced mass are algebraically equivalent.

Let me return now to the puzzle which I posed at the end of the last section.
Figure~\ref{fig:npart_ncoll} appears to require a term which grows \emph{faster} than $\Np$ with increasing density.
Is this really true?
No, it's not true.
At least it's not \emph{necessarily} true.
It turns out that that generalized mean ansatz \eqref{eq:gmean_ansatz} also describes the centrality dependence of charged-particle production, and it does so without invoking the problematic binary collision term.
Consider for instance, the entropy variant of the generalized mean ansatz
\begin{equation}
  \label{eq:gmean_entropy}
  s_0 \propto M_p(\T_A, \T_B),
\end{equation}
defined by equation~\eqref{eq:gmean_ansatz}.
In ideal boost-invariant hydrodynamics, the final charged-particle density scales linearly with the initial entropy density, i.e.\ $d\nch/d\eta \propto dS/d\eta_s$, so equation~\eqref{eq:gmean_entropy} implies that
\begin{equation}
  \label{eq:gmean_yield}
  d\nch/d\eta \propto \int d^2x_\perp T_R(\xv_\perp),
\end{equation}
where $T_R$ is the reduced thickness function defined by equation \eqref{eq:reduced_thickness}.
The participant thickness functions $\T_A, \T_B$ are easily simulated using equation~\eqref{eq:participant_thickness}, so I can test this prediction against experimental measurements.

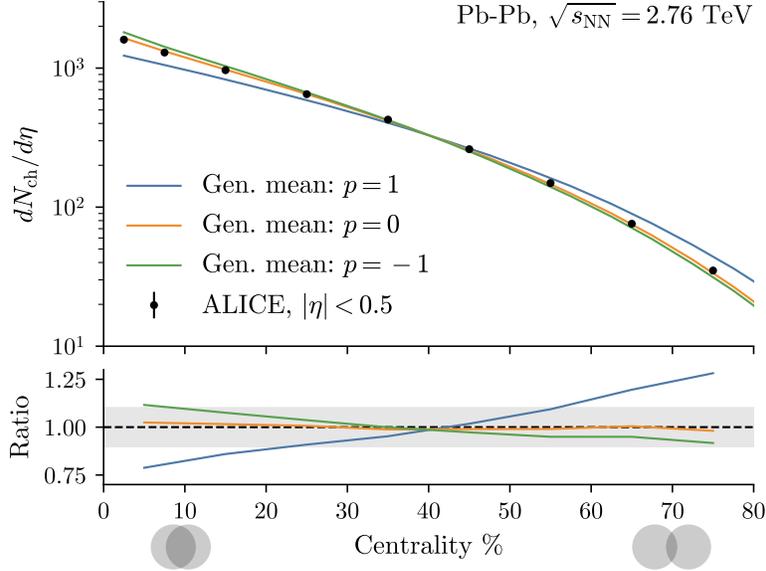
\begin{figure}[t]
  \centering
  \begin{tikzpicture}
    \node {\includegraphics{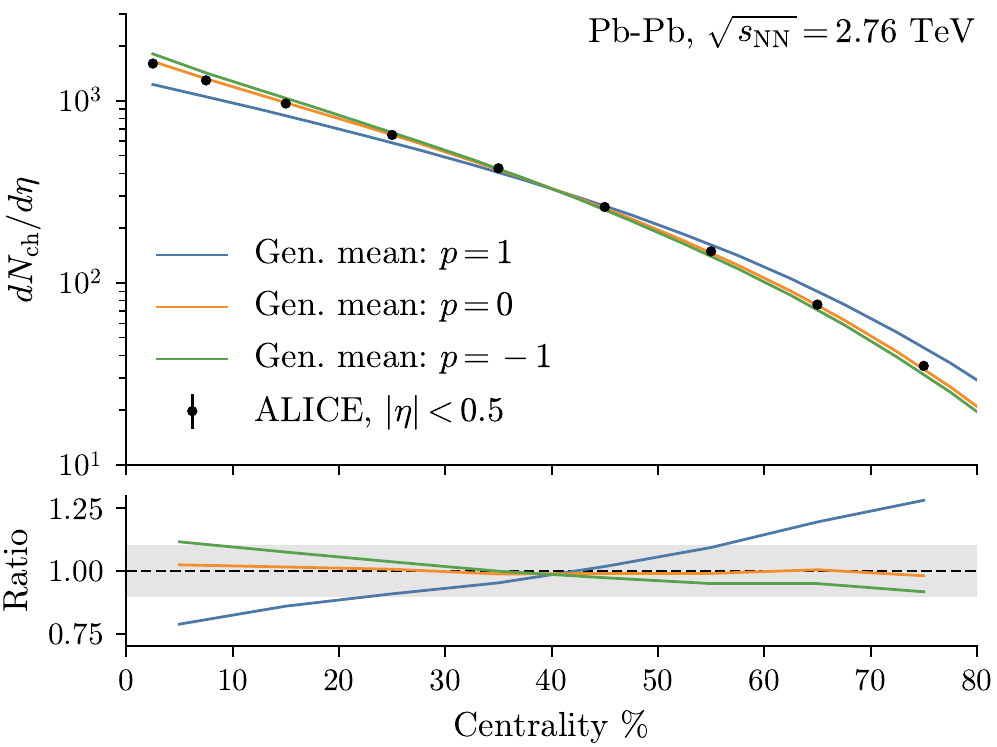}};
    \fill[black, fill opacity=0.2] (-2.85, -3.6) circle (.3);
    \fill[black, fill opacity=0.2] (-2.65, -3.6) circle (.3);
    \fill[black, fill opacity=0.2] (3.55, -3.6) circle (.3);
    \fill[black, fill opacity=0.2] (4, -3.6) circle (.3);
  \end{tikzpicture}
  \caption{
    \label{fig:gen_mean_model}
    Charged-particle density $d\nch/d\eta$ as a function of collision centrality for Pb-Pb collisions at $\sqrts=2.76$~TeV.
    Colored lines are model predictions based on the entropy variant of the generalized mean ansatz \eqref{eq:gmean_ansatz} using $p=-1, 0, 1$ and $d\nch/d\eta \propto dS/d\eta$.
    Black symbols show experimental data from the ALICE experiment \cite{Aamodt:2010cz}.
    The bottom plot shows the ratio of the model over the data (gray band $\pm 10\%$), and circles below the x-axis show the approximate nuclear overlap at two centralities.
  }
\end{figure}

Figure~\ref{fig:gen_mean_model} shows the centrality dependence of charged-particle production $d\nch/d\eta$ predicted by equation~\eqref{eq:gmean_entropy} for Pb-Pb collisions at $\sqrts=2.76$~TeV compared to experimental data from ALICE \cite{Aamodt:2010cz}.
The blue line shows an arithmetic mean $p=1$, the orange line shows a geometric mean $p=0$, and the green line shows a harmonic mean $p=-1$.
The participant thickness functions $\T_A, \T_B$ were calculated using a nucleon width $w=0.6$~fm and inelastic nucleon-nucleon cross section $\sigmann=6.4$~fm$^2$, although these details do not strongly affect the results as shown.

Notice how the generalized mean parameter $p$ varies the centrality dependence of charged-particle production, similar to the effect of $\alpha$ in the two-component model.
Smaller, more negative values of $p$ pull the generalized mean towards the minimum of the two thickness functions \eqref{eq:trento_p}.
This attenuates entropy deposition in asymmetric regions of the collision where $\T_\text{min} \ll \T_\text{max}$.
These asymmetric regions tend to be prevalent in peripheral collisions which occur at large impact parameters.
Hence decreasing $p$ (making it more negative) suppresses particle production in peripheral events at large centralities.
Conversely, the effect of the binary collision fraction in the two-component ansatz is to enhance particle production in central events.
The two ansatzes thus predict markedly different scaling behavior.

\begin{figure}[h]
  \centering
  \includegraphics{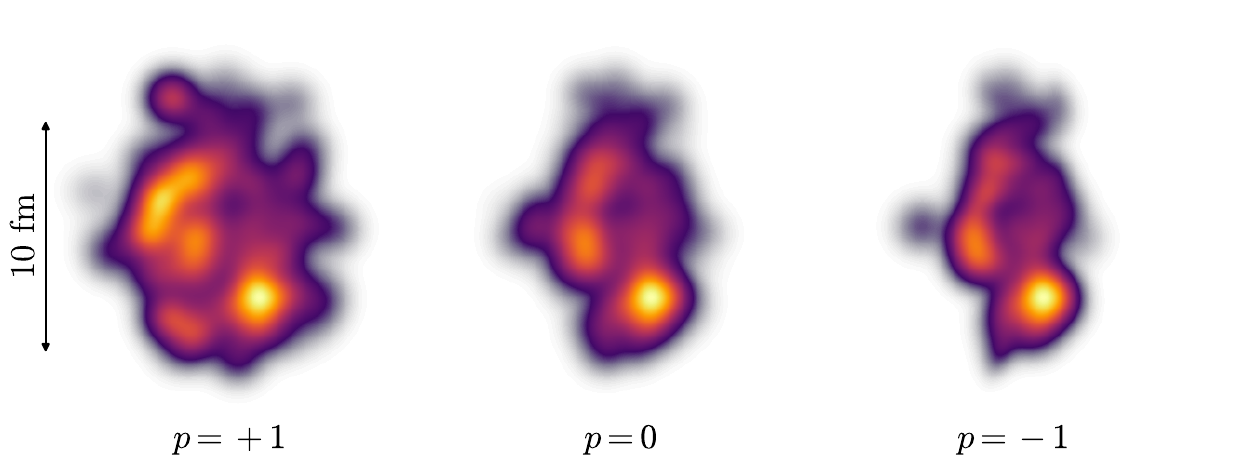}\\
  \includegraphics{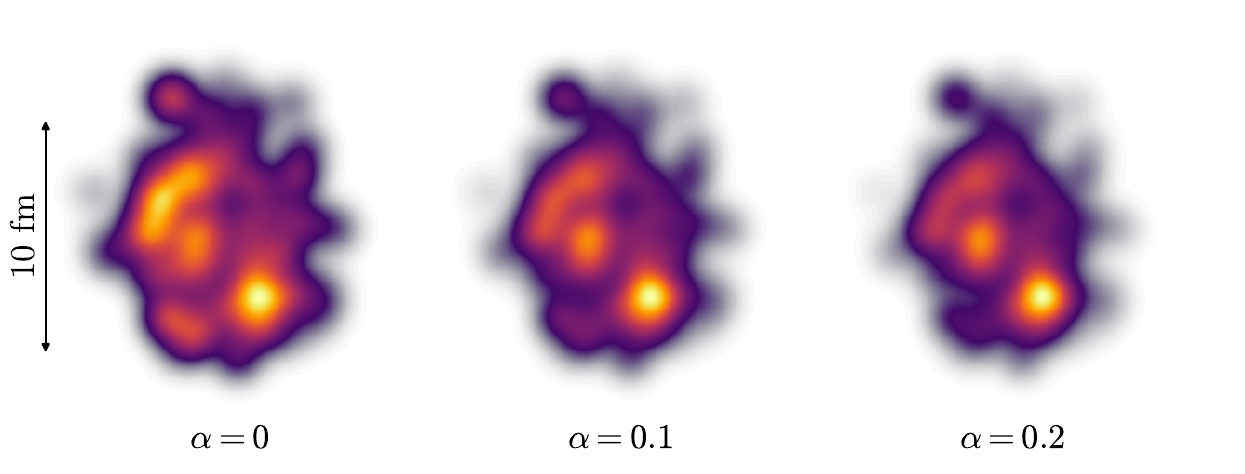}
  \caption{
    \label{fig:twocomp_vs_gmean}
    Realistic transverse entropy density profiles predicted by the generalized mean ansatz \eqref{eq:gmean_ansatz} with $p=+1$, $0$, and $-1$ (top row) compared to those predicted by the two-component ansatz \eqref{eq:two_comp_ansatz} with $\alpha = 0$, $0.1$, and $0.2$ (bottom row).
    Nucleon positions describe a peripheral Pb-Pb event at $\sqrts=2.76$~TeV.
    Each collision is identical except for the varied deposition ansatz.
  }
\end{figure}

These differences are imparted on the initial profiles of energy and entropy predicted by each model.
Figure~\ref{fig:twocomp_vs_gmean} compares the transverse profiles predicted by the two-component and generalized mean ansatzes---here assumed to be entropy densities---for a realistic Pb-Pb event at $\sqrts=2.76$~TeV.
The nucleon positions, cross sections and all other aspects of each calculation are identical, except for the chosen deposition ansatz.

First, look at the top row of the figure which shows the predictions of the generalized mean ansatz for $p=+1$, 0, and $-1$.
Smaller, more negative $p$ values compress the transverse profile along the impact parameter direction (here aligned with the $x$-axis), leading to smaller, more elliptic fireballs.
Meanwhile, the bottom row of the figure shows the predictions of the two-component ansatz using binary collision fractions $\alpha=0$, 0.1, and 0.2.
As the binary collision fraction $\alpha$ increases, the profile becomes more concentrated in the center of the collision, but the overall ellipticity of the fireball changes mildly.
Also notice, that for $p=+1$ and $\alpha=0$, both ansatzes predict wounded nucleon scaling.
Hence the far left profiles are identical.

The qualitative geometric scaling predicted by the generalized mean ansatz is of course nothing new.
Models based on Color Glass Condensate (CGC) effective field theory have predicted similar scaling behavior for quite some time; for example, see reference \cite{Drescher:2006pi}.
It explains why saturation-based models generally predict profiles of energy and entropy with larger ellipticities \cite{Drescher:2006pi, Schenke:2012wb}.
However, to the best of my knowledge, the generalized mean ansatz is the first model to effectively \emph{parametrize} this degree of freedom on an event-by-event basis.

\section{The \trento\ model}
\label{sec:trento_model}

\begin{figure}[t]
  \small
  \centering
  \makebox[\textwidth]{
    \begin{tabular}{ccccc}
      \includegraphics[scale=0.6]{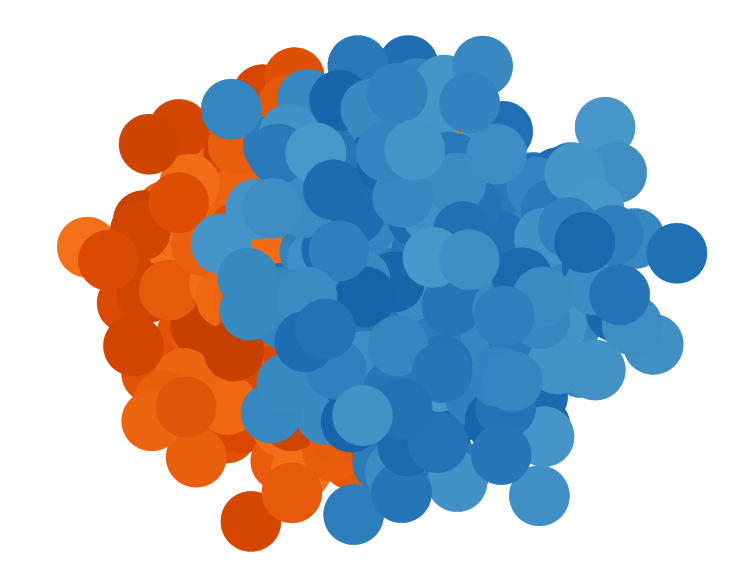} &
      \includegraphics[scale=0.6]{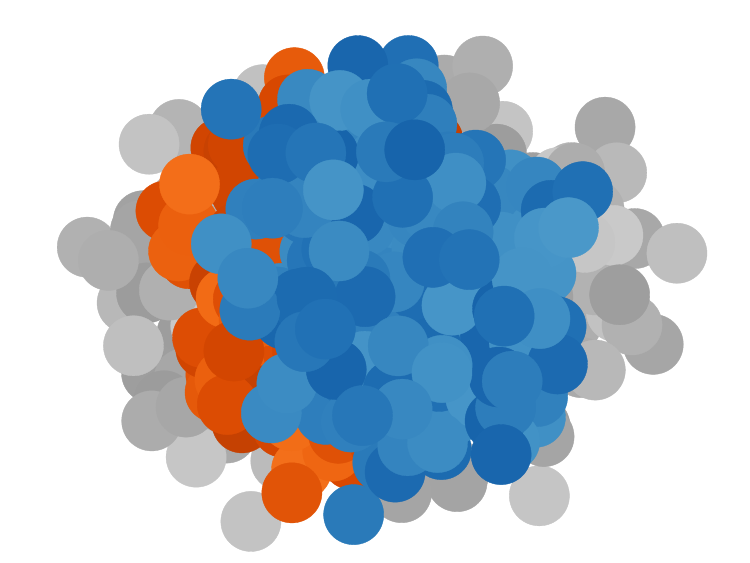} &
      \includegraphics[scale=0.6]{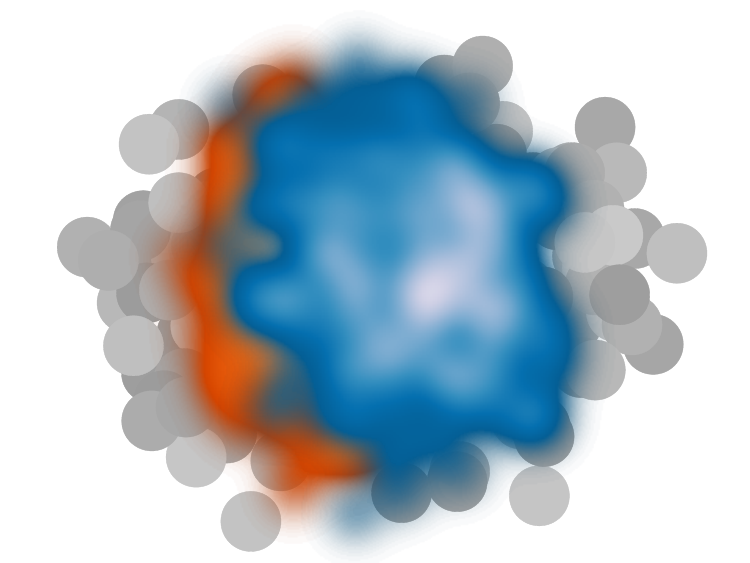} &
      \includegraphics[scale=0.6]{external/misc/trento_entropy} & \\
      \textbf{1.} Nucleon positions &
      \textbf{2.} Nucleon participants &
      \textbf{3.} Participant thicknesses &
      \textbf{4.} Reduced thickness
    \end{tabular}
  }
  \medskip
  \caption{
    \label{fig:trento_overview}
    Stages of the \trento\ initial condition model:
    1) sample nucleon positions,
    2) determine which nucleons participate inelastically,
    3) construct the participant thickness functions $\T_A$ and $\T_B$, and
    4) apply the generalized mean ansatz to calculate the reduced thickness function $T_R = M_p(\T_A, \T_B)$ which is renormalized to furnish the initial energy or entropy density.
  }
\end{figure}

I want to demonstrate now that the generalized mean ansatz is able to reproduce important features of the experimental data.
Before I proceed though, let me summarize each step described thus far
which I've illustrated in figure \ref{fig:trento_overview}.
These steps complete the specification of the so-called \trento\ model \cite{Moreland:2014oya} which stands for Reduced Thickness Event-by-event Nuclear Topology\footnote{Topology is a technical term that describes the study of geometric properties and spatial relations unaffected by continuous deformations. Geometry would be a more appropriate word for our acronym, but we liked the name \trento\, so we used the scale-invariant property of the generalized mean to justify our abuse of terminology.}.
The name pays homage to Trento, Italy where fellow graduate student Jonah Bernhard and I first conceived the formalism.
The model is summarized as follows:
\begin{enumerate}
  \item
    First, I sample nucleon positions in the rest frame of each nucleus.
    Heavy-ion nucleon positions are sampled using a deformed two-parameter Fermi distribution \eqref{eq:2pf}, while light-ion nucleon positions are modeled on a case-by-case basis.
    Optionally, when modeling heavy-ions, nucleon positions are sampled subject to a minimum distance criteria $|\xv_i - \xv_j| > d_\text{min}$ which mimics repulsive interactions between individual pairs of nucleons.
    I then randomly sample an impact parameter offset $b$ between the centroids of the two colliding nuclei and shift each set of nucleon positions by $\Delta \xv_\perp = \pm \mathbf{b}/2$.
  \item
    Once the nucleon positions are determined, inelastic nucleon-nucleon collisions are sampled using an impact parameter dependent nucleon-nucleon collision profile \eqref{eq:pcoll_pp} adapted from the analytic Glauber model, where each nucleon density $\rho_n(\xv)$ is assumed to be a three-dimensional Gaussian distribution \eqref{eq:nucleon_density} of one-sigma width $w$.
    This determines the subset of nucleons in each nucleus which collide inelastically.
  \item
    The density of each nucleon $\rho_n$ is then summed over $N_{A,B}$, the total number of participants in each nucleus, to produce a three-dimensional participant density field
    \begin{equation}
      \tilde{\rho}_{A,B}^\text{part}(\xv) = \sum\limits_{i=1}^{N_{A,B}} \gamma_i\, \rho_n(\xv - \xv_i \pm \mathbf{b}/2),
    \end{equation}
    which includes additional nucleon weights $\gamma_i$ sampled from a gamma distribution with unit mean and variance $1/k$.
    This participant density is finally projected onto the transverse plane $\xv_\perp$ by integrating over $z$ to construct two participant thickness functions
    \begin{equation}
      \T_{A,B}(\xv_\perp) = \int dz\, \tilde{\rho}_{A,B}^\text{part}(\xv_\perp, z).
    \end{equation}
    These participant thickness functions describe the fluctuated density of participant matter in each nucleus ``seen'' by a probe that moves parallel to the beam axis and pierces the transverse coordinate $\xv_\perp$.
  \item
    Finally, I set the initial energy density $e_0$ or entropy density $s_0$ proportional to the reduced thickness function
    \begin{equation}
      \label{eq:gmean_ansatz2}
      \left \{
      \begin{aligned}
        e_0\\
        s_0
      \end{aligned}
      \right \}
      = \text{Norm} \times T_R,
    \end{equation}
    defined as the generalized mean $T_R \equiv M_p(\T_A, \T_B)$ of the participant thickness functions $\T_A, \T_B$.
    The normalization prefactor is tuned once at a given beam energy $\sqrts$, after which it remains constant for all collision systems at the same energy.
\end{enumerate}

In principle, the only way to rigorously test the model is to evolve it through a realistic transport model such as viscous hydrodynamics.
The output of the simulation can then be used to assess the predictions of the model compared to the experimental data.
Indeed, I will do exactly this later in chapter \ref{ch:results}.
However, such calculations are laborious, and they often involve considerable computing resources.
Hence, before I perform such an analysis, let me first discuss properties of the initial state which \emph{do not} require a full-blown transport simulation to study.

\subsection{Mimicking existing models}
\label{subsec:mimic}

One of the primary strengths of the \trento\ model is its flexibility.
It can mimic a large number of initial condition models proposed in the literature, including (but not limited to) specific calculations in Color Glass Condensate effective field theory.
This section demonstrates this flexibility for several well-known initialization models by comparing each model's prediction for the initial entropy density $s_0$ as a function of $\T_A, \T_B$.
The following text and figures are adapted from my own contributions to
\begin{flushleft}
J.\ E.\ Bernhard, J.\ S.\ Moreland, S.\ A.\ Bass, J.\ Liu, and U.\ Heinz,
``Applying Bayesian parameter estimation to relativistic heavy-ion collisions:
simultaneous characterization of the initial state and quark-gluon plasma medium'', \href{https://arxiv.org/abs/1605.03954}{Phys.\ Rev.\ \textbf{C94}, 024907 (2016), arXiv:1605.03954 [nucl-th]},
\end{flushleft}
which I've lightly edited to conform to the notation used throughout this dissertation.

As discussed previously, one of the simplest and oldest models of heavy-ion initial conditions is the wounded nucleon model which deposits energy or entropy for each nucleon which engages in one or more inelastic collisions \cite{Bialas:1976ed}.
Consider, for example, the entropy variant of the wounded nucleon model $s_0 \propto \np$, which may be expressed in terms of the participant thickness functions $\T_A, \T_B$ in its Monte Carlo formulation \cite{Shor:1988vk, Wang:1991hta, Alver:2008aq, Broniowski:2007nz} as
\begin{equation}
  s_0 \propto \T_A + \T_B.
  \label{eq:wn}
\end{equation}
This form is exactly equivalent the generalized mean ansatz $s_0 \propto M_p(\T_A, \T_B)$ using $p=1$, so this model is a specific subcase of the \trento\ model.

More sophisticated calculations of the entropy density $s_0$ can be derived from Color Glass Condensate (CGC) effective field theory.
A common implementation of a CGC based saturation picture is the KLN model \cite{Kharzeev:2001yq, Kharzeev:2002ei, Kharzeev:2004if}, in which entropy deposition at the QGP thermalization time can be approximated from the produced gluon density, $s_0 \propto N_g$, where
\begin{equation}
  \frac{dN_g}{dy\,d^2r_\perp} \sim \Qs{min}^2 \biggl[
    2 + \log \biggl(\frac{\Qs{max}^2}{\Qs{min}^2} \biggr)
  \biggr],
  \label{eq:kln}
\end{equation}
and $\Qs{max}$ and $\Qs{min}$ denote the larger and smaller values of the two saturation momenta in opposite nuclei at any fixed position in the transverse plane \cite{Drescher:2006ca}.
In the original formulation of the KLN model, the two saturation scales are proportional to the local participant nucleon density in each nucleus, $Q^2_{s,A} \propto \T_A$, and the entropy density can be recast as
\begin{equation}
  s_0 \sim \T_\text{min} \bigl[ 2 + \log(\T_\text{max}/\T_\text{min}) \bigr],
\end{equation}
where $\T_\text{min}, \T_\text{max}$ are the minimum and maximum of the two participant thickness functions respectively.

Another saturation model which attracted interest after it successfully described an extensive list of experimental particle multiplicity and flow observables \cite{Niemi:2015qia, Paatelainen:2013eea} is the previously discussed EKRT model, which combines collinearly factorized pQCD minijet production with a simple conjecture for gluon saturation \cite{Eskola:1999fc, Eskola:2001bf}.
The energy density predicted by the model after a pre-thermal Bjorken free streaming stage is given by
\begin{equation}
  e_0 \sim \frac{K_\text{sat}}{\pi} p_\text{sat}^3(K_\text{sat}, \beta; T_A, T_B),
  \label{eq:ekrt_energy}
\end{equation}
where the saturation momentum $p_\text{sat}$ depends on the nuclear thickness functions $T_A$ and $T_B$, as well as two phenomenological model parameters $K_\text{sat}$ and $\beta$.
Calculating the saturation momentum in the EKRT formalism is computationally intensive, and hence---in its Monte Carlo implementation---the model parametrizes the saturation momentum $p_\text{sat}$ to facilitate efficient event sampling \cite{Niemi:2015qia}.
The energy density in equation~\eqref{eq:ekrt_energy} can then be recast as an entropy density using the thermodynamic relation ${s \sim e^{3/4}}$ to compare it with the previous models.

Note that equation~\eqref{eq:ekrt_energy} is expressed as a function of nuclear thickness $T$ which includes contributions from \emph{all} nucleons in the nucleus, as opposed to the participant thickness $\T$.
In order to express initial condition mappings as functions of a common variable one could, e.g.\ relate $\T$ and $T$ using an analytic wounded nucleon model.
The effect of this substitution on the EKRT model is small, as the mapping deposits zero entropy if nucleons are non-overlapping, effectively removing them from the participant thickness function.
We thus replace $T$ with $\T$ in the EKRT model and note that similar results are obtained by recasting the wounded nucleon, KLN, and \trento\ models as functions of $T$ using standard Glauber relations.

\begin{figure}
  \centering
  \makebox[\textwidth]{
    \includegraphics{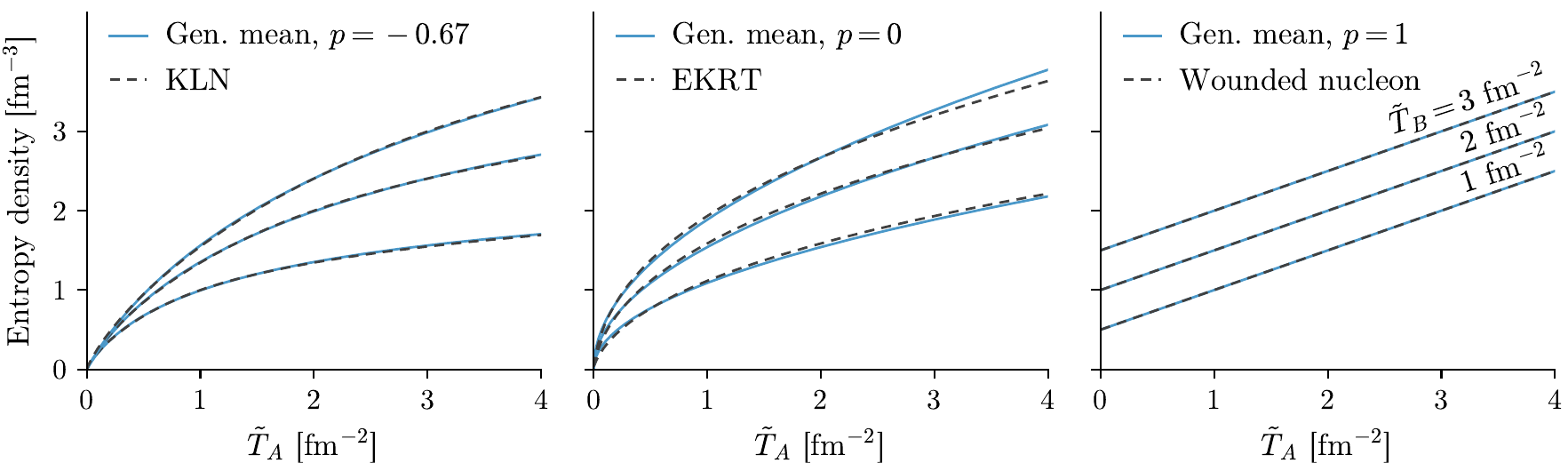}
  }
  \caption{
    \label{fig:cgc_compare}
    Profiles of the initial thermal distribution predicted by the KLN (left), EKRT (middle), and wounded nucleon (right) models (dashed black lines) compared to a generalized mean with different values of the parameter $p$ (solid blue lines).
    Staggered lines show different slices of the initial entropy density $s_0$ as a function of the participant nucleon density $\T_A$ for several values of $\T_B = 1, 2, 3$ [fm$^{-2}$].
    The EKRT mapping is shown with model parameters $K=0.64$ and $\beta=0.8$ \cite{Niemi:2015qia}.
    Entropy normalization is arbitrary.
    Figure and caption are from \cite{Bernhard:2016tnd}.
  }
\end{figure}

Figure~\ref{fig:cgc_compare} shows the midrapidity entropy density $s_0$ predicted by each of the aforementioned models as a function of $\T_A, \T_B$.
Each line is a one-dimensional cut through the two-dimensional surface $s_0(\T_A, \T_B)$.
We fix $\T_B=1, 2, 3$~fm$^{-2}$ and vary $\T_A$ from 0 to 4~fm$^{-2}$ to simulate typical values probed by a heavy-ion collision.
The dashed lines are the entropy densities $s_0$ of the KLN, EKRT and wounded nucleon (WN) models, and the solid lines are those predicted by the generalized mean ansatz, using $p$ values tuned to fit each model.
The figure illustrates the ability of the generalized mean ansatz to reproduce different initial condition calculations and quantifies differences among them in terms of the generalized mean parameter $p$.
The KLN model, for example, is well-described by $p\sim-0.67$, the EKRT model corresponds to $p \sim 0$, and the wounded nucleon model is precisely reproduced by $p=1$.
Smaller, more negative values of $p$ pull the generalized mean toward a minimum function and hence correspond to models with more extreme gluon saturation effects.

The three models considered in figure~\ref{fig:cgc_compare} are by no means an exhaustive list of proposed initial condition models, see e.g.\ references~\cite{Eremin:2003qn, Broniowski:2007nz, Pierog:2013ria, Drescher:2000ec, Chatterjee:2015aja, Zhang:1999bd}.
Notably absent, for instance, is the highly successful IP-Glasma model which combines IP-Sat CGC initial conditions with classical Yang-Mills dynamics to describe the full pre-equilibrium evolution of produced glasma fields \cite{Schenke:2012wb, Schenke:2012fw, Gale:2012rq}.
The IP-Glasma model lacks a simple analytic form for initial energy (or entropy) deposition at the QGP thermalization time and so it cannot be directly compared to the generalized mean ansatz.
In lieu of such a comparison, we examined the geometric properties of the IP-Glasma and \trento\ models through their eccentricity harmonics $\varepsilon_n$.

\begin{figure}
  \centering
  \includegraphics{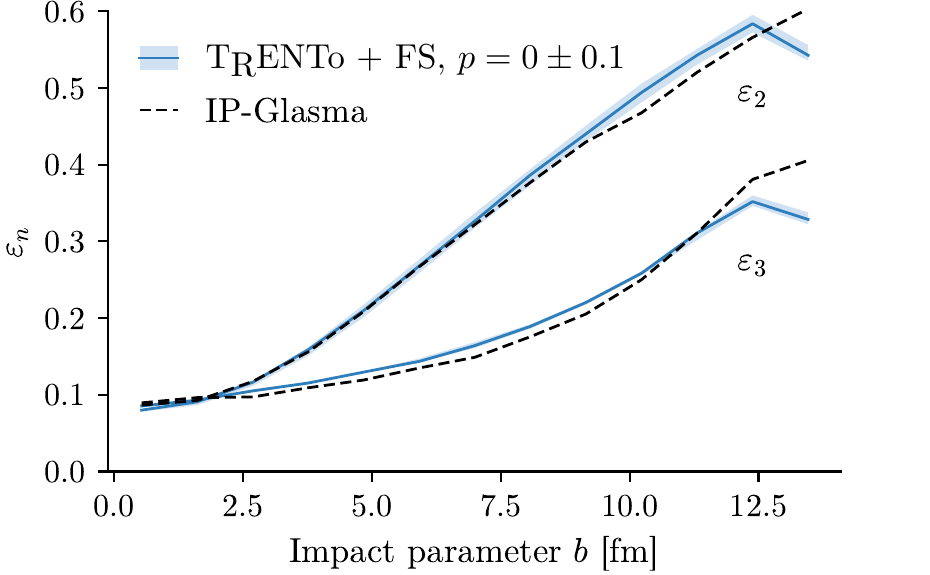}
  \caption{
    \label{fig:ipglasma}
    Eccentricity harmonics $\varepsilon_2$ and $\varepsilon_3$ as a function of impact parameter $b$ for Pb-Pb collisions at ${\sqrts=2.76}$~TeV calculated from IP-Glasma and \trento\ initial conditions.
    IP-Glasma events are evaluated after $\tau=0.4$~fm/$c$ classical Yang-Mills evolution \cite{Schenke:2012wb}; \trento\ events after $\tau=0.4$~fm/$c$ free streaming \cite{Liu:2015nwa, Broniowski:2008qk} and using parameters $p = 0 \pm 0.1$, $k = 1.6$, and nucleon width $w=0.4$~fm to match IP-Glasma \cite{Schenke:2013dpa}.
    Figure and caption are from \cite{Bernhard:2016tnd}.
  }
\end{figure}

We generated a large number of \trento\ events using entropy deposition parameter $p=0$, Gaussian nucleon width $w=0.4$~fm, and fluctuation parameter $k=1.6$, which were previously shown to reproduce the ratio of ellipticity and triangularity in IP-Glasma \cite{Moreland:2014oya}.
We then free streamed \cite{Liu:2015nwa, Broniowski:2008qk} the events for $\tau=0.4$~fm/$c$ to mimic the weakly coupled pre-equilibrium dynamics of IP-Glasma and match the evolution time of both models.
Finally, we calculated the eccentricity harmonics $\varepsilon_2$ and $\varepsilon_3$ weighted by energy density $e(x, y)$ according to the definition
\begin{equation}
    \varepsilon_n e^{i n \phi} = -\frac{\int dx\, dy\, r^n e^{i n \phi} e(x,y)}{\int dx\, dy\, e(x,y)},
\end{equation}
where the energy density is the time-time component of the stress-energy tensor after the free streaming phase, $T^{00}$.
The resulting eccentricities, pictured in figure~\ref{fig:ipglasma}, are in good agreement for all but the most peripheral collisions, where sub-nucleonic structure becomes important.
This similarity suggests that \trento\ with $p \sim 0$ can effectively reproduce the scaling behavior of IP-Glasma, although a more detailed comparison would be necessary to establish the strength of correspondence illustrated in figure~\ref{fig:cgc_compare}.

Needless to say, there are also models in the literature which \emph{cannot} be approximated by the generalized mean ansatz.
The generalized mean ansatz is scale-invariant, i.e.\
\begin{equation}
  \label{eq:scale_invariant}
  M_p(c \T_A, c \T_B) = c M_p(\T_A, \T_B)
\end{equation}
for any nonzero constant $c$, so the parametrization can only mimic initial condition models which scale accordingly.
Note, this property is clearly broken by the binary collision term $\nc$ in the two-component ansatz \eqref{eq:two_comp_ansatz}.
This property may seem overly restrictive.
However, I'll show somewhat later in the next section that even small deviations from scale invariance produce tension with the data.
The scale-invariant postulate thus appears to be a necessary requirement for physically reasonable models.\footnote{Recently, it has come to light that the IP-Glasma model predicts local energy density deposition which scales approximately like the product of nuclear thickness functions ${e_0 \propto T_A\, T_B}$ \cite{Nagle:2018ybc, Romatschke:2017ejr}. This functional form clearly breaks scale invariance. Understanding this feature of the model and its compatibility with the data is currently the subject of active investigation}

\subsection{Application to experimental data}

This section demonstrates \trento's ability to simultaneously describe a wide range of collision systems, using approximate scaling laws to compare the model predictions with experimental data.
The text and figures in this subsection section are adapted from one of my publications,
\begin{flushleft}
  J.\ S.\ Moreland, J.\ E.\ Bernhard, and S.\ A.\ Bass,
  ``Alternative ansatz to wounded nucleon and binary collision scaling in high-energy nuclear collisions'',
  \href{https://arxiv.org/abs/1804.06469}{Phys.\ Rev.\ \textbf{C92}, 011901 (2015) arXiv:1412.4708 [nucl-th]},
\end{flushleft}
which I've lightly edited for clarity and formatting.
Note, the \trento\ reduced thickness parameter $p$, gamma fluctuation parameter $k$, and nucleon profile $\rho_n$ used here are not rigorously constrained---doing so requires advanced Bayesian statistical machinery which I'll introduce in the next chapter.
Therefore, the results of this section do not necessarily represent the best possible fit of the model to data.

In this publication, we compared the \trento\ model with experimentally measured multiplicity distributions using a three-stage model for particle production, similar to that of reference \cite{Bozek:2013uha}, in which the final multiplicity arises from a convolution of the initial entropy deposited by the collision, viscous entropy production during hydrodynamic evolution, and statistical hadronization at freeze-out.
The average charged-particle multiplicity $\avg\nch$ after hydrodynamic evolution is roughly proportional to the total initial entropy \cite{Song:2008si} and hence to the integrated reduced thickness via equation~\eqref{eq:gmean_ansatz2}:
\begin{equation}
  \avg\nch \propto \int dx \, dy \, T_R.
\end{equation}
Then, assuming independent particle emission at freeze-out, the final number of charged particles is Poisson distributed \cite{Kisiel:2005hn, Chojnacki:2011hb}, i.e.\ $P(\nch) = \text{Poisson}(\avg\nch)$.
The folding of the Poisson fluctuations with the gamma weights for each participant yields a negative binomial distribution \cite{Bozek:2013uha}, which has historically been used to fit proton-proton multiplicity fluctuations (see subsection \ref{subsec:participant_thickness}).

\begin{figure}[t]
  \centering
  \makebox[\textwidth]{\includegraphics{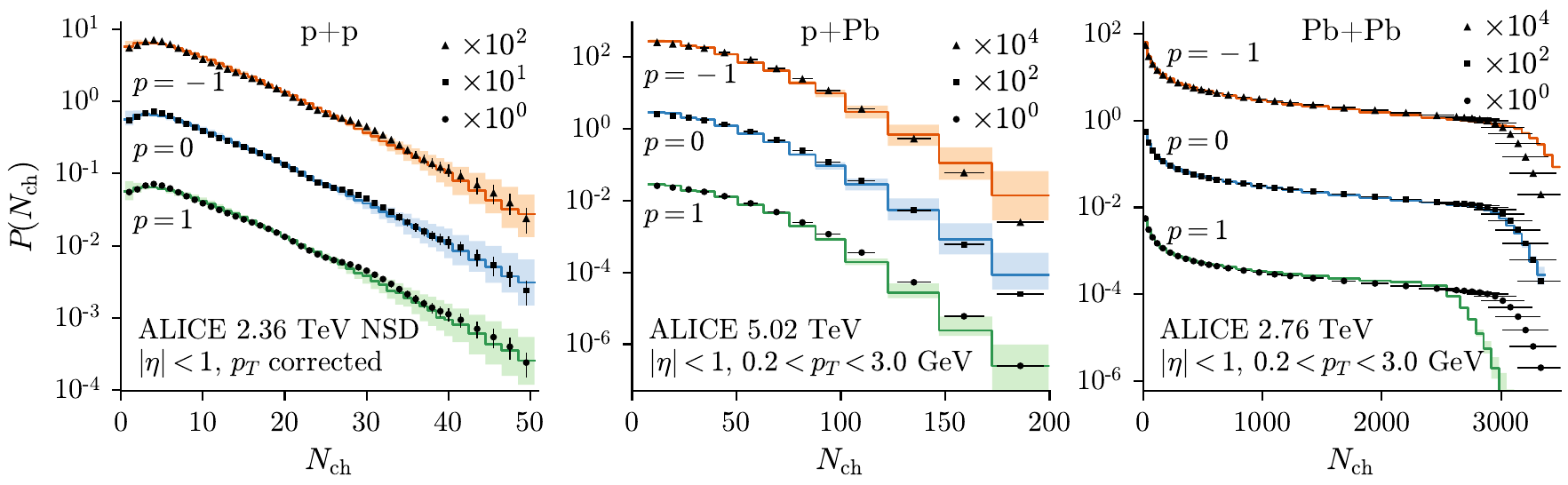}}
  \caption{
    \label{fig:nch}
    Multiplicity distributions for proton-proton, proton-lead, and lead-lead collisions.
    The histograms are \trento\ results for reduced thickness parameter $p = -1$ (top, orange), $p = 0$ (middle, blue), and $p = 1$ (bottom, green), with approximate best-fit fluctuation parameters $k$ and normalizations given in table~\ref{tab:nch}.
    The shaded bands show the sensitivity from varying $k$ by $\pm30\%$.
    Data points (triangles, squares, circles) are experimental distributions from ALICE \cite{Aamodt:2010ft, Abelev:2014mda} offset by powers of ten for comparison with the model.
  }
\end{figure}

\begin{table}[b]
  \centering
  \caption{
    \label{tab:nch}
    Approximate best-fit fluctuation parameters $k$ and normalizations for each $p$ value and collision system in figure~\ref{fig:nch}.
  }
  \begin{tabular}{rcccc}
    \toprule
    $p\;$  & $k$ & p-p norm & p-Pb norm & Pb-Pb norm \\
    \midrule
    $+1$   & 0.8 & 9.7      & 7.0       & 13.        \\
    $ 0$   & 1.4 & 19.      & 17.       & 16.        \\
    $-1$   & 2.2 & 24.      & 26.       & 18.        \\
    \bottomrule
  \end{tabular}
\end{table}

Following this prescription, we generated a large ensemble of minimum-bias events, integrated their $T_R$ profiles, rescaled them by an overall normalization constant, and sampled a Poisson number for the multiplicity of each event.
The left panel of figure~\ref{fig:nch} shows the $\nch$ distributions for proton-proton simulations with reduced thickness parameter $p = 1$, 0, $-1$, and Gaussian beam-integrated proton density
\begin{equation}
  \int dz \, \rho_n = \frac{1}{2\pi B} \exp\biggr( -\frac{x^2 + y^2}{2B} \biggr)
\end{equation}
with effective area $B = (0.6\;\text{fm})^2$.
We tuned the fluctuation parameter $k$ for each value of $p$ to qualitatively fit the experimental proton-proton distribution \cite{Aamodt:2010ft}, and additionally varied $k$ by $\pm30\%$ to explore the sensitivity of the model to the gamma participant weights.
For proton-lead and lead-lead collisions \cite{Abelev:2014mda} (middle and right panels), we used identical model parameters except for the overall normalization factor, which was allowed to vary independently across collision systems to account for differences in beam energy and kinematic cuts (annotated in the figure).
The $k$ values and normalizations are provided in table~\ref{tab:nch}.

\begin{figure}
  \centering
  \includegraphics{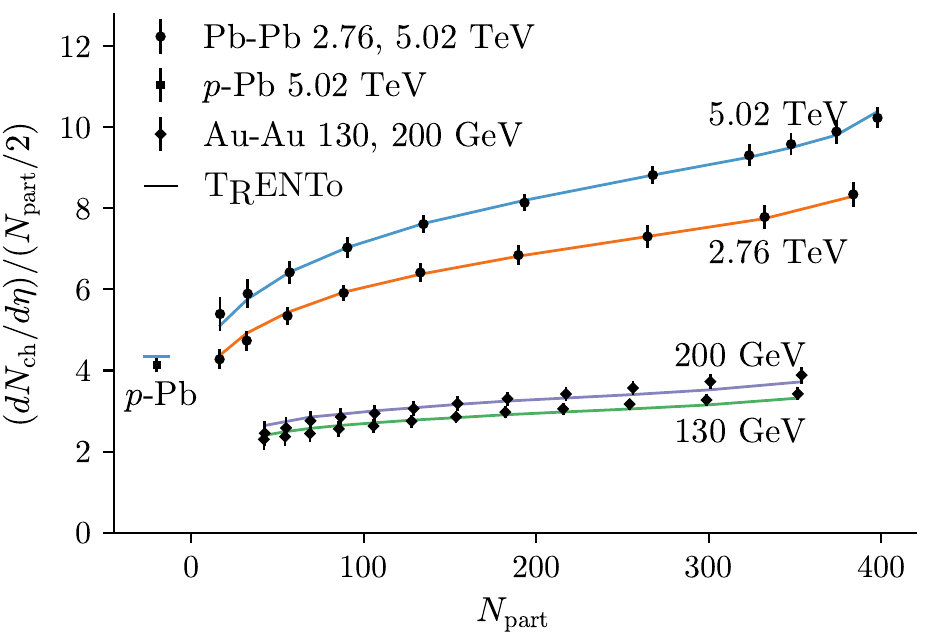}
  \caption{
    \label{fig:nch_per_npart}
    Average charged particle density per participant nucleon pair $(d\nch/d\eta)/(\Np/2)$ at midrapidity as a function of the participant number $\Np$ for Pb-Pb, p-Pb, and Au-Au systems at various collision energies.
    Lines are \trento\ calculations with generalized mean parameter $p = 0$, and symbols are data from PHENIX \cite{Adare:2015bua} and ALICE \cite{Aamodt:2010cz, Adam:2015ptt}.
    The average minimum bias participant number for p-Pb is shifted for clarity.
    Figure and caption are from \cite{Bernhard:2016tnd}.
  }
\end{figure}

The model is able to reproduce the experimental proton-proton distribution for each value of $p$, provided $k$ is appropriately tuned.
Varying the best-fit $k$ value (by $\pm30\%$) has a noticeable effect on proton-proton and proton-lead systems, especially in the high-multiplicity tails, but is less important in lead-lead collisions, where the gamma weights are averaged over many participant nucleons.
Each $p$ value also yields a reasonable fit to the shapes of the proton-lead and lead-lead distributions, although lead-lead appears to favor $p \approx 0$.
Note that the normalizations for $p = 1$ (wounded nucleon model) in proton-lead and lead-lead collisions (table~\ref{tab:nch}) are not self-consistent, since proton-lead requires roughly half the normalization as lead-lead, even though the experimental data were measured at a higher beam energy.

In a somewhat later publication \cite{Bernhard:2016tnd}, we extended the yield comparison for $p=0$ to include additional beam energies.
Figure \ref{fig:nch_per_npart} shows the charged-particle density per participant pair $(d\nch/d\eta)/(\Np/2)$ at midrapidity as a function of participant number for proton-lead, gold-gold, and lead-lead collisions with beam energies spanning several orders in magnitude \cite{Bernhard:2016tnd}.
The colored lines are \trento\ model calculations using $p=0$, and the black symbols are experimental data from PHENIX \cite{Adare:2015bua} and ALICE \cite{Aamodt:2010cz}.
The \trento\ calculations provides a superb fit to the experimental data, even fitting the average yield of the $p$-Pb data point using the same overall normalization as the Pb-Pb system at the same beam energy, consistent with previous observations.

\begin{figure}[t]
  \centering
  \makebox[\textwidth]{\includegraphics{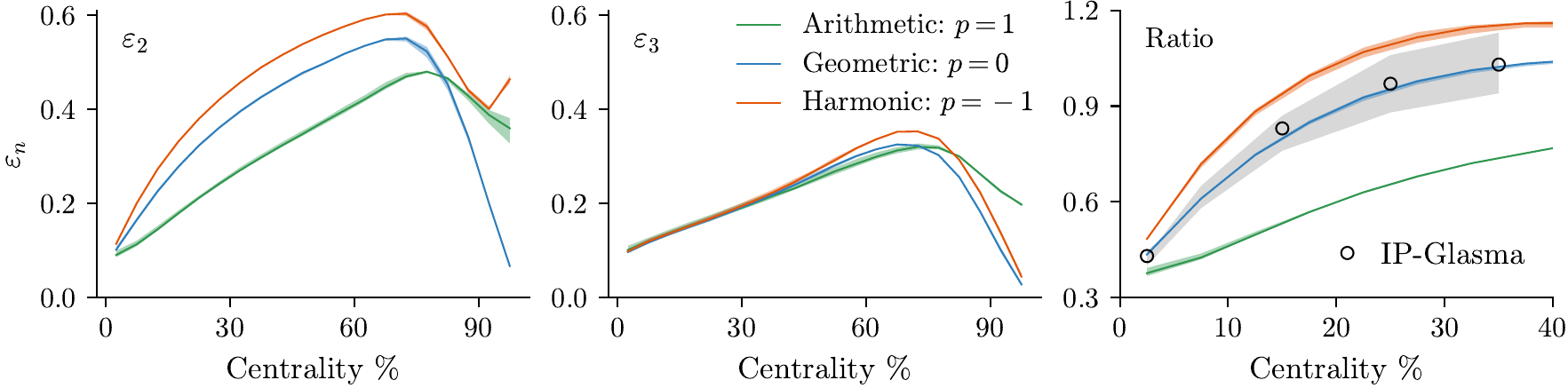}}
  \caption{
    \label{fig:eccen}
    Left and middle plots: Eccentricity harmonics $\varepsilon_2$ and $\varepsilon_3$ as a function of centrality for reduced thickness parameters $p = 1$, 0, $-1$ (green, blue, orange).
    The shaded bands show the sensitivity from varying $k$ by $\pm30\%$ from the values in table~\ref{tab:nch}.
    Right plot: Ratio of the rms eccentricities $\eccratio$ against the allowed region (grey band) and the ratio computed by IP-Glasma (circles) \cite{Retinskaya:2013gca}.
    Note that the axes have different ranges in the ratio plot.
  }
\end{figure}

In reference \cite{Moreland:2014oya}, we also examined the sensitivity of the eccentricity harmonics $\varepsilon_n$ to different values of the generalized mean parameter $p$ and fluctuation factor $k$.
The reduced thickness function eccentricity equals
\begin{equation}
  \varepsilon_n e^{i n\phi} = -\frac{\int dx \, dy\, r^n e^{i n \phi} \, T_R}{\int dx \, dy \, r^n \, T_R}.
\end{equation}
Figure~\ref{fig:eccen} shows ellipticity $\varepsilon_2$ and triangularity $\varepsilon_3$ as a function of centrality using the same lead-lead data as in figure~\ref{fig:nch}.
There is a clear trend of increasing eccentricity (particularly $\varepsilon_2$) with decreasing $p$.
As $p$ decreases, the generalized mean \eqref{eq:gmean_ansatz} attenuates entropy production in asymmetric regions of the collision, accentuating the elliptical overlap shape in non-central collisions and enhancing their eccentricity.
Meanwhile, varying the fluctuation parameter $k$ has limited effect.

In addition, we performed the test proposed by \cite{Retinskaya:2013gca}, which uses flow data and hydrodynamic calculations to determine an experimentally allowed band for the ratio of root-mean-square eccentricities $\eccratio$ as a function of centrality.
Among the initial condition models available at that time, only IP-Glasma consistently falls within the allowed region.
As shown in the right panel of figure~\ref{fig:eccen}, the \trento\ model with $p = 0$ (geometric mean) yields excellent agreement with the allowed band and is similar to IP-Glasma.

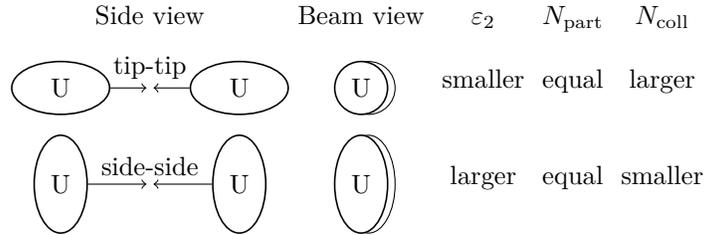
\begin{figure}[b]
  \centering \small
  \tikzsetnextfilename{uranium_diagram}
  \begin{tikzpicture}[
    uranium/.style={draw, semithick, ellipse, anchor=center},
    small width/.style={minimum width=20},
    large width/.style={minimum width=37},
    small height/.style={minimum height=20},
    large height/.style={minimum height=37}
  ]
    \matrix (m) [matrix of nodes] {
      &[-2.1ex] Side view &[-2.1ex] & Beam view & $\varepsilon_2$ & $\Np$ & $N_\text{coll}$ \\[.7ex]
      |[uranium, large width, small height] (ttl)| U &
      |[above]| tip-tip &
      |[uranium, large width, small height] (ttr)| U &
      \node[draw, circle, small height, small width, xshift=1mm] {};
      \node[uranium, circle, small width, small height, fill=white] {U}; &
      smaller & equal & larger \\[1.7ex]
      |[uranium, large height, small width] (ssl)| U &
      |[above]| side-side &
      |[uranium, large height, small width] (ssr)| U &
      \node[draw, ellipse, large height, small width, xshift=1mm] {};
      \node[uranium, large height, small width, fill=white] {U}; &
      larger & equal & smaller \\
    };
    \begin{scope}[->]
      \draw (ttl) -- ($(ttl)!.48!(ttr)$);
      \draw (ttr) -- ($(ttr)!.48!(ttl)$);
      \draw (ssl) -- ($(ssl)!.48!(ssr)$);
      \draw (ssr) -- ($(ssr)!.48!(ssl)$);
    \end{scope}
  \end{tikzpicture}
  \caption{
    \label{fig:uu-schematic}
    Comparison of tip-tip and side-side uranium-uranium collisions.
    Schematics are shown from a side view and looking down the beam axis, and the following quantities are compared: ellipticity $\varepsilon_2$, number of participating nucleons $\Np$, and number of binary nucleon-nucleon collisions $\Nc$.
  }
\end{figure}

As a final novel application, we explained how the generalized mean ansatz resolves an apparent puzzle in uranium-uranium collisions at RHIC.
Unlike e.g.~gold and lead, uranium nuclei have a highly deformed prolate spheroidal shape, so uranium-uranium collisions may achieve maximal overlap via two distinct orientations:
``tip-tip'', in which the long axes of the spheroids are aligned with the beam axis and the overlap area is circular;
or ``side-side'', where the long axes are perpendicular to the beam axis and the overlap area is elliptical, as shown in figure~\ref{fig:uu-schematic}.
Hence side-side collisions will in general have larger initial-state ellipticity $\varepsilon_2$ and final-state elliptic flow $v_2$ than tip-tip.

In the two-component Glauber model, tip-tip collisions produce more binary nucleon-nucleon collisions than side-side, so tip-tip collisions have larger charged-particle multiplicity $\nch$.
Therefore, the most central uranium-uranium events are dominated by tip-tip collisions with maximal $\nch$ and small $v_2$, while side-side collisions have a smaller $\nch$ and somewhat larger $v_2$.
This predicted drop in elliptic flow as a function of $\nch$ is known as the ``knee'' \cite{Voloshin:2010ut}.
Data from STAR on uranium-uranium collisions, however, exhibits no evidence of a knee \cite{Pandit:2013uiv,Wang:2014qxa}, at odds with the predictions of the two-component model.
It has been proposed that fluctuations could wash out the knee \cite{Rybczynski:2012av}, but a detailed flow analysis showed that it would still be visible \cite{Goldschmidt:2015qya}.

\begin{figure}[t]
  \centering
  \makebox[\textwidth]{\includegraphics{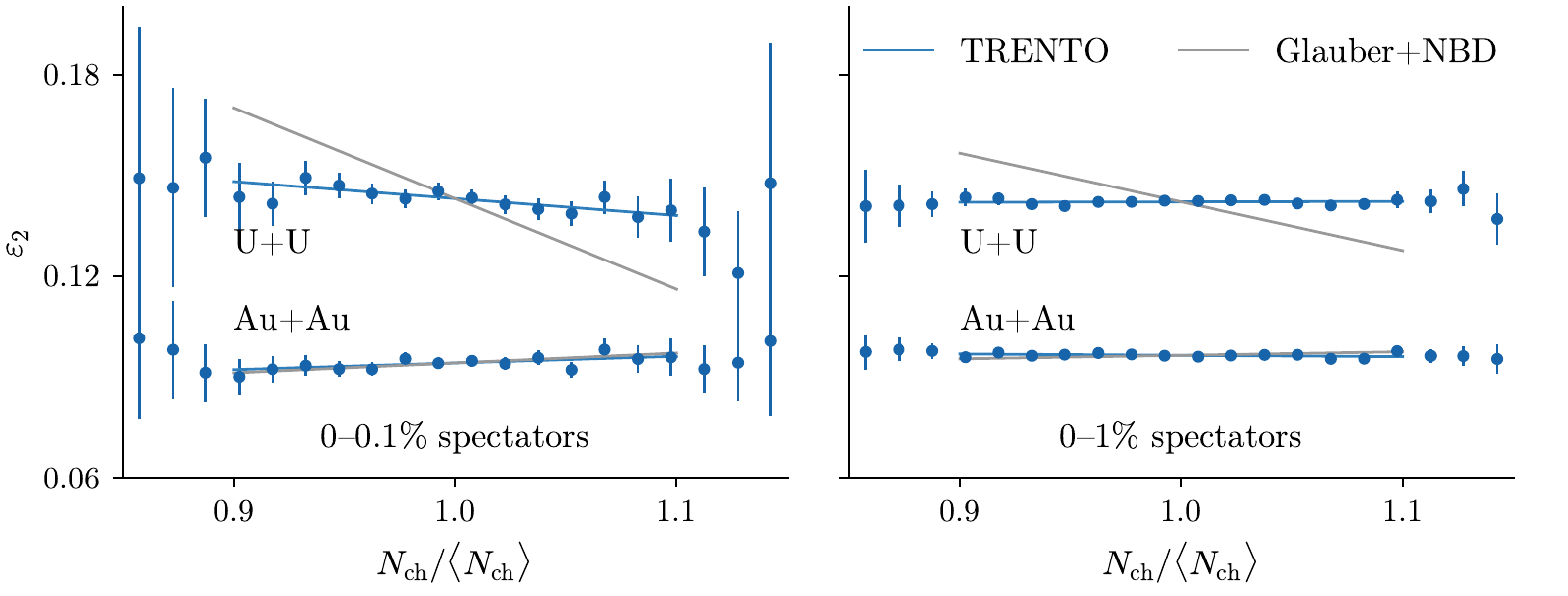}}
  \caption{
    \label{fig:uranium}
    Ellipticity $\varepsilon_2$ as a function of normalized charged-particle multiplicity $\nch/\avg\nch$ in ultra-central uranium-uranium and gold-gold collisions at RHIC.
    The top and bottom plots show the top 0.1\% and 1\% of collisions selected by number of spectators to mimic STAR's experimental ZDC selection \cite{Pandit:2013uiv}.
    Blue points with error bars are binned \protect\trento\ results with reduced thickness parameter $p = 0$ and best-fit fluctuation parameter $k = 1.4$.
    Blue lines are linear fits within $0.9~<~\nch/\avg\nch~<~1.1$.
    Grey lines represent the analogous Glauber+NBD slopes calculated in \cite{Pandit:2013uiv}.
  }
\end{figure}

The data therefore imply that multiplicity is independent of the number of binary collisions, justifying the scale-invariant condition \eqref{eq:scale_invariant},
postulated during the construction of the reduced thickness ansatz \eqref{eq:reduced_thickness}.
Due to this assumed scale invariance, \trento\ predicts roughly the same number of charged particles in tip-tip and side-side uranium-uranium collisions.
As shown in figure~\ref{fig:uranium}, the slope of $\varepsilon_2$ as a function of $\nch$ is approximately equal for uranium-uranium and gold-gold, in contrast to the two-component Glauber model which predicts a much steeper slope for uranium.
Short of conducting a full hydrodynamic analysis, \trento\ appears to be more consistent with STAR data than the two-component model, and behaves similarly to IP-Glasma \cite{Schenke:2014tga}.

\subsection{Proton-proton collision geometry}

The \trento\ model makes no distinction between small and large collision systems, so it can also be used to predict the profiles of energy (or entropy) produced by a single proton-proton collision.
The framework is admittedly strained---I crudely modeled each nucleon as a Gaussian blob---nevertheless, it should provide some qualitative insight into the macroscopic geometry of the produced fireball.

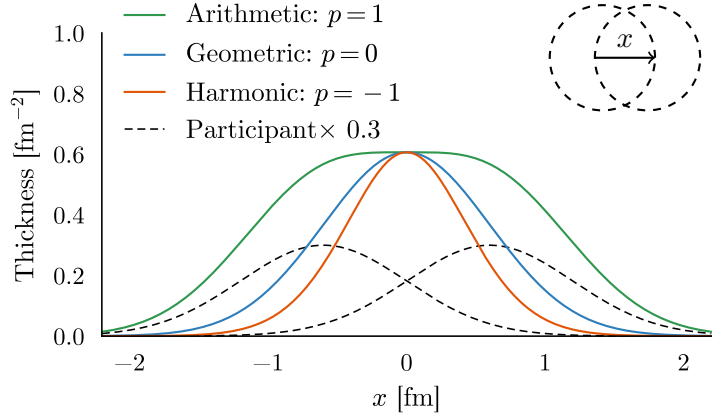
\begin{figure}
  \centering
  \begin{tikzpicture}
    \node at (0,0) {\includegraphics{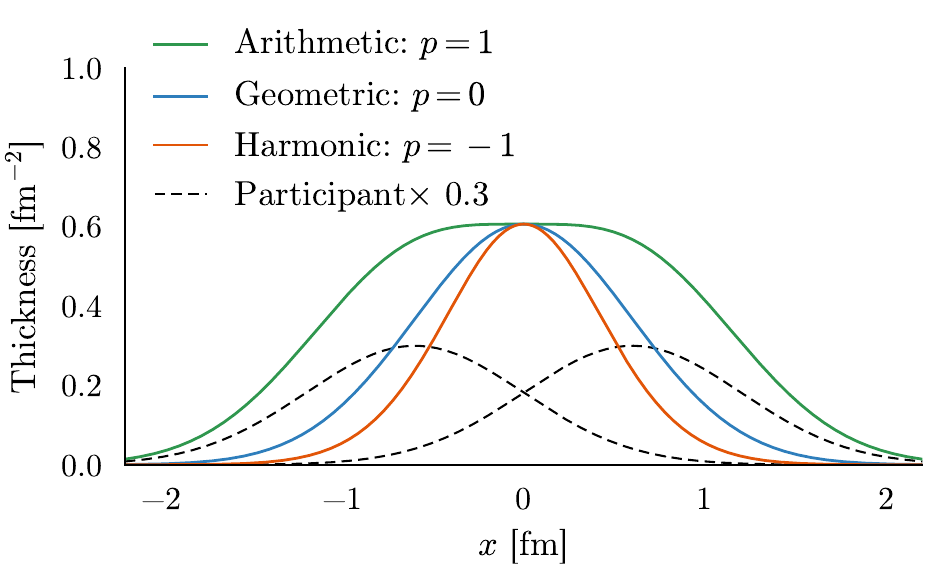}};
    \draw[thick, dashed] (3.8, 1.9) circle (.7);
    \draw[thick, dashed] (3.2, 1.9) circle (.7);
    \draw[thick, ->] (3.1, 1.9)--(3.9, 1.9) node[above, midway]{$x$};
  \end{tikzpicture}
  \caption{
    \label{fig:reduced_thickness}
    Cross section of the reduced thickness function for a pair of nucleon participants. The nucleons collide with a nonzero impact parameter along the $x$-direction as shown in the upper right.
    The black dashed lines are one-dimensional cross sections of the participant nucleon thickness functions $\T_A, \T_B$, and the colored lines are the reduced thickness $T_R$ for $p = 1$, $0$, $-1$ (green, blue, orange).
    Figure and caption are from \cite{Moreland:2014oya}.
  }
\end{figure}

\begin{figure}
  \centering
  \makebox[\textwidth]{\includegraphics{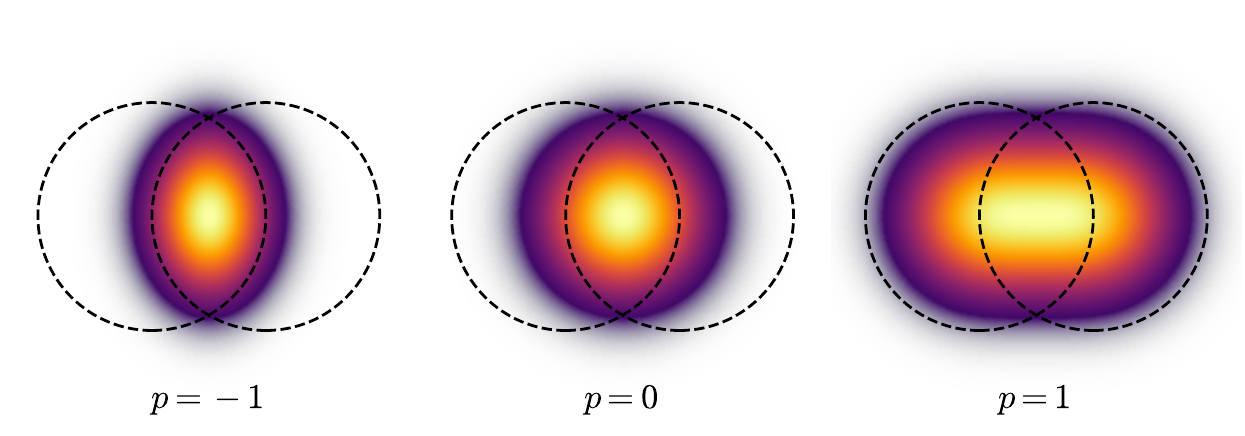}}
  \caption{
    \label{fig:proton_shapes}
    Reduced thickness for a pair of nucleon participants using a harmonic mean $p=-1$ (left), geometric mean $p=0$ (middle), and arithmetic mean $p=1$ (right).
    The black dashed circles are contours showing the location of each Gaussian nucleon participant.
    Geometry parallels figure \ref{fig:reduced_thickness}.
  }
\end{figure}

Figure \ref{fig:reduced_thickness} shows a cross section of the reduced thickness function $T_R$ for the collision of two Gaussian protons at non-zero impact parameter with the gamma fluctuations \eqref{eq:fluct_participant_density} turned off.
Each colored line is the generalized mean for a different value of $p$.
The green line is an arithmetic mean $p=1$, the blue line is a geometric mean $p=0$, and the orange line is a harmonic mean $p=-1$.
The black dashed lines outline each nucleon participant, scaled down by a factor of three for visibility.
Meanwhile, figure \ref{fig:proton_shapes} shows the same reduced thickness functions plotted as heatmaps in the transverse plane.
As before, the black dashed lines mark the outline of each nucleon participant.

Notice how the reduced thickness becomes more sharply peaked in the overlap region as $p \to -\infty$, producing a highly elliptic fireball.
Conversely, for $p \to +\infty$, the fireball grows in size and becomes oblong, reproducing the underlying density of each Gaussian participant.
For the intermediate value $p=0$, the energy (or entropy) produced by the collision is a Gaussian located midway between the two interacting protons.

Intriguingly, Bzdak \emph{et al.}\ pondered these exact three proton-proton collision scenarios two years before the generalized mean ansatz was even conceived; see figure 1 in reference \cite{Bzdak:2013zma}.
The \trento\ model explains these scenarios as certain subcases of the overall energy and entropy deposition mapping.
Hence, if we can rigorously constrain the parameter $p$, it might help to lift the veil on the distribution of energy and entropy deposited by individual proton-proton collisions.
The next chapter describes some advanced statistical machinery which will help rigorously pin down the parameters of the \trento\ initial condition model, enabling state-of-the-art extractions of the QGP transport coefficients.

\chapter[Bayesian parameter estimation]{Bayesian parameter estimation}
\label{ch:param_est}

\lettrine{Q}{uark-gluon plasma} produced in the laboratory is not directly observable.
Relativistic nuclear collisions produce small droplets of the substance, and these droplets expand and cool rapidly, disintegrating into showers of separate particles on the order of ${\sim}10^{-23}$ seconds.
All direct physical traces of the QGP are thus long gone by the time the collision is measured inside the detector.
Studying the QGP is hence a forensic endeavor.
Experimentalists measure final state properties of the collision, and theorists work to reconstruct the causal factors which explain their findings.
Such problems are called \emph{inverse problems} because the results of the process are known but its root causes are not.

Inverse problems are particularly challenging to solve when information is lost or scrambled during the process of interest.
For example, if someone kicks over a sandcastle, it's difficult to reconstruct the sandcastle from a picture of its demolished state.
The same is also true for a relativistic nuclear collision event.
These events generate considerable entropy, and hence---according to the second law of thermodynamics---their processes are irreversible.
This prevents one from simply running the dynamics of a nuclear collision simulation in reverse.

That's not to say, however, that \emph{all} information is lost during an irreversible process.
If someone kicked over a sandcastle, I could still estimate its total mass and get a rough idea of its initial shape.
If I wanted to be more precise, I could build lots of different sandcastles and kick each one over until I found one that resembled the demolished sandcastle I sought to reconstruct, i.e.\ I could attempt to solve the sandcastle inverse problem using brute force trial-and-error.

Trial-and-error is particularly useful for $A/B$ testing, i.e.\ testing whether causal factors are better explained by process $A$ or process $B$.
For example, it was originally unknown whether the QGP would behave like a strongly coupled liquid or a weakly coupled gas.
Researchers simulated the space-time dynamics of heavy-ion collisions using both descriptions and found that strong coupling provides a far superior description of the data \cite{Shuryak:2004cy, Molnar:2001ux}.

This is, of course, an over simplification of a more nuanced problem.
The QGP is neither infinitely strongly coupled nor infinitely weakly coupled but something inbetween.
The relevant problem-space is hence continuous, not discrete.
We can therefore frame the problem more rigorously if we imagine a meta-model of a single parameter $x$ which varies the simulation's coupling strength.
For example, suppose that for $x \sim 0$ the meta-model is weakly coupled, and for $x \gg 1$ it is strongly coupled.
Additionally, suppose there is some simulated observable $y$ which is sensitive to the coupling strength $x$.
Solving the inverse problem amounts to estimating the value of the coupling strength $x$, given some estimate (experimental measurement) for the simulated observable $y = f(x)$.

This one-parameter model is merely a pedagogical example.
Most nuclear collision models include multiple parameters $\xv = (x_1, x_2, \dots, x_n)$.
For example, one parameter might vary the QGP's average shear viscosity while another varies its average bulk viscosity.
They also generally predict more than one observable.
For instance, a model might simulate a large vector of observables $\yv = (y_1, y_2, \dots, y_m)$ describing, for example, the system's charged-particle yield in different centrality bins.
The model calibration problem hence seeks to estimate the model parameters $\xv$ using the simulation predictions $\yv = f(\xv)$ and their global fit to experimental measurements, described by some third vector $\yv_e$.

The model parameters $\xv$ typically correlate among each other and affect multiple observables.
Hence, if one wishes to obtain a global fit to the data, the only option is to fit all parameters simultaneously.
If there are only one or two model parameters, it is often possible to tune their values using a brute force approach or manually by hand, but when the number of parameters is large, this becomes infeasible.
A graduate student could easily spend an entire lifetime guessing and checking parameters if they live in a high-dimensional space.

The experimental data $\yv_e$ is also inherently \emph{uncertain}.
Measurement errors include, for example, statistical and systematic uncertainties such as finite event statistics and imperfect detector response.
The existence of these uncertainties naturally limits the ability to estimate $\xv$ precisely.
As a simple example, consider a ``physics model'' consisting of one parameter $x$ and one observable $y$ trivially related by
\begin{equation}
  y = c x,
\end{equation}
where $c = 0.1$ is a fixed coefficient that specifies the model.
Moreover, assume that $-1 < x < 1$ is bounded for physical reasons.
Suppose I measure the observable $y$ and find $y_e = 0.0_{-0.5}^{+0.5}$.
Can I conclude anything interesting about the parameter $x$?
Sadly, no I cannot.
The model maps the parameter range $-1 < x < 1$ to the observable range $-0.1 < y < 0.1$ which is well within the experimental uncertainty band.
Essentially all allowed values of $x$ are equally supported by the data!
However, if I blindly invert the function $y = f(x)$ ignoring this measurement uncertainty, I find a single preferred value $x = 0$.
This conclusion is of course \emph{nonsense}; there is nothing statistically special about this one value.
Measurement uncertainties are hence a crucial component of the inverse problem.

In this chapter, I describe a general statistical framework developed to address the aforementioned challenges of quantitative model-to-data comparison.
Commonly known as Bayesian parameter estimation, the procedure describes a powerful methodology to constrain the parameters of computationally intensive computer models.
I should emphasize that I did not develop the framework myself.
The groundwork was developed over time by numerous contributors \cite{OHagan:2006kll, Higdon:2008jqi, Higdon:2014tva}.
I also did not adapt the framework to heavy-ion collisions.
That effort was the subject of previous dissertations \cite{Coleman-Smith:2014hnz, Bernhard:2018hnz} and numerous published papers \cite{Novak:2013bqa, Pratt:2015zsa, Sangaline:2015isa, Bernhard:2015hxa}.
However, I use the framework extensively in this dissertation to study the properties of the QGP initial conditions \cite{Bernhard:2016tnd, Ke:2016jrd, Moreland:2018gsh}, and a firm understanding of Bayesian parameter estimation is essential to communicate these results.

The goal of this chapter is to introduce Bayesian parameter estimation using a simple example to capture the essential features of the methodology.
Disclaimer: in this chapter and throughout this dissertation, I use a suite of open source software tools developed by fellow Duke graduate student Jonah Bernhard.
His dissertation, \emph{Applying Bayesian parameter to heavy-ion collisions} \cite{Bernhard:2018hnz}, is the primary resource for this chapter, and many of my graphics are generated using his own original software.
For a more thorough discussion of Bayesian parameter estimation and its application to heavy-ion collisions, I recommend reading his dissertation.

\section{Problem statement}

The case study for this chapter is a simple vector-valued function ${\yv = f(\xv)}$ with three input parameters $\xv = (x_1, x_2, x_3)$ and ten output observables $\yv = (y_1, y_2, \dots, y_{10})$.
This mapping $f: \xv \mapsto \yv$ is assumed to model some physical process in nature.
The vector $\xv$ describes unknown attributes of the process, and the vector $\yv$ describes their measurable consequences.

Suppose that the function $f$ is described by the simple linear form
\begin{equation}
  \label{eq:linear_model}
  \yv = x_1 \mathbf{u} + (x_2 +  x_3) \vv,
\end{equation}
where $\mathbf{u}=(1, 1, \dots, 1)$ is a vector of ones, and $\vv = (0.05, 0.15, \dots, 0.95)$ is a vector that uniformly partitions the interval $[0, 1]$, both with ten elements.
Moreover, suppose that we are unable to evaluate this function exactly, but we can approximate it with some computer model $f_\text{model}: \xv \mapsto \yv + \delta \yv$, where $\delta \yv$ is a vector of uncorrelated random noise sampled from the normal distribution
\begin{equation}
  \label{eq:model_noise}
  \delta y_i = \mathcal{N}(\mu, \sigma) \quad\text{for $i=1,2,..., 10$},
\end{equation}
with mean $\mu=0$ and standard deviation $\sigma = 0.02$.
Finally, assume that this function $f_\text{model}$ is slow to evaluate, e.g.\ suppose it takes one processor hour of computation time to evaluate the function \emph{once}.

Now imagine the following scenario.
Someone secretly writes down
\begin{equation}
  x_1 = 0.3,~x_2=0.5,~\text{and}~x_3=0.7
\end{equation}
on a piece of paper.
This someone is the omniscient creator, and they feed these parameter values through the physical process modeled by the function $\yv = f(\xv)$.
An experimentalist then measures the output of this process and reports their measurement $\yv_e$ along with its uncertainties, quantified by the covariance matrix
\begin{align}
  (\Sigma_e)_{ij} &= \cov(y_i, y_j) \nonumber \\
                  &= E[(y_i - \langle y_i \rangle)(y_j - \langle y_j \rangle)],
\end{align}
where $y_i$ and $y_j$ are two observables from the vector $\yv_e$, and $E[y]$ is the expected value of $y$.
You are finally given the data $\yv_e$ and $\Sigma_e$ and tasked with estimating the true physical parameters $\xv_\text{true} = (0.3, 0.5, 0.7)$ using the computer model $\yv = f_\text{model}(\xv)$.
This chapter describes how Bayesian parameter estimation can be used to solve the problem.

\section{Parameter space}
\label{sec:parameter_space}

The aforementioned computer model maps a three-dimensional vector $\xv$ to a ten-dimensional vector $\yv$.
Here we've assume that $f_\text{model}$ is a linear equation, but imagine instead that it is something far more complicated, e.g.\ a large scale computer simulation that maps QGP initial condition parameters to simulated hadronic observables.
The function $f_\text{model}$ is thus assumed to be messy, non-algebraic, and slow to evaluate.
If we want to figure out what it does, we need to do so empirically.

With this in mind, imagine that you set out to explore the mapping $f_\text{model}: \xv \mapsto \yv + \delta \yv$.
You do not know the true parameters $\xv_\text{true}$, but you are told by the omniscient creator that each parameter $x_1, x_2, x_3 \in [0, 1]$.
You therefore decide to evaluate the function $f_\text{model}$ at lots of different parameter points $X = (\xv_1, \xv_2, \dots, \xv_d)$ uniformly distributed inside the unit-cube $[0, 1]^3$.
To distribute these points, you assign each parameter $x_1$, $x_2$, and $x_3$ to one of $k$ possible values uniformly distributed between 0 and 1.
You then take the Cartesian product of all possible parameter combinations.
This produces a matrix
\begin{equation}
  X =
  \begin{Bmatrix}
    (x_1)_1 \\
    (x_1)_2 \\
    \vdots \\
    (x_1)_k
  \end{Bmatrix}
  \times
  \begin{Bmatrix}
    (x_2)_1 \\
    (x_2)_2 \\
    \vdots \\
    (x_2)_k
  \end{Bmatrix}
  \times
  \begin{Bmatrix}
    (x_3)_1 \\
    (x_3)_2 \\
    \vdots \\
    (x_3)_k
  \end{Bmatrix},
\end{equation}
with $d = k^3$ rows and $n=3$ columns.
The row vectors $(\xv_1, \xv_2, \dots, \xv_d)$ are called \emph{design points}, and the matrix $X$ is called a \emph{design}.
This particular prescription for distributing the design points describes what's known as a \emph{factorial design}.

For the sake of our example, let's say that $k = 100$, a seemingly reasonable number for the problem at hand.
The resulting factorial design will have $d = 100 \times 100 \times 100 = 10^6$ design points.
If our function takes one processor hour to evaluate, the entire design will require ${\sim}100$ processor years!
Clearly, we need to cap the number of design points at something reasonable and find a way to distribute these points more sensibly throughout the design space.

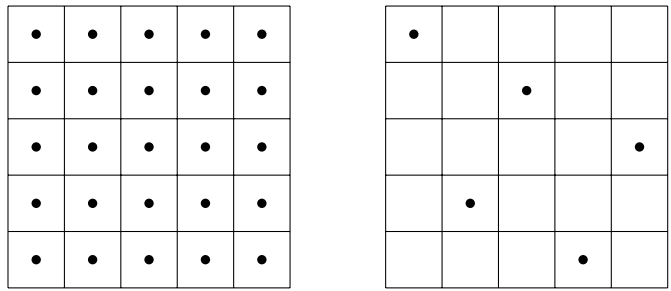
\begin{figure}
  \centering
  \begin{tikzpicture}[scale=.75]
    \draw[black, thin, step=1] (0, 0) grid +(5, 5);
    \foreach \x in {0,...,4}
      \foreach \y in {0,...,4}
      \fill[black] (\x,\y) ++(.5, .5) circle (.5ex);
  \end{tikzpicture}
  \hspace{1cm}
  \begin{tikzpicture}[scale=.75]
    \draw[black, thin, step=1] (0, 0) grid +(5, 5);
    \fill[black] (.5, 4.5) circle (.5ex);
    \fill[black] (1.5, 1.5) circle (.5ex);
    \fill[black] (2.5, 3.5) circle (.5ex);
    \fill[black] (3.5, 0.5) circle (.5ex);
    \fill[black] (4.5, 2.5) circle (.5ex);
  \end{tikzpicture}
  \medskip
  \caption{
    \label{fig:latin_hypercube}
    Left: $5 \times 5$ factorial design.
    Right: $5 \times 5$ Latin square.
    The Latin square contains one (and only one) point in each row and column.
    Both designs distribute their points uniformly across each dimension.
  }
\end{figure}

An algorithm known as Latin hypercube sampling is commonly used for this purpose \cite{Tang:1993yuu, Morris:1995tya}.
The Latin hypercube is a generalization of a Latin square to three or more dimensions.
A Latin square is a $d \times d$ grid filled with $d$ symbols (points) where each symbol appears once (and only once) in each row and column (it should be familiar to anyone who has played a game of Sudoku).
The algorithm hence distributes the symbols (points) uniformly across each row and column space.
Figure~\ref{fig:latin_hypercube} shows an example of a Latin square compared to a factorial design in two dimensions.

The Latin hypercube extends this idea for a square to an $n$-dimensional unit hypercube $[0, 1]^n$.
Unlike the factorial design where $d$ points-per-dimension requires $d^n$ total points, the Latin hypercube only requires $d$ total points.
It therefore scales linearly with the desired points-per-dimension, not exponentially.
At this point you might be wondering, ``Why not take the $n$-th root of the total number of design points and distribute them according to a factorial design? What's the advantage of the Latin hypercube algorithm?''

Often there exist model parameters that are inactive.
This means that a parameter $x$ has no affect on the observables $\yv$.
For example, let's imagine in our example that $x_1$ and $x_2$ are inactive and that $x_3$ is active, i.e.\ $x_3$ is the only parameter that affects the observables $\yv$.
Suppose that we create two separate designs: a $10 \times 10 \times 10$ factorial design with $d=10^3$ parameter points, and a Latin hypercube design with $d=10^3$ parameter points.

The parameters $x_1$ and $x_2$ are assumed to have no effect on the model so we can ignore their values.
This leaves 10 distinct values for $x_3$ in the factorial design and $10^3$ distinct values for $x_3$ in the Latin hypercube design.
This means that we wasted 99\% of our design points repeating the same 10 parameter values in the factorial design!
The Latin hypercube never repeats a parameter value so this redundancy is never an issue.

Returning to the problem at hand, let's generate a Latin hypercube design for the three model parameters $x_1$, $x_2$, and $x_3$ listed in equation~\eqref{eq:linear_model}.
I generate, for this purpose, $d=100$ parameter points using a publicly available Latin hypercube implementation written in R \cite{Carnel:2018web}.
In addition to satisfying the Latin hypercube requirements, it also attempts to optimize the sample by maximizing the minimum distance between every pair of design points (maximin criteria).
Note, the present parameters $x_1, x_2, x_3 \in [0, 1]$ lie inside a unit hypercube $[0, 1]^3$, but this is generally not the case.
It is therefore typically necessary to scale and shift the Latin hypercube design along each dimension to fit the desired parameter ranges.

\begin{figure}
  \centering
  \includegraphics{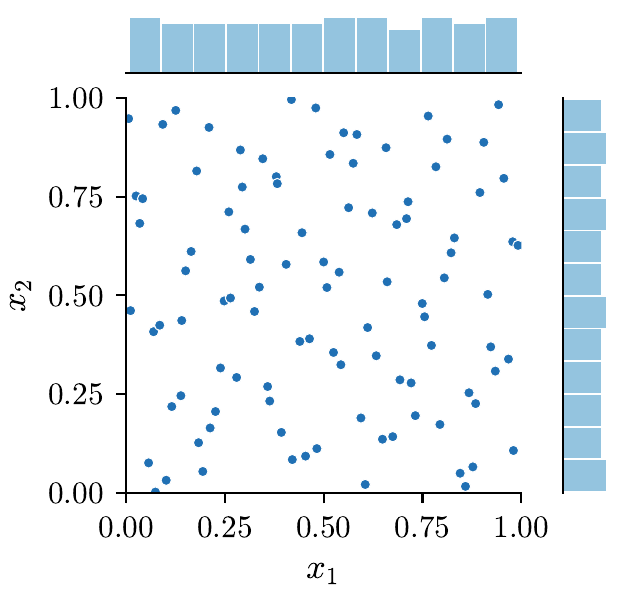}
  \caption{
    \label{fig:maximin_latin_hypercube}
    Latin hypercube design with $d=100$ design points and $n=3$ three parameters $x_1$, $x_2$, and $x_3$ distributed in the interval $x_i \in [0, 1]$.
    The figure shows the design projected onto two dimensions, $x_1$ and $x_2$, with the marginal distributions (histograms) for each dimension shown above and to the right of the design.
  }
\end{figure}

Figure~\ref{fig:maximin_latin_hypercube} shows the resulting three-dimensional design projected onto the two-dimensional subspace spanned by $x_1$ and $x_2$.
The blue histogram at the top of the figure shows the marginal distribution of parameter $x_1$, and the blue histogram on the right shows the marginal distribution of parameter $x_2$.
These distributions are uniform, illustrating a fundamental property of the Latin hypercube sampling procedure.
Note that $d=100$ points is overkill for this example; most physical problems only require \order{10} points per parameter dimension \cite{Loeppky:2009ahk}.

With the design in hand, I can then evaluate the computer model at each design point
\begin{equation}
  f_\text{model}: (\xv_1, \xv_2, \dots, \xv_d) \mapsto (\yv_1, \yv_2, \dots, \yv_d),
\end{equation}
to produce a $d \times m$ matrix $Y = (\yv_1, \yv_2, \dots, \yv_d)$ of simulated model observables.
Per specification, the computer model requires one processor hour to evaluate one point.
The entire Latin hypercube design therefore requires 100 processor hours.
Note, this number is actually reasonable compared to the 100 processor \emph{years} required by the factorial design.

\begin{figure}
  \centering
  \includegraphics{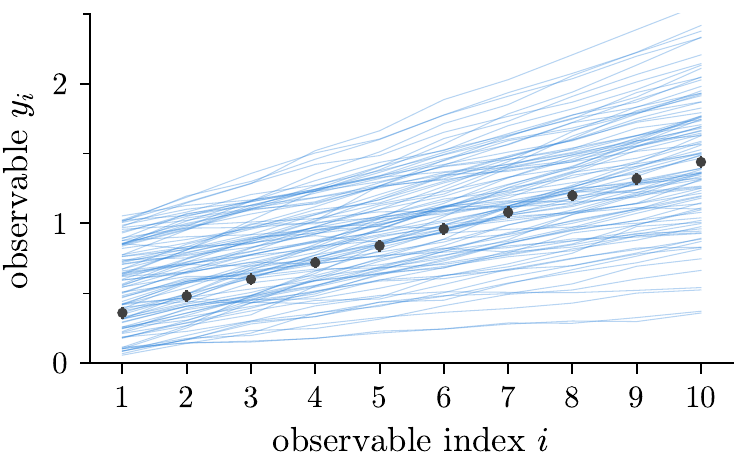}
  \caption{
    \label{fig:obs_design_example}
    Simulated observables compared to the experimental data.
    Blue lines are explicit model calculations $\yv = (y_1, y_2, \dots, y_{10})$ for each of the 100 design points, and black symbols with error bars are the mean $\yv_e$ and standard deviation $\boldsymbol{\sigma}_e = \diag(\Sigma_e)^{1/2}$ of the experimental measurement.
    The experimental covariance matrix $\Sigma_e$ is described later in the chapter, Eq.~\eqref{eq:expt_cov}.
  }
\end{figure}

Figure~\ref{fig:obs_design_example} shows the resulting computer model output at each design point (blue lines) compared to the experimental data (black symbols).
Recall that the experimental data is the output of the model using a certain set of \emph{true} parameters, $\xv_\text{true} = (0.3, 0.5, 0.7)$.
The black bars on each symbol (barely visible) are the experimental uncertainties, equal to $\boldsymbol{\sigma}_e = \diag(\Sigma_e)^{1/2}$, where $\Sigma_e$ is the experimental covariance matrix (I'll specify this matrix later in the chapter).
Notice the slight wiggle that is visible in each model calculation caused by the statistical noise term in equation~\eqref{eq:model_noise}.
The model outputs also have a large visual spread resulting from the varied parameter combinations sampled by the design matrix $X$.
Evidently, our design scaffolding nicely covers the data, and there appear to be some parameter values which describe the experimental data, although it is not yet clear what these values are.

\section{Computer model emulator}
\label{sec:emulator}

Seeing as we've already spent 100 processor hours evaluating the computer model at each design point, we can't afford to evaluate it any more times.
If we want to calculate $\yv' = f_\text{model}(\xv')$ at some new parameter point $\xv'$, we'll need to leverage our previous observations $f_\text{model}: X \mapsto Y$.
Emulators are essentially fancy interpolators that serve this purpose.

Broadly speaking, an emulator is a black box that accepts the same inputs and predicts the same outputs as the model that it emulates.
It is first trained or ``calibrated'' to reproduce a finite set of input-output observations.
Once trained, the emulator acts as a fast surrogate for the full model calculation, enabling essentially instantaneous predictions at arbitrary points in parameter space.

\subsection{Gaussian processes}

There are many different types of emulators in the literature, each having its own advantages and disadvantages.
When performing Bayesian parameter estimation, the convention is to use a specific type of emulator known as a Gaussian process (GP) emulator \cite{Rasmussen:2006gp}.

GP emulators have several strengths which make them popular for Bayesian parameter estimation:
\begin{itemize}
  \item
    They are non-parametric: GPs do not assume a specific algebraic form for the underlying function of interest.
  \item
    They readily generalize to functions with multiple input variables, i.e.\ they have multivariate support.
  \item
    They predict a \emph{distribution} of values. GPs return a mean prediction as well as an uncertainty estimate.
\end{itemize}
This last point is key.
Bayesian parameter estimation seeks to estimate the model parameters $\xv$ with quantitative uncertainty.
Emulators are naturally imperfect, and their prediction error biases parameter estimates.
In order to preserve the integrity of our parameter estimates, it's imperative that we account for our emulator error.
GPs also have one significant limitation:
\begin{itemize}
  \item They only support single-valued functions, $y = f(\xv)$.
\end{itemize}
There is, however, a simple procedure that can be used to circumvent this issue, enabling the application of GPs to vector-valued functions.
I'll postpone describing the details of this procedure till later in the chapter; for now, it suffices to say that such a work-around exists.
I'll therefore assume that we are only dealing with single-valued output for now, and I'll return to the issue of vector-valued output later.

A GP generalizes a multivariate normal distribution from the space of vectors to the space of functions.
We say that a $k$-dimensional random \emph{vector} $\yv = (y_1, y_2, \dots, y_k)$ is normally distributed if
\begin{equation}
  P(\yv) = \frac{1}{\sqrt{(2 \pi)^k | \Sigma|}} \exp \left ( -\frac{1}{2}(\yv - \boldsymbol{\mu})\tran \Sigma^{-1} (\yv - \boldsymbol{\mu}) \right ).
\end{equation}
This normal distribution is parametrized by a mean vector $\boldsymbol{\mu}$ and a positive semi-definite covariance matrix $\Sigma_{ij} = \cov(y_i, y_{j})$ where
where $y_i$ and $y_j$ are two elements of the vector $\yv$.
Hereafter, I'll write this equation more compactly using the shorthand notation
\begin{equation}
  \yv \sim \mathcal{N}(\boldsymbol{\mu}, \Sigma),
\end{equation}
to signify a normal random vector $\yv$ with mean $\boldsymbol{\mu}$ and covariance $\Sigma$.

Crudely speaking, we can also think of an arbitrary function as a vector.
For example, given some single-valued multivariate function $y = f(\xv)$, we can evaluate the function at a discrete number of points
\begin{equation}
  f: (\xv_1, \xv_2, \dots, \xv_k) \mapsto (y_1, y_2, \dots, y_k),
\end{equation}
to approximate its behavior over some finite region of its domain.
Taking the number (and density) of evaluation points to infinity, this discrete realization converges to the function $y = f(\xv)$, assuming it is continuous.
A function is thus something like a vector of infinite length.
Given enough evaluation points, it contains essentially all of the same information.

Abstractly, a GP is a multivariate normal distribution over \emph{functions}, i.e.\ it samples a continuous curve (or hypersurface) instead of a discrete vector.
This random curve is defined by a collection of random variables, and these variables are constructed such that every finite number of them shares a multivariate normal distribution.
Numerical implementations naturally cannot generate continuous functions, so in practice these curves are discretized.

Let $X = (\xv_1, \xv_2, \dots, \xv_k)$ be some points that discretize each curve.
Given these points, a GP predicts a random vector of corresponding function values $\yv_p = (y_1, y_2, \dots, y_k)$ that is distributed according to a multivariate normal distribution
\begin{equation}
  \label{eq:norm_dist}
  \yv_p \sim \mathcal{N}(\boldsymbol{\mu}, \Sigma).
\end{equation}
This property also implies that the GP output $y_p$ at a single point $\xv$ is a normal random variable with some mean value $\mu$ and variance $\sigma^2 = \cov(y_p, y_p)$.

Before a GP can be sampled to make predictions, it must be conditioned on some training data to determine the values of $\boldsymbol{\mu}$ and $\Sigma$ in equation~\eqref{eq:norm_dist}.
This conditioning process requires three additional ingredients:
\begin{enumerate}
  \item a list of training inputs, $X_t = (\xv_1, \xv_2, \dots, \xv_d)$, where each input is an $n$-dimensional vector,
  \item a list of training outputs, $\yv_t = (y_1, y_2, \dots, y_d)$, where each output is a single number, and
  \item an assumed covariance function $k(\xv, \xv') = \cov(y_p(\xv), y_p(\xv'))$
    which describes the similarity of the predicted emulator outputs $y_p(\xv)$ and $y_p(\xv')$ as a function of the emulator inputs $\xv$ and $\xv'$.
  \end{enumerate}
  Note, I've used a subscript $t$ to label training data and a subscript $p$ to label prediction data following the notation in reference \cite{Bernhard:2018hnz}, e.g.\ $y_t$ is a training point output, and $y_p$ is an emulator prediction.
This will help me distinguish between both types of variables when they appear side-by-side.
I'll stick to this convention throughout the remainder of the chapter.

Expressed as a single function call, we can now write the emulator prediction $y_p$ at the parameter point $\xv$ compactly using the following notation
\begin{equation}
  \label{eq:gp_func}
  y_p = \mathcal{GP}(\xv; X_t, \yv_t, k(\xv, \xv')).
\end{equation}
The arrays $X_t$ and $\yv_t$ on the right-side are the training data used to condition the emulator, while $k(\xv, \xv')$ is the GP covariance function.
The conditioning procedure that calculates $\boldsymbol{\mu}$ and $\Sigma$ for use in equation~\eqref{eq:norm_dist} given these variables is somewhat technical, and I won't describe it here, but a detailed discussion can be found in \cite{Rasmussen:2006gp, Bernhard:2018hnz}.
Rather, let me now pivot to some example visualizations which should make things more clear.

For the sake of simplicity, let me temporarily restrict our attention to single-variate GPs which are easier to visualize than multivariate GPs.
Suppose that you are given the training outputs $\yv_t=(y_1, y_2, \dots, y_d)$ at several training inputs $\xv_t=(x_1, x_2, \dots, x_d)$.
Moreover, suppose that you are also given the simple covariance function
\begin{equation}
  k(x, x') = \sigma_f^2 \exp \bigg ( {-}\frac{|x - x'|^2}{2 \ell^2} \bigg ) + \sigma_n^2 \delta(x - x'),
  \label{eq:example_covariance}
\end{equation}
for the similarity of $y_p(x)$ and $y_p(x')$ as a function of $x$ and $x'$.
The right-side of this equation consists of two separate terms.
The term on the left is a squared-exponential kernel with autocovariance $\sigma_f^2$ and correlation length $\ell$, and the term on the right is an uncorrelated white noise kernel with variance $\sigma_n^2$.
The squared exponential term asserts that the function is smoothly varying, and the white noise term allows for some additional emulator wiggle room at each training point.
This general two-component form is a common choice for real world applications, and I'll use a similar variant of it later in the text.

\begin{figure}[t]
  \centering
  \makebox[\textwidth]{
    \includegraphics{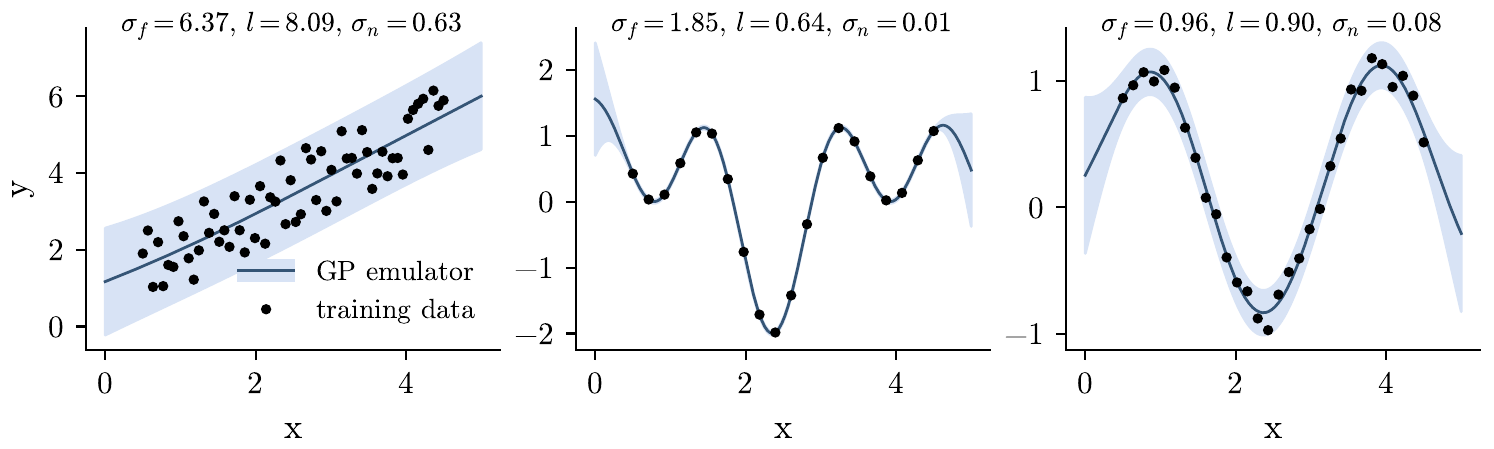}
  }
  \caption{
    \label{fig:example_gp}
    GPs trained on several different datasets (black symbols) using the covariance function \eqref{eq:example_covariance}.
    The dark blue line is the emulator mean prediction, and the shaded band is $\pm 2\sigma$ uncertainty.
    The maximum likelihood covariance hyperparameters $\sigma_f$, $\ell$, and $\sigma_n$ are listed above each figure.
  }
\end{figure}

Figure~\ref{fig:example_gp} shows several GP realizations conditioned to various datasets (black symbols) using the covariance function \eqref{eq:example_covariance}.
The emulators are trained using a GP module bundled with \textsc{scikit-learn}, a Python machine learning library \cite{sklearn}.
The emulator mean predictions are shown as solid blue lines and their $\pm 2 \sigma$ uncertainties as shaded blue bands.
The software automatically estimates maximum likelihood values for the covariance hyperparameters $\sigma_f$, $\ell$, and $\sigma_n$ which are annotated along the top of each figure.
For more information on the maximum likelihood hyperparameter calibration, see chapter 4 section 3.3 of reference~\cite{Bernhard:2018hnz}.

Notice how the covariance hyperparameters $\sigma_f$, $\ell$, and $\sigma_n$ affect the behavior of each GP.
For example, the left panel shows training data that's noisy, but slowly varying over the full range of the plot.
The preferred correlation length $\ell$ and noise term $\sigma_n$ are hence large.
Conversely, in the middle figure, the data points bend and twist rapidly over the plot range with very little apparent randomness.
The correlation length, $\ell$ and noise term $\sigma_n$ are therefore small.
The figure on the right, meanwhile, shows an intermediate example combining features of the first two examples.

I want to pause for a moment to explain how impressive this is.
I did not tell the GP what kind of function it was emulating.
Nor did I specify its covariance hyperparameters, e.g.\ the correlation length $\ell$ or statistical noise level $\sigma_n$.
I merely provided the \emph{form} of the GP covariance function and conservative \emph{bounds} for each hyperparameter.
The GP module was able to infer the optimal hyperparameters using its own internal maximum likelihood optimization routine.
Remarkably, it determined a sensible mean prediction and uncertainty band for three dramatically different datasets.
This is what people mean when they refer to GPs as flexible non-parametric functions.
Such flexibility is naturally vital if the framework is to generalize.

\subsection{Procedure for multiple correlated outputs}

GP emulators are scalar-valued functions.
Our toy-model in equation~\eqref{eq:linear_model}, however, is \emph{vector}-valued.
It has three input parameters $\xv = (x_1, x_2, x_3)$ and ten outputs $\yv = (y_1, y_2, \dots, y_{10})$.
This section describes a standard procedure which is used to apply GP emulators to vector-valued functions.

Consider, for the moment, the following vector-valued function
\begin{equation}
  \begin{pmatrix}
    y_1 \\
    y_2
  \end{pmatrix}
  =
  \begin{pmatrix}
    3\, x_1 \\
    2\, x_2
  \end{pmatrix}.
\end{equation}
This function consists of two independent equations, $y_1 = 3\, x_1$ and $y_2 = 2\, x_2$.
These equations are decoupled, so we are free to train two independent GP emulators, one for each scalar-valued equation.
Now consider the slightly more complicated function
\begin{equation}
  \label{eq:coupled_eqns}
  \begin{pmatrix}
    y_1 \\
    y_2
  \end{pmatrix}
  =
  \begin{pmatrix}
    x_1  + x_2 \\
    x_1 - x_2
  \end{pmatrix}.
\end{equation}
This example consists of two \emph{coupled} equations $y_1 = x_1 + x_2$ and $y_2 = x_1 - x_2$.
Both equations depend on the parameters $x_1$ and $x_2$ so they cannot be emulated independently.

If we want to model this function with two independent emulators, we need to first transform the function into a new basis representation where the resulting equations are decoupled.
Let's define two orthogonal variables, $z_1 = (y_1 + y_2)/2$ and $z_2 = (y_1 - y_2)/2$ expressed as functions of $y_1$ and $y_2$.
This transformation describes a linear operation
\begin{equation}
  \begin{pmatrix}
    z_1 \\
    z_2
  \end{pmatrix}
  =
  \begin{pmatrix}
    1/2 & 1/2 \\
    1/2 & -1/2
  \end{pmatrix}
  \begin{pmatrix}
    y_1 \\
    y_2
  \end{pmatrix},
  \label{eq:linear_opp}
\end{equation}
on the vector $\yv = (y_1, y_2)$.
In the transformed basis representation, equation~\eqref{eq:coupled_eqns} becomes
\begin{equation}
  \begin{pmatrix}
    z_1 \\
    z_2
  \end{pmatrix}
  =
  \begin{pmatrix}
    x_1 \\
    x_2
  \end{pmatrix}.
\end{equation}
The variables $z_1$ and $z_2$ are now \emph{decoupled}, i.e.\ they vary independently as functions of $x_1$ and $x_2$.
This allows us to train two separate GPs
\begin{equation}
  \label{eq:gp_emu}
  \begin{pmatrix}
    {z_p}_1 \\
    {z_p}_2
  \end{pmatrix}
  =
  \begin{pmatrix}
    \mathcal{GP}_1(x_1, x_2) \\
    \mathcal{GP}_2(x_1, x_2)
  \end{pmatrix},
\end{equation}
to emulate the function inputs and outputs.
Recall that the subscript $p$ here means that the variable is an emulator prediction.

Suppose now that we want emulate the vector $\yv = (y_1, y_2)$.
We simply emulate $z_1$ and $z_2$ using equation~\eqref{eq:gp_emu} then transform the emulated variables ${z_p}_1$ and ${z_p}_2$ back to the original basis
\begin{equation}
  \begin{pmatrix}
    {y_p}_1 \\
    {y_p}_2
  \end{pmatrix}
  =
  \begin{pmatrix}
    1  &  1  \\
    1  & -1
  \end{pmatrix}
  \begin{pmatrix}
    {z_p}_1 \\
    {z_p}_2
  \end{pmatrix}.
\end{equation}
When expanded, this yields
\begin{align}
  {y_p}_1 &= \mathcal{GP}_1(x_1, x_2) + \mathcal{GP}_2(x_1, x_2), \\
  {y_p}_2 &= \mathcal{GP}_1(x_1, x_2) - \mathcal{GP}_2(x_1, x_2).
\end{align}
Needless to say, realistic vector-valued functions are far more complicated than this simple example.
When a function has many correlated outputs, we need to use a more sophisticated procedure to decouple the output variables.

\subsubsection{Principal component analysis}

Before proceeding, let me briefly define some additional notation which will be important for the discussion that follows.
Throughout this dissertation I've used regular type lowercase letters to denote scalars, bolded lowercase letters to denote vectors, and regular type capital letters to denote matrices.
This means that $y$ is a scalar, $\mathbf{y}$ is a vector, and $Y$ is a matrix.
When two objects share the same letter, e.g.\ $\mathbf{y}$ and $Y$, it's been assumed that they are related, i.e.\ $\mathbf{y}$ is some vector from $Y$.

In this section, I'll also need to distinguish between row vectors and column vectors.
If \emph{not otherwise specified}, it should be assumed that $i$ is a row index and $j$ is a column index.
Moreover, indices that are not specified are assumed to be arbitrary.
For example, given some matrix $Y$:
\begin{itemize}
  \item $\yv_i$ is a row vector,
  \item $\yv_j$ is a column vector,
  \item $y_i$ is the element in the $i$th row of some column $j$,
  \item $y_j$ is the element in the $j$th column of some row $i$, and
  \item $y_{ij}$ is the element in row $i$ and column $j$.
\end{itemize}
With this comment about notation out of the way, let's return to the topic at hand, functions with many correlated output variables.

Bayesian parameter estimation commonly uses a statistical procedure known as principal component analysis (PCA) to decouple correlated model outputs.
Suppose that you are given a $d \times m$ matrix $Y$.
Let the variables $y_j$ and $y_{j'}$ denote the matrix elements in columns $j$ and $j'$ of two random row vectors $\yv_i$ and $\yv_{i'}$.
Now assume that these variables are non-trivially correlated:
\begin{equation}
  \corr(y_j, y_{j'}) \neq 0, \quad\text{for}~j \neq j'.
\end{equation}

The PCA transformation changes the basis of $Y$ to produce a new $d \times m$ matrix $Z$.
It sends each row vector $\yv_i = (y_{i1}, y_{i2}, \dots, y_{im})$ to a new row vector $\zv_i = (z_{i1}, z_{i2}, \dots, z_{id})$, where each element $z_{ij}$ is a linear combination of the elements of $\yv_i$.
The transformation is constructed such that it removes all linear correlations between the different columns of $Z$.
For example, if $z_j$ and $z_{j'}$ are the values in columns $j$ and $j'$ of two random row vectors, $\zv_i$ and $\zv_{i'}$, then
\begin{equation}
  \corr(z_j, z_{j'}) = 0, \quad\text{for}~j \neq j'.
\end{equation}

This linear transformation is described by an $m \times m$ matrix $V$ which multiplies $Y$ from the right,
\begin{equation}
  \label{eq:pca}
  Z = Y V.
\end{equation}
The coefficients $(z_1, z_2, \dots, z_d)$ of each row vector $\zv$ are called the \emph{principal components} (PCs) of the transformation, and they are ranked in order of explained variance---that is to say that each PC describes the maximal variance possible subject to the constraint that its basis vector remains orthogonal to the basis vectors of the preceding components.
The first variable $z_1$ is called the first principal component, the second variable $z_2$ the second principal component, and so on.

The transformation matrix $V$ is calculated using the singular value decomposition (SVD) of the matrix $Y$.
Since we are only concerned with real-valued output, let's assume that $Y$ is real.
Taking the SVD of the real matrix $Y$ then yields
\begin{equation}
  \label{eq:svd}
  Y = U \Sigma V\tran,
\end{equation}
where $U$ is a $d \times d$ orthogonal matrix, $\Sigma$ is a $d \times m$ rectangular diagonal matrix, and $V$ is an $m \times m$ orthogonal matrix.
The diagonal entries of $\Sigma$ are called the \emph{singular values} of $Y$, while the columns of $U$ and the columns of $V$ are called the left and right \emph{singular vectors}.
Multiplying both sides of this equation by $V$, we see that
\begin{equation}
  Y V = U \Sigma,
\end{equation}
which performs the desired PCA decomposition.
The columns of the matrix $Z = U \Sigma$ are uncorrelated and sorted in order of decreasing variance.
The matrix $V$ of right singular vectors is hence the desired PCA transformation matrix.

This is all very nice, but it's not particularly illuminating.
In order to see why PCA is \emph{useful}, let's apply the transformation to a simple dataset in order to visualize its effect.
Consider, for this purpose, the bivariate normal distribution
\begin{equation}
  P(\yv) = \frac{1}{2 \pi \sqrt{|\Sigma_y|}} \exp \left ( -\frac{1}{2} (\yv - \boldsymbol{\mu})\tran \Sigma_y^{-1} (\yv - \boldsymbol{\mu}) \right ),
  \label{eq:mult_gauss}
\end{equation}
of a random two-component vector $\yv=(y_1, y_2)$.
This distribution is parametrized by a mean vector $\boldsymbol{\mu}$ and a covariance matrix
\begin{equation}
  \Sigma_y =
  \begin{pmatrix}
    \cov(y_1, y_1) & \cov(y_1, y_2) \\
    \cov(y_2, y_1) & \cov(y_2, y_2)
  \end{pmatrix}.
\end{equation}
If you are confused by the subscript $y$ on the covariance matrix, its purpose will become clear in a moment.
Consider now some arbitrary parameters for this distribution, namely
\begin{equation}
  \boldsymbol{\mu} = (0, 0)\quad\text{and}\quad
  \Sigma_y =
  \begin{pmatrix}
    1 & 0.8 \\
    0.8 & 1
  \end{pmatrix}.
  \label{eq:cov_form}
\end{equation}
Let's sample the distribution $1000$ times and concatenate the samples into a $1000 \times 2$ matrix $Y = (\yv_1, \yv_2, \dots, \yv_{1000})$.

\begin{figure}[t]
  \centering
  \makebox[\textwidth]{
    \includegraphics{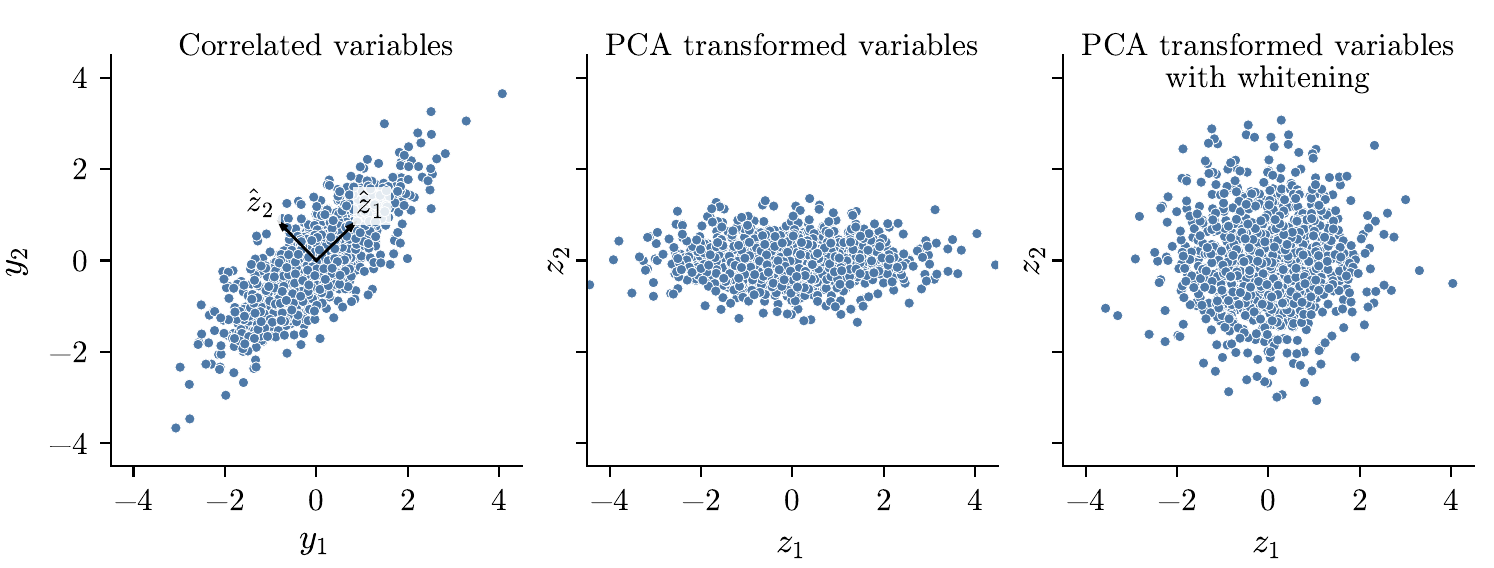}
  }
  \caption{
    \label{fig:pca}
    Left: Scatter plot of $1000$ vectors $Y = (\yv_1, \yv_2, \dots, \yv_{1000})$ randomly sampled from the bivariate normal distribution \eqref{eq:mult_gauss} with parameters $\boldsymbol{\mu}$ and $\Sigma_y$ from Eq.~\eqref{eq:cov_form}.
    The black arrows $\hat{\zv}_1$ and $\hat{\zv}_2$ are the basis vectors of the PC transformation. Middle: Scatter plot of $1000$ vectors $Z = (\zv_1, \zv_2, \dots, \zv_{1000})$ obtained by applying the PCA transformation to the vectors of $Y$.
    Right: Same as the middle figure, but with each PC vector whitened to unit variance.
  }
\end{figure}

The left-side of figure~\ref{fig:pca} shows the scatter plot of these samples.
Each blue symbol is a random vector $\yv= (y_1, y_2)$ sampled from the distribution.
The covariance matrix \eqref{eq:cov_form} has non-zero off-diagonal elements, so $y_1$ and $y_2$ are linearly correlated.
Let's apply the PCA transformation now to the matrix $Y$ and plot the result.
The resulting PC basis vectors, $\hat{\zv}_1$ and $\hat{\zv}_2$, are plotted as black arrows on top of the original vectors $(\yv_1, \yv_2, \dots, \yv_{1000})$.
The PC basis is orthogonal by construction, so $\hat{\zv}_1 \perp \hat{\zv}_2$.
The first PC vector points in the direction of largest variance, while the second PC vector points in the direction of second largest variance.

Now look at the middle-panel of figure~\ref{fig:pca}.
This panel shows the result of the PCA transformation applied to the matrix $Y$.
Each vector $\yv=(y_1, y_2)$ is sent to a new vector $\zv=(z_1, z_2)$, where linear correlations between $z_1$ and $z_2$ vanish.
The PCA transformation is linear, so we can easily reexpress the covariance matrix $\Sigma_y$ in the basis of the transformed variables.
It equals
\begin{equation}
  \Sigma_z = V\tran \Sigma_y V,
\end{equation}
where $V$ is the PCA transformation matrix defined by equation~\eqref{eq:svd}.
The PCs of different order are uncorrelated, so this matrix is diagonal.
Specifically,
\begin{equation}
  \Sigma_z = \diag(\sigma_{z_1}^2, \sigma_{z_2}^2),
\end{equation}
where $\sigma_{z_1}^2=1.8$ and $\sigma_{z_2}^2=0.2$ are the variances along the directions of PCs $z_1$ and $z_2$ respectively.
Notice that $\sigma_{z_1}^2 \geq \sigma_{z_2}^2$ as required: the PCs of the transformation are always sorted in order of decreasing variance.

Often, when applying PCA to real world problems, it is common to \emph{whiten} the PC vectors.
Whitening divides each PC by its standard deviation to produce a covariance matrix that is unit diagonal.
Specifically, it sends the $d \times m$ matrix of PCs
\begin{equation}
  Z \to Z W,
\end{equation}
where $W$ is an $m \times m$ whitening matrix
\begin{equation}
  W = \diag(\sigma_{z_1}^{-1}, \sigma_{z_2}^{-1}, \dots, \sigma_{z_m}^{-1}),
\end{equation}
and $\sigma_{z_j}$ is the standard deviation of the $j$th column of $Z$.
Combining this with equation~\eqref{eq:pca}, the full PCA transformation with whitening is
\begin{equation}
  Z = Y V W.
\end{equation}
The right-side of figure~\ref{fig:pca} shows the result of the whitened PCA transformation applied to the matrix $Y$.
Notice how the whitened PCs now have unit variance.
Whitening is thus commonly used to standardize the input of a downstream machine learning estimator.

\begin{remark}
  Whitening is applied internally by the \textsc{sklearn} PCA transformer.
  I always work with the whitened PCs in this dissertation, and it is cumbersome to keep track of the whitened and non-whitened PCs separately.
  Therefore, I define $Z$ to be the matrix of whitened PCs and $V$ to be the matrix which performs the PCA transformation with whitening.
\end{remark}

\subsection{Emulator calibration}

Now that I've described how PCA can be used to decouple our model outputs, let me return to the original task at hand.
I've digressed from this problem, so let me briefly jog your memory.
We've been tasked with estimating the parameters of a vector-valued function $\yv = f_\text{model}(\xv)$ described by equation~\eqref{eq:linear_model}.
The function is a computer model with three inputs $\xv = (x_1, x_2, x_3)$ and ten outputs $\yv = (y_1, y_2, \dots, y_{10})$.
Several experimentalists have measured the physical process $f: \xv \mapsto \yv$ modeled by the function $f_\text{model}$, and they've reported their measurement $\yv_e$ along with a covariance matrix $\Sigma_e$.
The physical process has some \emph{true} parameters $\xv_\text{true} = (0.3, 0.5, 0.7)$ which have been kept secret, and we've been tasked with estimating these unknown parameters using the experimental data.

In section \ref{sec:parameter_space} we evaluated the computer model at $d=100$ parameter points $X = (\xv_1, \xv_2, \dots, \xv_d)$ selected to fill the space of reasonable parameter combinations.
This required significant computing resources (the model is slow), so we've decided to emulate the discrete mapping $f_\text{model}: X \mapsto Y$.
Once this emulator is trained, we'll be able to rapidly predict the model output $\yv = f_\text{model}(\xv)$ at any point $\xv$ in the parameter space.
This will enable a thorough exploration of the three-dimensional parameter space using Bayesian methods.
I'll now describe how this emulator is assembled and calibrated using the existing training data.

First, let's take our $d \times m$ matrix $Y$ of simulated model outputs, whose $d=100$ rows correspond to design points and $d=10$ columns to model observables, and center the data by subtracting off the mean of each column
\begin{equation}
  \yv_j \to \yv_j - \langle \yv_j \rangle.
\end{equation}
This is a necessary preprocessing step for PCA, and it is applied automatically by the \textsc{scikit-learn} PCA transformer \cite{sklearn}.
Next, let's divide out the standard deviation $\sigma_{y_j}$ of each observable column $\yv_j$
\begin{equation}
  \yv_j - \langle \yv_j \rangle \to \frac{\yv_j - \langle \yv_j \rangle}{\sigma_{y_j}},
\end{equation}
so that every column has unit variance.
This is another common machine learning procedure that places every observable (column) on the same general scale.
Without it, I would be able to change the form of the PCA decomposition simply by changing the units of each observable which is clearly undesirable.
Now, let's define $\hat{Y}$ to be the matrix of scaled observables
\begin{equation}
  \label{eq:scaler}
  \hat{Y} = (Y - \bar{Y}) S^{-1},
\end{equation}
constructed by applying the aforementioned transformation to each of the columns of $Y$.
Here I use the $d \times m$ matrix $\bar{Y}_{ij} = \langle \yv_j \rangle$ to recenter the observables, and the $m \times m$ diagonal matrix
\begin{equation}
  S = \diag(\sigma_{y_1}, \sigma_{y_2}, \dots, \sigma_{y_m}),
\end{equation}
to divide out their standard deviations $\sigma_{y_j}$ and rescale them.

The next step is to use PCA to transform the $d \times m$ matrix $\hat{Y}$ into a new $d \times m$ matrix $Z$ of whitened PCs
\begin{equation}
  \label{eq:pca_transform}
  Z = \hat{Y} V.
\end{equation}
The first column $\zv_1$ of $Z$ is the first PC at every design point, the second column $\zv_2$ is the second PC at every design point, and so on.
Each PC column vector $\zv_j$ is linearly uncorrelated with the other PC column vectors $\zv_{j'} \neq \zv_j$, so we can train a single GP to emulate each column separately.
Collectively, this forms a probabilistic vector-valued emulator:
\begin{equation}
  \label{eq:pca_emu}
  \zv_p \sim
  \begin{pmatrix}
    {z_p}_1 \\
    {z_p}_2 \\
    \vdots \\
    {z_p}_{10}
  \end{pmatrix}
  =
  \begin{pmatrix}
    \mathcal{GP}_1(\xv; X_t, {\zv_t}_1, k(\xv, \xv')) \\
    \mathcal{GP}_2(\xv; X_t, {\zv_t}_2, k(\xv, \xv')) \\
    \vdots \\
    \mathcal{GP}_{10}(\xv; X_t, {\zv_t}_{10}, k(\xv, \xv'))
  \end{pmatrix},
\end{equation}
for the PC vector $\zv_p = ({z_p}_1, {z_p}_2, \dots, {z_p}_{10})$ as a function of the model input parameters $\xv = (x_1, x_2, x_3)$.
Here $X_t$ is the $d \times m$ matrix of training inputs, ${\zv_t}_j$ is the $d \times 1$ vector of training outputs of the $j$th PC, and $k(\xv, \xv')$ is the GP covariance function.
Note, technically $\zv_p$ is an $1 \times 10$ row vector here.
I've simply written $\zv_p$ as a column vector for notational convenience.

In principle, we could train ten independent GP emulators---one for each PC---but ten PCs would be overkill for the problem at hand.
The first two PCs describe 91.728\% and 8.008\% of the model's total output variance respectively, while PCs 3--10 describe the remaining 0.264\%.
Emulating the first two PCs therefore reproduces essentially all of the model variance.
While the first two PCs include meaningful variance, e.g.\ variance that arises from sensitivity to the model input parameters, PCs 3--10 include \emph{meaningless} variance, i.e.\ they are essentially random statistical noise.
This noise results from the statistical noise of the model, introduced by equation~\eqref{eq:model_noise}.
PCA is an information preserving transformation so this statistical noise has to go somewhere.
PCs 3--10 are essentially a projection of this noise onto random orthogonal vectors which complete the basis of the linear transformation.

GP implementations will occasionally fail to determine the correct hyperparameters when fitting noisy data, so it is generally safer to model the especially noisy PCs using the sample mean and sample variance of each PC over the entire design.
Hence, if a PC is pure noise, it samples a normal distribution $\mathcal{N}(\mu, \sigma^2)$ with mean $\mu=0$ and variance $\sigma^2=1$, the mean and variance of each PC after whitening.
This is a conservative approximation that over predicts the emulator and model uncertainties.
Equation~\eqref{eq:pca_emu} may then be written as
\begin{equation}
  \label{eq:pca_emu_approx}
  \zv_p \sim
  \begin{pmatrix}
    {z_p}_1 \\
    {z_p}_2 \\
    {z_p}_3 \\
    \vdots \\
    {z_p}_{10}
  \end{pmatrix}
  =
  \begin{pmatrix}
    \mathcal{GP}_1(\xv; X_t, {\zv_t}_1, k(\xv, \xv')) \\
    \mathcal{GP}_2(\xv; X_t, {\zv_t}_2, k(\xv, \xv')) \\
    \mathcal{N}(0, 1) \\
    \vdots \\
    \mathcal{N}(0, 1)
  \end{pmatrix},
\end{equation}
using $\mathcal{N}(0, 1)$ to replace each noise-dominated GP emulator prediction.

Suppose now that I want to sample the emulator prediction $\yv_p$ at some parameter point $\xv$.
The first step is to sample the emulated PC vector $\zv_p(\xv)$ defined by equation~\eqref{eq:pca_emu_approx}.
I can then revert the PC transformation \eqref{eq:pca_transform} and the scaling and centering transformations \eqref{eq:scaler} in sequence yielding,
\begin{equation}
  \label{eq:emu_pred}
  \yv_p = \zv_p V\tran S + \bar{\yv},
\end{equation}
where $\bar{\yv}$ is now a vector since we are recentering a single model output.

Similarly, if I want to sample the predicted emulator output at $k$ parameter points, I can sample $\zv_p$ at those points and concatenate the samples into a $k \times 10$ matrix $Z_p$, where each row of $Z_p$ is the PC vector at a single point.
I can then transform all of the PC vectors in one pass,
\begin{equation}
  Y_p = Z_p V\tran S + \bar{Y}.
\end{equation}

Moreover, I can also use this procedure to directly calculate the \emph{mean} emulator output $\langle \yv_p \rangle$ at a certain parameter point $\xv$.
The mean vector in PC space is
\begin{equation}
  \langle \zv_p(\xv) \rangle = (\mu_1(\xv), \mu_2(\xv), 0, \dots, 0),
\end{equation}
where $\mu_1(\xv)$ and $\mu_2(\xv)$ are the means predicted by GP1 and GP2 respectively.
The other components are zero since they sample a normal distribution with zero mean and unit variance.
Equation~\eqref{eq:emu_pred} can then be used to transform the mean vector $\langle \zv_p(\xv) \rangle$ into the mean vector $\langle \yv_p(\xv) \rangle$, equal to the average emulator prediction at the point $\xv$.

\subsection{Emulator uncertainty}
\label{subsec:emu_uncertainty}

Additionally, we can also transform the emulator uncertainty on $\zv_p(\xv)$ to calculate the emulator uncertainty on $\yv_p(\xv)$.
The variance of the PC vector $\zv_p(\xv)$ is given by
\begin{equation}
  \label{eq:stdz}
  \mathrm{var}(\zv_p(\xv)) = (\sigma_{z_1}^2(\xv), \sigma_{z_2}^2(\xv), 1, \dots, 1),
\end{equation}
where $\sigma_{z_1}^2(\xv)$ and $\sigma_{z_2}^2(\xv)$ are the predictive variances of PCs ${z_p}_1$ and ${z_p}_2$ respectively at the point $\xv$.
Here I've dropped the subscript $p$ from the right-side of equation~\eqref{eq:stdz} because I'm running out of room.
The other PCs have zero mean and unit variance across the entire design, so their variances are set to one.
The individual components of $\zv_p$ are all linearly uncorrelated, so the resulting covariance matrix is diagonal:
\begin{equation}
  \label{eq:pc_cov}
  \Sigma_z(\xv) = \diag[\sigma_{z_1}^2(\xv), \sigma_{z_2}^2(\xv), 1, \dots, 1].
\end{equation}
Again, $\sigma_{z_i}^2(\xv)$ is the \emph{predictive} variance that's returned by the GP emulator at a single parameter point, not to be confused with $\sigma_{z_i}^2$, the variance of that PC over all design points.
The former is generally smaller than unity, while the latter is unity by construction; recall that each PC is whitened to unit variance.

The PCA transformation is linear, so the covariance matrix \eqref{eq:pc_cov} is easily reexpressed in the original basis of the model observables $\yv$.
Performing this change of coordinates yields
\begin{equation}
  \label{eq:cov_transform}
  \Sigma_y(\xv) = Q \Sigma_z(\xv) Q\tran,
\end{equation}
where $Q = S^{-1} V$ is the pair of transformations performed by the scaling matrix $S$ \eqref{eq:scaler} and the whitened PCA transformation $V$ \eqref{eq:pca_transform}.
The centering matrix $\bar{Y}$ does not contribute here since the overall mean has no effect on the covariance matrix.
Note, while $\Sigma_z$ is diagonal, the matrix $\Sigma_y$ is generally \emph{non}-diagonal.
Applying the inverse PCA transformation reintroduces all the observable correlations that were originally removed by the forward PCA transformation.

\begin{figure}
  \centering
  \makebox[\textwidth]{
    \includegraphics{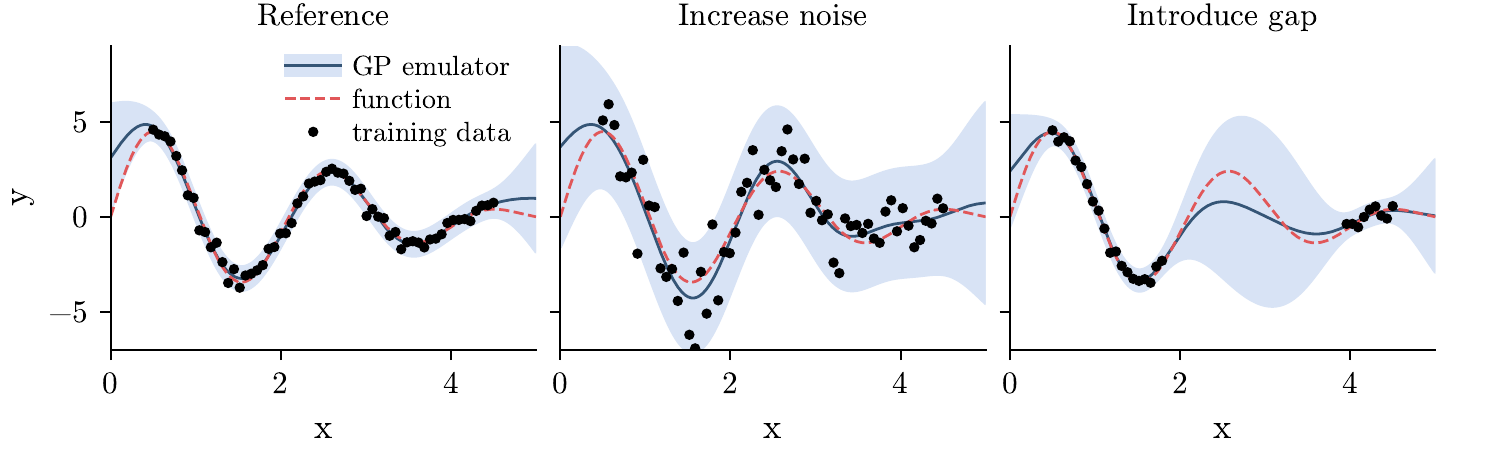}
  }
  \caption{
    \label{fig:gp_interp}
    The reference figure on the left shows a GP (blue line and error band) conditioned on some training points (black symbols) sampled from a function (red line) with added statistical noise.
    The setup in the middle figure is identical to that of the reference figure, but with larger statistical noise added to the training data.
    The figure on the right, meanwhile, is also identical to the reference figure, but this time with a large chunk of training points removed from the middle.
  }
\end{figure}

So where does the emulator uncertainty come from?
Generally speaking, the emulator is only as good as its training data.
If the training data is noisy, or there are too few parameter points, then the emulator uncertainty will be large.
To get a better feel for why this happens, let's look at a specific example using a GP emulator in one dimension.
The left panel of figure~\ref{fig:gp_interp} shows some training data (black points) sampled from a function (red dashed line) with added statistical noise.
A GP is then trained to emulate the black points using the covariance function from equation~\eqref{eq:example_covariance}.
The blue line is the GP's mean prediction, and the blue band is the region containing $\pm 2 \sigma$ uncertainty.

Using this training data and GP as a reference point, let's increase the statistical noise of the training data.
The middle panel of figure~\ref{fig:gp_interp} shows the qualitative effect of this change.
The emulator uncertainty clearly grows to accommodate the larger variance of the training data.
In a similar fashion, we can also investigate what happens when the emulator is forced to interpolate over a larger distance.
The right panel of figure~\ref{fig:gp_interp} shows the effect of removing a large chunk of training points from the middle of the dataset to create a gap.
This causes the GP uncertainty to swell in the middle of the gap where its predictions are furthest from the neighboring training points.

The overall emulator uncertainty is therefore reduced by either running the computer model more times to reduce statistical noise in the model output at each design point, or by using more design points to produce a denser scaffolding of the parameter space.
Both require more computing resources, so there is generally a trade off to be made in optimizing each variable separately.
I do not analyze these trade offs in the present dissertation, but it would nevertheless an interesting topic for future study.

\subsection{Emulator validation}
\label{subsec:emu_validation}

The emulator will serve as a stand-in for the physics model so it is imperative to check that it is working correctly.
The emulator is validated by checking its predictions against new inputs and outputs that were excluded from the calibration process.
In the parlance of machine learning, we say that this validation data is \emph{out of sample}.
Generally speaking, the emulator is working properly if it makes statistically accurate statements.
This is somewhat loaded terminology, so let me elaborate with a concrete example.

Imagine that you have a model which generates probabilistic predictions for the value of some variable $y$.
Suppose, for example, that this variable is the point-total of two teams in a basketball game.
Moreover, assume that each prediction is a normal random variable $\mathcal{N}(\mu, \sigma)$, parametrized by some mean value $\mu$ and standard deviation $\sigma$.

Now imagine that you run this basketball model to predict the outcome of $k$ different games which are each part of some future tournament.
These predictions form a list of $k$ normal distributions
\begin{equation}
  \text{predictions} = [\mathcal{N}(\mu_1, \sigma_1), \mathcal{N}(\mu_2, \sigma_2), \dots, \mathcal{N}(\mu_k, \sigma_k)],
\end{equation}
where $\mathcal{N}(\mu_i, \sigma_i)$ is the probabilistic prediction of the $i$th game.
After the tournament ends, you record the point-total of each game and construct a list of corresponding observations
\begin{equation}
  \text{observations} = [y_1, y_2, \dots, y_k],
\end{equation}
where each observation is the outcome of a single prediction.

If the basketball point-total prediction model is truthful, then each outcome $y_i$ is a ``sample'' of its predicted distribution $\mathcal{N}(\mu_i, \sigma_i)$.
Moreover, if $y_i$ is a sample of $\mathcal{N}(\mu_i, \sigma_i)$, then the transformed variable
\begin{equation}
  z_i = \frac{y_i - \mu_i}{\sigma_i},
\end{equation}
is normally distributed with zero mean and unit variance.
In the literature, this variable is commonly referred to as a $z$-score.

Hence, if I calculate the $z$-score of every model prediction, the $z$-scores should populate a normal distribution with zero mean and unit variance.
I can easily check this property by calculating the $z$-scores and histogramming their distribution.
If the $z$-scores deviate strongly from a normal distribution, then the predictions are statistically inaccurate.
Note, \emph{statistical} accuracy has a very specific meaning.
I can make a prediction that claims to know nothing at all, but with very large error bars, and have it be statistically accurate.
In other words, it's ok if the emulator is imperfect so long as it returns a reasonable estimate of its own uncertainty.

\begin{figure}
  \centering
  \makebox[\textwidth]{
    \includegraphics{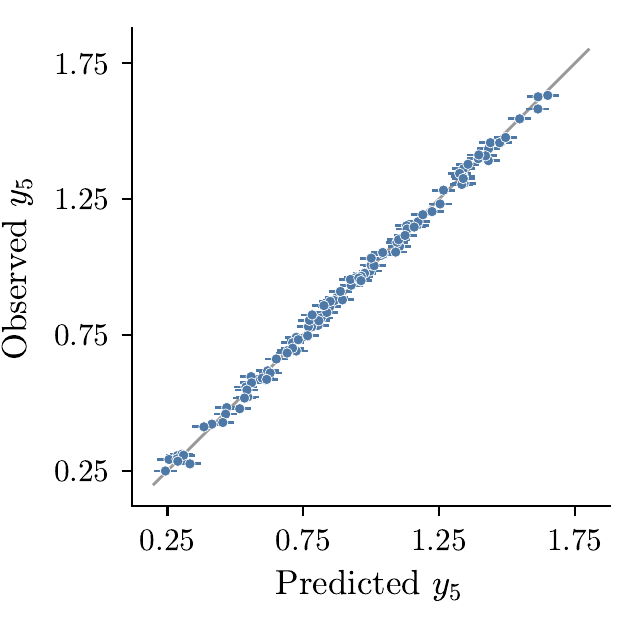}
    \includegraphics{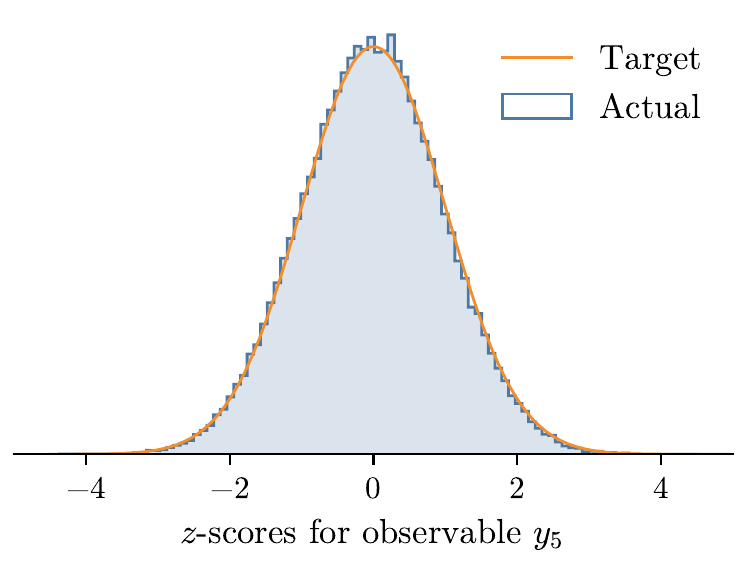}
  }
  \caption{
    \label{fig:emu_validation}
    Left: Emulator predictions for observable $y_5$ scatter plotted against the model value for $y_5$ using random parameter values $x_1, x_2, x_3 \in [0, 1]$.
    The error bar on the emulator prediction is $\pm 2\sigma$.
    Right: Histogram of the predicted $z$-scores equal to $z = (y_\text{pred} - y_\text{obs})/\sigma_\text{pred}$ for the same observable, but using a much larger number of parameter samples.
    The orange line is the target line for a perfectly accurate emulator.
  }
\end{figure}

Returning to the example that's the subject of this chapter, let's apply the aforementioned $z$-score test to the emulator constructed in the previous section.
Unfortunately, there's no simple analogue for a $z$-score test in multiple dimensions, so I'll have to validate each model output $(y_1, y_2, \dots, y_{10})$ separately.
This is somewhat tedious and repetitive, so let me demonstrate the procedure for just one observable.
For no particular reason, let's look at the fifth emulated model output $y_5$.

Figure~\ref{fig:emu_validation} shows two complementary tests of the emulator prediction accuracy.
The panel on the left shows a scatter plot of the model and emulator predictions at 100 random validation points $\xv \in [0, 1]^3$.
Each blue symbol is the model's observed value plotted against the emulator's predicted value calculated using the same input parameters.
The emulator predictions are close to the true model values, so each point falls close to the thin gray line which indicates perfect agreement.
The horizontal error bars on the predicted values of $y_5$ are $\pm 2\sigma$ emulator uncertainties, so we'd expect ${\sim}95\%$ of these error bars to touch the gray line.
This is of course difficult to eye ball, so let's plot the distribution of $z$-scores which serves as a more sensitive test.

The right panel of figure~\ref{fig:emu_validation} shows the distribution of $z$-scores (blue histogram) for observable $y_5$ calculated from the emulator and model predictions.
Here, I've used significantly more validations points, $\xv \in [0, 1]^3$, to produce a smooth distribution.
This histogram should be compared to the orange curve which is a normal distribution with zero mean and unit variance.
The agreement here is \emph{excellent}.
Evidently, the emulator uncertainties are correct estimated.
This means that we will be able to properly account for this uncertainty when using Bayesian parameter estimation to calculate the unknown model parameters.

\section{Applying Bayes' theorem}

This section describes how Bayesian statistics is used to solve the inverse problem, enabling statistically accurate statements to be made about the three unknown model parameters $(x_1, x_2, x_3)$ which are the subject of this chapter.
The following explanation of Bayesian parameter estimation is adapted from one of my publications
\begin{flushleft}
J.\ S.\ Moreland, J.\ E.\ Bernhard, and S.\ A.\ Bass, ``Bayesian calibration of a hybrid nuclear collision model using $p$-Pb and Pb-Pb data from the LHC,'' \href{https://arxiv.org/abs/1808.02106}{Submitted for publication, arXiv:1808.02106 [nucl-th]},
\end{flushleft}
which I've edited for both content and clarity.

\subsection{Bayesian posterior}

Ultimately, the goal of this chapter is to estimate the \emph{true} parameters of our model $\xv_\text{true}$, provided some evidence that the model predictions describe a set of experimental measurements $\yv_e$.
The problem involves three distinct components:
\begin{enumerate}
  \item $H_f$: the hypothesis that the assumed model $\yv = f(\xv)$ provides a realistic description of the physical process that it simulates,
  \item $H_\xv$: the hypothesis that $\xv$ are the \emph{true} parameters $\xv_\text{true}$ of the physical process modeled by the function $f$, and
  \item $E$: the evidence provided by the model, the experimental data, and their associated uncertainties.
\end{enumerate}
As a practical matter, one always asserts the truth of hypothesis $H_f$.
This means that there are no glaring flaws in the chosen theoretical model framework, i.e.\ there exist \emph{some} model parameters $\xv$ where the model provides a sensible description of reality.
This is a significant assumption, and all of our results are predicated on it.
If the model is completely bogus, the constraints on the parameters are meaningless as well.

Subject to this assumption, we can apply Bayes' theorem to evaluate hypothesis $H_\xv$ \emph{given} the evidence provided by $E$.
Simplifying my notation and writing $H_\xv$ as just $\xv$, Bayes' theorem yields
\begin{equation}
  \label{eq:bayes}
  P(\xv | E) \propto P(E | \xv)\, P(\xv).
\end{equation}
The left-side of this expression is the \emph{posterior}: the probability of $\xv = \xv_\text{true}$ given the experimental evidence $E$.
On the right-side there are two separate terms.
The first term $P(E | \xv)$ is the \emph{likelihood} function: the probability of observing the evidence $E$ provided that $\xv = \xv_\text{true}$, and the second term $P(\xv)$ is the \emph{prior}: an estimate of the probability of $\xv = \xv_\text{true}$ in the absence of evidence $E$.

\subsubsection{Likelihood function}
\label{subsec:likelihood}

Let's assume that the likelihood function $P(E | \xv)$ in equation~\eqref{eq:bayes} is described by a multivariate normal distribution:
\begin{equation}
  \label{eq:likelihood}
  P(E | \xv) = \frac{1}{\sqrt{(2\pi)^m \det \Sigma}} \exp \left [ -\frac{1}{2}\Delta\yv(\xv)\tran \Sigma^{-1}(\xv) \Delta\yv(\xv) \right ],
\end{equation}
where $\Delta\yv = \yv_m(\xv) - \yv_e$ is a vector of length $m$ equal to the discrepancy of the model and experiment, and
\begin{equation}
  \label{eq:tot_cov}
  \Sigma = \Sigma_m(\xv) + \Sigma_e,
\end{equation}
is a \emph{total} covariance matrix, equal to the sum of a modeling component $\Sigma_m(\xv)$ and an experimental component $\Sigma_e$ which account for all known sources of uncertainty in the simulated and measured observables.

\begin{remark}
  When two uncertainties are uncorrelated their variances add. The same general rule also applies when combining two covariance matrices.
\end{remark}

Our model is agonizingly slow to run directly, so we've trained an emulator to replace the model calculation $\yv_m(\xv)$ with an emulator prediction $\yv_m^\text{emu}(\xv)$.
The covariance matrix $\Sigma_m(\xv)$ is therefore the covariance matrix of our emulator, $\Sigma_m^\text{emu}(\xv)$, which subsumes all \emph{quantifiable} sources of model and emulator uncertainty.
Namely, it accounts for the interpolation uncertainty introduced by using a finite number of emulator training points as well as the statistical uncertainty introduced by the noise in the simulation outputs.

To complete the specification of equation~\eqref{eq:tot_cov}, I'll also need to specify an experimental covariance matrix $\Sigma_e$ for use in our example problem.
Generally speaking, the experimental covariance consists of separate statistical and systematic contributions,
\begin{equation}
  \label{eq:expt_cov}
  \Sigma_e = \Sigma_e^\text{stat} + \Sigma_e^\text{sys}.
\end{equation}
The statistical errors in $\Sigma_e^\text{stat}$ are uncorrelated, so its covariance matrix is diagonal:
\begin{equation}
  \label{eq:sig_stat}
  \Sigma_e^\text{stat} = \diag[(\sigma_{y_1}^\text{stat})^2, (\sigma_{y_2}^\text{stat})^2, \dots, (\sigma_{y_m}^\text{stat})^2].
\end{equation}
Here $\sigma_{y_j}^\text{stat}$ is the statistical uncertainty of observable $y_j$ in the experimental observable vector $\yv_e = (y_1, y_2, \dots, y_m)$.
For the purpose of our example, let's assume that $\sigma_{y_j}^\text{stat} = 0.03$ for every observable component.

The systematic errors in $\Sigma_e^\text{sys}$, meanwhile, are correlated so the matrix is generally non-diagonal.
Given an arbitrary covariance, we can decompose it into the form
\begin{equation}
  \cov(y_i, y_j) = \rho_{ij} \sigma_i \sigma_j,
\end{equation}
where $\sigma_i$ and $\sigma_j$ are the standard deviations of $y_i$ and $y_j$ respectively, and $\rho_{ij}$ is their Pearson correlation coefficient:
\begin{equation}
  \rho_{ij} = \frac{\cov(y_i, y_j)}{\sigma_i \sigma_j},
\end{equation}
satisfying $\rho_{ij}=1$ for $i=j$ and $|\rho_{ij}| \le 1$ for $i \neq j$.
I'll use this decomposition now to define a systematic covariance matrix for our example problem.

Let's assume that the systematic standard deviation $\sigma_{y_i}^\text{sys} = 0.03$ for every component $y_i$.
Let's also assume that each pair of observables, $y_i$ and $y_j$, has a systematic error correlation coefficient
\begin{equation}
  \label{eq:eg_corr}
  \rho_{ij}^\text{sys} = \exp \left [ -\frac{1}{2} \left ( \frac{v_i - v_j}{l} \right )^2 \right ],
\end{equation}
where $v=(0.05, 0.15, \dots, 0.95)$ is a ten component vector that evenly partitions the interval $[0, 1]$, and $l=0.5$ is a fixed correlation length.
This functional form strongly correlates two observables $y_i$ and $y_j$ if they occupy proximate elements of the vector $\yv_e$.
For example, the observables $y_1$ and $y_2$ are strongly correlated, while $y_1$ and $y_{10}$ are not.
The correlation length $l$ controls the extent of the correlation.
As $l \to \infty$, the systematic errors become perfectly correlated, and as $l \to 0$ the systematic errors become perfectly uncorrelated.

\begin{figure}
  \centering
  \includegraphics{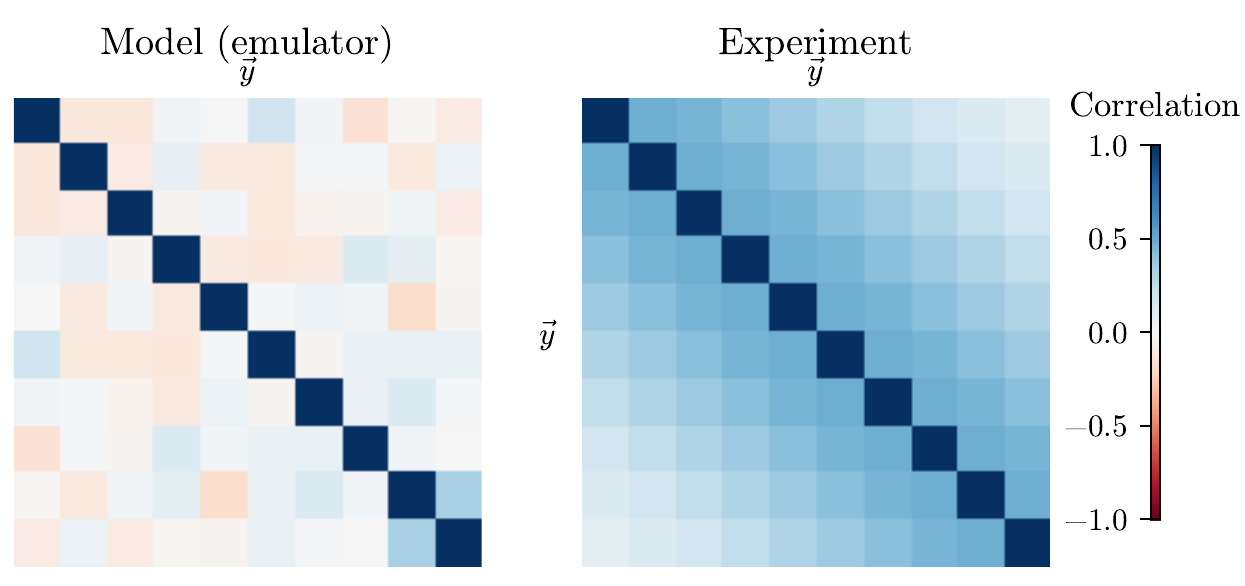}
  \caption{
    \label{fig:corr_matrices}
    Correlation matrix, $\corr(y_i, y_j) = \cov(y_i, y_j)/(\sigma_i \sigma_j)$, visualized for the model (left-side) and the experimental data (right-side).
    Each colored cell is the correlation coefficient of a pair of observables $y_i$ and $y_j$.
    Cooler colors indicate positive correlations and warmer colors indicate negative correlations.
    Observables are trivially correlated with themselves, so the diagonal is unity.
  }
\end{figure}

Figure~\ref{fig:corr_matrices} compares the correlation structure of the model emulator (left-side) to that of the experimental data (right-side).
Each colored cell is the correlation coefficient of a pair of observables:
\begin{equation}
  \corr(y_i, y_j) = \frac{\cov(y_i, y_j)}{\sigma_i \sigma_j},
\end{equation}
defined as the covariance of $y_i$ and $y_j$ divided by $\sigma_i$ and $\sigma_j$, the standard deviation of each observable.
Note, the correlation matrix is shown here instead of the covariance matrix because the correlation coefficient is bounded between -1 and 1.
This makes it significantly easier to visualize the matrix elements using a single heatmap.

Looking at the left-side of figure~\ref{fig:corr_matrices}, we see that the emulator's correlation matrix is essentially diagonal.
This makes sense.
When I constructed the model, I added some additional statistical noise to the model outputs.
This statistical noise is uncorrelated, so it only appears in the diagonal elements of the correlation matrix.
Every model output is perfectly correlated with itself, so the diagonal elements are unity by construction.
Presumably the emulator also includes some interpolation uncertainty which is likely correlated, but evidently this interpolation uncertainty is small compared the model's overall statistical uncertainty.
Generally speaking, this will not always be the case.

Compare this emulator correlation matrix to the experimental correlation matrix depicted on the right-side of figure~\ref{fig:corr_matrices}.
The experimental correlation matrix includes contributions from both a statistical component and a systematic component.
The statistical errors are naturally uncorrelated so they only contribute to the matrix diagonal.
The systematic errors, meanwhile, are correlated, so they introduce nonzero off-diagonal elements in the correlation matrix, producing the soft blue gradient that is visible in the figure.
This systematic correlation uncertainty was modeled using a finite correlation length, so the correlation fades as the element-wise distance between the components increases, i.e.\ $y_1$ and $y_2$ are more strongly correlated than $y_1$ and $y_{10}$.

\subsubsection{Specifying a prior}

Now that I've specified the likelihood function $P(E | \xv)$, I need to specify the prior $P(\xv)$ to complete the right-side of equation~\eqref{eq:bayes}.
The Bayesian prior describes the initial uncertainty on the model parameters $\xv$, absent the evidence provided by the likelihood function.
Specifically, it is a probability distribution of the model inputs parameters.
Regions of parameter space where the prior probability density is large indicate likely regions for the true parameters $\xv_\text{true}$.
Conversely, if the prior probability density is small, then those regions of parameter space are unlikely to contain the true model parameters $\xv_\text{true}$.

Prior specification is a hotly contested topic in Bayesian statistics.
If a prior is too restrictive, i.e.\ specific, it can bias the posterior in harmful ways.
For example, suppose I choose a prior that's a Dirac delta function
\begin{equation}
  P(\xv) = \delta(\xv - \xv'),
\end{equation}
where $\xv'$ is some specific set of parameter values.
The Bayesian posterior is proportional to the product of the likelihood function and the prior, so the resulting posterior $P(\xv | E)$ will be \emph{zero} for all values of $\xv \neq \xv'$.
This essentially forces the posterior to equal the prior.
In other words, I learn nothing from the likelihood function at all because I've purported to know everything there is to know with my choice of prior.

Alternatively, there are times when it makes sense to use a restrictive prior.
For example, if it's known that a parameter cannot be negative, then the prior distribution should be zero for all negative parameter values.
This asserts that there is 0\% chance the true parameter value $x_\text{true}$ is negative.

In the present example, we've been told that each parameter $(x_1, x_2, x_3)$ lies inside a unit cube, $[0, 1]^3$ so the relevant prior distribution is
\begin{equation}
  P(x_1, x_2, x_3) =
  \begin{cases}
    1 \quad \text{if $0 \leq x_i \leq 1$ for all $x_i$},\\
    0 \quad \text{otherwise.}
  \end{cases}
  \label{eq:example_prior}
\end{equation}
This is, by no accident, the same range of values that I used to construct the design matrix $X$.
There is no need to train the emulator outside this region, since the prior is zero there anyhow.
If the prior is zero, the posterior also will be zero regardless of the value of the likelihood function.

\subsection{Importance Sampling}

Now that I've specified the likelihood $P(E | \xv)$ and the prior $P(\xv)$, I can proceed to calculate the Bayesian posterior
\begin{equation}
  P(\xv | E) \propto P(E | \xv) P (\xv).
\end{equation}
Written in this way, I can only calculate the Bayesian posterior up to an overall normalization factor.
However, only relative probabilities will matter in the steps that follow, so I can disregard the value of this factor.

In the present example, the Bayesian posterior distribution $P(\xv | E)$ is three-dimensional, i.e.\ it has one dimension for each parameter $x_1$, $x_2$, and $x_3$.
The plane of this page is two-dimensional, so I'll ultimately have to project this distribution down onto a lower number of dimensions.
One common procedure to do this is to sample the distribution a large number of times.
Once the distribution is sampled, I can histogram the samples using a single parameter value $x_i$ to project the joint posterior distribution onto that dimension.
In a similar fashion, I can also histogram a pair of parameters $x_i$ and $x_j$ to visualize their correlations.

Sampling a multidimensional probability distribution is a difficult task in and of itself.
Bayesian posterior distributions may include as many as ten parameters or more, so accept-reject sampling is out of the question; this basic sampling algorithm fails spectacularly when the number of dimensions is large.
The canonical procedure for sampling a multi-dimensional probability distribution is to use Markov chain Monte Carlo importance sampling, commonly referred to as MCMC for short.

Given a target probability distribution, the MCMC sampling procedure constructs a sequence of random values known as a \emph{chain}.
When this chain is sufficiently long, one can obtain a sample of the target distribution by observing the random values at the end of the chain.
The simplest and most commonly used MCMC algorithm is the Metropolis-Hastings (M-H) algorithm.
M-H samples the probability distribution using a series of proposed updates which are either accepted or rejected to produce the next sample in the chain.
The algorithm has a number of \emph{walkers} which wander the parameter space.
At any given moment, each walker occupies a single parameter point $\xv$.
The algorithm then proposes a new parameter point $\xv'$ which is obtained by perturbing the parameter $\xv$ in a random direction.
The walker then moves to the proposed parameter point $\xv'$ with probability
\begin{equation}
  \alpha = \min\left(1, \frac{P(\xv')}{P(\xv)} \right),
\end{equation}
referred to as the proposal acceptance fraction.
If the proposal is rejected, the walker stays in its current location.
If it is accepted, the walker moves to the new location.
Since this update step only depends on the relative probability $P(\xv') / P(\xv)$, it is not necessary that the target probability distribution is normalized.
Every time the walker position is updated, the new position is appended to the end of the MCMC chain.
The update step is then repeated many times to build up the length of the chain.
Eventually, the samples at the end of the chain form a random sample of the target probability distribution.

The Bayesian parameter estimation framework \cite{Bernhard:2018hnz} used in this dissertation uses \textsc{emcee} \cite{emcee}, an affine-invariant MCMC ensemble sampler implemented in Python that uses a large number of interdependent walkers \cite{Foreman-Mackey:2013, Goodman:2010en}.
The algorithm is qualitatively similar to the original M-H algorithm, but it generally converges much faster to a stationary set of samples from the target distribution.
This reduces the amount of ``burn-in'' steps which must be discarded from the beginning of the MCMC chain to ensure the same level of statistical accuracy.

\section{Visualizing the posterior distribution}
\label{sec:visualize_posterior}

The final step of Bayesian parameter estimation is to visualize the multidimensional posterior distribution defined by equation~\eqref{eq:bayes}.
Let's use the \textsc{emcee} ensemble sampler to draw \order{10^7} parameter samples $\{\xv_i\}$ from the posterior distribution of our example problem.
Using these samples, we'll be able to visualize the constraints provided by the model, the experimental data, and all associated uncertainties.

\begin{figure}
  \centering
  \makebox[\textwidth]{
    \includegraphics{external/misc/observables_design_example}
    \includegraphics{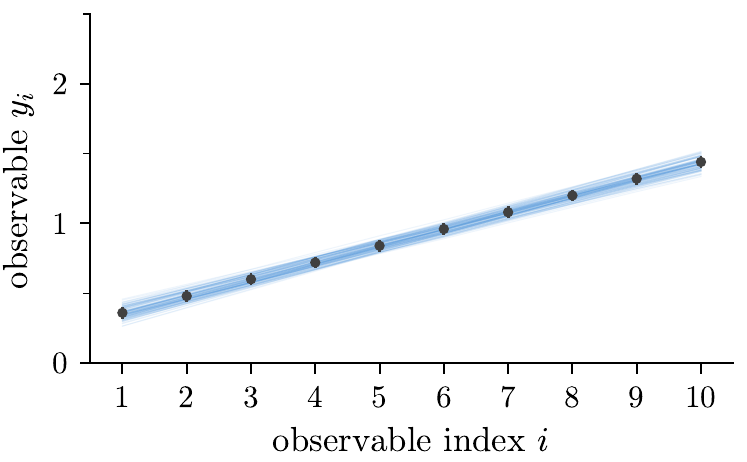}
  }
  \caption{
    \label{fig:post_obs_example}
    Simulated observables (blue lines) compared to the experimental data (black symbols with errors bars).
    Left: Explicit model calculations (no emulator) for each of the $d=100$ design points.
    Right: Emulator predictions for $100$ random samples drawn from the posterior.
  }
\end{figure}

Figure~\ref{fig:post_obs_example} shows various model calculations (blue lines) compared to the experimental data (black symbols with error bars).
The plot on the left shows the model output $\yv = f_\text{model}(\xv)$ at each of the $d=100$ design points (same as figure~\ref{fig:obs_design_example}), while the plot on the right shows the mean emulator output $\yv = f_\text{emu}(\xv)$ at $100$ different parameter points randomly selected from the Bayesian posterior.

We initially assumed that each parameter $\xv \in [0, 1]^3$, so the model outputs on the left span a wide range of values.
This visual spread is therefore the uncertainty of the assumed prior \eqref{eq:example_prior}.
The plot on the right, meanwhile, shows the constraining power of the evidence, i.e.\ the refinement to the prior provided the model, the experimental data, and all uncertainties.
These calculations are now tightly clustered around the experimental data, giving us confidence that the model is in fact a good representation of the ground truth.
Of course, the experimental data $\yv_e$ was constructed by running the model using a certain set of true parameters, $\yv_e = f_\text{model}(\xv_\text{true})$, so it should be no surprise that this is the case.

\begin{figure}[t]
  \centering
  \includegraphics{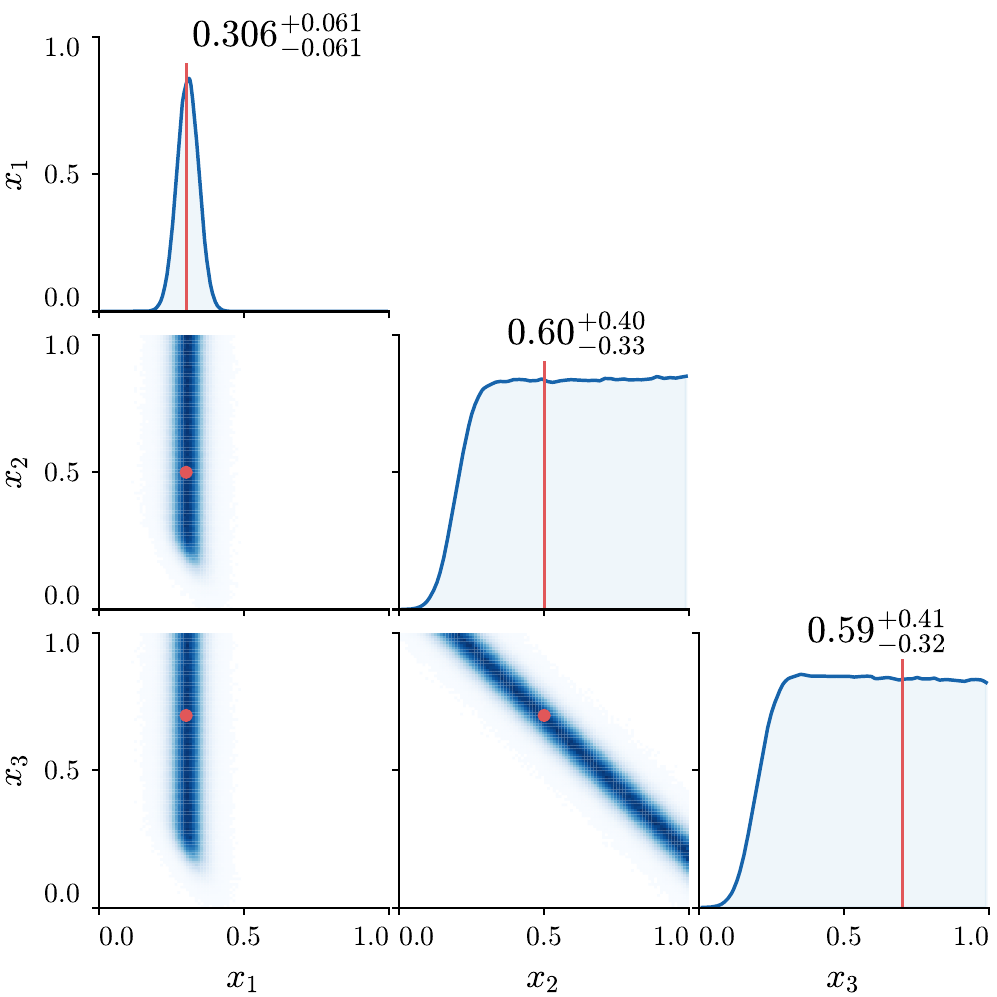}
  \caption{
    \label{fig:posterior_example}
    Bayesian posterior distribution for the three model input parameters $x_1$, $x_2$, and $x_3$.
    The diagonal panels show the marginalized distributions of individual model parameters, while off-diagonal panels show the joint distributions for pairs of model parameters, visualizing their correlations.
    The marginalized distribution medians and 90\% credible intervals are annotated along the diagonal.
    Red symbols and lines indicate the \emph{true} value of each model parameter.
  }
\end{figure}

Finally, let's address the original problem that we were tasked with solving at the beginning of this chapter.
I'll now present the posterior estimate for the model parameters $\xv$.
Figure~\ref{fig:posterior_example} visualizes the three-dimensional Bayesian posterior distribution $P(\xv | E)$.
Each diagonal panel is the marginal distribution of a single model parameter constructed by integrating the posterior distribution over all other parameters.
For example, the marginal distribution for parameter $x_1$ is
\begin{equation}
  P(x_1) = \int dx_2\, dx_3\, P(x_1, x_2, x_3).
\end{equation}
The off-diagonal panels, meanwhile, are the joint posterior distribution for a \emph{pair} of model parameters, visualizing their correlations.
They are similarly constructed by marginalizing (integrating) over all but two model parameters.
Note, the probability distributions in figure~\ref{fig:posterior_example} are histograms, so there's no need to actually integrate, we simply bin the variables of interest.

The black text above each diagonal panel is the marginalized posterior distribution's median value and 90\% highest posterior density (HPD) credible interval.
This latter quantity is defined as the narrowest parameter interval containing 90\% of the posterior density.
Roughly speaking, this means that we expect $\xv_\text{true}$ to land inside the interval 90\% of the time.
The estimates are presented as a median value plus-minus the distance to each edge of the credible interval.
For example, the estimate $x_1 = 0.306_{-0.061}^{+0.061}$ corresponds to a median value $\tilde{x}_1 = 0.306$ and 90\% credible interval $0.245 < x_1 < 0.367$.
Also shown, for reference, are the true model parameters $x_1=0.3$, $x_2=0.5$, and $x_3=0.7$.
These values are plotted as a red dot on each joint posterior distribution and as a red line on each marginal distribution.

There are several important features that should be noticed when looking at this figure.
First, the posterior distribution of $x_1$ is tightly constrained, while the estimates for $x_2$ and $x_3$ span a large range of values.
This result is easily understood if we look at the original function that was modeled, equation~\eqref{eq:linear_model}, which I've written below:
\begin{equation}
  \label{eq:linear_model_ref}
  \yv = x_1 \mathbf{u} + (x_2 + x_3) \mathbf{v}.
\end{equation}
Recall that $\mathbf{u}=(1, 1, \dots, 1)$ is a vector of ones, and $\mathbf{v}=(0.05, 0.15, \dots, 0.95)$ is a vector that uniformly partitions the interval $[0, 1]$.
Inspecting this function, we see that parameter $x_1$ shifts the vector $\yv$ vertically up and down, while the parameters $x_2$ and $x_3$ vary its slope.
The effect of parameter $x_1$ is unique, while $x_2$ and $x_3$ are degenerate; all combinations of $x_2$ and $x_3$ have the same effect if their sum is held constant.

This illustrates a very important property of the inverse problem, namely \emph{information loss}.
If two parameters produce the same effect, then their values cannot be constrained independently.
Look now at the joint posterior distribution of parameters $x_2$ and $x_3$.
This joint distribution is highly correlated: $x_2$ can be small if $x_3$ is large and vice versa, but both parameters cannot be large (or small) at the same time.
This is, of course, exactly what we'd expect from looking at equation~\eqref{eq:linear_model_ref}.
Clearly, it is their \emph{sum}, $x_2 + x_3$, which is constrained by the data.

Before I conclude this chapter, I want to make one more statement about the Bayesian posterior distribution.
Evidently, we did a pretty good job determining $x_1$, but our estimates were considerably less certain about the individual values of $x_2$ and $x_3$.
This is not a failure!
This is merely a fact of life.
Our goal here was obviously to say as much as humanly possible about each parameter, but we also wanted to do so \emph{truthfully}.
Sometimes the correct answer is ``I don't know'' or ``I can't tell'', and that's perfectly acceptable.
In fact, in this case, it was the only correct answer.

In the next chapter, I will use the Bayesian parameter estimation framework developed in \cite{Bernhard:2018hnz} to study the \trento\ initial condition model which is the focus of this dissertation.
When I constructed the model, I tried to assume as little as possible.
The model is intentionally flexible, constructed to interpolate a subspace of \emph{all} initialization models including (but not limited to) specific calculations
in Color Glass Condensate effective field theory.
This flexibility is parametrized by a handful of model parameters whose values are uncertain.
Bayesian parameter estimation is useful, in large part, because it allows one to marginalize over this uncertainty, i.e.\ the constraint on any one parameter accounts for the uncertainty of all other parameters.
This enables robust statements about the initial conditions and QGP medium parameters which would not otherwise be possible.

\chapter[Calibration and comparison to data]{Model calibration and comparison to data}
\label{ch:results}

\lettrine{U}{ltimately}, the goal of this dissertation is to study the QGP initial conditions through the lens of the inverse problem.
While most calculations seek to derive the QGP initial conditions from first-principles or approximations thereof, I want to approach the problem from the opposite direction.
Namely, I want to use the experimental data and the predictions of relativistic fluid dynamics to place robust constraints on the QGP initial conditions without assuming a narrow theoretical formalism for their derivation.
This constitutes what is broadly referred to as a top-down or data-driven approach.

Section \ref{sec:trento_model} described a parametric model of the QGP initial conditions called \trento\ developed for this purpose.
The model was designed to be maximally flexible, enabling future analyses to determine the correct form of the initial conditions from their resulting description of the data and not the other way around.
It therefore describes a sort of meta-model that encompasses a large subspace of reasonable theoretical descriptions.

The method that I will use to study this initial condition model is hypothesis testing.
First, I fix the parameters of the \trento\ model and propose a certain parametric description of the initial state.
Then, I evolve the system forward in time using relativistic viscous hydrodynamics and microscopic Boltzmann transport to simulate the final state of the system as it is observed inside the detector.
Finally, I evaluate my initial hypothesis by comparing the model's simulated output to experimental data.

The problem is challenging for several reasons.
First, the initial conditions are but one part of a multi-stage simulation.
In addition to the handful of parameters needed to describe the initial state, I'll need several more to account for uncertainties in the QGP medium properties.
These parameters typically correlate among each other and affect multiple observables so I cannot tune them individually by hand.
And second, the model is computationally intensive.
Evaluating just a single set of model parameters requires hundreds of CPU hours.
Therefore, I cannot afford to brute force the problem by running the model countless times.

In chapter \ref{ch:param_est}, I described a general statistical framework known as Bayesian parameter estimation designed to handle these problems.
In this chapter, I present several Bayesian studies of the \trento\ initial condition model using a software package for Bayesian parameter estimation developed by fellow graduate student Jonah Bernhard \cite{Bernhard:2018hnz}.
I should emphasize again that I did not write the Bayesian framework.
Each project is also a collaborative effort involving multiple contributors.
My own specific contribution level correspondingly varies from project to project.

Generally speaking, each study applies Bayesian parameter estimation using the same sequence of steps:
\begin{enumerate}[leftmargin=1\parindent]
  \item
    First, we construct a dynamical model of a relativistic nuclear collision event. The model components change slightly from study to study, but the overall framework is more or less the same.
    We use relativistic hydrodynamics to model the hot and dense early stage of the collision and microscopic Boltzmann transport to simulate individual particle interactions once the QGP freezes out into hadrons.
  \item
    Next, we identify a number of free parameters in the model framework which are uncertain.
    These parameters are the objects of interest in each study.
    We place a conservative prior distribution on each model parameter and run the computer model at lots of different parameter points uniformly selected from the prior distribution.
  \item
    For each analysis, we identify a set of physical observables which we use to build evidence for (or against) the aforementioned parameter values.
    We calculate these observables using the model predictions at every evaluation point and use these predictions to train an emulator.
    Once the emulator is trained to reproduce the input-output relationship of the model at each evaluation point, it can be used to make essentially instantaneous predictions at new points in parameter space.
  \item
    With the trained emulator in hand, we calculate an overall likelihood function for each set of model parameters that quantifies the global compatibility of the experimental data with the model predictions at that point.
    This likelihood function is then multiplied by the prior to furnish the Bayesian posterior probability for the model parameters up to an overall normalization constant.
  \item
    Finally, we use MCMC to sample and visualize the Bayesian posterior distribution.
    This posterior is the central result of each analysis, and it includes a wealth of information about the unknown model parameters.
    We therefore conclude each study with a detailed discussion of the posterior distribution.
\end{enumerate}

The following chapter results are divided into three sections which address different aspects of the QGP initial condition problem.
In the first section, I examine the QGP initial conditions at midrapidity using boost-invariant approximations.
Specifically, I describe and discuss the first ``proof of concept'' Bayesian analysis of the \trento\ initial condition model, calibrated to Pb-Pb collision data at $\sqrts=2.76$~TeV.
I also briefly summarize a greatly improved analysis of Pb-Pb collisions at $\sqrts=2.76$ and 5.02 TeV, and use the results of the improved analysis to generate predictions for Xe-Xe collisions at $\sqrts=5.44$~TeV.

In the second section, I relax the boost-invariant approximation to explore the three-dimensional structure of the collision.
My coauthors and I construct a flexible parametrization for the rapidity-dependence of the QGP initial conditions and use this parametrization to extend the \trento\ model beyond midrapidity.
The model is then matched to a dynamical simulation consisting of three-dimensional viscous hydrodynamics and microscopic Boltzmann transport.
Using the model, we perform a Bayesian analysis of $p$-Pb and Pb-Pb collisions at the LHC and present posterior results for the three-dimensional structure of the produced plasma.

Finally in the third section, I investigate the effect of nucleon substructure on hydrodynamic bulk observables.
I model this substructure using three parameters to vary the number, size, and distribution of constituent sources inside each nucleon.
This parametric nucleon substructure is then added to a dynamical model consisting of \trento\ initial conditions, pre-equilibrium free streaming, boost-invariant viscous hydrodynamics, and microscopic Boltzmann transport.
Free parameters of the model which describe the initial state and QGP medium are then simultaneously calibrated to fit bulk observables in $p$-Pb and Pb-Pb collisions at $\sqrts=5.02$~TeV.
Posterior estimates for the nucleon substructure parameters are obtained, and implications for hydrodynamic flow in small collision systems discussed.
I'll now describe each of these projects in detail, following the general sequence of steps outlined on the previous page.

\section{Bulk matter at midrapidity}
\label{sec:midrapidity}

The following describes the first Bayesian analysis using \trento\ initial conditions.
It is based off the publication
\begin{flushleft}
J.\ E.\ Bernhard, J.\ S.\ Moreland, S.\ A.\ Bass, J.\ Liu, and U.\ Heinz,
``Applying Bayesian parameter estimation to relativistic heavy-ion collisions:
simultaneous characterization of the initial state and quark-gluon plasma medium'', \href{https://arxiv.org/abs/1605.03954}{Phys.\ Rev.\ \textbf{C94}, 024907 (2016), arXiv:1605.03954 [nucl-th]},
\end{flushleft}
which combines my own contributions in initialization theory \cite{Moreland:2014oya} with the Bayesian parameter estimation framework developed by coauthor Jonah Bernhard \cite{Bernhard:2018hnz}.
He wrote the Bayesian parameter estimation software used for this project, ran the hydrodynamic events, and performed the primary data analysis.
I contributed to the design of the analysis (parameters, observables, model components, etc.), co-developed the \trento\ model used in the analysis, and was involved in multiple aspects of the data analysis and manuscript preparation.

Prior to this publication, we had already found compelling evidence that \trento\ would be able to simultaneously and self-consistently describe hydrodynamic bulk observables.
For instance, in figure~\ref{fig:nch}, we showed that model was able to describe $p$-$p$, $p$-Pb, and Pb-Pb charged particle multiplicity distributions using a single set of model parameters.
Additionally, in figure~\ref{fig:eccen} we showed that these \emph{same} parameters were also those that were needed to describe the relative magnitude of the second and third eccentricity harmonics, a quantity that was found to be strongly constrained by ALICE flow data \cite{Retinskaya:2013gca}.
Moreover, unlike the two-component ansatz \eqref{eq:two_comp_ansatz} which predicts a knee-shaped structure in ultra-central U-U elliptic flow data \ref{fig:uranium}, the generalized mean ansatz \eqref{eq:gmean} predicts that $v_2$ should flat-line (e.g.\ see figure \ref{fig:uranium}) in qualitative agreement with measurements by the STAR collaboration \cite{Pandit:2013uiv}.
These bread crumbs all suggested that our parametric approach was sensible.

Nevertheless, these indicators were really just that, indicators.
The only reliable way to validate the initial condition model is to run it through a full hydrodynamic simulation.
The \trento\ model is parametric, so naturally we didn't want to just check a single set of model parameters, we wanted to check \emph{every} set of model parameters.
Thus, we used the Bayesian parameter estimation framework \cite{Bernhard:2015hxa} recently developed by Jonah Bernhard \emph{et al.}\ to estimate the \trento\ initial condition and QGP medium parameters using the predictions of a realistic hybrid transport model \cite{Shen:2014vra}.

\subsection{Boost-invariant nuclear collision model}
\label{subsec:nuclear_collision_model_v1}

This study was conducted using on an older version of our nuclear collision model that's missing many of the bells and whistles included in our later studies.
The study's parameter estimates are therefore outdated, superseded by the parameter estimates of our more recent work.
I show them here purely for historical context.
The nuclear collision model consisted of the following components.

\subsubsection{Initial conditions}

We used the \trento\ model to initialize the hydrodynamic simulations.
The version of the model used at the time was largely identical to the current version described in chapter \ref{ch:initialization} with one small difference.
It did not yet include the inter-nucleon minimum distance algorithm described in subsection~\ref{subsec:nucleon_corr} to sample correlated nucleon positions.
The nucleon positions were therefore sampled from a standard uncorrelated Woods-Saxon distribution \cite{Loizides:2014vua}.
Hence the nucleon minimum distance parameter $d_\text{min}$ was absent from the analysis.

In this particular study, we initialized the hydrodynamic medium using the entropy variant of the generalized mean ansatz, $s_0 \propto T_R$, where $T_R$ is the reduced thickness function \eqref{eq:reduced_thickness}.
The initial flow velocity $u^\mu$ was set to zero at the hydrodynamic starting time as well as all viscous correction terms, $\pi^{\mu\nu}$ and $\Pi$.
We therefore assumed instant thermalization at the hydrodynamic starting time $\tau_0$ and parametrized the initial \emph{entropy} density $s = dS/(d^2x_\perp \tau_0\,d\eta_s)$ according to
\begin{equation}
  \label{eq:entropy_deposition1}
  s(\xv_\perp; \eta_s=0, \tau_0) = \frac{\text{Norm}}{\tau_0} \times \left[ \frac{\T_A(\xv_\perp)^p + \T_B(\xv_\perp)^p}{2} \right]^{1/p},
\end{equation}
where Norm is a dimensionless normalization constant, $\tau_0=0.6~\fmc$ is the hydrodynamic starting time, and $\T_A$ and $T_B$ are the participant thickness functions defined by equation~\eqref{eq:participant_thickness}.

\subsubsection{Boost-invariant viscous hydrodynamics}
\label{subsec:boostinv_hydro}

This study also debuted an upgraded version of VISH2+1 \cite{Song:2007ux}, a boost-invariant viscous hydrodynamics code developed by our collaborators at Ohio State.
These upgrades included support for fluctuating event-by-event initial conditions \cite{Shen:2014vra} and bulk viscous corrections with shear-bulk coupling \cite{Liu:2015bik}.
The code has been extensively validated, and it reproduces semi-analytic solutions for ideal hydrodynamics to excellent precision \cite{Shen:2014vra}.

It's numerical implementation solves the second-order Israel-Stewart equations \cite{Israel:1979wp, Israel:1976aa} in the so-called 14-moment approximation.
This formalism produces a set of relaxation-type equations \cite{Denicol:2014vaa, Ryu:2015vwa}
\begin{subequations}
  \label{eq:relaxation}
  \begin{align}
    \tau_\Pi \Pi + \dot{\Pi} &=
      - \zeta \theta - \delta_{\Pi\Pi} \Pi\theta
      + \lambda_{\Pi\pi} \pi^{\mu\nu} \sigma_{\mu\nu}, \\[1ex]
    \tau_\pi \dot{\pi}^{\langle \mu\nu \rangle} + \pi^{\mu\nu} &=
      2\eta\sigma^{\mu\nu} - \delta_{\pi\pi} \pi^{\mu\nu} \theta
      + \phi_7 \pi_\alpha^{\langle \mu} \pi^{\nu \rangle \alpha} \nonumber \\
      &\qquad {} - \tau_{\pi\pi} \pi_\alpha^{\langle \mu}\sigma^{\nu \rangle \alpha}
      + \lambda_{\pi\Pi} \Pi \sigma^{\mu\nu},
  \end{align}
\end{subequations}
for the shear viscosity $\eta$ and bulk viscosity $\zeta$ which are parametrized below.
All other transport coefficients were fixed using analytic results derived from the Boltzmann equation near the conformal limit \cite{Denicol:2014vaa}.

The hydrodynamic equations of motion were closed using a modern EoS based on lattice calculations at zero baryon density published by the HotQCD collaboration \cite{Bazavov:2014pvz}.
These calculations were then blended with a hadron resonance gas EoS in the temperature interval $110 \le T \le 130$~TeV using a smoothstep interpolation function \cite{Moreland:2015dvc}.
This matching procedure was done differently in subsequent studies \cite{Moreland:2018gsh, Bernhard:2019ntr}, although the effect of the difference is likely small.
See figure 3 in \cite{Moreland:2015dvc} and figure 3.9 in \cite{Bernhard:2018hnz} for a direct comparison of the two different methods.

The study sought to estimate both the properties of the initial conditions as well as the properties of the produced QGP medium.
So we parametrized the temperature dependence of the QGP shear and bulk viscosities expressed as dimensionless ratios $\etas$ and $\zetas$ where $s$ is the entropy density.
We used a piecewise linear parametrization for the shear viscosity
\begin{equation}
  \label{eq:shear_param1}
  (\etas)(T) =
  \begin{cases}
    \etasmin + \etasslope(T - T_c) & T > T_c, \\
    \etashrg & T \leq T_c,
  \end{cases}
\end{equation}
which is constant below the temperature $T_c$ and linearly rising above it.
The constant $T_c=0.154$~GeV was then fixed using the pseudocritical transition temperature of the HotQCD EoS, motivated by studies which demonstrate that $\etas$ has a minimum near the QCD transition temperature \cite{Prakash:1993bt, Arnold:2003zc, Csernai:2006zz}.
The constants $\etas$ hrg, min, and slope, meanwhile, were treated as variable model input parameters to be determined by the Bayesian analysis.

\begin{figure}[t]
  \centering
  \makebox[\textwidth]{\includegraphics{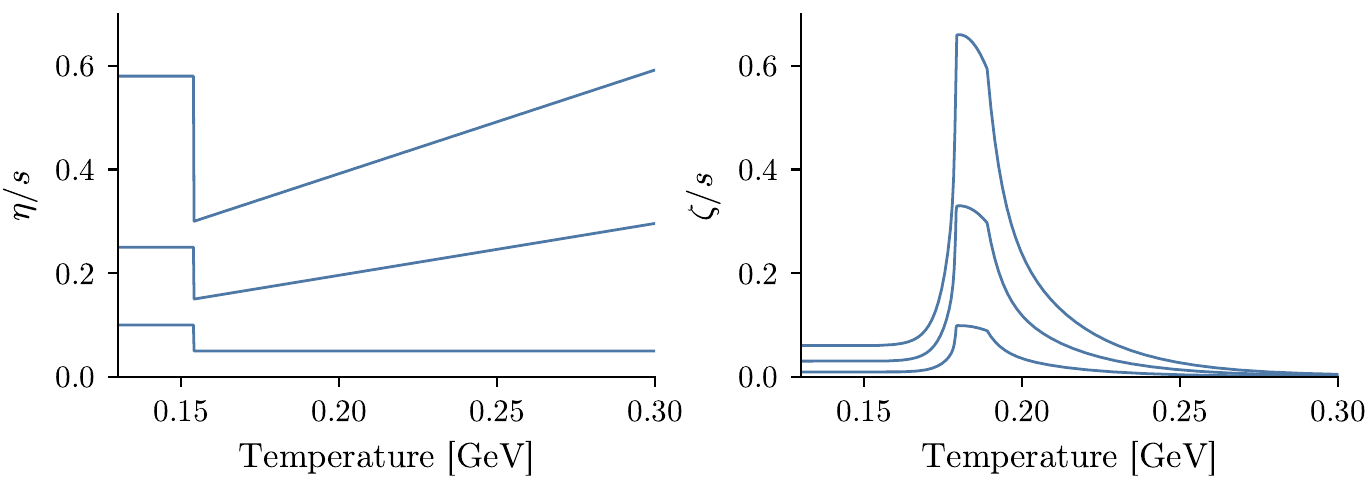}}
  \caption{
    \label{fig:viscosity_dof1}
    Some example curves demonstrating the variability of the temperature-dependent shear viscosity parametrization \eqref{eq:shear_param1} and the bulk viscosity parametrization \eqref{eq:bulk_param1}.
    Lines are chosen for illustrative purposes only and do not represent all possible variability.
  }
\end{figure}

For the specific bulk viscosity $\zetas$, we used the parametrization from references \cite{Ryu:2015vwa, Denicol:2009am}
\begin{equation}
  \label{eq:bulk_param1}
  (\zetas)(T) =
  \begin{cases}
    \begin{aligned}
      C_1 &+ \lambda_1 \exp [(x-1)/\sigma_1]  \\ &+ \lambda_2 \exp [ (x-1)/\sigma_2]
    \end{aligned}
    &T < T_a \\[3ex]
    A_0 + A_1 x + A_2 x^2 &T_a \le T \le T_b \\[2ex]
    \begin{aligned}
      C_2 &+ \lambda_3 \exp [-(x-1)/\sigma_3]  \\ &+ \lambda_4 \exp [-(x-1)/\sigma_4]
    \end{aligned}
    &T > T_b
  \end{cases},
\end{equation}
with $x = T/T_0$ and coefficients:
\begin{align*}
  &C_1=0.03,\quad C_2=0.001, \\
  &A_0=-13.45,\quad A_1=27.55,\quad A_2=-13.77, \\
  &\sigma_1=0.0025,\quad \sigma_2=0.022,\quad \sigma_3=0.025,\quad \sigma_4=0.13, \\
  &\lambda_1=0.9,\quad \lambda_2=0.22,\quad \lambda_3=0.9,\quad \lambda_4=0.25, \\
  &T_0 = 0.18 \text{ GeV},\quad T_a = 0.995\, T_0,\quad T_b = 1.05\, T_0.
\end{align*}
We fixed the peak temperature $T_0=180$~MeV chosen to lie somewhat above the EoS critical temperature and scaled the overall magnitude of the curve using a tunable normalization factor $\zetasnorm$.
Hence for $\zetasnorm = 1$, our bulk viscosity parametrization reproduced the curve used in reference \cite{Ryu:2015vwa, Denicol:2009am}.
Figure \ref{fig:viscosity_dof1} shows some of the temperature-dependent specific viscosities $(\etas)(T)$ and $(\zetas)(T)$ describable using each parametrization.

\subsubsection{Particlization and Boltzmann transport}

The study employed a hybrid transport model that simulated the hot and dense phase of the collision using viscous hydrodynamics and the cooler, more dilute regions of the fireball using microscopic Boltzmann transport \cite{Bass:2000ib, Nonaka:2006yn, Petersen:2008dd}.
The two phases were separated by a pre-specified switching temperature $\Tsw$.
Matter above this temperature was simulated hydrodynamically, and matter below this temperature was simulated using the microscopic transport model.
To preserve the continuity of energy and momentum at the interface between the two regimes, the hydrodynamic medium was converted into particles as it cooled below the $\Tsw$ isotherm.

Particles were sampled from this isotherm using the standard Cooper-Frye algorithm \cite{Cooper-Frye:1974frz}
\begin{equation}
  E \frac{dN_i}{d^3p} = \frac{g_i}{(2 \pi)^3} \int_\sigma f_i(p) p^\mu d^3 \sigma_\mu,
\end{equation}
where $g_i$ and $f_i$ are the degeneracy and distribution function of particle species $i$, and $d^3\sigma_\mu$ is a volume element of the isothermal hypersurface $\sigma$ defined by $\Tsw$.

Following convention, we decomposed the distribution function into an ideal part $f_0$ and a viscous correction $\delta f$.
We modeled the shear contribution to $\delta f$ using the common form \cite{Teaney:2003kp}
\begin{equation}
  \label{eq:shear_corrections}
  \delta f_\text{shear} = f_0 (1 \pm f_0) \frac{1}{2 T^2 (e + P)} p^\mu p^\nu \pi_{\mu\nu},
\end{equation}
where $f_0$ is the ideal Bose-Einstein or Fermi-Dirac distribution, $T$ is the fluid cell temperature, $e$ is its energy density, $P$ is its ideal pressure, and $\pi_{\mu\nu}$ is its shear stress tensor in the fluid rest frame.

The bulk viscous correction to the distribution function, meanwhile, was completely neglected in this study.
Our justification for this choice at the time was two fold.
First, our chosen bulk viscosity parametrization \eqref{eq:bulk_param1} predicted a small value for $\zetas$ at most reasonable particlization temperatures $0.135 \leq \Tsw \leq 0.165$~GeV, so we expected this correction to be small.
And second, the various algorithms used to implement bulk viscous corrections at the time predicted different behavior when either the bulk pressure $\Pi$ or the momentum $p$ are large \cite{Dusling:2011fd, Noronha-Hostler:2013gga}.
Therefore, we decided to neglect bulk viscous corrections until a better algorithm could be implemented.
This choice precluded a quantitative study of the QGP bulk viscosity which we emphasized in the publication at the time \cite{Bernhard:2016tnd}.
Indeed, I will show somewhat later in this chapter that estimates for the temperature dependence of $\zetas$ are quite different if the bulk viscous corrections at freeze-out are appropriately accounted for.

These particles were then fed into the Ultrarelativistic Quantum Molecular Dynamics (UrQMD) model \cite{Bass:1998ca, Bleicher:1999xi} which simulates the particles microscopically until they stop interacting.
UrQMD models individual particle collisions using Monte Carlo techniques to solve the Boltzmann equation
\begin{equation}
  \frac{df_i(x, p)}{dt} = \mathcal{C}_i(x, p),
\end{equation}
where $f_i(x, p)$ is the distribution function of particle species $i$, and $\mathcal{C}_i(x, p)$ is its collision kernel.
The model propagates particles along classical trajectories and simulates their scatterings, resonance formations, and decays.
This produces a list of final particle data, where each particle has an identification number, four-momentum, and four-position at the moment of last interaction.
The particle data for each event was finally processed to calculate the model observables used in the analysis.

\subsection{Parameter design and observables}

This study estimated the joint posterior distribution for nine model parameters used in the construction of the aforementioned nuclear collision model.
Four parameters varied the QGP initial conditions modeled by \trento:
\begin{enumerate}
  \item
    the overall normalization factor for initial entropy deposition,
  \item
    the generalized mean parameter $p$ for the scaling of initial entropy deposition as a function of nuclear thickness,
  \item
    the gamma shape parameter $k$ controlling nucleon multiplicity fluctuations, and
  \item
    the Gaussian nucleon width $w$ determining initial state granularity.
\end{enumerate}
Meanwhile, another five parameters varied the properties of the hybrid model simulation:
\begin{enumerate}
  \item[5--7.]
    three parameters ($\etas$ hrg, min, and slope) for the temperature dependence of the QGP shear viscosity,
  \setcounter{enumi}{7}
  \item
    one parameter $\zetas$ norm for the overall normalization and magnitude of the QGP bulk viscosity, and
  \item
    a particlization temperature \Tsw\ that defined the isotherm for Cooper-Frye particle emission.
\end{enumerate}

We assigned each parameter the conservative range of prior values listed in table \ref{tab:design1} and sampled 300 parameter points inside the resulting nine-dimensional parameter space using a maximin Latin hypercube design.
We then ran \order{10^4} minimum-bias Pb-Pb collisions at $\sqrts=2.76$~TeV at each parameter point.
Each nuclear collision event consisted of a single fluctuated initial condition profile and hydrodynamic simulation followed by numerous hadronic afterburner oversamples, i.e.\ repeated iterations of the Cooper-Frye emission and UrQMD evolution.
The completed events were then partitioned into centrality classes according to their final charged-particle multiplicity at midrapidity.

\begin{table}[h]
  \centering
  \captionsetup{width=\textwidth}
  \caption{
    Input parameter ranges for the nuclear collision model.
  }
  \label{tab:design1}
  \small
  \begin{tabular}{lll}
    \toprule
    Parameter                  & Description                            & Prior range \\
    \midrule
    Norm                       & Normalization factor                   & 100--250 \\
    $p$                        & Entropy deposition parameter           & $-1$ to $+1$ \\
    $k$                        & Multiplicity fluct.\ shape             & 0.8--2.2 \\
    $w$ [fm]                   & Gaussian nucleon width                 & 0.4--1.0 \\
    $\etas$ hrg                & Const.\ shear viscosity below $T_c$    & 0.3--1.0 \\
    $\etas$ min                & Minimum value of $\etas$ (at $T_c$)    & 0--0.3 \\
    $\etas$ slope [GeV$^{-1}$] & Slope of $\etas$ above $T_c$           & 0--2 \\
    $\zetas$ norm              & Prefactor for $(\zetas)(T)$            & 0--2 \\
    $\Tsw$ [GeV]               & Switching / particlization temperature & 0.135--0.165 \\
    \bottomrule
  \end{tabular}
\end{table}

Using this event data, we calculated the charged-particle yield $d\nch/d\eta$, identified-particle yields $dN/dy$, identified-particle mean $p_T$, and two-particle flow cumulants $\vnk{n}{2}$ for $n=2$, 3, and 4.
Each observable was calculated for a number of different centrality bins selected to match the experiment.
Table \ref{tab:observables1} summarizes these Pb-Pb $\sqrts=2.76$~TeV observables, including the kinematic cuts, centrality bins, and experimental data, which were taken from the ALICE experiment \cite{ALICE:2011ab, Abelev:2013vea}.

\begin{table}
  \centering
  \caption{
    Experimental data to be compared with model calculations.
  }
  \label{tab:observables1}
  \small
  \makebox[\textwidth]{
    \begin{tabular}{lcccc}
      \toprule
      Observable & Particle species & Kinematic cuts & Centrality classes & Ref. \\
      \midrule
      Yields $dN/dy$                       & $\pi^\pm$, $K^\pm$, $p\bar p$ &
      $|y| < 0.5$ & 0--5, 5--10, 10--20, \ldots, 60--70 & \cite{Abelev:2013vea} \\
      \noalign{\smallskip}
      Mean transverse momentum $\avg{p_T}$ & $\pi^\pm$, $K^\pm$, $p\bar p$ &
      $|y| < 0.5$ & 0--5, 5--10, 10--20, \ldots, 60--70 & \cite{Abelev:2013vea} \\
      \noalign{\smallskip}
      Two-particle flow cumulants $\vnk n 2$ & \multirow{2}{*}{all charged} &
      $|\eta| < 1$ & 0--5, 5--10, 10--20, \ldots, 40--50 &
      \multirow{2}{*}{\cite{ALICE:2011ab}} \\
      $n = 2$, 3, 4 & & $0.2 < p_T < 5.0$ GeV & $n = 2$ only: 50--60, 60--70 & \\
      \bottomrule
    \end{tabular}
  }
\end{table}

These measurements are of course a small cross section of the data that ALICE and the other experiments have to offer.
We chose these specific observables as a starting point because they are statistically cheap to compute and because nuclear collision models often struggle to describe them simultaneously, e.g.\ see reference \cite{Bernhard:2015hxa}.
I'll describe a far more grandiose study later in this chapter which used a much larger cross section of the available experimental data.

\subsection{Bayesian parameter estimation}

\begin{figure}[t]
  \centering
  \makebox[\textwidth]{\includegraphics{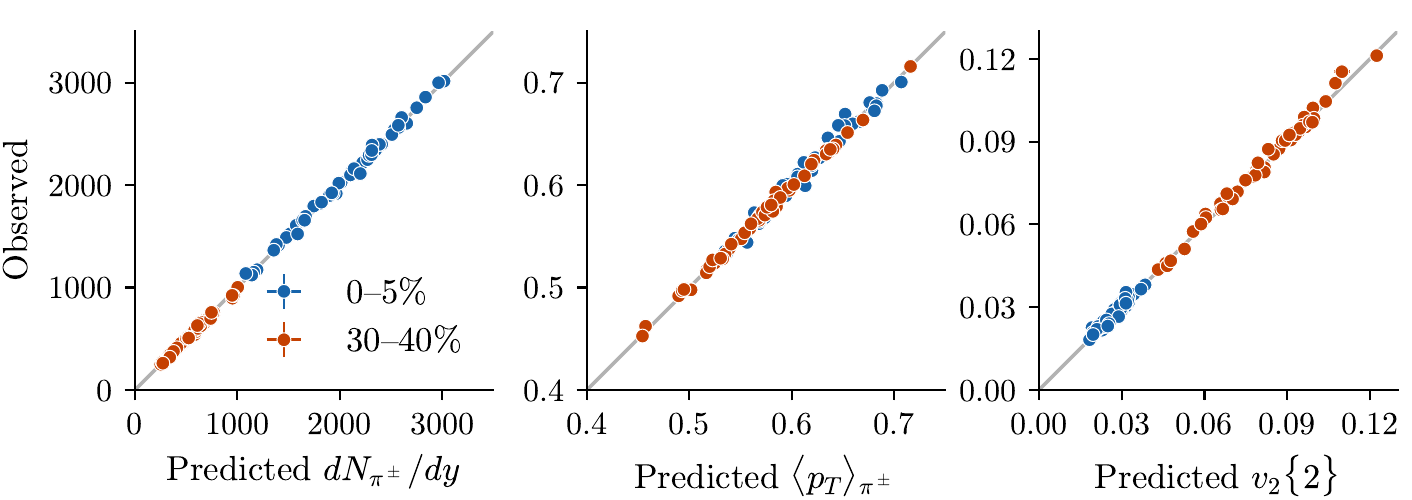}}
  \caption{
    \label{fig:validation1}
    Scatter plot of the predicted (emulated) model output versus the observed (true) model output for the pion yield (left), mean $p_T$ (middle) and elliptic flow cumulant $\vnk{2}{2}$ (right).
    Diagonal line indicates perfect agreement.
  }
\end{figure}

Bayesian parameter estimation was then used to estimate the values of the nine model parameters listed in table \ref{tab:design1} using the experimental data in table \ref{tab:observables1} and the predictions of our nuclear collision model.
We trained, for this purpose, an emulator using the simulated observables predicted by the nuclear collision model at each design point (see section \ref{sec:emulator} for details).
Once the emulator was trained, we validated the emulator results by comparing the emulator predictions to explicit model calculations using an independent set of parameter points which were excluded from the calibration.
Figure \ref{fig:validation1} shows this validation test applied to several of the model observables used in the analysis.
The results demonstrate the accuracy of the emulator, as evidenced by the proximity of each point to the diagonal gray line indicating perfect emulator and model agreement.

Before I proceed to present the results of the analysis, a few comments are in order about its treatment of uncertainties.
Generally speaking, there are two sources of quantifiable uncertainty which should be accounted for in the analysis.
First, there is the experimental uncertainty quantified by the covariance matrix $\Sigma_e$, consisting of individual statistical and systematic contributions.
And second, there is the model (emulator) uncertainty quantified by the covariance matrix $\Sigma_m$.
It includes statistical uncertainty in the model outputs and interpolation uncertainty arising from limited training data.
The overall likelihood function is then calculated using the total covariance matrix \eqref{eq:tot_cov}, equal to their sum, $\Sigma = \Sigma_m + \Sigma_e$.

Previously, in subsection \ref{subsec:likelihood}, I described how to calculate the likelihood function equation \eqref{eq:likelihood} in the physical observable basis, i.e.\ in the basis of $\yv$, following the derivation outlined in \cite{Bernhard:2018hnz}.
In the study I am discussing now, the likelihood function was calculated in the basis of the PCA transformed observables, i.e.\ in the basis of $\zv$.
Expressed in this form, the likelihood function becomes
\begin{equation}
  \label{eq:likelihood_pc_space}
  P(E | \xv) \propto \exp\left \{-\frac{1}{2} [\zv_m(\xv) - \zv_e]\tran \Sigma_z^{-1} [\zv_m(\xv) - \zv_e] \right\},
\end{equation}
where $\zv_m(\xv)$ and $\zv_e$ are the PCA transformed model (emulator) observables and experimental observables respectively, and $\Sigma_z$ is their total covariance matrix.
Here I've discarded the overall normalization prefactor when writing equation \eqref{eq:likelihood_pc_space} because it is not necessary for MCMC sampling.
A 10\% fractional uncertainty was then placed on the PCs following \cite{Novak:2013bqa,Bernhard:2015hxa}
\begin{equation}
  \label{eq:frac_uncertainty1}
  \Sigma_z = \diag(\sigma^2_z\,\zv_e),
\end{equation}
using $\sigma_z = 0.10$.
This was a rough approximation intended to conservatively account for various sources of uncertainty in the experimental data, model calculations, and emulator predictions.
Recently, more advanced methods have been developed to rigorously estimate the likelihood covariance matrix \cite{Bernhard:2019ntr}.
At the time, approximation \eqref{eq:frac_uncertainty1} was merely a stop-gap.

We placed a uniform prior on each model parameter which was constant within the design ranges listed in table \ref{tab:design1} and zero outside them.
We then applied Bayes' theorem \eqref{eq:bayes} to calculate the posterior probability density at each parameter point using the likelihood function and the prior.
Finally, we used the affine-invariant MCMC ensemble sampler \textsc{emcee} \cite{emcee} to draw samples from the posterior distribution, first generating \order{10^6} burn-in samples to allow the chain to equilibrate followed by \order{10^7} production samples to visualize the posterior.

\subsection{Posterior parameter estimates}

\begin{figure}[!b]
  \centering
  \makebox[\textwidth]{\includegraphics{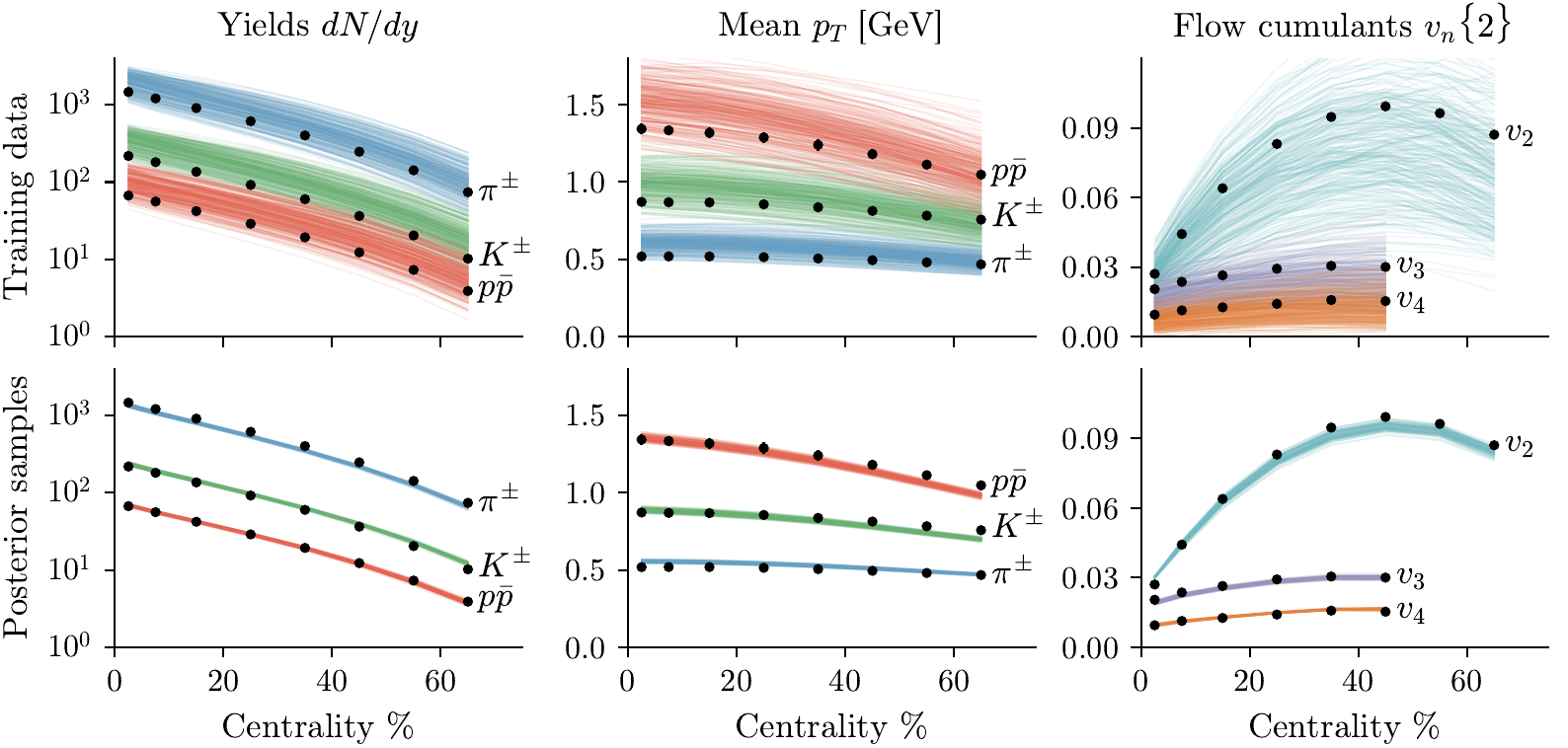}}
  \caption{
    \label{fig:obs_samples1}
    Model and emulator predictions (colored lines) compared to experimental data (black symbols and error bars).
    Top row: explicit model calculations at each of the 300 design points used to train the emulator.
    Bottom row: emulator predictions using 100 parameter combinations randomly drawn from the posterior.
    Left column shows identified particle yields $dN/dy$ for pions, kaons, and protons, middle column shows mean $p_T$ for pions, kaons, and protons, and right column shows the two-particle flow cumulants $\vnk{n}{2}$ for charged particles.
    Experimental data is from the ALICE collaboration \cite{ALICE:2011ab, Abelev:2013vea}.
  }
\end{figure}

Figure \ref{fig:obs_samples1} shows simulated model observables (colored lines) compared to the experimental data (black symbols).
The top row of the figure shows explicit model calculations at each of the 300 design points.
Their large visual spread is a result of each parameter's initial design range (see table \ref{tab:design1}) reflecting the prior uncertainty in the true parameter values.
Now direct your attention to the bottom row of the figure.
These lines are emulator predictions using 100 parameter combinations randomly sampled from the posterior.
The lines are now tightly clustered around the experimental data, reflecting the uncertainty of the posterior distribution and the trade-offs that are made when fitting all of the observables simultaneously.

The agreement of the posterior sampled emulator predictions is really quite impressive, matching the accuracy of the most advanced dynamical models at the time \cite{Niemi:2015qia, Schenke:2012wb, Gale:2012rq, Ryu:2015vwa}.
The result is testament to the remarkable success of the hydrodynamic standard model which is the foundation of our model-to-data comparison.
Moreover, it validates the general parametric initial condition approach used throughout this dissertation.
The agreement of the calibrated model with the data is evidence that the \trento\ model is able to well reproduce the data using appropriately chosen parameters.
Of course, this does not necessarily mean the model is \emph{correct} or anything of that nature---our inference could still be contaminated by errors lurking elsewhere in the framework.
Rather, it demonstrates that the model survives a significant test of its prediction accuracy.

\begin{figure}[p]
  \centering
  \makebox[\textwidth]{\includegraphics{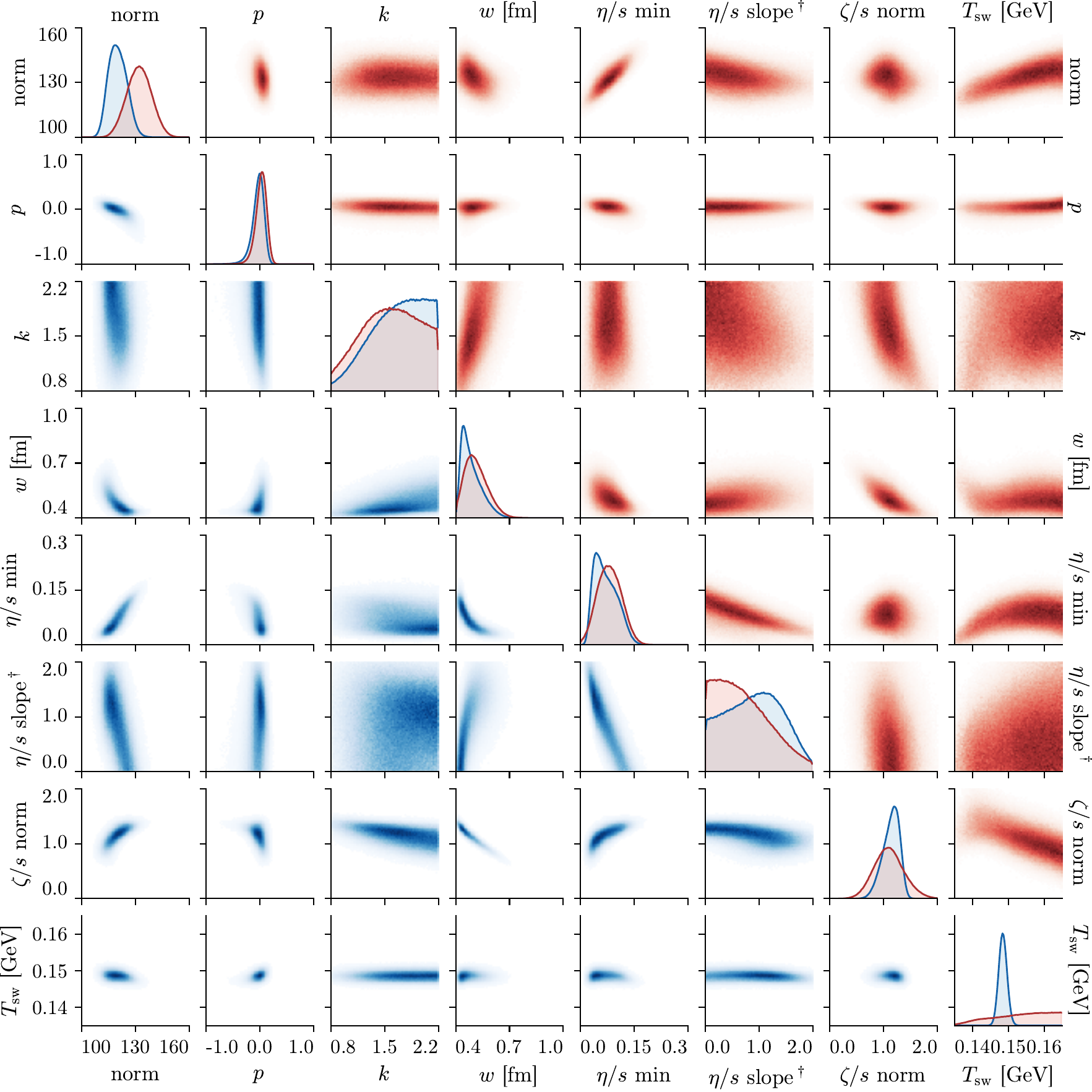}}
  \captionsetup{width=.75\paperwidth}
  \caption{
    \label{fig:posterior1}
    Posterior distribution for the model parameters, calibrated including identified particle yields (blue, lower triangle) and charged particle yields (red, upper triangle).
    Each diagonal panels is the marginal distribution of a single parameter, and each off-diagonal panel is the joint distribution of a pair of parameters visualizing their correlations.
    The parameter $\etashrg$ has been dropped from the posterior distribution for reasons which are explained in the text.
  }
\end{figure}

The larger scope of this work is of course not to merely validate the model, but to learn about the underlying physics of the collision using the constraints on each parameter imposed by the experimental data and the assumptions of our framework.
Figure \ref{fig:posterior1} presents the primary result of this study, a visualization of the Bayesian posterior distribution for the model input parameters.
Each diagonal panel is the marginal distribution of a single model parameter, and each off-diagonal panel is the joint distribution of a pair of model parameters visualizing their correlations.
There are, in fact, \emph{two} posterior distributions in this figure.
The blue lower-triangle is the posterior that's obtained when pion, kaon, and proton yields $dN/dy$ are included in the calibration, and the red upper-triangle is the posterior that's obtained when these identified yields are replaced by the charged particle yield $d\nch/d\eta$ instead.
The reason for showing both posteriors will be explained in a moment.

\begin{table}
  \centering
  \caption{
    Estimated parameter values (medians) and uncertainties (90\% credible intervals) from the posterior distributions calibrated to identified and charged particle yields (middle and right columns, respectively).
    The distribution for $\Tsw$ based on charged particles is essentially flat, so we do not report a quantitative estimate.
  }
  \small
  \label{tab:posterior1}
  \begin{tabular}{lll}
    \toprule
    & \multicolumn{2}{c}{Calibrated to:} \\
    \noalign{\smallskip}\cline{2-3}\noalign{\smallskip}
    Parameter & \multicolumn{1}{c}{Identified} & \multicolumn{1}{c}{Charged} \\
    \midrule
    Normalization & \parbox{3.5em}{\hfill $120.$}$_{-8.}^{+8.}$ & \parbox{3.5em}{\hfill $132.$}$_{-11.}^{+11.}$ \\ \noalign{\smallskip}
    $p$ & \parbox{3.5em}{\hfill $-0.02$}$_{-0.18}^{+0.16}$ & \parbox{3.5em}{\hfill $0.03$}$_{-0.17}^{+0.16}$ \\ \noalign{\smallskip}
    $k$ & \parbox{3.5em}{\hfill $1.7$}$_{-0.5}^{+0.5}$ & \parbox{3.5em}{\hfill $1.6$}$_{-0.5}^{+0.6}$ \\ \noalign{\smallskip}
    $w$ [fm] & \parbox{3.5em}{\hfill $0.48$}$_{-0.07}^{+0.10}$ & \parbox{3.5em}{\hfill $0.51$}$_{-0.09}^{+0.10}$ \\ \noalign{\smallskip}
    $\etas$ min & \parbox{3.5em}{\hfill $0.07$}$_{-0.04}^{+0.05}$ & \parbox{3.5em}{\hfill $0.08$}$_{-0.05}^{+0.05}$ \\ \noalign{\smallskip}
    $\etas$ slope [GeV$^{-1}$] & \parbox{3.5em}{\hfill $0.93$}$_{-0.92}^{+0.65}$ & \parbox{3.5em}{\hfill $0.65$}$_{-0.65}^{+0.77}$ \\ \noalign{\smallskip}
    $\zetas$ norm & \parbox{3.5em}{\hfill $1.2$}$_{-0.3}^{+0.2}$ & \parbox{3.5em}{\hfill $1.1$}$_{-0.5}^{+0.5}$ \\ \noalign{\smallskip}
    $T_\mathrm{switch}$ [GeV] & \parbox{3.5em}{\hfill $0.148$}$_{-0.002}^{+0.002}$ & \hspace{3em}--- \\ \noalign{\smallskip}
    \bottomrule
  \end{tabular}
\end{table}

Table \ref{tab:posterior1} also lists the median and 90\% highest posterior density (HPD) credible interval for each model parameter.
This interval is defined in section \ref{sec:visualize_posterior} as the smallest interval containing 90\% of the parameter's posterior density.
Roughly speaking, we expect the true parameter values to fall within these ranges 90\% of the time assuming the model is exact and that all uncertainties are accounted for.
There is a lot of information to unpack here, so I'll start with the marginal distributions in figure \ref{fig:posterior1}.
These are the red and blue histograms located on the diagonal.
I'll begin with the top-left corner and work my way to the bottom-right.

\subsubsection{Initial condition parameters}

The initial entropy normalization is unsurprisingly well constrained.
Here we see that the normalization is different if we calibrate to fit the identified particle yields (blue histograms) or the charged particle yields (red histograms).
This is because the model cannot fit the pion, kaon, and proton yields simultaneously.
So it generally tends to underpredict the pion yield in order to better fit the kaons and protons.
This was our original motivation for calibrating with and without the identified particle yields included.

The next diagonal panel shows the marginal distribution for the generalized mean parameter $p$ which controls the scaling of initial entropy deposition as a function of nuclear thickness, equation \eqref{eq:entropy_deposition1}.
Here we see a peak centered at $p=0$.
Recall that this value corresponds to a geometric mean
\begin{equation}
  s(\xv_\perp; \eta_s=0, \tau_0) = \frac{\text{Norm}}{\tau_0} \times \sqrt{\T_A\, \T_B},
\end{equation}
where $\T_A$ and $\T_B$ are the participant thickness functions.
Note, this is the exact same $p$ value supported by our preliminary examination of charged particle multiplicity distributions, figure \ref{fig:nch}, and eccentricity harmonics, figure \ref{fig:eccen}.
This consensus is strong indication of universal scaling.

\begin{figure}[t]
  \centering
  \includegraphics{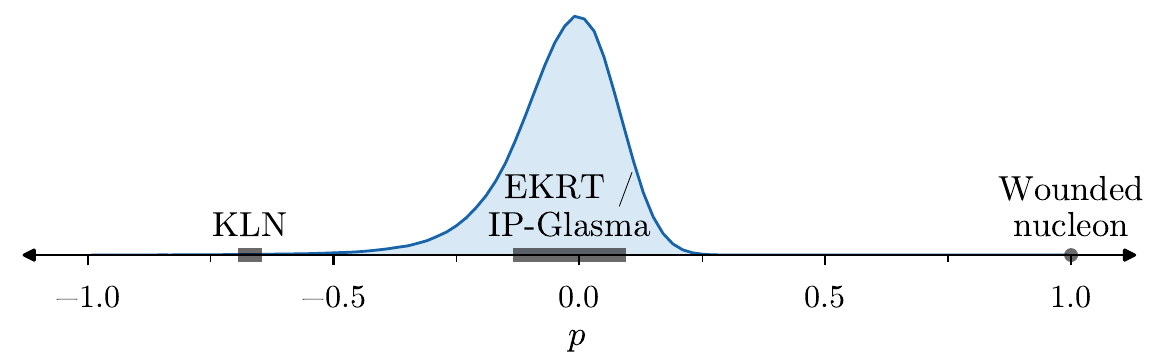}
  \caption{
    \label{fig:posterior_p_arrows1}
    Posterior distribution on the \trento\ generalized mean parameter $p$ compared to the effective values of $p$ needed to mimic the KLN, EKRT, IP-Glasma, and Wounded nucleon models.
    These values are explained in subsection \ref{subsec:mimic}.
  }
\end{figure}

Previously in subsection \ref{subsec:mimic}, I showed that the $p$ parameter smoothly interpolates different classes of initial condition models; the wounded nucleon model is equivalent to arithmetic mean scaling $p = 1$, the EKRT and IP-Glasma models behave similarly to geometric mean scaling $p \sim 0$, and the KLN model is closely fit by $p = -0.67$.
Figure \ref{fig:posterior_p_arrows1} shows an expanded view of the Bayesian posterior distribution on $p$ with black symbols and bands marking the effective $p$ values needed to describe each of the aforementioned models.
The EKRT and IP-Glasma models lie squarely in the peak of the posterior distribution, while the KLN and wounded nucleon models are considerably outside, corroborating the findings of several previous model validation studies \cite{Niemi:2015qia, Bernhard:2015hxa, Song:2011hk}.

The posterior distribution on the parameter $k$, meanwhile, is inconclusive.
Recall that this parameter is the inverse variance of the Gamma distribution used to sample nucleon-nucleon fluctuations; see equation \eqref{eq:participant_thickness}.
In hindsight, we've realized there is a problem with this parametrization.
The only way to turn the fluctuations off is to send $k$ to infinity.
Our prior tests a finite range of $k$ values, so it never tests the model predictions with the fluctuations turned completely off.
We've fixed this issue in later studies by replacing the parameter $k$ with the standard deviation of the fluctuations $\sigma_\text{fluct}=1/\sqrt{k}$.
We'll use this modified parametrization in the next study.

Continuing down the diagonal, the Gaussian nucleon width $w$ is narrowly peaked mostly within the range 0.4--0.6~fm.
This estimate is quantitatively consistent with the value for the gluonic nucleon width extracted from deep inelastic scattering data at HERA \cite{Chekanov:2004mw, Kowalski:2006hc, Rezaeian:2012ji}, and it is also similar to the values
commonly used in studies of the IP-Glasma and EKRT models \cite{Schenke:2012wb, Niemi:2015qia}.
However, this estimate is also somewhat puzzling in hindsight because we obtained much larger estimates for $w$ in our subsequent Bayesian analyses.
Evidently the preferred nucleon width is sensitive to the details of the dynamical model.
I'll return to this puzzle later in the chapter.

\subsubsection{QGP medium parameters}

The next three parameters $\etasmin$, $\etasslope$, and $\zetasnorm$ all control the temperature dependence of the QGP viscosity.
Here, we've dropped the parameter $\etashrg$ from the posterior distribution because it has no effect on the model, i.e.\ its posterior distribution is completely flat.
The $\etashrg$ parameter controls the specific shear viscosity of the hydrodynamic simulation below the temperature $T_c = 0.154$~GeV, so it only affects the model output if we particlize the fluid well below that temperature.
The model generally prefers a particlization (switching) temperature near $T_c$ (see below), so the $\etashrg$ parameter ultimately has little effect.

The posterior on $\etasmin$ is mostly peaked between 0 and 0.15.
Its temperature dependent slope $\etasslope$, meanwhile, is far broader and extends to either edge of its prior; only large slopes $\etasslope > 2$~GeV appear to be excluded.
These two marginal distributions, however, paint an incomplete picture of the posterior on $(\etas)(T)$.
The joint posterior distribution of $\etasmin$ and $\etasslope$ reveals that both parameters are inversely correlated.
The posterior allows for a small value of $\etasmin$ and a large value of $\etasslope$ or \emph{vice versa}, but both cannot be large (or small) at the same time.
Hence, most of the uncertainty in $\etasslope$ arises from this degeneracy.
This is exactly the type of relationship that is difficult to discover without using Bayesian parameter estimation.

\begin{figure}[t]
  \centering
  \includegraphics{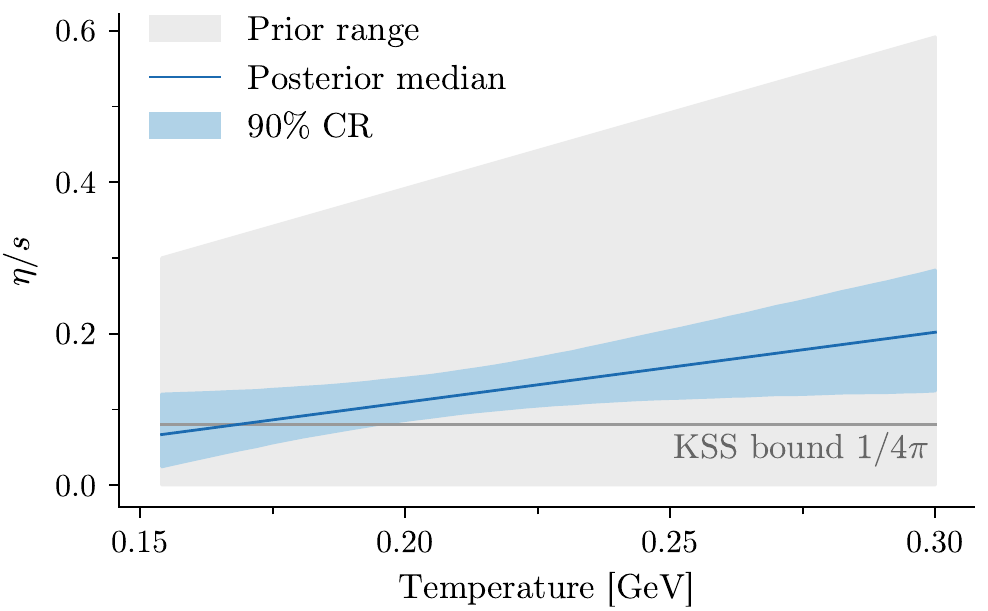}
  \caption{
    \label{fig:etas_estimate1}
    Posterior estimate for the temperature dependence of the QGP specific shear viscosity $\etas$.
    The trapezoidal gray region is the prior range explored by the parametrization \eqref{eq:shear_param1_repeat}.
    The blue band is the parametrization's posterior 90\% HPD credible region, and the blue line its median.
    The horizontal gray line is the KSS bound $\etas \geq 1/4 \pi$ \cite{Kovtun:2004de, Policastro:2001yc, Danielewicz:1984ww}.
  }
\end{figure}

There is, alternatively, a better way to visualize the posterior estimate for the temperature dependence of the specific shear viscosity $(\etas)(T)$.
We modeled this function above the critical temperature $T_c$ using the linear ansatz
\begin{equation}
  \label{eq:shear_param1_repeat}
  (\etas)(T) = \etasmin + \etasslope (T - T_c),
\end{equation}
with two parameters $\etasmin$ and $\etasslope$.
Now that we have the joint posterior distribution for these parameters, we can sample the function defined by equation \eqref{eq:shear_param1_repeat}.
Concatenating these samples into a single list, we can then calculate a credible interval at each value of the temperature to form a credible region (CR).
Figure \ref{fig:etas_estimate1} applies this method to visualize our estimate for the temperature dependence of the specific shear viscosity for $T \geq T_c$.
The gray band is the region spanned by the prior distribution on the shear viscosity parameters, the blue band is the posterior's 90\% CR, and the blue line is its median.
We've also superimposed the well-known KSS bound \cite{Kovtun:2004de, Policastro:2001yc, Danielewicz:1984ww} from AdS-CFT holography for reference purposes.

Perhaps the most interesting feature of this figure is its characteristic bow-tie shape.
The posterior distribution narrows slightly above ${\sim}200$~MeV and fans out again on either side.
Presumably this is the temperature that the medium evolution is most sensitive to.
It suggests that we may be able to get a better handle on the temperature dependence of the QGP viscosity by including more beam energies and hence more handles at different temperature points on this graph.
Indeed, I will summarize the results of a study which does exactly that in the next section.

The posterior on the bulk viscosity $\zetasnorm$ parameter (scaling prefactor for equation~\eqref{eq:bulk_param1}), meanwhile, is clearly peaked just above one.
Therefore, the results agree with leaving the bulk viscosity parametrization unscaled, as in reference \cite{Ryu:2015vwa}.
Given the previously mentioned short-comings in our treatment of bulk viscosity, namely neglecting bulk corrections at particlization and the lack of a dynamical pre-equilibrium phase, we refrained from making any quantitative statements about the temperature dependence of bulk viscosity.
Suffice to say, the analysis clearly prefers non-zero bulk viscosity.

The last model parameter is the Cooper-Frye particlization temperature \Tsw.
This parameter sets the temperature of thermal particle emission, and it strongly affects the relative particle abundances.
Therefore it is primarily constrained by the pion, kaon, and proton yields $dN/dy$.
This is easily verified by removing these yields from the calibration (red posterior distribution).
Once the yields are removed, the constraint on the particlization temperature vanishes.
Evidently, there was no single particlization temperature which could fit the pion, kaon, and proton yields simultaneously, but $\Tsw \sim 148$~MeV provided the best overall compromise.

\subsection{Verification of high-probability parameters}
\label{subsec:verification_trento3d}

\begin{figure}[t]
  \centering
  \makebox[\textwidth]{\includegraphics{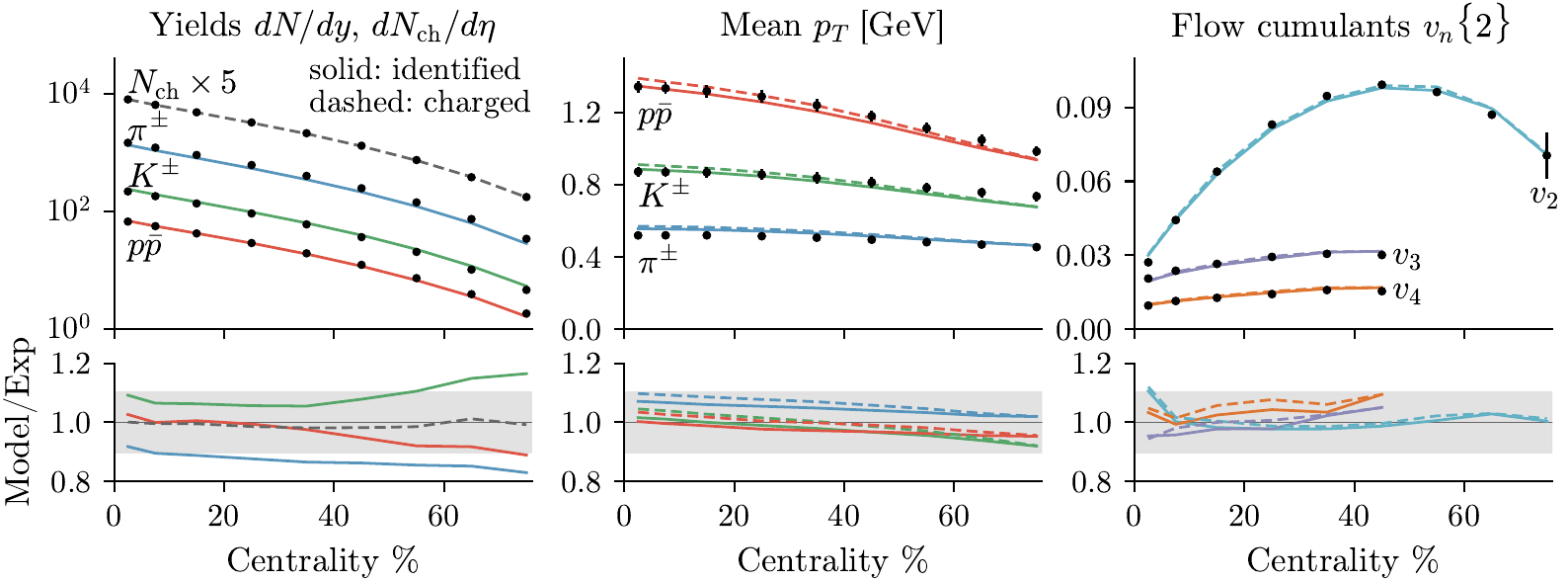}}
  \caption{
    \label{fig:mode_observables1}
    Explicit model calculations using the high-probability parameters listed in table \ref{tab:mode_params1}.
    Solid lines are calculations using parameters based on the identified particle posterior, dashed lines are calculations using parameters based on the charged particle posterior, and black symbols with errors are experimental data from ALICE \cite{Abelev:2013vea, ALICE:2011ab}.
    Top row: Charged or identified particle yields $dN/dy$ or $d\nch/d\eta$ (left), identified particle mean $p_T$ (middle), and two-particle flow cumulants $\vnk{n}{2}$ (right).
    Bottom row: Ratio of the model to the experimental data.
  }
\end{figure}

As a final test of the emulator prediction accuracy and our model's veracity, we ran \order{10^6} events using a single combination of high-probability parameters selected from the mode of the posterior probability distribution.
Roughly speaking, these values demonstrate the single best-fit attainable by the model.
Table \ref{tab:mode_params1} lists these parameter values and figure \ref{fig:mode_observables1} shows the resulting model predictions compared to the experimental data.
The identified and charged particle yields $dN/dy$ and $d\nch/d\eta$ are shown on the left, the identified mean $p_T$ in the middle, and the two-particle flow cumulants $\vnk{n}{2}$ on the right.
The inset below each figure shows the ratio of the model calculations to the data, and the gray band indicates $\pm10\%$ error.

\begin{table}[h]
  \centering
  \caption{
    \label{tab:mode_params1}
    High-probability parameters chosen based on the posterior distributions and used to generate Fig.~\ref{fig:mode_observables1}.
    Pairs of values separated by slashes are based on identified / charged particle yields, respectively.
    Single values are the same for both cases.
  }
  \small
  \begin{tabular}{ll@{\hspace{2em}}ll}
    \toprule
    \multicolumn{2}{c}{Initial condition / Pre-eq}  & \multicolumn{2}{c}{QGP medium} \\
    \cmidrule(r){1-2}                                 \cmidrule(l){3-4}
    norm & 120. / 129.                              & $\etas$ min      & 0.08 \\
    $p$  & 0.0                                      & $\etas$ slope    & 0.85 / 0.75 GeV$^{-1}$ \\
    $k$  & 1.5  / 1.6                               & $\zetas$ norm    & 1.25 / 1.10 \\
    $w$  & 0.43 / 0.49 fm                           & $T_\text{switch}$ & 0.148 GeV \\
    \bottomrule
  \end{tabular}
\end{table}

The agreement of all observables is ${\sim}10\%$ or better with two notable exceptions.
The pion, kaon, and proton yields $dN/dy$ are off by ${\sim}10$--$20\%$, particularly in the most peripheral centrality bins, and the two-particle flow cumulants $\vnk{n}{2}$ deviate from the data in the 0--5\% centrality range.
Both issues are known problems for this class of model.
All nitpicking aside, the model performance is really quite remarkable.
For instance, the total charged particle production is essentially perfect all the way out to 80\% centrality, suggesting that the tension in the identified particle yields is a problem with the medium chemistry, not with initial entropy deposition.
It's also impressive that the model correctly describes the nuanced shape of each mean $p_T$ and flow cumulant curve.
For example, the mean $p_T$ of pions, kaons, and protons is more separated in 0--10\% centrality range than it is in the 60--70\% centrality range for both the model calculation and experiment.
The model also nails the centrality dependence of the anisotropic flow harmonics, observables that are notoriously sensitive to the geometry of the initial conditions and the temperature dependence of the hydrodynamic transport coefficients.

I should reemphasize, however, that the aforementioned results include several caveats.
We did not account for any uncertainty in the pre-equilibrium dynamics that describe the first moments of the collision prior to the onset of hydrodynamic expansion, asserting instead instant thermalization at the hydrodynamic starting time $\tau_0$.
We also chose a rather restrictive parametrization for the temperature dependence of the specific bulk viscosity, and we completely neglected the bulk viscous corrections to the particle distribution at freeze-out.
Moreover, we took several short-cuts when estimating the likelihood covariance matrix which degrade the strict interpretation of our uncertainty estimates.
These defects are the reason why I initially referred to the study as a proof of concept.

\subsection{Improved boost-invariant nuclear collision model}
\label{subsec:nuclear_collision_model_v2}

This subsection describes several improvements to the boost-invariant nuclear collision model used in the previous study.
The most significant changes to the model are the addition of a pre-equilibrium free streaming stage \cite{freestream} and a newly developed particle sampler \cite{frzout} which were added by Jonah Bernhard.
The following describes these upgrades among others which distinguish the dynamical model from its previous version.
I'll also describe some computational tricks which greatly reduced the time needed to run the model.
These modifications are relevant to results which I will present later in the chapter.

\subsubsection{Initial conditions}

The updated model includes two important changes to \trento:
\begin{enumerate}[leftmargin=1\parindent]
  \item
    The nucleon minimum distance algorithm described in subsection \ref{subsec:nucleon_corr} was added to study the effect of nucleon correlations on our sampled lead nuclei.
  The algorithm imposes a minimum distance criterion
  \begin{equation}
    |\xv_i - \xv_j| > d_\text{min}
  \end{equation}
  between all pairs of nucleons $i, j$, where $d_\text{min}$ is a variable model parameter.
  This algorithm is constructed to preserve the desired Woods-Saxon radial distribution so there is no artificial swelling of the nucleus with increasing $d_\text{min}$.
  \item
    Second, we reinterpreted the output of the \trento\ model as an initial \emph{energy} density profile $e = dE/(d^2x_\perp \tau_0\, d\eta_s)$, written as
  \begin{equation}
    \label{eq:energy_deposition}
    e(\xv_\perp; \eta_s=0, \tau_0) = \text{const} \times \left[ \frac{\T_A(\xv_\perp)^p + \T_B(\xv_\perp)^p}{2} \right]^{1/p},
  \end{equation}
  where const is an overall normalization factor with units of GeV/fm.
\end{enumerate}

Here we opted to parametrize the initial energy density (rather than the initial entropy density) since its dynamics are exactly calculable in the weakly-coupled non-interacting limit, subject to certain simplifying assumptions \cite{Broniowski:2008qk,Liu:2015nwa}.
This so-called free streaming approximation provides a more realistic starting point for the hydrodynamic equations of motion, as I'll explain below.

\subsubsection{Pre-equilibrium evolution}

The nuclear collision model was updated to free stream the initial energy density modeled by equation \eqref{eq:energy_deposition} to the hydrodynamic starting time using a procedure developed in \cite{Broniowski:2008qk,Liu:2015nwa}.
The free streaming evolution was implemented using \textsc{freestream}, a publicly available Python code written by J.\ Bernhard \cite{freestream}.
The following excerpt summarizes the free streaming procedure in my own words.
This text also appears in reference \cite{Bernhard:2019ntr}.

\begin{figure}[!b]
  \centering
  \includegraphics{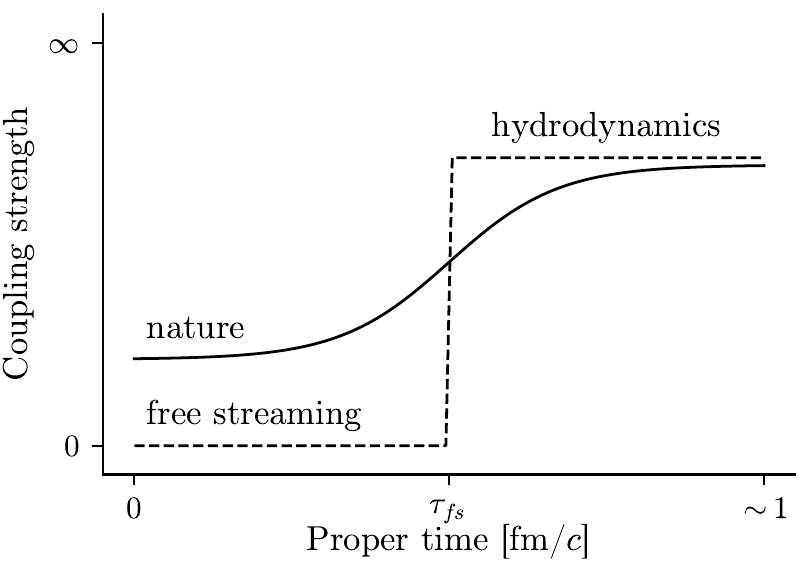}
  \caption{
    \label{fig:coupling}
    Cartoon of the coupling strength for the first ${\sim}1~\fmc$ of the collision.
    The solid line shows a reasonable time-evolution for the coupling strength in nature, and the dashed line shows the free streaming approximation used in this work.
  }
\end{figure}

The initially deposited matter modeled by equation \eqref{eq:energy_deposition} is expected to rapidly approach the conditions of hydrodynamic applicability over a timescale of $\tau \sim 1~\fmc$, although the details of this evolution are the subject of ongoing investigation \cite{Heinz:2015arc,Kurkela:2018vqr}.
There are, of course, two natural limiting cases for the initial strength of the medium interactions: infinitely weak coupling, where the matter free streams without interacting; and infinitely strong coupling, where the inter-particle mean free path effectively vanishes.
Realistically, one expects the initially produced medium to lie somewhere between these two extremes.
The updated model therefore free streams the energy density $e(\xv_\perp, \eta_s=0, \tau_0)$ until a variable hydrodynamic starting time $\tfs > \tau_0$ to modulate the time-averaged strength of medium interactions prior to the onset of hydrodynamic expansion.
Figure \ref{fig:coupling} shows a cartoon of this evolution.

The free streaming evolution is easily calculable for a boost-invariant gas of massless partons with locally isotropic transverse momenta \cite{Broniowski:2008qk,Liu:2015nwa}.
Under these conditions, the energy-momentum tensor at transverse position $\xv_\perp$ and time $\tau > \tau_0$ is given by
\begin{equation}
  T^{\mu\nu}(\xv_\perp, \eta_s=0, \tau) = \\
  \frac{\tau_0}{\tau} \int_{-\pi}^{\pi} d\phi_p\,\hat{p}^\mu \hat{p}^\nu e(\xv_\perp - (\tau - \tau_0) \hat{\mathbf{p}}_\perp, \eta_s=0, \tau_0),
  \label{eq:freestream}
\end{equation}
where $\hat{p}^\mu = (1, \cos\phi_p, \sin\phi_p, 0)$ and $\hat{\mathbf{p}}_\perp = (\cos\phi_p, \sin\phi_p)$ are momentum unit vectors.
At top RHIC and LHC energies, the nuclear interpenetration time $\tau_0 \ll 1$~fm/$c$ so the free streaming time $\tfs \equiv \tau - \tau_0 \approx \tau$.
One can also combine the normalization constant in equation \eqref{eq:energy_deposition} with the prefactor $\tau_0$ in equation \eqref{eq:freestream} to yield a single normalization factor, $\text{Norm} = \text{const} \,\tau_0$, with units of energy, which varies the overall magnitude of energy deposition for all collisions at a given beam energy.

At time $\tfs$, the energy-momentum tensor can be decomposed into hydrodynamic form
\begin{equation}
  T^{\mu\nu} = e u^\mu u^\nu - (P + \Pi) \Delta^{\mu\nu} + \pi^{\mu\nu}
  \label{eq:Tuv-hydro}
\end{equation}
in order to match the free streamed energy-momentum tensor with viscous relativistic hydrodynamics.
Here $e$ and $P$ are the energy density and pressure in the local fluid rest frame, $u^\mu$ is the local fluid velocity, $\Delta^{\mu\nu} = g^{\mu\nu} - u^\mu u^\nu$ is the projector onto the space orthogonal to $u^\mu$, and $\pi^{\mu\nu}$ and $\Pi$ are the shear and bulk viscous pressures, respectively.
The Landau matching condition furnishes the energy density $e$ and fluid velocity $u^\mu$ as the timelike eigenvalue and eigenvector of the energy-momentum tensor:
\begin{equation}
  T^{\mu\nu} u_\nu = e u^\mu.
\end{equation}
The equilibrium pressure $P = P(e)$ can then obtained from the equation of state (see below), and the bulk pressure from the difference with the total pressure
\begin{equation}
  P + \Pi = -\frac{1}{3} \Delta_{\mu\nu} T^{\mu\nu}.
\end{equation}
Finally, the shear pressure tensor can be obtained by rearranging equation \eqref{eq:Tuv-hydro}
\begin{equation}
  \pi^{\mu\nu} = T^{\mu\nu} - e u^\mu u^\nu + (P + \Pi) \Delta^{\mu\nu},
\end{equation}
since everything on the right-hand side is now known.
This provides all the information necessary to initialize the hydrodynamic equations of motion.

\subsubsection{Boost-invariant viscous hydrodynamics}

The upgraded transport model uses the exact same boost-invariant VISH2+1 hydrodynamics code with shear-bulk coupling \cite{Song:2007ux, Shen:2014vra, Liu:2015bik} described in subsection \ref{subsec:nuclear_collision_model_v1} with a slight modification to the equation of state (EoS).
As before, the EoS was constructed by blending lattice QCD calculations at high-temperature \cite{Bazavov:2014pvz} with a hadron resonance gas calculation at low temperatures.
However, in the updated model, the blending procedure is performed somewhat differently.
It matches the trace anomalies $I(T) = (e - 3p)/T^4$ of each calculation in the interval between 165 and 200 MeV using a Krogh polynomial, which ensures continuity of the functions and their first several derivatives; see reference \cite{Bernhard:2018hnz}.

The updated model also uses new parametrizations for the temperature dependent shear and bulk viscosities \cite{Bernhard:2018hnz}.
For the shear viscosity, it uses
\begin{equation}
  \label{eq:shear_param2}
  (\etas)(T) =
  \begin{cases}
    \etasmin + \etasslope (T - T_c) \cdot (T/T_c)^{\etascrv} & T > T_c, \\
    \etashrg & T \leq T_c,
  \end{cases}
\end{equation}
which upgrades equation \eqref{eq:shear_param1} by adding a dimensionless curvature parameter $\etascrv$ to vary the second derivative of the $(\etas)(T)$ above the critical temperature $T_c = 0.154$~GeV.
Values $\etascrv > 1$ introduce positive curvature, while values $\etascrv < 1$ introduce negative curvature.
Meanwhile, for $\etascrv = 1$, the curvature is turned off, reproducing equation \eqref{eq:shear_param1}.

\begin{figure}[t]
  \centering
  \makebox[\textwidth]{\includegraphics{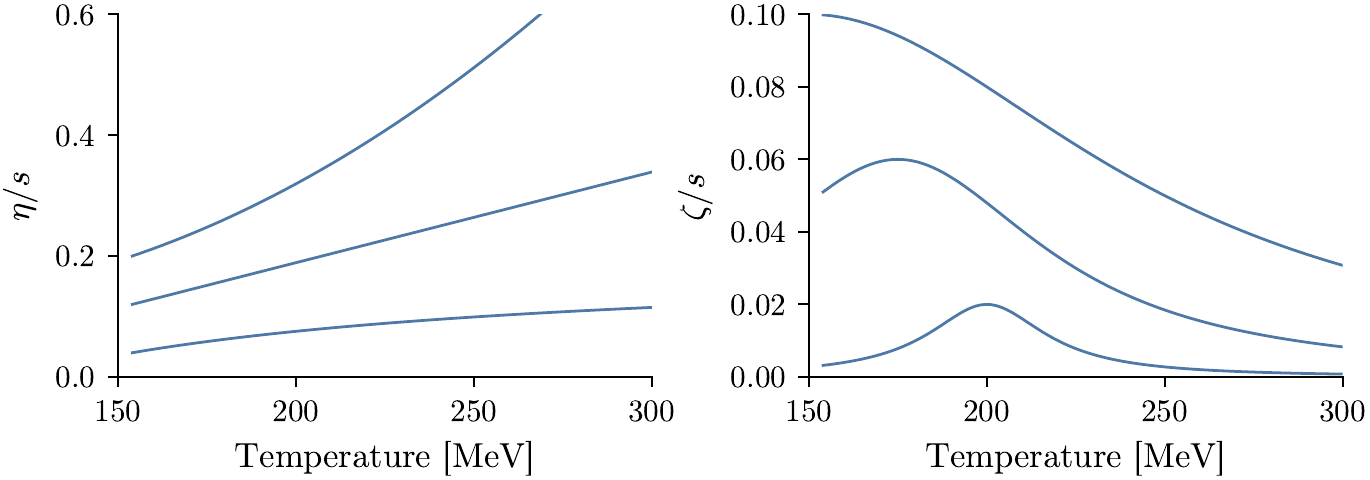}}
  \caption{
    \label{fig:viscosity_dof_v2}
    Degrees of freedom in the temperature dependent shear and bulk viscosity parametrizations.
    Lines are chosen for illustrative purposes only and do not represent all possible variability.
    For instance, $\eta/s$ could have a large slope and negative curvature, or $\zeta/s$ could have a large max and narrow width, neither of which are depicted above.
    Note, $y$-axis limits are different for the shear and bulk viscosity plots.
  }
\end{figure}

The bulk viscosity parametrization was completely overhauled using an unscaled Cauchy distribution
\begin{equation}
  \label{eq:bulk_param2}
  (\zetas)(T) = \dfrac{\zetasmax}{1 + \left( \dfrac{T - \zetasloc}{\zetaswidth} \right)^2},
\end{equation}
where $\zetasmax$ is the maximum value of $\zetas$, $\zetasloc$ is the location of the peak (units of temperature), and $\zetaswidth$ is the width of the peak (also units of temperature).
Figure \ref{fig:viscosity_dof_v2} shows several of the possible $(\etas)(T)$ and $(\zetas)(T)$ curves parametrized by equations \eqref{eq:shear_param2} and \eqref{eq:bulk_param2}.

Additionally, we modified the simulation's hydrodynamic grid.
In the previous version of the transport model, we used a square transverse grid $|x| < x_\text{max}$ of fixed maximum extent $x_\text{max} = 13$~fm and step width $dx=0.1$~fm along each dimension.
These are the VISH2+1 defaults.
In general, the grid should be made large enough to contain the entire event but no larger.
If the grid is too large, precious computation time is wasted, and if the grid is too small, the space-time evolution of the event will be clipped by the edge of the grid.

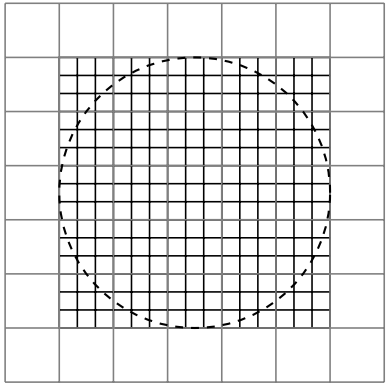
\begin{figure}[t]
  \centering
  \begin{tikzpicture}[scale=.8]
    \draw [black, semithick, step=0.3cm] (0.9, 0.9) grid +(4.5, 4.5);
    \draw [gray, semithick, step=0.9cm] (0, 0) grid +(6.3, 6.3);
    \draw [black, dashed, thick] (3.15, 3.15) circle (2.25cm);
  \end{tikzpicture}
  \caption{
    \label{fig:adaptive_grid}
    Diagram of the adaptive grid resizing algorithm (not to scale).
    Each initial condition event is first run on a very large coarse-grained mesh (large gray grid) of one-third the spatial resolution otherwise required to measure hydrodynamic observables.
    We then measure the maximum transverse radius $R_\text{max}$ (dashed circle) of the hypersurface defined by the temperature isotherm $T = T_\text{min}$.
    The temperature $T_\text{min}$ is a prespecified isotherm, below which the event can be truncated without modifying the simulated observables.
    Finally, the initial condition event is rerun on a smaller and finer mesh (smaller black grid) with three-times the cell density of the pre-run event and a smaller transverse extent $x_\text{max} = R_\text{max}$.
    Figure and caption are adapted from \cite{Moreland:2018gsh}.
  }
\end{figure}

The updated model uses a newly developed procedure, visualized in figure \ref{fig:adaptive_grid}, to vary the maximum grid size $x_\text{max}$ event-by-event, allowing large events to run with large grids and small events with small grids.
This procedure is as follows:
\begin{enumerate}
  \item
    Generate a minimum-bias \trento\ event using a square transverse grid with $x_\text{max}=15$~fm and $dx = 0.15 \times w$, where $w$ is the nucleon width (I'll explain this choice shortly).
    Call this the fine grid.
  \item
    Free stream the event to time $\tfs$ to calculate the energy density $e$, flow velocity $u^\mu$, and viscous corrections $\pi^{\mu\nu}$ and $\Pi$ on the fine grid.
  \item
    Enlarge these fine grids to $x_\text{max} = 27$~fm by padding each with zeros.
    Then take every third grid cell along each dimension to resolve the same event on a thrice coarser grid.
    Call this the coarse grid.
  \item
    Run the large coarse grid through the hydrodynamic simulation, setting the viscosities to zero, and calculate the \emph{smallest} transverse radius $R_\text{max}$ that fully encloses the prespecified isotherm $T = T_\text{min}$ for the full lifetime of the event.
  \item
    Trim the original fine grid using $x_\text{max} \to R_\text{max}$.
    Then rerun the event (with viscosity on) using the trimmed fine grid.
\end{enumerate}
This procedure works because the coarse ideal hydro event runs dramatically faster than the thrice finer viscous hydro event.
The numerical computation time for VISH2+1 scales like ${\sim}n_x^3$, where $n_x$ is the number of grid cells along either spatial dimension, so the thrice coarser grid runs ${\sim}27$ times faster.
Obviously, if the spatial step width $dx$ is too large, numerical viscosity will spoil the simulation accuracy.
However, the maximum truncation radius $R_\text{max}$ can be estimated using grids which are far coarser than those needed to calculate typical observables.
For the truncation isotherm, we found that $T_\text{min}=110$~MeV leads to a good compromise between numerical accuracy and speed, resulting in numerical errors of a few percent or less for most observables at LHC energies.
Of course, we could lower $T_\text{min}$ further still, but it would likely hurt our simulation accuracy since it would limit our simulation statistics.

In a similar fashion, the hydrodynamic grid should be fine enough to calculate simulated observables accurately but no finer.
Generally speaking, smaller nucleon widths $w$ require finer grids, since they produce smaller transverse structures in the initial energy density profile.
Thus, we decided to replace the fixed  grid step width $w=0.1$~fm in the previous version of the model with a variable step width $dx = 0.15 \times w$, equal to 15\% of the chosen nucleon width.
This leads to slight improvements in the simulation accuracy and faster event times when the nucleon width $w$ is large.

While somewhat technical and mundane, these grid modifications cut the CPU time needed to run the model in \emph{half}.
The projects discussed in this dissertation required \order{10} million CPU hours to execute, so these savings are significant.
This grid resizing method is also completely general and should work for a number of different applications.
Indeed, I'll use this exact same method in my final study where I study nuclear collisions with nucleon substructure.

\subsubsection{Particlization and Boltzmann transport}

One issue with the previous model described in \ref{subsec:nuclear_collision_model_v1} was that it completely neglected bulk viscous corrections when converting the hydrodynamic fluid into particles using the Cooper-Frye algorithm.
This precluded quantitative statements about the QGP bulk viscosity, and it also put a caveat on other parameter estimates that are strongly correlated with bulk viscosity.
For example, figure \ref{fig:posterior1} shows that the Gaussian nucleon width $w$ and bulk viscosity normalization $\zetasnorm$ are inversely correlated.
Hence, one should interpret our previous estimate for the nucleon width $w$ with care.

Collaborator Jonah Bernhard addressed this issue by developing \mbox{\textsc{frzout}}, a new computer model for Cooper-Frye particle emission \cite{frzout}.
His model includes, among other features, a novel method to implement bulk viscous corrections to the particle distribution functions.
It addresses many of the issues which led us to neglect bulk corrections in the first place, such as their stability against large bulk pressures, enabling a more realistic treatment of the interface between the hydrodynamic and microscopic transport models.
The following is a short summary of his model \cite{Bernhard:2018hnz}.

The \textsc{frzout} model starts by sampling particles from the isothermal space-time hypersurface $\sigma$ defined by the hydro-to-micro switching temperature $\Tsw$.
Particles are emitted from the hypersurface $\sigma$ with momentum distributions described by the Cooper-Frye formula \cite{Cooper-Frye:1974frz}
\begin{equation}
  \label{eq:cooper-frye2}
  E \frac{dN_i}{d^3p} = \frac{g_i}{(2 \pi)^3} \int_\sigma f_i(p) p^\mu d^3 \sigma_\mu,
\end{equation}
where $g_i$ and $f_i$ are the degeneracy and distribution function of particle species $i$, and $d^3\sigma_\mu$ is a volume element on the hypersurface $\sigma$.

The particles are sampled in the local rest frame of each fluid cell, using either a Bose-Einstein or Fermi-Dirac thermal distribution
\begin{equation}
  \label{eq:fmp}
  f(m, p) = \frac{1}{\exp(\sqrt{m^2 + p^2}/T) \mp 1},
\end{equation}
where $m$ is the mass of the particle, $p$ is its momentum in the rest frame of the fluid cell, and $T$ is the fluid cell temperature.
If the particle is a resonance, it samples a \emph{distribution} of masses $P(m)$.
The distribution $f(p)$ is then calculated by integrating out this mass dependence, yielding
\begin{equation}
  f(p) = \int dm \mathcal{P}(m) f(m, p).
\end{equation}
Typically, computational implementations neglect the mass-width of resonances, using instead $\mathcal{P}(m) = \delta(m - m_0)$ for every particle where $m_0$ is the mass where the resonance peaks.
Indeed, our previous study did not include the effect of each resonance width.
However, such approximations are crude and are known to affect the particle yields, particularly at low $p_T$ \cite{Sollfrank:1991xm, Huovinen:2016xxq}.

The \textsc{frzout} model therefore samples resonances with a distribution of masses, modeled by a Breit-Wigner distribution
\begin{equation}
  \mathcal{P}(m) \propto \frac{\Gamma(m)}{(m - m_0)^2 + \Gamma(m)^2/4},
\end{equation}
where $m_0$ is the usual Breit-Wigner mass, and $\Gamma$ is its mass dependent width.
This width is then parametrized using the simple form
\begin{equation}
  \Gamma(m) = \Gamma_0 \sqrt{\frac{m - m_\mathrm{min}}{m_0 - m_\mathrm{min}}},
\end{equation}
where $\Gamma_0$ is the usual Breit-Wigner width and $m_\text{min}$ is a minimum mass threshold equal to the sum of the masses of the lightest decay products.
The particle properties $\Gamma_0$, $m_0$, $m_\text{min}$ are provided by the Particle Data Group \cite{Tanabashi:2018oca}.

In hydrodynamic theory, the energy-momentum tensor is described by equation \eqref{eq:Tuv-hydro}, while in kinetic theory it equals
\begin{equation}
  T^{\mu\nu} = \sum\limits_i g_i \int \frac{d^3p}{(2 \pi)^3} \frac{p^\mu p^\nu}{E} f_i(p),
\end{equation}
where the summation runs over all particle species.
Comparing these two forms, it is clear that the distribution function $f_i$ must be modified as a function of the shear pressure $\pi^{\mu\nu}$ and bulk pressure $\Pi$, to preserve the continuity of $T^{\mu\nu}$ as the system transitions from hydrodynamics to Boltzmann transport.

The \textsc{frzout} model applies these viscous corrections using a general procedure developed by Pratt and Torrieri to transform the sampled particle momenta \emph{inside} the distribution function \cite{Pratt:2010jt}
\begin{align}
  \label{viscous_correction}
  p_i \rightarrow p'_i = p_i + \sum\limits_j \lambda_{ij}\, p_j,
\end{align}
using a linear transformation matrix
\begin{equation}
  \lambda_{ij} = (\lambda_\mathrm{shear})_{ij} + \lambda_\mathrm{bulk}\, \delta_{ij},
\end{equation}
composed of a traceless shear contribution $\lambda_\text{shear}$ and a bulk contribution $\lambda_\text{bulk}$ that's proportional to the identity matrix $\delta_{ij}$.

When the shear pressure is small relative to the ideal pressure, the shear transformation equals
\begin{equation}
  (\lambda_\mathrm{shear})_{ij} = \frac{\tau}{2 \eta} \pi_{ij},
\end{equation}
where $\pi_{ij}$ is the shear stress in the local fluid rest frame, and $\tau / \eta$ is the ratio of the relaxation time to the shear viscosity.
In the non-interacting hadron resonance gas model, the inverse of this later quantity equals
\begin{equation}
  \frac{\eta}{\tau} = \frac{1}{15 T} \sum\limits_i g_i \int \frac{d^3p}{(2\pi)^3}\frac{p^4}{E^2} f_i (1 \pm f_i),
\end{equation}
where the summation runs over all species in the gas, and $f_i$ is the equilibrium Fermi-Dirac or Bose-Einstein distribution of that species.

For the bulk viscous corrections, the \textsc{frzout} model applies a novel numerical algorithm developed in \cite{Bernhard:2018hnz}.
The total kinetic pressure of the gas in kinetic theory is
\begin{equation}
  \label{eq:kinetic_pressure}
  P + \Pi = \sum\limits_i g_i \int \frac{d^3p}{(2\pi)^3} \frac{p^2}{3E} f_i(p),
\end{equation}
where $P$ is the fluid's ideal pressure, and $\Pi$ is its bulk pressure.
For a given fluid bulk pressure $\Pi$, the algorithm rescales the momentum inside the distribution function $f(p) \to f(p + \lambda_\text{bulk} p)$ and adjusts the parameter $\lambda_\text{bulk}$ to match the total pressure on the left-side of the equation.
This modification also alters the fluid's energy density
\begin{equation}
  \label{eq:kinetic_energy_density}
  e = \sum\limits_i g_i \int \frac{d^3p}{(2 \pi)^3} E f_i(p),
\end{equation}
so the algorithm introduces a second parameter $z_\text{bulk}$---assumed to be the same for all particle species---that rescales the particle yield $f(p) \to z_\text{bulk} f(p)$ to compensate.
The parameters $z_\text{bulk}$ and $\lambda_\text{bulk}$ are determined numerically for each value of the bulk pressure to match the fluid cell's pressure and energy density.

As before, the sampled particles are passed to the UrQMD transport model which simulates their individual microscopic interactions until the last interactions cease and the medium ``freezes out'', producing a list of particle IDs and four-momenta for each simulated event.
There were no significant changes to our implementation of UrQMD compared to the previous iteration of the model.

\subsection{Predictions for Xe-Xe collisions at 5.44 TeV}

To this point, I have only demonstrated that the \trento\ initial condition model is \emph{postdictive}, i.e.\ given some calibration data, the model can be tuned to fit that data.
Naturally, our hope is that once the model is calibrated, it will also be able to predict observables that are out of sample, i.e.\ measurements not included in the calibration data.
This subsection presents some previously unpublished \trento\ model predictions for Xe-Xe collisions at $\sqrts=5.44$~TeV using the updated version of our nuclear collision model which I just described.

\subsubsection{Constraints from Pb-Pb collisions at 2.76 and 5.02 TeV}

I fix the \trento\ and QGP medium parameters for the Xe-Xe predictions using the results of a recently published Bayesian analysis of Pb-Pb bulk observables at $\sqrts=2.76$ and 5.02 TeV \cite{Bernhard:2016tnd}.
This reference builds upon the previously described analysis \cite{Bernhard:2016tnd} and is the capstone of J.\ Bernhard's dissertation work on Bayesian parameter estimation \cite{Bernhard:2018hnz}.
My own contributions to this updated global analysis are meaningful, but they are not significant enough to warrant a full description of the project in its entirety here.
Rather, let me highlight some key differences compared to the previous Bayesian analysis which I just described:
\begin{itemize}
  \item
    The study combines Pb-Pb data from two beam energies at the LHC, $\sqrts=2.76$ and 5.02 TeV.
  \item
    It adds transverse energy and mean $p_T$ fluctuations to the list of calibration observables.
  \item
    It uses the significantly upgraded nuclear collision model discussed in subsection \ref{subsec:nuclear_collision_model_v2}.
    Notable additions include a pre-equilibrium free streaming evolution \cite{freestream} and a new particle sampler \cite{frzout}.
  \item
    It significantly improves the treatment of uncertainties in \cite{Bernhard:2016tnd}.
\end{itemize}

\begin{figure}[p]
  \centering
  \makebox[\textwidth]{
    \includegraphics[width=.95\paperwidth]{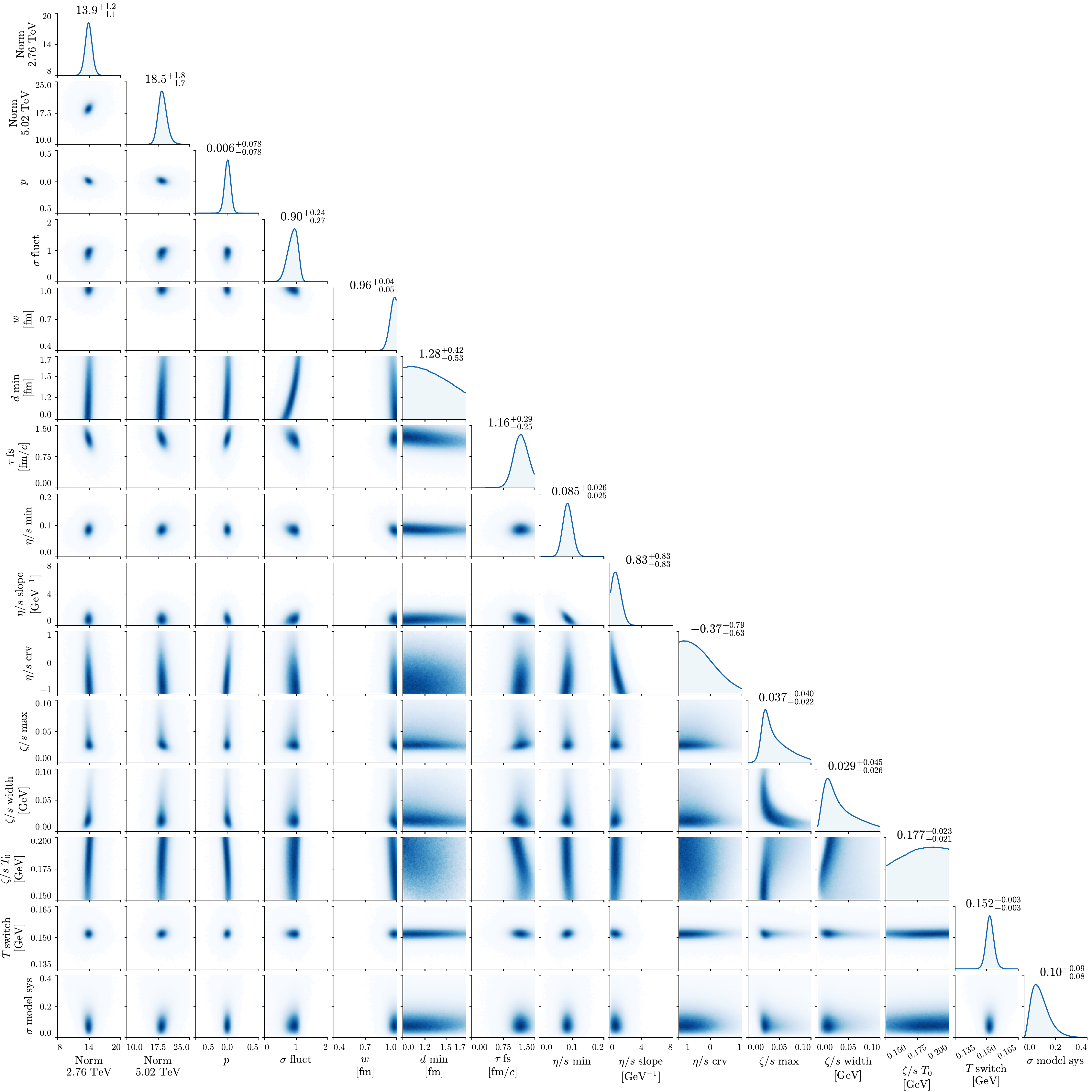}
  }
  \captionsetup{width=.85\paperwidth}
  \caption{
    \label{fig:posterior_v2}
    Posterior distribution for the model input parameters of Ref.~\cite{Bernhard:2019ntr}.
    Diagonal panels are the marginal distributions of individual parameters and off-diagonal panels are the joint distributions for pairs of parameters.
  }
\end{figure}

\begin{figure}[p]
  \centering
  \makebox[\textwidth]{
    \includegraphics{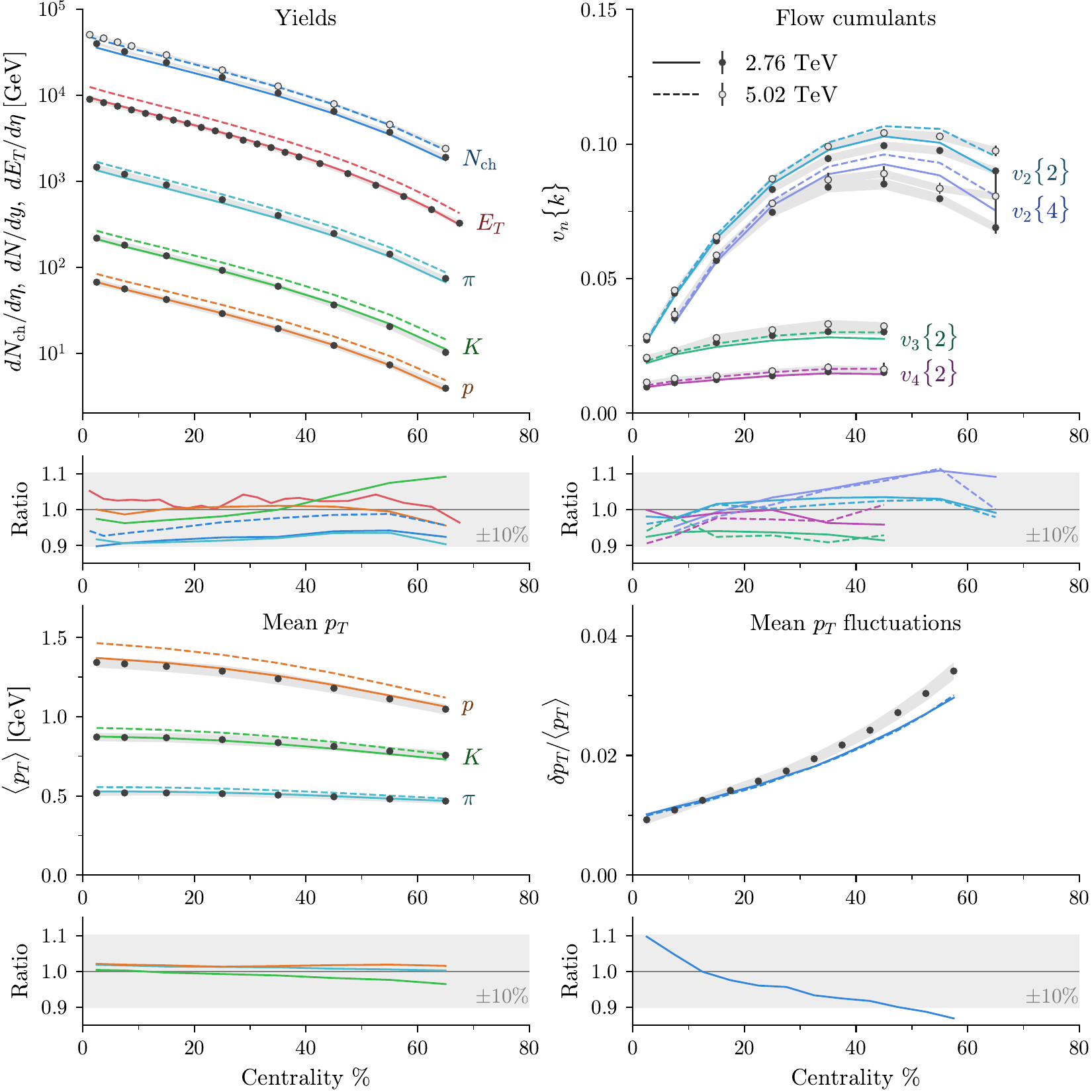}
  }
  \captionsetup{width=.6\paperwidth}
  \caption{
    \label{fig:observables_map_v2}
    Model calculations (colored lines) for Pb-Pb collisions at two beam energies, $\sqrts=2.76$~TeV (solid lines) and $\sqrts=5.02$~TeV (dashed lines), using the MAP parameters from Ref.~\cite{Bernhard:2019ntr}, listed in table \ref{tab:map_param_v2}.
    All experimental data (symbols) are from ALICE \cite{Aamodt:2010cz,Adam:2016thv,Abelev:2013vea,Abelev:2014ckr,ALICE:2011ab,Adam:2015ptt,Adam:2016izf}.
    The subpanel below each plot shows the ratio of the model over the experimental data.
    Figure is from Ref.~\cite{Bernhard:2019ntr}.
  }
\end{figure}

Figure \ref{fig:posterior_v2} shows the fifteen dimensional posterior distribution obtained from the analysis, and figure \ref{fig:observables_map_v2} shows the calibrated model predictions compared to experimental data, using parameters selected from the mode of the posterior distribution which are listed in table \ref{tab:map_param_v2}.
The mode parameter point maximizes the posterior probability density and characterizes the best possible agreement of the model with the data.
It is also commonly referred to as the \emph{maximum a posteriori} (MAP) estimate.

\begin{table}[h]
  \centering
  \caption{
    \label{tab:map_param_v2}
    \emph{Maximum a posteriori} (MAP) model parameters from Ref.~\cite{Bernhard:2019ntr}.
  }
  \small
  \begin{tabular}{ll@{\hspace{2em}}ll}
    \toprule
    \multicolumn{2}{c}{Initial condition / Pre-eq} & \multicolumn{2}{c}{QGP medium} \\
    \cmidrule(r){1-2}                                \cmidrule(l){3-4}
    Norm 2.76 TeV & 13.94 GeV & $\etas$ min          & 0.081 \\
    Norm 5.02 TeV & 18.38 GeV & $\etas$ slope        & 1.11 GeV$^{-1}$ \\
    $p$  & 0.007              & $\etas$ crv          & -0.48  \\
    $\sigf$  & 0.918          & $\zetas$ max         & 0.052 \\
    $w$  & 0.956 fm           & $\zetas$ width       & 0.022 GeV \\
    $\dmin$  & 1.27 fm        & $\zetas$ $T_0$       & 0.183 GeV \\
    $\tfs$  & 1.16 \fmc       & $\Tsw$               & 0.151 GeV \\
    \bottomrule
  \end{tabular}
\end{table}

In principle, these MAP parameters should also describe Xe-Xe collisions at LHC energies as well, with one small caveat.
The \trento\ model has several parameters that scale strongly with the collision energy, so we'll need to extrapolate these parameters from the Pb-Pb energy $\sqrts=5.02$ TeV to the Xe-Xe energy $\sqrts=5.44$ TeV.
I'll describe this extrapolation in a moment.
First, I want to comment briefly on figures \ref{fig:posterior_v2} and \ref{fig:observables_map_v2}.

There are two results in figure \ref{fig:posterior_v2} which are particularly striking: the marginal posterior distribution on the generalized mean parameter $p$, and the joint posterior distribution of the QGP viscosity parameters.

\begin{figure}[t]
  \centering
  \makebox[\textwidth]{
    \includegraphics[width=.9\textwidth]{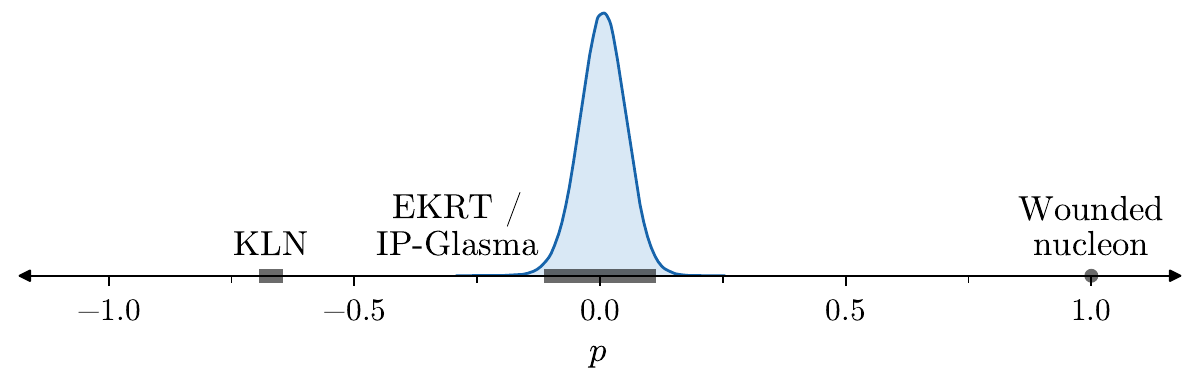}
  }
  \caption{
    \label{fig:posterior_p_alt}
    Posterior on the \trento\ generalized mean parameter $p$ compared to the effective values needed to mimic the KLN, EKRT, IP-Glasma, and wounded nucleon models.
    Note, each effective $p$ value was determined in subsection \ref{subsec:mimic} assuming that \trento\ deposits entropy density.
    However, the present analysis reinterprets this profile as an energy density.
    The free streamed energy density after Landau matching is roughly proportional to the entropy density at the hydrodynamic starting time, so the two can be directly compared.
    Figure from Ref.~\cite{Bernhard:2019ntr}.
  }
\end{figure}

The analysis obtains a sharply peaked posterior on the \trento\ generalized mean parameter $p$ centered essentially at zero.
Figure \ref{fig:posterior_p_alt} shows this posterior distribution compared to the effective $p$ values needed to fit the KLN, EKRT, IP-Glasma, and wounded nucleon models determined in subsection \ref{subsec:mimic}.
Note, our interpretation of this parameter is somewhat different than before.
Originally in reference \cite{Bernhard:2016tnd}, we parametrized the initial \emph{entropy} density using the generalized mean ansatz \eqref{eq:gmean_ansatz}
\begin{equation}
  s_0 \propto M_p(\T_A, \T_B),
\end{equation}
assuming static profiles at the hydrodynamic starting time with $u^\mu=0$.
However in this study, we parametrized the initial \emph{energy} density
\begin{equation}
  e_0 \propto M_p(\T_A, \T_B).
\end{equation}
This energy density was free streamed to the hydrodynamic starting time and matched to viscous hydrodynamics using the Landau matching procedure.
Evidently, both prescriptions prefer $p \sim 0$, leading to similar geometric mean scaling
\begin{equation}
   \left \{
  \begin{aligned}
    e_0\\
    s_0
  \end{aligned}
  \right \}
  \propto \sqrt{\T_A\, \T_B},
\end{equation}
as a function of nuclear thickness.
Suffice to say, our strict interpretation of the parameter $p$ depends on the assumed matching procedure.

The study also presents the first Bayesian estimate for the temperature dependence of the QGP specific shear and bulk viscosities extracted from hydrodynamic simulations with rigorously calculated uncertainties.
These uncertainties account for the model's finite statistical error, the emulator's systematic interpolation error, and the experiment's statistical and systematic errors.
Moreover, by marginalizing over the \trento\ model parameters, the analysis also accounts for the correlated effect of the QGP initial condition uncertainties.

\begin{figure}[t]
  \centering
  \makebox[\textwidth]{
    \includegraphics{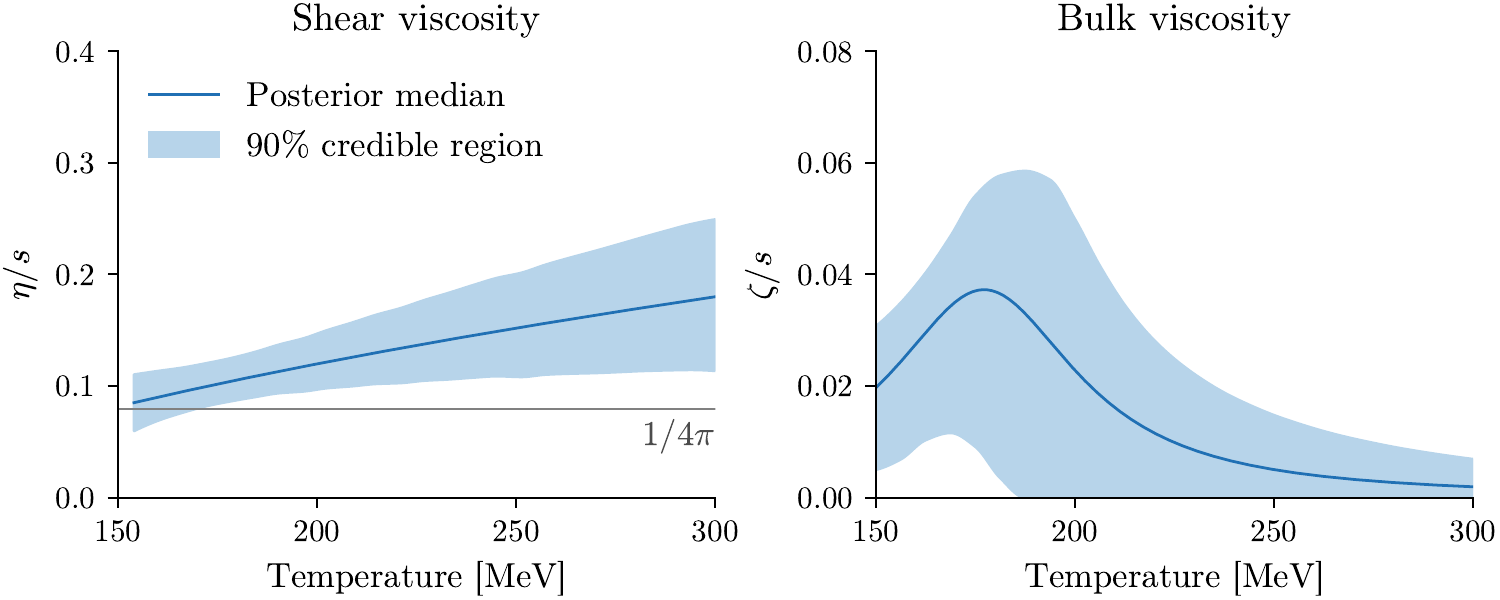}
  }
  \caption{
    \label{fig:region_shear_bulk_v2}
    Posterior estimates for the temperature dependence of the QGP specific shear viscosity $\etas$ (left) and bulk viscosity $\zetas$ (right).
    The blue band is the parametrization's posterior 90\% HPD credible region, and the blue line is its median.
    Here the prior covers essentially the entire region of the figure so it is not explicitly shown.
    The horizontal gray line (left) is the KSS bound $\etas \ge 1/4\pi$ \cite{Kovtun:2004de, Policastro:2001yc, Danielewicz:1984ww}.
    Figure is from Ref.~\cite{Bernhard:2018hnz, Bernhard:2019ntr}.
  }
\end{figure}

Figure \ref{fig:region_shear_bulk_v2} visualizes the posterior estimates for the specific shear viscosity $(\etas)(T)$ and specific bulk viscosity $(\zetas)(T)$.
The QGP shear viscosity prefers a minimum value $\etasmin=0.085_{-0.025}^{+0.026}$ and a gently rising or flat slope $\etasslope=0.83_{-0.83}^{+0.83}$~GeV$^{-1}$ in agreement with reference \cite{Bernhard:2016tnd}.
Note, we restricted the slope parameter to positive values so we excluded negative slopes \emph{a priori}.
The bulk viscosity, meanwhile, can be either tall and narrow, or short and broad.
Evidently, the important quantity is something like the integral of the bulk viscosity curve which scales like the product of the $\zetaswidth$ and $\zetasmax$ parameters.

Finally, I want to reflect on the global agreement of the model's MAP estimate with the experimental data from ALICE \cite{Aamodt:2010cz,Adam:2016thv,Abelev:2013vea,Abelev:2014ckr,ALICE:2011ab,Adam:2015ptt,Adam:2016izf} which is plotted figure \ref{fig:observables_map_v2}.
The hybrid nuclear collision model consisting of \trento\ initial conditions, free streaming, VISH2+1 boost-invariant viscous hydrodynamics, \textsc{frzout}, and UrQMD microscopic hadronic transport describes nearly all calibration observables at the 10\% level or better.
Indeed many of the observables agree at the level of \emph{a few percent}.
This is a truly remarkable agreement of theory with experiment.

\subsubsection{Extrapolating predictions to 5.44 TeV}

Let me return now to the topic of this subsection, generating hydrodynamic model predictions for Xe-Xe collisions at $\sqrts=5.44$~TeV.
I'll use the Bayesian MAP parameters (table \ref{tab:map_param_v2}) determined by reference \cite{Bernhard:2019ntr} to fix the parameters of our Xe-Xe predictions.
The \trento\ model's inelastic nucleon-nucleon cross section $\sigmann$ and overall normalization factor vary as a function of the beam energy, so we'll need to scale these values to $\sqrts=5.44$~TeV to account for the slight difference in beam energy compared to our Pb-Pb calibration data.

\begin{figure}[h]
  \centering
  \includegraphics{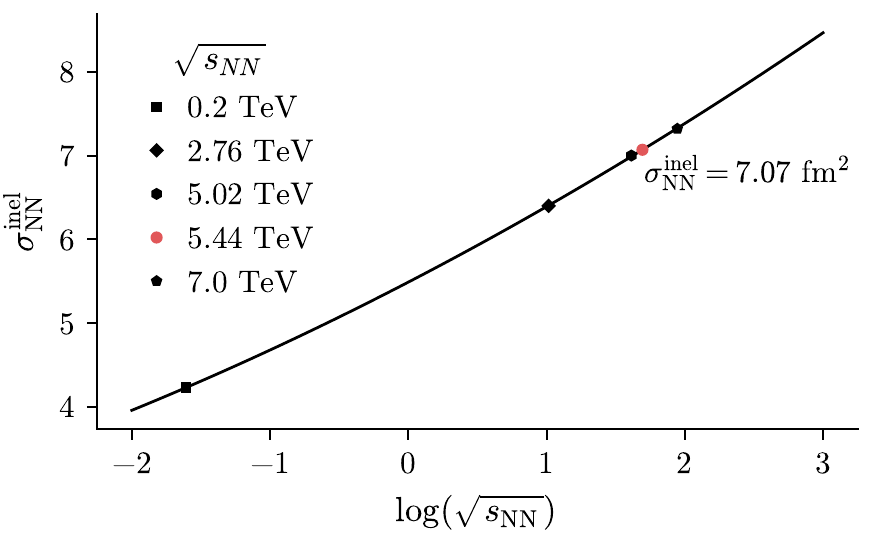}
  \caption{
    \label{fig:xenon_cross_section}
    Experimentally measured values for the inelastic nucleon-nucleon cross section $\sigmann$ at $\sqrts=0.2$, 2.76, 5.02, and 7~TeV \cite{Adare:2015bua, ATLAS:2011ag, ALICE:2012xs, Abelev:2012sea} plotted as a function of $\log(\sqrts)$ (black symbols) and fit with a second-order polynomial (black line).
    The extrapolated value for the inelastic nucleon-nucleon cross section at $\sqrts=5.44$~TeV (red circle) equals $\sigmann=7.07$~fm$^2$.
  }
\end{figure}

\begin{figure}[h]
  \centering
  \includegraphics{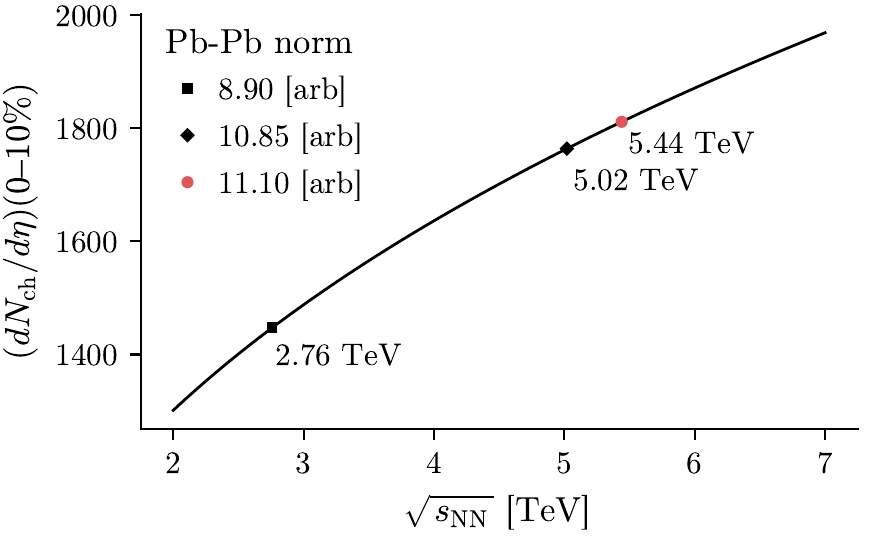}
  \caption{
    \label{fig:xenon_entropy_norm}
    Pb-Pb charged-particle density $d\nch/d\eta$ for 0--10\% central events at $\sqrts=2.76$ and 5.02 TeV (black square and diamond) measured by ALICE \cite{Aamodt:2010cz, Adam:2015ptt}, fit with a power law (black line) and used to predict the same charged-particle density at $\sqrts=5.44$~TeV (red circle).
    Legend lists the normalizations, equation \eqref{eq:yield_scaling}, needed to fit the charged-particle density at each collision energy.
  }
\end{figure}

We fix the inelastic nucleon-nucleon cross section $\sigmann$ at 5.44 TeV by interpolating previous cross section measurements at $\sqrts=0.2$, 2.76, 5.02 and 7 TeV \cite{Adare:2015bua, ATLAS:2011ag, ALICE:2012xs, Abelev:2012sea}.
Figure \ref{fig:xenon_cross_section} shows the value of the cross section at each energy fit to a second-order polynomial as a function of $\log(\sqrts)$.
Using this fit, we estimate that the inelastic nucleon-nucleon cross section at 5.44 TeV is $\sigmann=7.07$~fm$^2$.
Then, we estimate the relative increase of the \trento\ normalization factor from 5.02 to 5.44 TeV as follows:

\begin{enumerate}[leftmargin=1\parindent]
  \item
    Fit the experimental Pb-Pb charged-particle density $d\nch/d\eta$, 0--10\% centrality, at $\sqrts=2.76$ and 5.02 TeV \cite{Aamodt:2010cz, Adam:2015ptt} with a power law.
  \item
    Predict the Pb-Pb charged-particle density $d\nch/d\eta$, 0--10\% centrality, at $\sqrts=5.44$ TeV using the power law fit.
  \item
    Run \trento\ Pb-Pb events at 5.02 and 5.44 TeV and calculate the normalization factor which fits the target charged-particle density $d\nch/d\eta$, 0--10\% centrality, at each beam energy, using the approximate scaling relation
    \begin{equation}
      \label{eq:yield_scaling}
      d\nch/d\eta = \text{norm} \times \int d^2x_\perp T_R(\xv_\perp),
    \end{equation}
    as depicted in figure \ref{fig:xenon_entropy_norm}.
    Finally, divide the normalization factor at 5.44 TeV by the normalization factor at 5.02 TeV to calculate their ratio.
    This predicts a normalization that's ${\sim}2\%$ larger for the Xe-Xe events.
\end{enumerate}

\begin{figure}[p]
  \centering
  \makebox[\textwidth]{
    \includegraphics{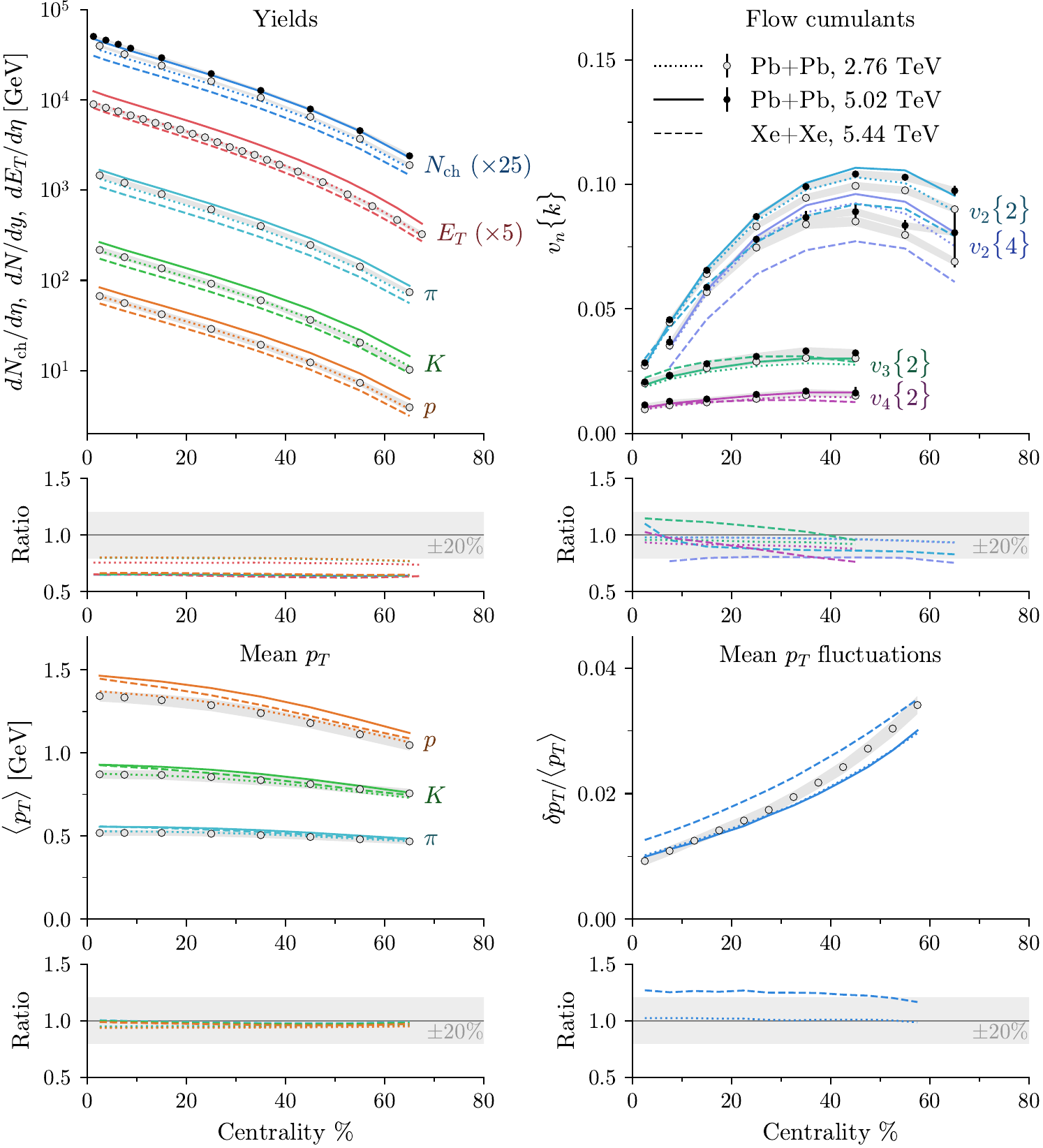}
  }
  \captionsetup{width=.7\paperwidth}
  \caption{
    \label{fig:xenon_observables_map}
    Model calculations (colored lines) using the MAP parameters from Ref.~\cite{Bernhard:2019ntr} compared to experimental data (black and white symbols) from ALICE \cite{Aamodt:2010cz,Adam:2016thv,Abelev:2013vea,Abelev:2014ckr,ALICE:2011ab,Adam:2015ptt,Adam:2016izf}.
    The Pb-Pb model calculations at 2.76 TeV (dotted) and 5.02 TeV (solid) are postdictions, calibrated to fit the experimental data, while the Xe-Xe model calculations at 5.44 TeV (dashed) are predictions based on the calibrated parameters.
  }
\end{figure}

After adjusting the 5.02 TeV normalization in table \ref{tab:map_param_v2} for the slight increase in beam energy, we run \order{10^6} minimum-bias Xe-Xe events using the MAP parameters from reference \cite{Bernhard:2019ntr} and calculate the same observables as before.
Figure \ref{fig:xenon_observables_map} shows our MAP estimate (prediction) for Xe-Xe collisions at 5.44 TeV (dashed lines) plotted on top of our MAP estimate (postdiction) for Pb-Pb collisions at 2.76 TeV (dotted lines) and 5.02 TeV (solid lines).
We also show, for reference purposes, the Pb-Pb experimental data (filled and open symbols) used to calibrate the MAP parameters  \cite{Aamodt:2010cz,Adam:2016thv,Abelev:2013vea,Abelev:2014ckr,ALICE:2011ab,Adam:2015ptt,Adam:2016izf}.
The subpanel below each figure shows the ratio of each model calculation to the Pb-Pb model calculation at 5.02 TeV.
Caution, this is a ratio of two model calculations, \emph{not} the ratio of the model to data as previously plotted in figure \ref{fig:observables_map_v2}.
These ratio plots predict several interesting features for the Xe-Xe collision system.

We see that changing the collision configuration from Xe-Xe to Pb-Pb essentially rescales the centrality dependence of charged-particle production by a constant overall factor.
Comparing the Xe-Xe system at 5.44 TeV to the Pb-Pb system at 5.02 TeV, we also predict a large enhancement for the triangular flow cumulant $\vnk{3}{2}$, particularly at small centralities, and a similarly large suppression for $\vnk{2}{4}$.
The identified mean $p_T$'s, meanwhile, change very little as a function of beam energy and system size.
Finally, we predict a striking enhancement of the mean $p_T$ fluctuations $\delta p_T / \langle p_T \rangle$ for the Xe-Xe system at 5.44 TeV compared to the Pb-Pb systems at 2.76 and 5.02 TeV.
Presumably, this observable is sensitive to event-by-event fluctuations which depend on the number of nucleons inside each nucleus.

\section{Bulk matter far from midrapidity}
\label{sec:trento3d}

In the previous section, I described several Bayesian studies of the initial condition and QGP medium properties at midrapidity using boost-invariant approximations.
This section describes an exploratory study which relaxes these approximations to analyze the full three-dimensional structure of the produced fireball.
The following content is based on the publication
\begin{flushleft}
W.\ Ke, J.\ S.\ Moreland, J. E.\ Bernhard, and S.\ A.\ Bass,
``Constraints on rapidity-dependent initial conditions from charged particle pseudorapidity densities and two-particle correlations'', \href{https://arxiv.org/abs/1610.08490}{Phys.\ Rev.\ \textbf{C96}, 044912 (2017), arXiv:1610.08490 [nucl-th]},
\end{flushleft}
which presents a study that I co-developed with primary author Weiyao Ke.
I helped conceive, design, and execute the study, but I did not write the software or perform the primary data analysis.

While \emph{ab initio} theoretical calculations have made tremendous progress determining the QGP initial conditions at midrapidity \cite{Schenke:2012wb, Niemi:2015qia}, the search for a comprehensive theory which describes the full three-dimensional structure of the produced plasma remains an outstanding challenge \cite{Sirunyan:2019cgy}.
Consider, for example, the average charged-particle yield $d\nch/d\eta$ which measure's the collision's single-particle distribution.
Multiple theories nicely describe the centrality dependence of this observable at midrapidity \cite{Adam:2015ptt}, but there exists (to date) no compelling explanation for its rapidity dependence.
Theoretical calculations of the QGP initial conditions at nonzero rapidity therefore tend to be somewhat speculative.

Data-driven Bayesian methods provide an attractive, complementary approach to bottom-up theory calculations \cite{Bernhard:2018hnz}.
In this work, we parametrized the three-dimensional structure of the QGP initial conditions and constrained the model's parameters using multiplicity observables measured in $p$-Pb and Pb-Pb collisions at the LHC \cite{Abbas:2013bpa, Adam:2015kda, Aad:2015zza, Radhakrishnan:2015eqq}.
The optimized initial condition model was then compared to a number of experimental observables outside our calibration dataset---pseudorapidity-dependent anisotropic flows, event-plane decorrelations, and flow correlations---to assess its veracity.
Our results shed light on the three-dimensional structure of the QGP initial conditions and provide useful guidance for \emph{ab initio} theory calculations.
They also determine realistic three-dimensional profiles of QGP entropy density which can be used to calculate hydrodynamic backgrounds for hard-probe and jet energy-loss calculations.

\subsection{Nuclear collision model in 3+1 dimensions}
\label{subsec:model3d}

This study employed a hybrid transport model qualitatively similar to the model described in subsection \ref{subsec:nuclear_collision_model_v1}.
We used relativistic hydrodynamics to simulate the hot and dense QGP phase of the collision and microscopic Boltzmann transport to simulate the cooler hadron resonance gas.
However, unlike the boost-invariant model described in subsection \ref{subsec:nuclear_collision_model_v1}, we modeled the collision dynamics in all three spatial dimensions.
This required two significant changes to the nuclear collision model: we parametrized the rapidity dependence of local entropy deposition to extend the \trento\ initial condition model from two to three spatial dimensions, and we replaced the boost-invariant VISH2+1 hydrodynamic code \cite{Song:2007ux} with the 3+1 dimensional hydrodynamic code vHLLE \cite{Karpenko:2013wva}.
The specifics of this three-dimensional hybrid nuclear collision model are summarized below.

\subsubsection{Initial conditions}

We modeled the three-dimensional entropy density $s(\xv_\perp, \eta_s, \tau_0)$ at the hydrodynamic starting time $\tau_0$ as the product of two functions
\begin{equation}
  \label{eq:trento3d_ansatz}
  s(\xv_\perp, \eta_s, \tau_0) = f(\xv_\perp) \times g(\xv_\perp, \eta_s),
\end{equation}
where the second function satisfies $g(\xv_\perp, 0) = 1$.
The function $f(\xv_\perp)$ models the initial transverse entropy density at midrapidity $\eta_s=0$, and the function $g(\xv_\perp, \eta_s)$ models the \emph{shape} of the initial entropy density profile as a function of space-time rapidity $\eta_s$.

Following previous work \cite{Bernhard:2016tnd}, we parametrized the midrapidity transverse entropy density $f(\xv_\perp)$ using the entropy variant of the generalized mean ansatz
\begin{equation}
  \label{eq:midrapidity_scaling}
  f(\xv_\perp) \propto \left [ \frac{\T_A(\xv_\perp)^p + \T_B(\xv_\perp)^p}{2} \right]^{1/p},
\end{equation}
where $\T_A, \T_B$ are the participant thickness functions of nucleus $A,B$ respectively.
Meanwhile, for the rapidity-dependent mapping $g(\xv_\perp, \eta_s)$, we constructed a flexible parametrization using a cumulant generating function characterized by a variable mean, standard deviation, and skewness.
I'll describe this rapidity parametrization in a moment.
First, let me clarify our use of space-time rapidity $\eta_s$, pseudorapidity $\eta$, and rapidity $y$.

Assuming, to first approximation, that the initially produced particles are massless and free streaming along the $\hat{z}$ direction, then
\begin{equation}
  \frac{z}{t} = \frac{p_z}{|\mathbf{p}|}.
\end{equation}
This approximation allows us to identify the space-time rapidity $\eta_s$ with the pseudorapidity $\eta$,
\begin{equation}
  \eta_s = \frac{1}{2} \log \left( \frac{t + z}{t - z} \right) \quad\sim\quad \eta = \frac{1}{2} \log \left( \frac{|\mathbf{p}| + p_z}{|\mathbf{p}| - p_z} \right ),
\end{equation}
from which it follows that $g(\xv_\perp, \eta_s) \approx g(\xv_\perp, \eta)$.
Furthermore, we can perform a change of variables from pseudorapidity $\eta$ to rapidity $y$ using the relations
\begin{subequations}
  \label{eq:dydeta}
  \begin{align}
    g(\xv_\perp, \eta)\, d\eta &= g(\xv_\perp, y)\, dy, \\[1ex]
    \frac{dy}{d\eta} &= \frac{J \cosh \eta}{\sqrt{1 + J^2 \sinh^2{\eta}}},
  \end{align}
\end{subequations}
where $J \approx \langle p_T \rangle / \langle m_T \rangle$ is a free parameter that characterizes the entropy density's effective particle composition.
It follows that the space-time rapidity profile $g(\xv_\perp, \eta_s)$ can be written as
\begin{equation}
  \label{eq:mult_dydeta}
  g(\xv_\perp, \eta_s) \approx g(\xv_\perp, y)\, \frac{dy}{d\eta},
\end{equation}
with $dy/d\eta$ provided by equation \eqref{eq:dydeta}.

We modeled the rapidity dependence of the function $g(\xv_\perp, y)$ on the right-side of equation \eqref{eq:mult_dydeta} at each transverse coordinate $\xv_\perp$ by parametrizing its cumulants as functions of $\T_A(\xv_\perp)$ and $\T_B(\xv_\perp)$.
Once these cumulants are known, we can reconstruct the function $g(\xv_\perp, y)$ by taking the inverse Fourier transform of the cumulant generating function
\begin{subequations}
  \label{eq:generating_func}
  \begin{align}
    g(\xv_\perp, y) &= \mathcal{F}^{-1} \{\tilde{g}(\xv_\perp, k)\}, \\
    \log \tilde{g} &= i \mu k - \frac{1}{2} \sigma^2 k^2 - \frac{1}{6} i \gamma \sigma^3 k^3 + \dots.
  \end{align}
\end{subequations}
The function is then normalized, $g(\xv_\perp, 0) = 1$, so the midrapidity entropy density is unmodified.

Different rapidity dependent initial condition models are described by different parametrizations of the generating function cumulants.
For example, the authors in reference \cite{Bozek:2010bi} investigated two models which ``shifted'' and ``tilted'' the rapidity profile $g$.
The shifted model varied the mean of the rapidity profile, and the tilted model strongly varied its skewness.
Generally speaking, however, all of the cumulants of the rapidity profile could be nonzero, and all of them could vary as functions of $\T_A, \T_B$.

\begin{table}[t]
  \centering
  \caption{
    \label{tab:cumulants}
    Generating function cumulant parametrizations used in the present analysis. The variable $y_\text{cm} = \frac{1}{2}\log(\T_A/\T_B)$, and the constant $\T_0 = 1$~fm$^{-2}$.
  }
  \small
  \begin{tabular}{lccc}
    \toprule
    & \multicolumn{3}{c}{Distribution cumulant:} \\
    \noalign{\smallskip}\cline{2-4}\noalign{\smallskip}
    Model variant & \multicolumn{1}{c}{mean $\mu$} & \multicolumn{1}{c}{std.\ $\sigma$} & \multicolumn{1}{c}{skewness $\gamma$} \\
    \midrule
    Relative skewness  & $\mu_0 \,y_\text{cm}$ & $\sigma_0$ & $\gamma_0 \left( \dfrac{\T_A - \T_B}{\T_A + \T_B} \right )$ \smallskip\\[2ex]
    Absolute skewness & $\mu_0 \,y_\text{cm}$  & $\sigma_0$ & $\gamma_0 \left ( \dfrac{\T_A - \T_B}{\T_0} \right )$ \\
    \bottomrule
  \end{tabular}
\end{table}

We therefore parametrized the first three cumulants of the rapidity profile as functions of $\T_A, \T_B$.
These cumulants describe the distribution's mean $\mu$, standard deviation $\sigma$, and skewness $\gamma$.
We parametrized the mean $\mu$ as
\begin{equation}
  \mu = \mu_0 \,y_\text{cm}
\end{equation}
where $\mu_0$ is a dimensionless parameter, and $y_\text{cm} = \frac{1}{2} \log(\T_A/\T_B)$ is the center-of-mass rapidity of the colliding matter.
For the distribution's standard deviation $\sigma$, we asserted a constant $\sigma = \sigma_0$, and for its skewness $\gamma$, we explored two parametrizations: a relative skewness model
\begin{equation}
  \gamma = \gamma_0 \left ( \frac{\T_A - \T_B}{\T_A + \T_B} \right ),
\end{equation}
and an absolute skewness model
\begin{equation}
  \label{eq:regulator}
  \gamma = \gamma_0 \left ( \frac{\T_A - \T_B}{\T_0} \right ),
\end{equation}
with $\T_0 = 1$~fm$^{-2}$.
Additional cumulants can be added to the generating function, but they increase the model complexity.
Table \ref{tab:cumulants} summarizes the parametrizations used for each generating function cumulant.

\begin{figure}[t]
  \centering
  \includegraphics{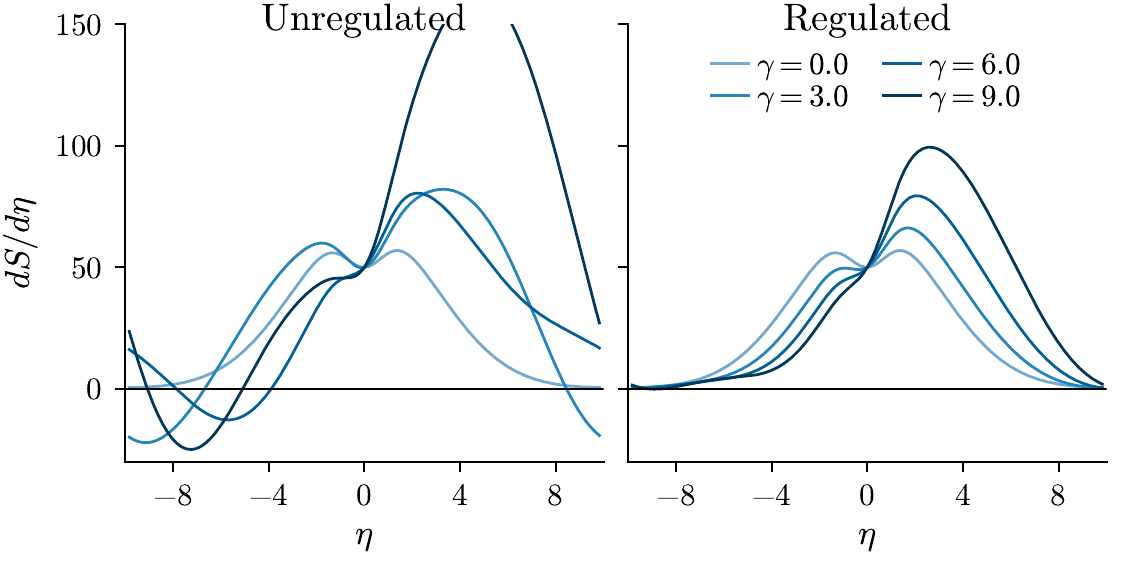}
  \caption{
    \label{fig:regulate}
    Rapidity-dependence of initial entropy deposition shown for different values of the skewness parameter $\gamma = 0, 3, 6, 9$.
    The left panel shows the unmodified parametrization, and the right panel shows the parametrization with the regulator, equation \eqref{eq:regulator}.
    The regulation term suppresses negative regions while maintaining monotonic scaling of the skewness.
  }
\end{figure}

There is, unfortunately, one significant problem with the aforementioned generating function approach.
The left panel of figure \ref{fig:regulate} shows several example functions $g$ generated by equation \eqref{eq:generating_func} using different values of the skewness $\gamma$.
When $\gamma$ is large, the distribution is ill behaved and turns negative at large rapidities.
The conditions to ensure a positive-definite Fourier transform are involved, so instead we introduced a regulation term
\begin{equation}
  \gamma \to \gamma \exp \left ( -\frac{1}{2} \sigma^2 k^2 \right ),
\end{equation}
to suppress spurious behavior at large rapidities.
The right panel of figure \ref{fig:regulate} shows the function with the regulator applied.
It suppresses negative regions while maintaining monotonic scaling of the skewness.
In realistic calculations, we set all negative entropy densities to zero.

\begin{figure}[t]
  \centering
  \includegraphics{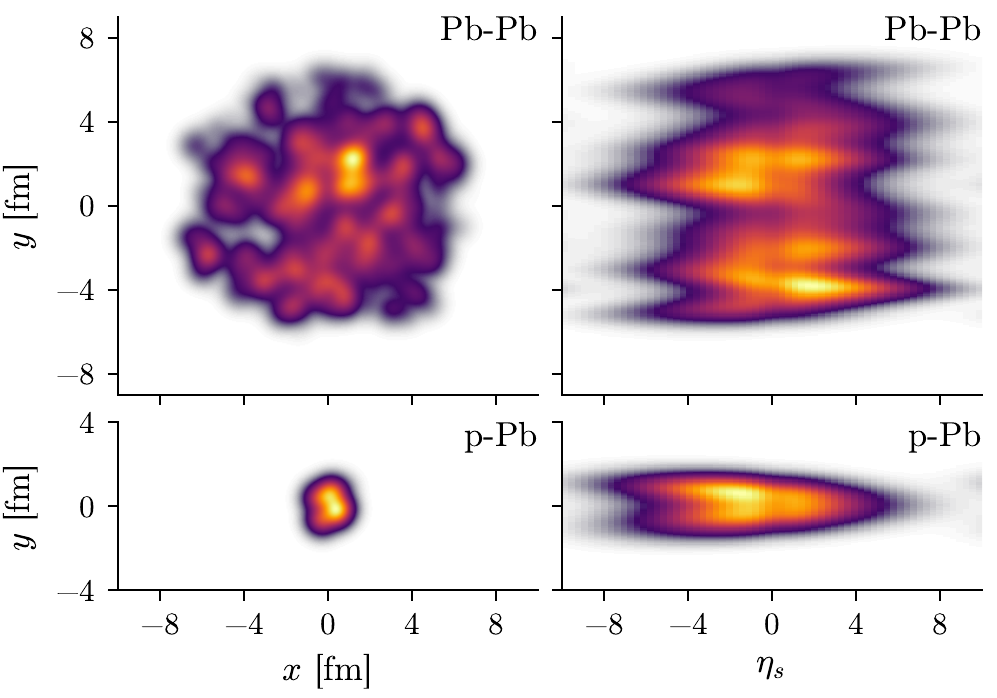}
  \caption{
    \label{fig:trento3d_example}
    Initial entropy density for an example Pb-Pb event (top) and $p$-Pb event (bottom) sliced along $\eta_s=0$ (left) and $x=0$ (right).
    Events are simulated using the relative skewness model in table \ref{tab:cumulants} with $\mu_0=1$, $\sigma_0=3$, and $\gamma_0 = 6$ along with midrapidity parameters from \cite{Bernhard:2016tnd}.
  }
\end{figure}

Figure \ref{fig:trento3d_example} shows the initial entropy density profile $s(\xv_\perp, \eta_s, \tau_0)$ at the hydrodynamic starting time $\tau_0$ generated by the model for an example Pb-Pb event (top) and $p$-Pb event (bottom) sliced along $\eta_s = 0$ (left) and $x=0$ (right), using typical parameter values for $\mu_0, \sigma_0$, and $\gamma_0$ annotated in the figure caption.
There are large entropy density fluctuations in the $(x, y)$ plane resulting from local nucleon density fluctuations.
These fluctuations generate momentum density anisotropies and hence significant forward-backward rapidity fluctuations which are visible in the $(\eta_s, y)$ plane.
Note, these parameters are not yet optimized, but I'll show somewhat later that they are in fact close to the model's best fit parameters.

\subsubsection{Hybrid transport model}

We replaced the boost-invariant VISH2+1 hydrodynamic code \cite{Song:2007ux} with vHLLE \cite{Karpenko:2013wva}, a 3+1D viscous hydrodynamics code with shear and bulk corrections.
The vHLLE code was initialized using the entropy density $s(\xv_\perp, \eta_s, \tau_0)$ in Milne coordinates at the hydrodynamic starting time $\tau_0$.
Following reference \cite{Bernhard:2016tnd}, we assumed instant thermalization at the hydrodynamic starting time $\tau_0$, setting the initial flow velocity $u^\mu$, shear correction $\pi^{\mu\nu}$, and bulk correction $\Pi$ identically to zero.
We also implemented the modern lattice based equation of state described in subsection \ref{subsec:nuclear_collision_model_v1}.

Unlike the boost-invariant studies in section \ref{sec:midrapidity}, we opted to fix the model's hydrodynamic medium parameters.
Ideally, we would parametrize and constrain these parameters concurrently with the model's initial condition parameters.
However, three-dimensional viscous hydrodynamic simulations require an order of magnitude more computing resources than boost-invariant hydrodynamic simulations.
This makes it difficult to calibrate on statistically intensive observables such as the two-particle flow cumulants $\vnk{n}{2}$ which are generally needed to constrain the QGP medium parameters.
In light of this issue, we calibrated the model using azimuthally integrated quantities, e.g.\ charged-particle densities $d\nch/d\eta$, which are statistically cheap to compute.

At the time of this study, the vHLLE code did not implement terms in the hydrodynamic equations of motion which couple the shear correction $\pi^{\mu\nu}$ to the bulk correction $\Pi$.
There were also a number of challenges which limited our ability to apply bulk viscous corrections to the fluid's distribution function at particlization; e.g.\ see section \ref{subsec:nuclear_collision_model_v1}.
These issues precluded a realistic treatment of bulk viscosity, so we opted to turn the bulk viscosity completely off, i.e.\ we set $\zeta/s = 0$ for all calculations.
For the QGP shear viscosity $\etas$, we assumed a constant (temperature-independent) value, hand tuned in the range 0.17--0.28 for each calculation to compensate for our simplistic treatment of viscous corrections.

All other aspects of the model follow the specification of subsection \ref{subsec:nuclear_collision_model_v1}:
we converted the fluid into particles along a fixed hydro-to-micro switching isotherm $\Tsw = 0.154$~GeV, we applied shear viscous corrections to the particle distribution function using equation \eqref{eq:shear_corrections}, and we modeled the hadronic evolution below the switching temperature using the UrQMD microscopic transport model.

\subsection{Parameter design and observables}

The study estimated the joint posterior distribution for nine initial condition parameters.
Five parameters modeled the entropy density at midrapidity:
\begin{enumerate}
  \item[1--2.]
    two overall normalization factors for initial entropy deposition; one for Pb-Pb collisions at $\sqrts=2.76$~TeV and one for $p$-Pb collisions at $\sqrts=5.02$~TeV,
  \setcounter{enumi}{2}
  \item
    the generalized mean parameter $p$ for the scaling of initial entropy deposition at midrapidity,
  \item
    the gamma shape parameter $k$ equal to the inverse variance of the random weights used to fluctuate each nucleon density, and
  \item
    the Gaussian nucleon width $w$ to specify initial state granularity.
\end{enumerate}
Meanwhile, another four parameters modeled its rapidity dependence:
\begin{enumerate}
  \item[6--8.]
    three coefficients $\mu_0$, $\sigma_0$, and $\gamma_0$ to parametrize the local rapidity distribution's mean, standard deviation, and skewness, and
  \setcounter{enumi}{8}
  \item
    one parameter $J$ to specify the pseudorapidity Jacobian.
\end{enumerate}

We assigned each parameter the conservative range of prior values listed in table \ref{tab:design3d} and sampled $d=100$ space filling parameter points within these ranges using a maximin Latin hypercube design.
We then executed $4 \times 10^3$ minimum bias Pb-Pb events at $\sqrts=2.76$~TeV and $10^4$ minimum bias $p$-Pb events at $\sqrts=5.02$~TeV at each parameter point.
Each nuclear collision event was run on a coarse space-time grid in ideal (non-viscous) mode to greatly reduce computational overhead.
Finally, we divided our minimum-bias events into centrality bins according to the charged-particle multiplicity $\nch$ in a given acceptance $\eta_\text{min} < \eta < \eta_\text{max}$.
We used $|\eta| < 0.8$ for Pb-Pb collisions and $-4.9 < \eta < -3.1$ for $p$-Pb collisions, selected to match our calibration data.

\begin{table}[h]
  \centering
  \captionsetup{width=\textwidth}
  \caption{
    Input parameter ranges for the nuclear collision model.
  }
  \label{tab:design3d}
  \small
  \begin{tabular}{lll}
    \toprule
    Parameter                   & Description                                & Prior range \\
    \midrule
    Norm Pb-Pb                  & Normalization factor $\sqrts=2.76$~TeV     & 140--190 \\
    Norm $p$-Pb                 & Normalization factor $\sqrts=5.02$~TeV     & 150--200 \\
    $p$                         & Generalized mean parameter                 & $-0.3$--$0.3$ \\
    $k$                         & Multiplicity fluct.\ shape                 & 1.0--5.0 \\
    $w$ [fm]                    & Gaussian nucleon width                     & 0.4--0.6 \\
    $\mu_0$                     & Rapidity shift mean coeff.\                & 0.0--1.0 \\
    $\sigma_0$                  & Rapidity width std.\ coeff.\               & 2.0--4.0 \\
    \multirow{2}{*}{$\gamma_0$} & \multirow{2}{*}{Rapidity skewness coeff.}  & 0.0--10.0 (rel) \\
                                &                                            & 0.0--3.6 (abs) \\
    $J$                         & Pseudorapidity Jacobian param.\            & 0.6--0.9 \\
    \bottomrule
  \end{tabular}
\end{table}

Running the hydrodynamic simulations on a coarse space-time grid in ideal mode precluded a meaningful comparison to anisotropic flow observables, so we calibrated the model using azimuthally integrated multiplicity observables which are less sensitive to the hydrodynamic viscosity and grid resolution.
We calculated the rapidity dependence of:
\begin{itemize}
  \item
    the charged-particle density $d\nch/d\eta$ measured in various centrality bins for Pb-Pb collisions at $\sqrts=2.76$~TeV and $p$-Pb collisions at $\sqrts=5.02$~TeV, and
  \item
    the root-mean-square of the two-particle pseudorapidity correlation coefficient $a_1$ for Pb-Pb collisions at $\sqrts=2.76$~TeV.
    I'll define this observable momentarily.
\end{itemize}
The first observable is sensitive to the rapidity dependence of the ensemble-averaged entropy density profile, while the second is sensitive to its event-by-event rapidity fluctuations.
Table \ref{tab:observables_trento3d} lists the experimental data used to calibrate each of these observables, taken from the ALICE \cite{Abbas:2013bpa, Adam:2015kda} and ATLAS collaborations \cite{Aad:2015zza, Radhakrishnan:2015eqq}.
Let me explain now what I mean by the two-particle pseudorapidity correlation coefficient $a_1$.

\begin{table}[b]
  \centering
  \caption{
    \label{tab:observables_trento3d}
    Experimental data used to calibrate the model parameters.
  }
  \small
  \makebox[\textwidth]{
    \begin{tabular}{ccccc}
      \toprule
      Collision system & Observable & Centrality bins & Ref. \\
      \midrule
      \multirow{2}{*}{Pb-Pb, 2.76 TeV} & Charged-particle pseudorapidity dist.\ $d\nch/d\eta$ & [0, 5, 10, 20, 30, 40, 50, 60, 70, 80, 90] & \cite{Abbas:2013bpa, Adam:2015kda}\\
                                       & Two-particle pseudorapidity corr.\ $\sqrt{\langle a_1^2 \rangle}$ & 5--10, 20--25, 35--40, 50--55, 65--70, 80--86 & \cite{Radhakrishnan:2015eqq} \\[1ex]
      $p$-Pb, 5.02 TeV & Charged-particle pseudorapidity dist.\ $d\nch/d\eta$ & [0, 1, 5, 10, 20, 30, 40, 60, 90] & \cite{Aad:2015zza} \\
      \bottomrule
    \end{tabular}
  }
\end{table}

Previously in subsection \ref{subsec:qgp_signatures}, I described how to project the event-by-event azimuthal particle distribution $dN/d\phi$ onto a basis of orthogonal harmonics.
In a similar fashion, we can project the event-by-event pseudorapidity distribution $dN/d\eta$ within the acceptance $[-Y, Y]$ onto a basis of orthonormal polynomials \cite{Bzdak:2012tp}
\begin{equation}
  \label{eq:legendre}
  \frac{dN}{d\eta} = \left \langle \frac{dN}{d\eta} \right \rangle \left [ 1 + \sum\limits_{n=0}^{\infty} a_n T_n \left ( \frac{\eta}{Y} \right ) \right ],
\end{equation}
using, for example, the normalized Legendre polynomials \cite{Jia:2015jga, Radhakrishnan:2015eqq}
\begin{equation}
  T_n(x) = \sqrt{n + 1/2}\, P_n(x).
\end{equation}
These normalized Legendre polynomials can then be used to expand the two-particle correlation function
\begin{equation}
  C(\eta_1, \eta_2) = \frac{ \langle N(\eta_1) N(\eta_2) \rangle }{\langle N(\eta_1) \rangle \langle N(\eta_2) \rangle},
\end{equation}
where $N(\eta) \equiv dN/d\eta$ is the multiplicity density at $\eta$, and $\langle N(\eta) \rangle$ is the average multiplicity at $\eta$ for a given event class.

Expressed as a sum of symmetrized Legendre polynomials, the two-particle correlation function $C(\eta_1, \eta_2)$ equals
\begin{subequations}
  \label{eq:sym_legendre}
  \begin{align}
    C(\eta_1, \eta_2) &= 1 + \sum\limits_{m,n=0}^\infty \langle a_m a_n \rangle T_{mn}(\eta_1, \eta_2), \\[1ex]
    T_{mn}(\eta_1, \eta_2) &= \frac{T_m(\eta_1) T_n(\eta_2) + T_m(\eta_2) T_n(\eta_1)}{2}.
  \end{align}
\end{subequations}
Rearranging terms, this expansion can be written as
\begin{multline}
  \label{eq:legendre_exp}
  C(\eta_1, \eta_2) = 1 + \frac{1}{2} \langle a_0 a_0 \rangle + \frac{1}{\sqrt{2}} \sum\limits_{n=1}^\infty \langle a_0 a_n \rangle (T_n(\eta_1) + T_n(\eta_2)) \\
  + \sum\limits_{m,n=1}^\infty \langle a_m a_n \rangle \frac{T_m(\eta_1) T_n(\eta_2) + T_m(\eta_2) T_n(\eta_1)}{2}.
\end{multline}
The last term containing coefficients $\langle a_m a_n \rangle$ with $m,n > 0$ describes the system's dynamical rapidity fluctuations at fixed centrality.
It can be isolated from the preceding terms by dividing the correlation function $C(\eta_1, \eta_2)$ by its projection onto the $\eta_1$ and $\eta_2$ dimensions:
\begin{subequations}
  \begin{align}
    C_N(\eta_1, \eta_2) &= \frac{C(\eta_1, \eta_2)}{C_1(\eta_1) C_2(\eta_2)}, \\[1ex]
    C_{1,2}(\eta_{1,2}) &= \frac{1}{2 Y} \int_{-Y}^{Y} C(\eta_1, \eta_2)\, d\eta_{1,2}.
  \end{align}
\end{subequations}
This rescaling sends the $\langle a_0 a_0 \rangle$ and $\langle a_0 a_n \rangle$ coefficients to zero, yielding
\begin{equation}
  C_N(\eta_1, \eta_2) = 1 + \sum\limits_{m,n=1}^\infty \langle a_m a_n \rangle \frac{T_m(\eta_1) T_n(\eta_2) + T_m(\eta_2) T_n(\eta_1)}{2}.
\end{equation}
Finally, the terms $\langle a_m a_n \rangle$ with $m,n > 0$ are projected out of the renormalized correlation function using the expansion
\begin{equation}
  C_N(\eta_1, \eta_2) = 1 + \frac{3}{2} \langle a_1^2 \rangle \frac{\eta_1 \eta_2}{Y^2} + \dots.
\end{equation}

Each coefficient $\langle a_m a_n \rangle$ receives contributions from long-range correlations, such as those introduced by initial state rapidity fluctuations, as well as short-range correlations, such as those caused by resonance decays.
Recent work has shown that short-range correlations from resonance decays are a significant contribution to the $\langle a_m a_n \rangle$ signal, while the QGP viscosity has a much smaller effect \cite{Denicol:2015bnf}.
Generally speaking, one can either model the short-range correlations or subtract their effect from both the model and the experimental data \cite{Jia:2016jlg}.
The hadronic transport model UrQMD properly accounts for short-range correlations, so we opted to compare to experimental data with the short-range correlations included \cite{Radhakrishnan:2015eqq}.

\subsection{Bayesian parameter estimation}

Bayesian parameter estimation was then applied to estimate the values of the nine initial condition parameters listed in table \ref{tab:design3d} using the predictions of our nuclear collision model and the experimental data listed in table \ref{tab:observables_trento3d}.
We followed the procedure described in section \ref{sec:emulator} and trained an emulator to reproduce the model's input-output mapping using the model calculations at each design point.
This emulator was trained to reproduce the first $q = 6$ principal components (PCs) of the $d\nch/d\eta$ data, and the first $q=4$ PCs of the rms $a_1$ data.
Collectively, these PCs explain 99.5\% of the observed variance across the computer experiment design.

We constructed the Bayesian likelihood from the product of three terms
\begin{equation}
  P(E | \xv) = P \big ( E_{d\nch/d\eta}^\text{$p$-Pb} \big | \xv \big) \cdot P \big (E_{d\nch/d\eta}^\text{Pb-Pb} \big | \xv \big ) \cdot P \big (E_\text{rms $a_1$}^\text{Pb-Pb} \big | \xv \big ),
\end{equation}
where each term quantifies the experimental evidence provided by a certain class of observable.
The first two terms on the right quantify the evidence provided by the $p$-Pb and Pb-Pb $d\nch/d\eta$ data, while the last term quantifies the evidence provided by the Pb-Pb rms $a_1$ observable.

Following reference \cite{Bernhard:2016tnd}, we calculated each likelihood function $P(E| \xv)$ in the PCA transformed observable basis
\begin{equation}
  P(E | \xv) \propto \exp\left \{-\frac{1}{2} [\zv_m(\xv) - \zv_e]\tran \Sigma_z^{-1} [\zv_m(\xv) - \zv_e] \right\},
\end{equation}
where $\zv_m(\xv)$ is the vector of PCA transformed model (emulator) observables, $\zv_e$ is the vector of PCA transformed experimental observables, and $\Sigma_z$ is their total covariance matrix.
We used a covariance matrix in the PC space proportional to the identity matrix $\Sigma_z = \sigma I$, corresponding to 5\%, 10\%, and 20\% relative error on the total variance of the $p$-Pb $d\nch/d\eta$, Pb-Pb $d\nch/d\eta$, and Pb-Pb rms $a_1$ observables.
This was a rough approximation, similar to reference \cite{Bernhard:2016tnd}, designed to
conservatively account for various sources of uncertainty in the experimental data, model calculations, and emulator predictions.
We elected to give more weight to charged-particle yields $d\nch/d\eta$ relative to the rms $a_1$ coefficient because they measure the single-particle distribution which is more fundamental than the two-particle distribution.
We also gave more weight to the $p$-Pb yields relative to the Pb-Pb yields, since they are more sensitive to the asymmetry parameters of the model.

In this study, we placed an informative prior $P(\xv)$ on the generalized mean parameter $p$, equal to the posterior distribution obtained from reference \cite{Bernhard:2016tnd}.
For all other model parameters, we used a uniform prior which was constant within the design ranges listed in table \ref{tab:design3d} and zero outside them.
The Bayesian posterior distribution was then calculated from the likelihood and the prior using equation \eqref{eq:bayes}.
Finally, we used the affine-invariant MCMC sampler \textsc{emcee} to draw samples from the Bayesian posterior distribution.
We ran \order{10^5} burn-in steps to allow the chain to equilibrate followed by \order{10^6} production steps.

\subsection{Posterior parameter distribution}

\begin{figure}[p]
  \centering
  \makebox[\textwidth]{
    \includegraphics{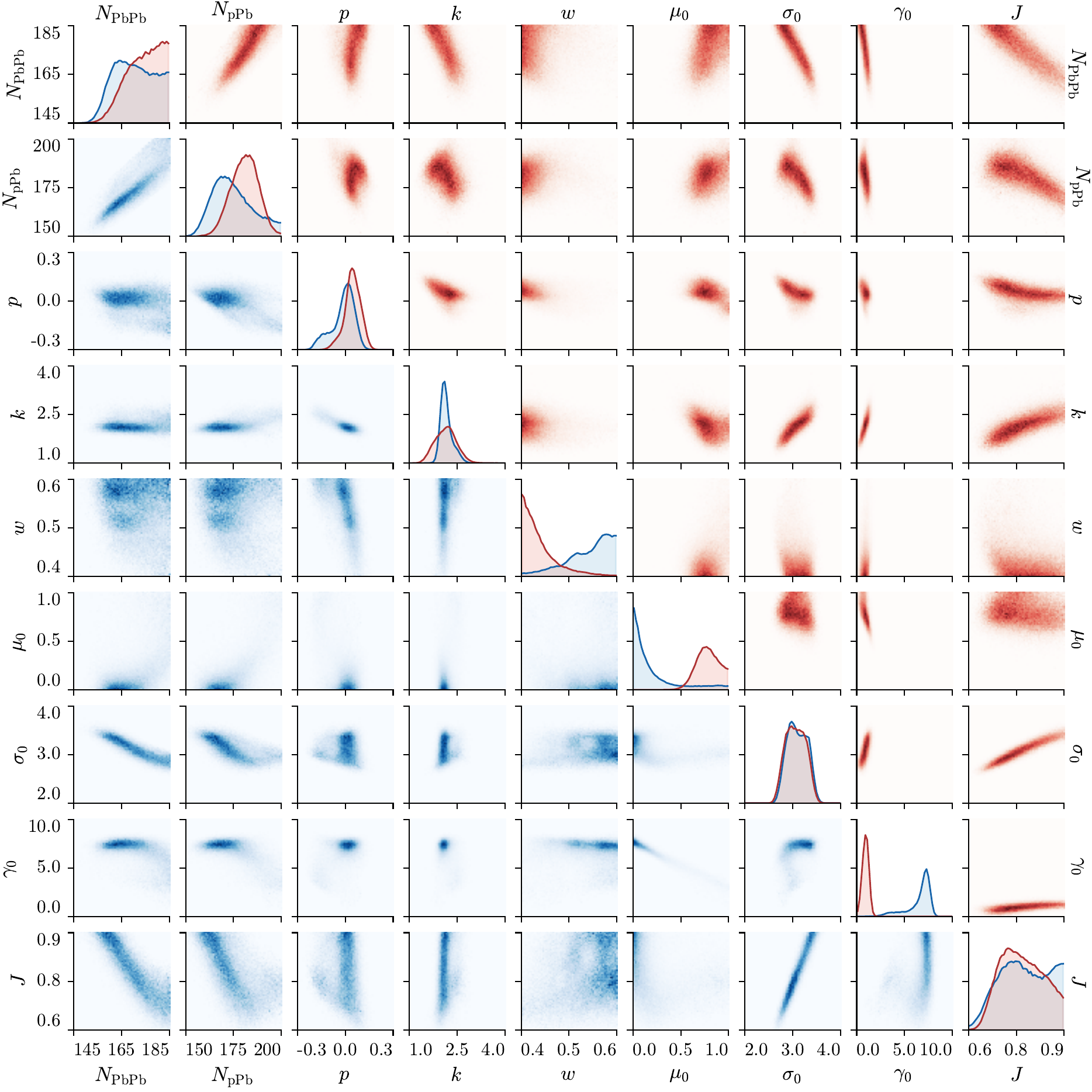}
  }
  \captionsetup{width=.8\paperwidth}
  \caption{
    \label{fig:posterior_trento3d}
    Posterior distribution for the model input parameters of Ref.~\cite{Ke:2016jrd}.
    Diagonal panels are the marginal distributions of individual parameters and off-diagonal panels are the joint distributions for pairs of parameters.
    The blue lower triangle is the posterior for the relative skewness model, and the red upper triangle is the posterior for the absolute skewness model.
  }
\end{figure}

Figure \ref{fig:posterior_trento3d} shows the Bayesian posterior distribution for the relative skewness model (blue lower triangle) and absolute skewness model (red upper triangle) parametrized by table \ref{tab:cumulants}.
I'll start, as before, by describing the marginal distributions shown on the figure diagonal, starting in the upper-left corner and working my way to the bottom-right.
As I traverse these marginal distributions, I'll also comment on some of the interesting correlations contained in the joint posterior distributions visualized by the off-diagonal elements.

The posterior on the Pb-Pb normalization $N_\text{Pb-Pb}$ is broad for both the relative and absolute skewness models.
Evidently, we should have increased the upper bound on the $N_\text{Pb-Pb}$ prior since its posterior is clipped.
Unsurprisingly, the Pb-Pb normalization is strongly correlated with the $p$-Pb normalization.
The $p$-Pb normalization $N_\text{$p$-Pb}$, meanwhile, peaks in the middle of the design range, with the absolute skewness model preferring somewhat larger values than the relative skewness model.

There's not much to say about the generalized mean parameter $p$ that has not already been said.
Recall that this parameter controlled the scaling of initial entropy deposition as a function of nuclear thickness at midrapidity \eqref{eq:gmean_ansatz}.
We assigned this parameter an informative prior peaked near zero using the results of reference \cite{Bernhard:2016tnd}, so the peak on our posterior is not particularly informative; it merely reflects our chosen prior.

Interestingly, the nucleon fluctuation parameter $k$ is tightly constrained.
This parameter specifies the inverse variance of the gamma random variables used to fluctuate each nucleon's contribution to the participant thickness function \eqref{eq:fluct_participant_density}.
Hence, small $k$ values enhance local nuclear density fluctuations, while large $k$ values suppress them.
Both the relative and absolute skewness models prefer $k \sim 2$, although it is unclear what drives this constraint.

The posterior on the Gaussian nucleon width $w$, meanwhile, is less clear.
The relative skewness model prefers significantly larger nucleons than the absolute skewness model.
The Gaussian nucleon width strongly affects the average asymmetry of nuclear density fluctuations, so it is not surprising that this parameter correlates with the skewness of the rapidity distribution.
This suggests that one should be careful when interpreting the posterior on the Gaussian nucleon width as it is highly sensitive to our modeling assumptions.

\begin{figure}[t]
  \centering
  \makebox[\textwidth]{
    \includegraphics{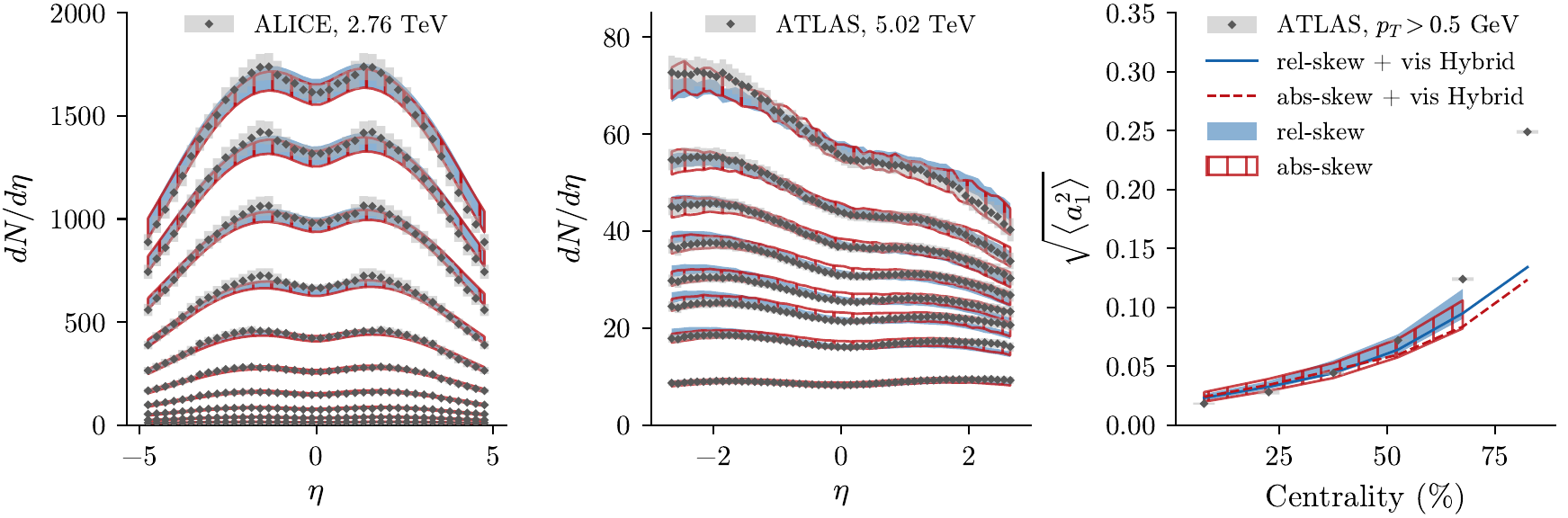}
  }
  \caption{
    \label{fig:post_obs_trento3d}
    Left and middle: Pseudorapidity dependence of the charged-particle density $d\nch/d\eta$ for Pb-Pb collisions at $\sqrts=2.76$~TeV (left) and $p$-Pb collisions at $\sqrts=5.02$~TeV (middle).
    Colored bands cover the model emulator's mean posterior prediction $\pm 2$ standard deviations, and black symbols are experimental data from ALICE \cite{Abbas:2013bpa, Adam:2015kda} and ATLAS \cite{Aad:2015zza}.
    Right: Two-particle pseudorapidity correlations quantified by $\sqrt{\langle a_1^2 \rangle }$ plotted versus collision centrality.
    Colored bands are same as before; black symbols are data from ATLAS \cite{Radhakrishnan:2015eqq}.
    The colored lines, meanwhile, are results from full event-by-event viscous hybrid model simulations using selected parameters from the Bayesian posterior.
  }
\end{figure}

The next three parameters $\mu_0$, $\sigma_0$, and $\gamma_0$ parametrize the cumulants of the rapidity distribution as functions of $\T_A, \T_B$.
Their precise values are not particularly meaningful, but we do observe some interesting features.
For example, the relative skewness model prefers a small rapidity shift, while the absolute skewness model prefers a large rapidity shift close to the nucleon center-of-mass rapidity $y_\text{cm}$.
We also observe a strong correlation between the width of the rapidity distribution $\sigma_0$ and the value of the Jacobian parameter $J$.
Finally, we observe that the relative and absolute skewness models prefer sharply peaked distributions for $\gamma_0$.
It's not interesting that these distributions peak in different locations; each model's parametrization of the skewness is different.
However, it \emph{is} interesting that the skewness is tightly constrained for both parametrizations.
Evidently, the data appears to prefer rapidity profiles with nonzero skewness.

Figure \ref{fig:post_obs_trento3d}, meanwhile, compares both models (colored bands) to the experimental calibration data (black symbols) using parameters sampled from the Bayesian posterior distribution.
Each colored band covers the model emulator's mean posterior prediction $\pm 2$ standard deviations.
First, direct your attention to the left two panels which show the Pb-Pb and $p$-Pb charged-particle densities $d\nch/d\eta$.
Both models nicely describe the centrality and rapidity dependence of $d\nch/d\eta$, illustrating the flexibility of the cumulant generating function.
Now, look at the far right panel which shows the centrality dependence of two-particle pseudorapidity correlations, quantified by the $\sqrt{\langle a_1^2 \rangle}$ observable.
Both models describe this observable to 20\% accuracy in the 0--50\% centrality range, but they underestimate its value at larger centralities.
Recently, it was shown that the microscopic model \mbox{HIJING} \cite{Gyulassy:1994ew} reproduces rms $a_1$ for $\Np < 80$, while it overestimates its value for more central collisions \cite{Radhakrishnan:2015eqq}.
This suggests that microscopic and hadronic models are complementary in understanding the longitudinal rapidity fluctuations.

\subsection{Verification of high-probability parameters}

This subsection evaluates our three-dimensional hybrid model's ``best fit'' predictions for new observables not included in the calibration process, using \emph{maximum a posteriori} (MAP) parameters (table \ref{tab:map_param_3d}) selected from the mode of the posterior distribution.
In the parlance of machine learning, the following predictions are said to be \emph{out of sample}.
These verification tests will help us assess the hidden assumptions of our model framework.

\begin{table}[t]
  \centering
  \caption{
    \label{tab:map_param_3d}
    High-probability \emph{maximum a posterior} (MAP) parameters selected from the mode of the posterior distribution.
    Values are approximate.
  }
  \small
  \begin{tabular}{lcc}
    \toprule
    & \multicolumn{2}{c}{Model variant:} \\
    \noalign{\smallskip}\cline{2-3}\noalign{\smallskip}
    Parameter         & Relative skewness & Absolute skewness \\
    \midrule
    $N_\text{Pb-Pb}$  & 150                     & 154 \\
    $p$               & 0.0                     & 0.0 \\
    $k$               & 2.0                     & 2.0 \\
    $w$ [fm]          & 0.59                    & 0.42 \\
    $\mu_0$           & 0.0                     & 0.75 \\
    $\sigma_0$        & 2.9                     & 2.9 \\
    $\gamma_0$        & 7.3                     & 1.0 \\
    $J$               & 0.75                    & 0.75 \\
    \bottomrule
  \end{tabular}
\end{table}

\subsubsection{Anisotropic flow cumulants}

First, we checked the centrality and rapidity dependence of the model's two-particle flow cumulants $\vnk{n}{k}$, calculated using the Q-cumulant method described in \cite{Bilandzic:2010jr}, which are highly sensitive to the initial distribution of matter in the $(x, y)$ plane.
Fitting the anisotropic flow cumulants is therefore a sensitive test of the QGP initial condition geometry.

\begin{figure}[t]
  \centering
  \includegraphics{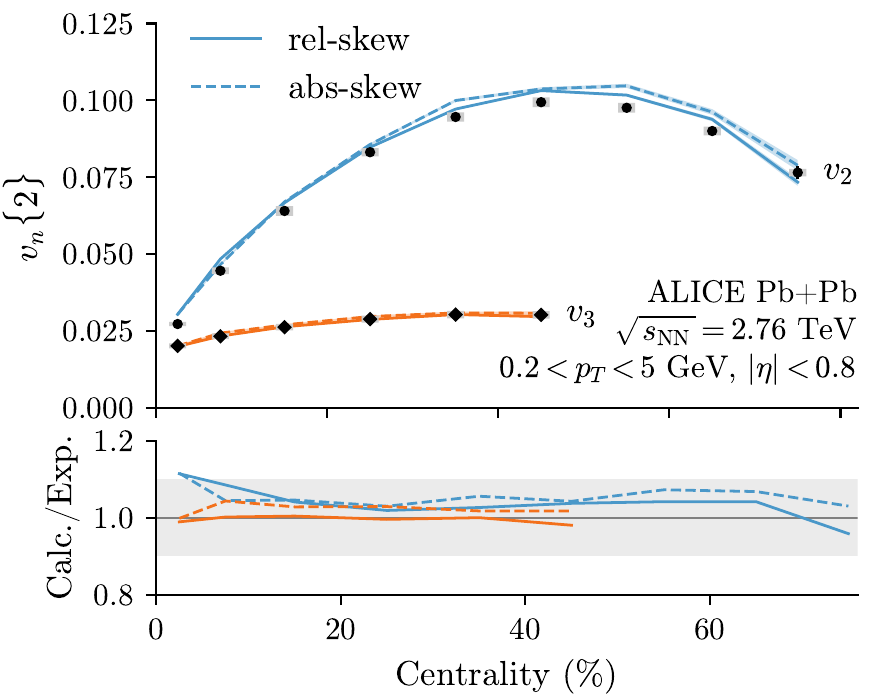}
  \caption{
    \label{fig:vn_cen}
    Anisotropic flow cumulants $\vnk{2}{2}$ and $\vnk{3}{2}$ at midrapidity $|\eta| < 0.8$ plotted as a function of collision centrality for Pb-Pb collisions at $\sqrts=2.76$~TeV.
    Colored lines are three-dimensional viscous hybrid model calculations using the relative (solid) and absolute (dashed) skewness models for initial entropy deposition, and symbols are experimental data from ALICE \cite{ALICE:2011ab}.
  }
\end{figure}

\begin{figure}[t]
  \centering
  \makebox[\textwidth]{
    \includegraphics{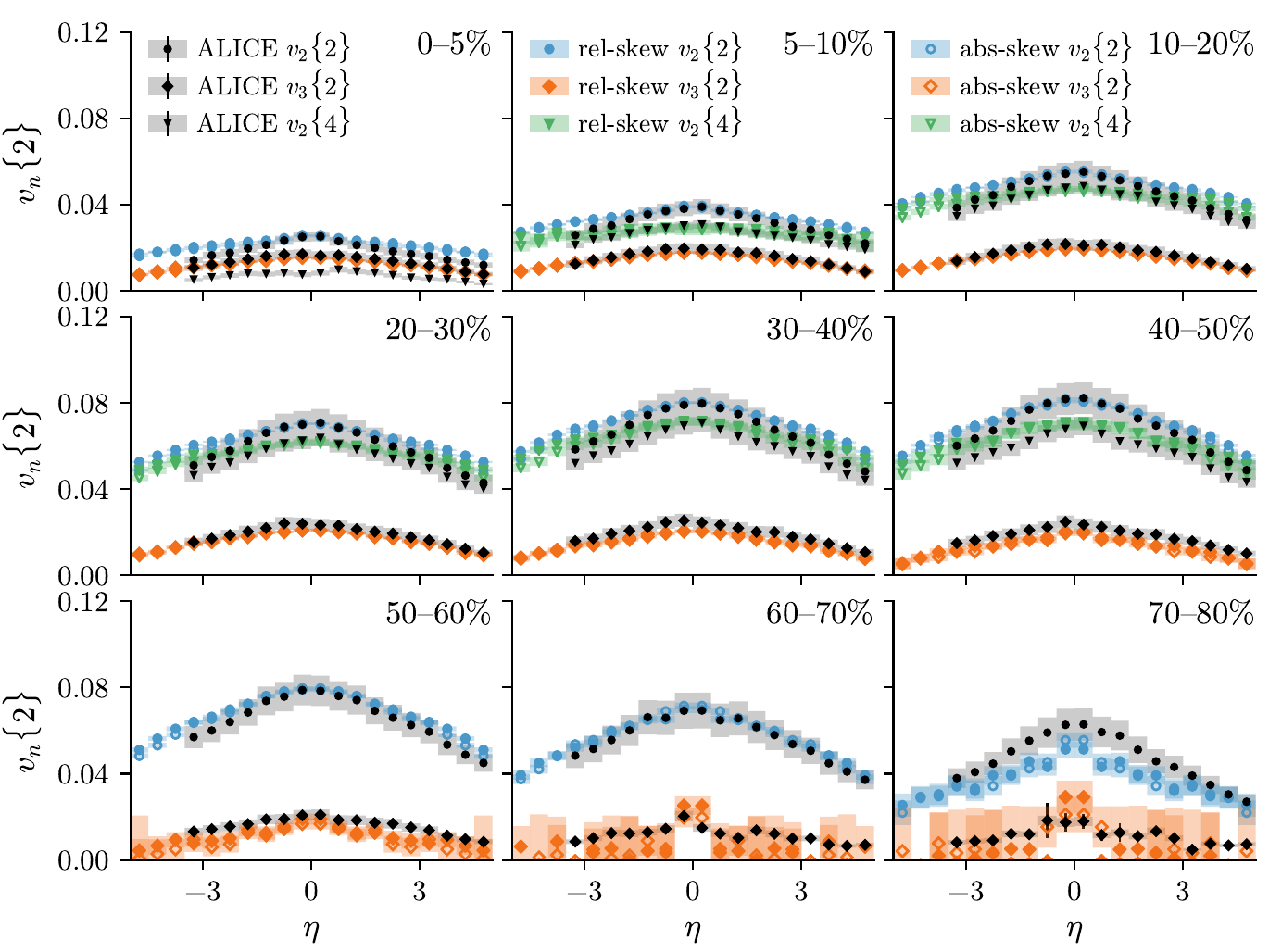}
  }
  \caption{
    \label{fig:vn_eta}
    Rapidity-dependence of the flow cumulants $\vnk{2}{2}$, $\vnk{3}{2}$, and $\vnk{2}{4}$ plotted for various centrality intervals.
    Colored symbols are three-dimensional viscous hybrid model calculations using high-probability parameters from the Bayesian posterior, and the bands are their $\pm2\sigma$ uncertainties.
    The black symbols and gray bands are experimental data from ALICE \cite{Adam:2016ows}.
  }
\end{figure}

Previously, in section \ref{sec:midrapidity}, I showed that the boost-invariant \trento\ model nicely describes the two-particle flow cumulants $\vnk{n}{2}$ for $n=2$, 3, and 4 at midrapidity $|\eta| < 0.8$.
Our rapidity-dependent initial condition model includes the boost-invariant \trento\ model as a specific subcase, so the present analysis should \emph{also} describe the experimentally measured two-particle flow cumulants at midrapidity provided suitably chosen parameters.

Figure \ref{fig:vn_cen} shows the present study's \emph{three-dimensional} hybrid model predictions (colored lines) for the two-particle flow cumulants $\vnk{2}{2}$ and $\vnk{3}{2}$ at midrapidity $|\eta| < 0.8$ plotted as a function of centrality for \mbox{Pb-Pb} collisions at $\sqrts=2.76$~TeV compared to experimental data (black symbols) from \mbox{ALICE} \cite{ALICE:2011ab}.
As expected, our three-dimensional model well describes the experimental data at midrapidity in agreement with similar model results based on boost-invariant approximations \cite{Bernhard:2016tnd}.

Now that we are confident that our three-dimensional model reproduces the results of our boost-invariant model at midrapidity, we can leverage the posterior constraints of the present analysis to predict each observable's rapidity dependence.
Figure \ref{fig:vn_eta} shows the rapidity dependence of the flow cumulants $\vnk{2}{2}$, $\vnk{3}{2}$, and $\vnk{2}{4}$ calculated in various centrality bins using the same particle selection criteria applied by the experiment.
The colored lines are the calculations of our three-dimensional hybrid model, and the black symbols are experimental data from ALICE \cite{Adam:2016ows}.

The flow cumulants peak at midrapidity and decrease at forward/backward rapidity, producing a triangular shape similar to the ALICE data.
Generally speaking, the model describes the data quite well, although the description is somewhat worse at large rapidities where the model tends to over predict the magnitude of each flow harmonic.
Recently, it was shown that the pseudorapidity dependence of $v_n(\eta)$ is highly sensitive to the hadronic shear viscosity \cite{Denicol:2015bnf}, so figure \ref{fig:vn_eta} corroborates the effective shear viscosity determined by UrQMD transport dynamics.

It's also worth noting that we have neglected the effects of nonzero baryon density.
For example, the initial conditions impart a rapidity-dependent baryon current which has been shown to affect final state observables \cite{Shen:2017bsr, Denicol:2018wdp}.
The QCD equation of state also varies as a function of baryochemical potential which we have completely omitted \cite{Denicol:2018wdp, Hasenfratz:1992ys, Barbour:1997ej, Fodor:2001au, Allton:2002zi, deForcrand:2002hgr, DElia:2002tig}.
Presumably, these effects would improve our model's description of the data at large rapidities where baryon density effects are most important.

\subsubsection{Event-plane decorrelation}

\begin{figure}[t]
  \centering
  \includegraphics[scale=.5]{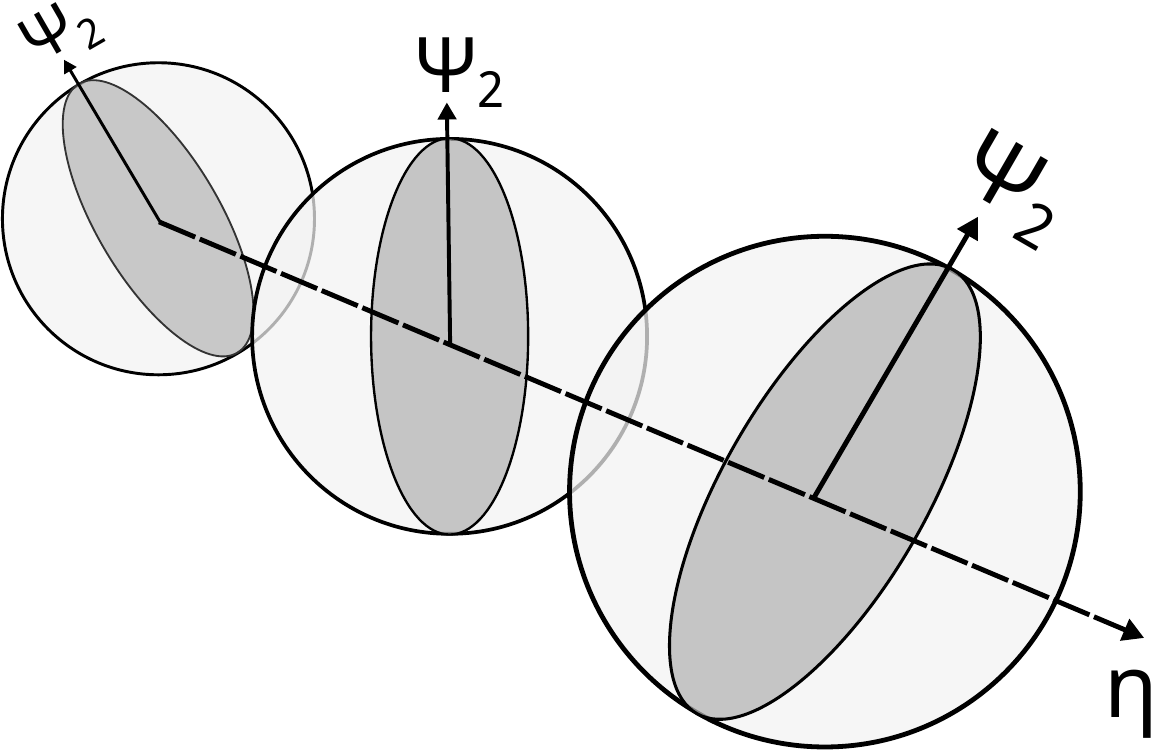}
  \caption{
    \label{fig:evt_plane_decorr}
    Twisting of the second-order event plane angle $\Psi_2$ as a function of pseudorapidity $\eta$.
    This effect decorrelates particles emitted with different pseudorapidities $\eta_1 \neq \eta_2$.
    Based on diagram by Xiang-Yu Wu \cite{Wu:2018ath}.
  }
\end{figure}

Recall that the anisotropic flow harmonics $v_n$ describe the modulation of the azimuthal single-particle distribution \cite{Ollitrault:1993ba, Voloshin:1994mz, Poskanzer:1998yz}
\begin{equation}
  \frac{dN}{d\phi} \propto 1 + 2 \sum\limits_{n=1}^\infty v_n \cos[ n(\phi - \Psi_n)],
\end{equation}
relative to the phase
\begin{equation}
  \Psi_n = \frac{1}{n} \textrm{arctan2}( \langle \sin n \phi \rangle, \langle \cos n \phi \rangle),
\end{equation}
commonly known as the event-plane angle.
In general, the event-plane angle $\Psi_n$ may change as a function of rapidity $\eta$ in a single event due to longitudinal rapidity fluctuations and finite particle effects.
Consequently, the event-plane angles $\Psi_n(\eta_1\pm \delta\eta)$ and $\Psi_n(\eta_2 \pm \delta\eta)$ are expected to decorrelate as the gap between $\eta_1$ and $\eta_2$ increases; e.g.\ see figure \ref{fig:evt_plane_decorr}.
This effect produces energy and entropy density profiles that are twisted or ``torqued'' along the pseudorapidity direction \cite{Bozek:2010vz}.

One method to study the single-particle anisotropic flow harmonics $v_n$ experimentally is to measure two-particle azimuthal correlations \cite{Ollitrault:1993ba, Voloshin:1994mz, Poskanzer:1998yz}
\begin{equation}
  \frac{dN^\text{pair}}{d\Delta\phi} \propto 1 + 2 \sum\limits_{n=1}^\infty V_{n\Delta} \cos( n \Delta \phi),
\end{equation}
where $\Delta\phi = \phi_a - \phi_b$ is the relative angle between each particle pair ($a$ and $b$) in the event.
If the particles are independently correlated with a common source, e.g.\ a hydrodynamic flow field, then the two-particle distribution factorizes into a product of single-particle distributions, so that
\begin{equation}
  V_{n\Delta} = v_n^a v_n^b.
\end{equation}
If, however, the particles $a$ and $b$ sample different underlying particle distributions, e.g.\ different flow fields, then the factorization breaks.
For example, if the fireball is torqued, then particles emitted with different rapidities $\eta^a \neq \eta^b$ will sample different flow fields so that $V_{n\Delta} \neq v_n^a v_n^b$.

The CMS collaboration measured the pseudorapidity dependence of this event-plane decorrelation effect in Pb-Pb collisions at $\sqrts=2.76$~TeV using the $\eta$-dependent factorization ratio $r_n$ \cite{Khachatryan:2015oea}, defined as
\begin{subequations}
  \begin{align}
    r_n(\eta^a, \eta^b) &\equiv \frac{V_{n\Delta}(-\eta^a, \eta^b)}{V_{n\Delta}(\eta^a, \eta^b)}, \\[1ex]
    V_{n\Delta}(\eta^a, \eta^b) &= \langle \langle \cos (n \Delta \phi) \rangle \rangle,
  \end{align}
\end{subequations}
where the inner average means averaging over all particle pairs in a given event, and the outer average means averaging over all events in a given centrality class.
Here, three rapidity bins, $\pm \eta^a$ and $\eta^b$, are used to remove the contamination of short-range jet-like two-particle correlations.
The resulting $\eta$-dependent factorization ratio $r_n(\eta^a, \eta^b)$ equals unity if the factorization holds, and is expected to be smaller than unity in the presence of rapidity fluctuations.

\begin{figure}[t]
  \centering
  \makebox[\textwidth]{
    \includegraphics{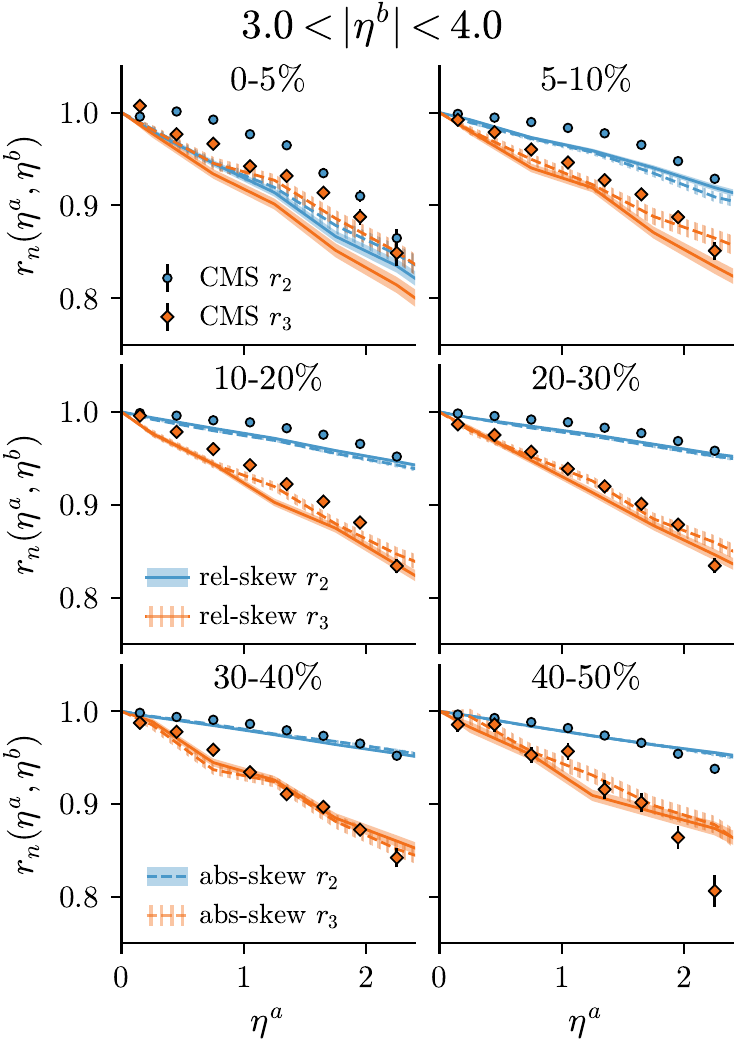}
    \hspace{2ex}
    \includegraphics{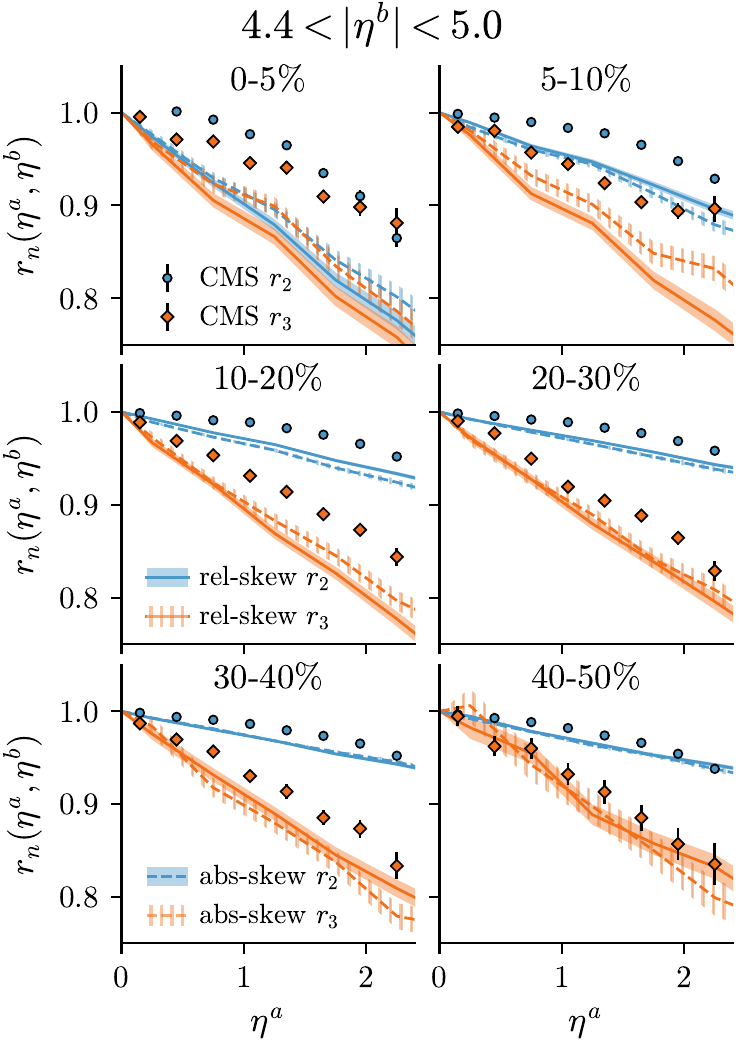}
  }
  \caption{
    \label{fig:evt_pln_decorr}
    Two-particle factorization ratio $r_n(\eta^a, \eta^b)$ for Pb-Pb collisions at $\sqrts=2.76$~TeV plotted as a function of the pseudorapidity $\eta^a$, for particle pairs selected with $0.3 < p_T^a < 3.0$~GeV and $p_T^b > 0$~GeV using $3.0 < |\eta^b| < 4.0$ (left figure) and $4.4 < |\eta^b| < 5.0$ (right figure).
    Colored lines are viscous hybrid model calculations and black symbols are experimental data from CMS \cite{Khachatryan:2015oea}.
  }
\end{figure}

Figure \ref{fig:evt_pln_decorr} shows our calibrated model calculations for the factorization ratio $r_n(\eta^a, \eta^b)$ plotted as a function of $\eta^a$ using $0.3 < p_T^a < 3.0$~GeV and $p_T^b > 0$~GeV, with two different rapidity cuts for $\eta^b$ selected to match the experiment: $3.0 < \eta^b < 4.0$ (left figure) and $4.4 < \eta^b < 5.0$ (right figure).
The factorization ratios are shown for the second harmonic $n=2$ (blue lines), the third harmonic $n=3$ (orange lines) and for six centrality classes (figure panels).
We show calculations for both the relative skewness model (solid lines) and absolute skewness model (dashed lines).

The model reproduces the general shape and approximate magnitude of the factorization breaking $r_n(\eta^a, \eta^b)$ for $n=2,3$ plotted as a function of the pseudorapidity $\eta^a$ for multiple centrality bins.
It correctly describes, for instance, the hierarchy of $r_2$ and $r_3$, and it reproduces the magnitude of the factorization breaking for non-central events provided $3.0 < \eta^b < 4.0$.
However, the agreement with the data also leaves room for improvement.
The model fails to describe the factorization breaking for 0--5\% centrality events, and it over predicts the strength of the factorization breaking when one of the particles is selected from $4.4 < |\eta^b| < 5.0$, i.e.\ far from midrapidity

Needless to say, some discrepancy at large rapidities is to be expected.
Our model is constructed to extrapolate well-developed initial conditions at midrapidity to nonzero rapidity using the constraints of azimuthally integrated multiplicity observables.
Naturally, this extrapolation gradually loses its predictive power for fine-structure flow observables as one moves far from midrapidity.
The observed tension indicates that our extrapolation is breaking down.
Clearly, future improvements to the model are needed at large rapidities.

\subsubsection{Symmetric cumulants}

As a final test of the calibrated model, we investigated the correlations between anisotropic flow harmonics of different order which have been shown to provide additional constraints on the QGP initial conditions \cite{Niemi:2012aj}.
Experimentally, these correlations can be quantified by the symmetric cumulants $\SC(m, n)$ \cite{Bilandzic:2013kga}, defined as
\begin{align}
  \SC(m, n) &= \langle\langle \cos[m(\phi_1 - \phi_3) + n(\phi_2-\phi_4)]\rangle\rangle \nonumber \\
  \nonumber &- \langle\langle\cos[m(\phi_1-\phi_2)]\rangle\rangle\langle\langle\cos[n(\phi_1-\phi_2)]\rangle\rangle \label{eq:scmn}\\
  &= \langle v_m^2 v_n^2 \rangle - \langle v_m^2\rangle\langle v_n^2\rangle,
\end{align}
where the double average means averaging over particles in each event and then averaging over all events in a given centrality class.
This quantity is positive if $v_m$ and $v_n$ are correlated, zero if they are uncorrelated, and negative if they are anti-correlated.
We also calculated the normalized symmetric cumulants
\begin{equation}
  \NSC(m, n) = \frac{\SC(m, n)}{\langle v_m^2 \rangle \langle v_n^2 \rangle},
\end{equation}
which divide out the magnitudes of $\langle v_m^2 \rangle$ and $\langle v_n^2 \rangle$.
Previous studies show that the $\NSC(3,2)$ observable is sensitive mainly to the initial conditions, while $\NSC(4,2)$ is sensitive to both the initial conditions and QGP medium properties \cite{ALICE:2016kpq, Niemi:2012aj}.

\begin{figure}[t]
  \centering
  \makebox[\textwidth]{
    \includegraphics{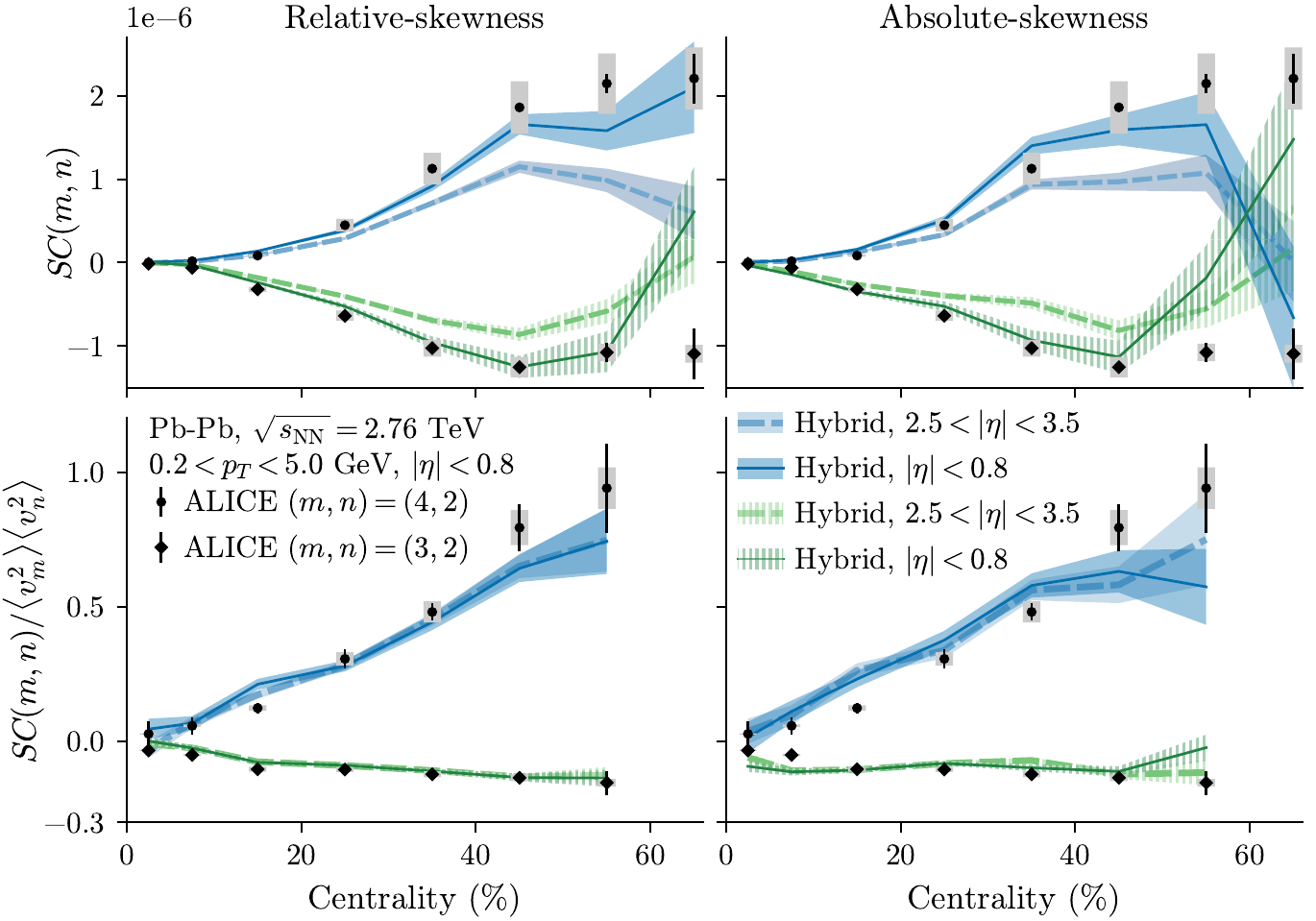}
  }
  \caption{
    \label{fig:sc_trento3d}
    Top panels: Symmetric cumulants $\SC(4, 2)$ (blue) and $\SC(3,2)$ (green) plotted as a function of collision centrality for the relative skewness model (left) and absolute skewness model (right) using $|\eta| < 0.8$ (solid lines) and ${2.5 < \eta < 3.5}$ (dashed lines) calculated using the same hybrid model setup as figure \ref{fig:vn_cen}.
    Experimental data (black symbols) are from ALICE \cite{ALICE:2016kpq}.
    Bottom panels: Same as top but for the normalized symmetric cumulants $\NSC(m,n)$.
  }
\end{figure}

Figure \ref{fig:sc_trento3d} shows the symmetric cumulants $\SC(m, n)$ (top row) and normalized symmetric cumulants $\NSC(m, n)$ (bottom row) calculated for the Pb-Pb system at $\sqrts=2.76$~TeV using the relative skewness model (left column) and absolute skewness model (right column).
The blue lines/bands are $(m, n) = (4, 2)$ and the green lines/bands are $(m, n) = (3, 2)$.
In addition, we also calculated each observable using two different sets of kinematic cuts (solid and dashed lines) described below.

First, we calculated each observable at midrapidity using the same kinematic cut $|\eta| < 0.8$ used by the ALICE experiment \cite{ALICE:2016kpq}.
These model calculations are the solid lines in figure \ref{fig:sc_trento3d}.
They nicely reproduce the centrality dependence of the symmetric and normalized symmetric cumulants measured by ALICE (black symbols), although we underpredict the magnitudes of the measured symmetric cumulants at large centralities.
Curiously, we obtained a worse description of this observable in reference \cite{Bernhard:2019ntr}, although it is unclear why.
Determining the culprit which explains this discrepancy would be an interesting project for future work.

Next, we calculated each observable using two reference particles sampled from $|\eta| < 0.8$ and two particles of interest sampled from $2.5 < |\eta| < 3.5$.
This new observable quantifies correlations between different flow harmonics separated by a large rapidity gap.
Therefore we expect it to be sensitive to the rapidity dependence of the initial conditions.
The dashed lines in figure \ref{fig:sc_trento3d} show our predictions for this novel observable which has not yet been measured.
Selecting the particles of interest from $2.5 < |\eta| < 3.5$ suppresses the magnitude of the symmetric cumulants at large centralities, but it has no meaningful effect on the normalized symmetric cumulants.
The invariance of the normalized symmetric cumulant to the rapidity of the particles of interest is therefore a testable prediction which can be used to validate the assumptions of our framework.

\section{Bulk matter in small collision systems}
\label{sec:nucleon_substructure}

Clearly, hydrodynamics well describes heavy-ion collisions at ultrarelativistic energies, but what about smaller collision systems, e.g.\ $p$-$p$ and $p$-Pb collisions, which generate nucleon sized QGP droplets and fewer particles per unit rapidity?
Do these small fireballs expand hydrodynamically as well?
This section presents the results of an exploratory study of hydrodynamic behavior in small collision systems using a modified version of the \trento\ initial condition model described in section \ref{sec:trento_model}.
The following text and figures appear in one of my publications,
\begin{flushleft}
J.\ S.\ Moreland, J.\ E.\ Bernhard, and S.\ A.\ Bass,
``Bayesian calibration of a hybrid nuclear collision model using $p$-Pb and Pb-Pb data from the LHC'', \href{https://arxiv.org/abs/1808.02106}{Submitted for publication, 1808.02106 [nucl-th]},
\end{flushleft}
which I've lightly edited to fit the format and layout of this dissertation.

Relativistic heavy-ion collisions produce long-range multiparticle correlations which are commonly explained by the existence of hydrodynamic flow \cite{deSouza:2015ena}.
This narrative is evidenced by the global, self-consistent and highly non-trivial quantitative agreement of hydrodynamic models with a large number of heavy-ion bulk observables \cite{Niemi:2015qia, Bernhard:2016tnd, Gale:2012rq}.
Naturally, such descriptions rely on the validity of hydrodynamic approximations, and these approximations begin to break down in the so-called dilute limit where discrete particle degrees-of-freedom dominate and continuous field descriptions of the medium cease to make sense.
Tell-tale signatures of hydrodynamic collectivity were thus always expected to vanish in smaller nuclear collision systems, e.g.\ $p$-$p$ and $p$-Pb collisions, where the number of produced particles is orders of magnitude smaller than a typical Pb-Pb collision.

These expectations were upended, however, when long-range multiparticle correlations were detected in high-multiplicity $p$-Pb collisions and found to be similar in magnitude to those observed in Pb-Pb collisions \cite{CMS:2012qk, Abelev:2012ola, Aad:2012gla}.
Nuclear collision systems which were previously thought to be too small for hydrodynamic flow, were subsequently found to generate the same collectivity used to justify hydrodynamic flow in heavy-ion collisions.
It is thus natural to wonder if a single unified hydrodynamic framework might be able to describe $p$-Pb and Pb-Pb bulk observables simultaneously.

In this work, we performed a semi-exhaustive search for a unified description of $p$-Pb and Pb-Pb collisions at $\sqrts=5.02$~TeV using Bayesian methods to rigorously calibrate and constrain free parameters of a flexible nuclear collision model based on viscous hydrodynamics.
The goal of our study was two fold.
First, we aimed to establish whether or not our hydrodynamic framework was able to describe both collision systems simultaneously.
And second, in the event that the former was true, we wished to obtain estimates for the \emph{true} parameters of our model given the assumptions of our framework and the evidence provided by the experimental data.

\subsection{Boost-invariant model with nucleon substructure}
\label{subsec:nucleon_substructure}

This study used an updated version of our nuclear collision model largely identical to the model which I described in subsection \ref{subsec:nuclear_collision_model_v2}.
It consisted of:
\begin{itemize}[leftmargin=1\parindent]
  \item
    \trento\ initial conditions,
  \item
    pre-equilibrium free streaming and Landau matching,
  \item
    boost-invariant viscous hydrodynamics with shear and bulk coupling,
  \item
    a modern lattice-based QCD equation of state,
  \item
    Cooper-Frye particlization with shear and bulk viscous corrections using the \textsc{frzout} sampler, and
  \item
    UrQMD to simulate microscopic hadronic interactions below the pre-specified hydro-to-micro switching temperature.
\end{itemize}
The study also made one important addition to the nuclear collision model.
We replaced Gaussian nucleons in the \trento\ model with ``lumpy'' nucleons characterized by several free parameters to vary the fluctuating size and shape of constituent degrees of freedom inside each nucleon.
I'll explain the specifics of this nucleon substructure extension shortly.
First, let me motivate why this additional physics is relevant in the first place.

\subsubsection{Case for nucleon substructure}
\label{subsec:case_for_substruct}

IP-Glasma \cite{Schenke:2012wb} is one of the most theoretically sound and phenomenologically successful models for the initial conditions of ultrarelativistic nuclear collisions.
The model describes, for instance, numerous bulk observables in heavy-ion collisions at RHIC and LHC energies \cite{Gale:2013da}.
Thus, when it was first observed that high-multiplicity $p$-Pb collisions generate flow-like signatures---not definitive \emph{proof} of flow, but tantalizing indicators---it was natural to check the predictions of hydrodynamic simulations using IP-Glasma initial conditions against said measurements.

\begin{figure}[t]
  \centering
  \includegraphics[scale=.9]{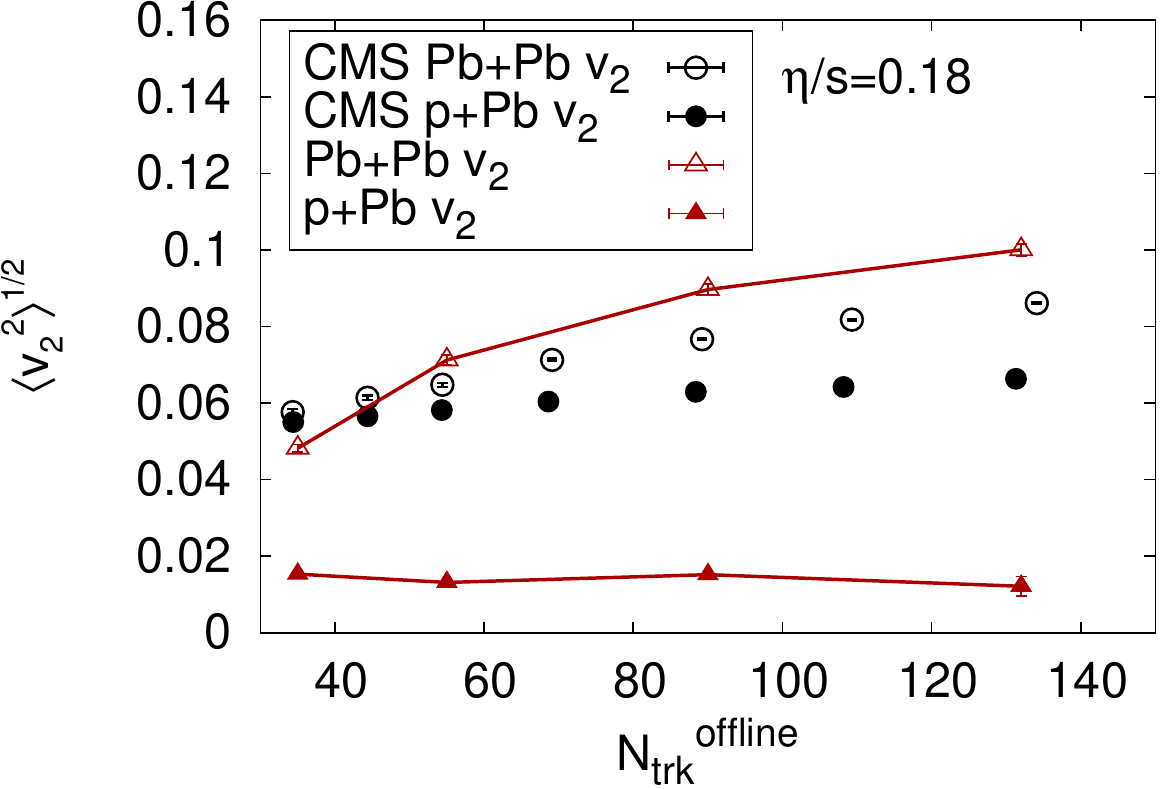}
  \caption{
    \label{fig:ipglasma_ppb_flow}
    Root-mean-square elliptic flow coefficient $\langle v_2^2 \rangle^{1/2}$ plotted as a function of the number of detector tracks offline $\ntrk$ (multiplicity) for $p$-Pb collisions (closed symbols) and Pb-Pb collisions (open symbols) at the LHC.
    Red triangles are the predictions of the MUSIC hydrodynamic model with IP-Glasma initial conditions, and black circles are experimental measurements from CMS \cite{Chatrchyan:2013nka}.
    Figure is from reference \cite{Schenke:2014zha}.
  }
\end{figure}

Schenke and Venugopolan performed this test in reference \cite{Schenke:2014zha} using the MUSIC viscous hydrodynamics code \cite{Schenke:2010nt, Schenke:2010rr, Ryu:2015vwa}.
While IP-Glasma coupled to MUSIC well reproduces numerous heavy-ion bulk observables, they found that the model significantly underpredicts the elliptic flow $v_2$ observed in $p$-Pb collisions at $\sqrts=5.02$~TeV as shown in figure \ref{fig:ipglasma_ppb_flow}.
This observed tension could mean one of several things:
\begin{enumerate}[leftmargin=1\parindent]
  \item
    Hydrodynamics is the correct framework to study $p$-Pb collisions, but IP-Glasma is flawed.
  \item
    IP-Glasma is essentially correct, and the application of hydrodynamics to $p$-Pb collisions is flawed.
    Perhaps some additional non-hydrodynamic correlations are required to describe the data.
  \item
    IP-Glasma coupled to hydrodynamics is a sensible framework to study $p$-Pb collisions, but the IP-Glasma model is \emph{incomplete}.
\end{enumerate}

Confronted with this list of possibilities, the authors suggested that eccentric nucleons might offer an attractive solution to the problem.
In its original formulation, the IP-Glasma model uses round, spherically symmetric nucleons.
This symmetry produces roundish blobs of initial energy density in $p$-$p$ and $p$-Pb collisions \cite{Bzdak:2013zma} and hence small initial state eccentricities.
Conversely, deformed nucleons generate significantly larger $p$-$p$ and $p$-Pb eccentricities, enhancing the anisotropic flow predicted for small systems.
Since Pb-Pb collisions are only weakly sensitive to nucleon substructure \cite{Moreland:2012qw}, this solution could potentially resolve the $p$-Pb discrepancy without spoiling the model's simultaneous description of heavy-ion data.
Indeed, recent studies of the IP-Glasma initial condition model using nucleon substructure show promising descriptions of $p$-Pb bulk observables \cite{Mantysaari:2017cni} and work is on-going to extend these comparisons to heavy-ion observables which were previously fit by IP-Glasma model calculations without nucleon substructure  \cite{Schenke:2018fci}.

Much like IP-Glasma, the calibrated \trento\ model predicts roundish blobs of energy (or entropy) in $p$-$p$ and $p$-Pb collisions.
Indeed, when the generalized mean parameter $p=0$, the \trento\ model predicts energy (or entropy) density profiles for $p$-$p$ collisions that are \emph{exactly} Gaussian; see figure \ref{fig:proton_shapes}.
Certainly, such similarities are to be expected if the bottom-up approach of IP-Glasma and the top-down approach of \trento\ are describing the same underlying physics.
Therefore, it stands to reason that \trento\ will also need nucleon substructure if our hydrodynamic model is to have any chance of describing $p$-Pb and Pb-Pb bulk observables simultaneously.

\subsubsection{Nucleon substructure in the literature}

Unlike heavy-ions for which there exists an established procedure to randomly sample nuclear density configurations (see section \ref{sec:nuclear_structure}), the nucleon's fluctuating structure is poorly understood from first principles and difficult to measure experimentally.
Nucleon substructure implementations in the literature have thus often employed simplistic models, replacing round protons with composite protons described by a few salient model parameters, in order to investigate the effect of each model parameter on simulated observables \cite{Adler:2013aqf, Mitchell:2016jio, Welsh:2016siu, Broniowski:2016pvx, Bozek:2017jog}.
Sensitivity studies such as these have been able to identify cause and effect relationships between model inputs and outputs, but they lack the ability to constrain nucleon substructure parameters in any kind of global or systematic fashion.
Multiple nucleon substructure descriptions appear to be \emph{qualitatively} consistent with the usual list of hydrodynamic bulk observables (yields, mean $p_T$'s and flows), so new observables are needed to discriminate mutually incompatible model assumptions.

Several such observables have been identified in proton-proton and proton-lepton scattering data.
Measurements by the TOTEM collaboration at $\sqrt{s}=7$~TeV, for instance, found an unexpected dip in the inelasticity density of $p$-$p$ collisions at zero impact parameter \cite{Antchev:2011zz}.
It was later realized that this depression, or so-called hollowness effect in the $p$-$p$ inelastic collision profile \cite{Arriola2016}, can be explained by the existence of correlated domains inside the proton, and that aspects of these domains, such as their size and correlation strength, may be constrained by comparing model predictions to inelastic $p$-$p$ measurements \cite{Albacete:2016gxu, Albacete:2016pmp}.

Independently, studies of coherent and incoherent $J/\psi$ production based on a color dipole picture of vector meson production were used to simultaneously constrain both the average color charge density of the proton as well as its event-by-event fluctuations in a saturation based framework \cite{Mantysaari:2016ykx, Mantysaari:2016jaz, Aaron:2009aa, Abramowicz:2015mha}.
Initial condition studies using the IP-Glasma model of Color Glass Condensate effective field theory \cite{Schenke:2012wb} simultaneously demonstrated that these color charge fluctuations leave a lasting imprint on the \mbox{small-x} gluon distribution of the proton and hence the initial geometry of QGP energy deposition \cite{Schlichting:2014ipa}.
In addition, it was recently shown that hydrodynamic simulations using IP-Glasma initial conditions with color charge fluctuations calibrated to fit coherent and incoherent $J/\psi$ diffraction measured by the H1 and Zeus experiments at HERA \cite{Aaron:2009aa, Abramowicz:2015mha} provide a good description of collectivity in small and large collision systems \cite{Schenke:2018fci}.

Model parameters, such as those calibrated by the aforementioned studies, are of course always in some degree of tension.
For instance, fitting one observable may require parameter values that degrade the quantitative description of some other observable.
Similarly, parameters which provide an optimal description of small-system observables may lead to a sub-optimal description of heavy-ion observables or \emph{vice versa}.
It is thus import to look at the experimental data holistically, and to use model calibration methods which
\begin{enumerate*}[label=(\arabic*)]
  \item explore all parameter combinations and
  \item compare model predictions to all experimental measurements in a statistically rigorous fashion.
\end{enumerate*}

\subsubsection{Parametric nucleon substructure}
\label{sec:initial_state}

From a Bayesian perspective, there is nothing unique about nucleon substructure uncertainty compared to the other sources of initial condition uncertainty already parametrized by the \trento\ model.
Indeed, the study which I will discuss now is essentially identical to that of reference \cite{Bernhard:2019ntr} with one significant change to the nuclear collision model.
We replaced one-parameter round nucleons with three-parameter lumpy nucleons.

\begin{figure}[t]
  \centering
  \begin{tikzpicture}[scale=.9]
    \node {\includegraphics[scale=.9]{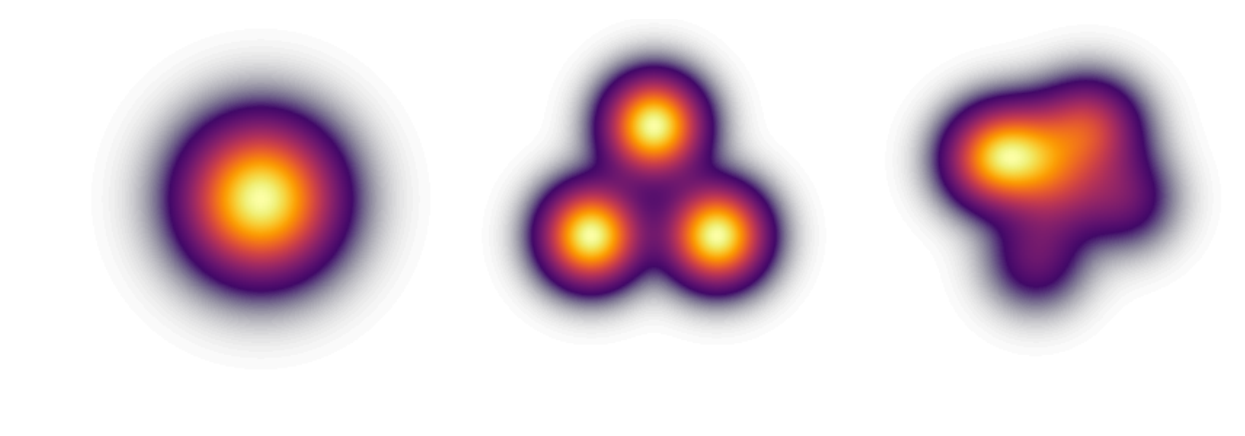}};
    \node[label=above:{2 fm}] (a) at (-5.75, 1.1) {};
    \node (b) at (-5.75, -0.85) {};
    \draw [<->, thick] (a) to (b);
  \end{tikzpicture}
  \vspace{-1cm}
  \caption{
    \label{fig:substructure}
    Thickness function $T(\xv_\perp) = \int dz \rho_n(\xv_\perp,z)$ for three example nucleon densities $\rho_n$ generated by the parametric substructure model.
    These example nucleons show a small subset of all possible variability.
  }
\end{figure}
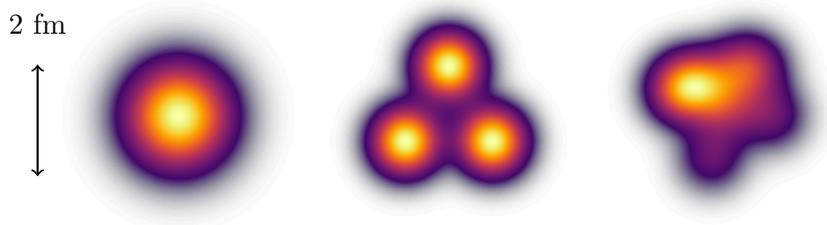

Specifically, we modeled each lumpy nucleon density as a sum of $n_c$ independent constituent densities:
\begin{equation}
  \label{eq:rho}
  \rho_n(\xv) = \frac{1}{n_c} \sum\limits_{i=1}^{n_c} \gamma_i\, \rho_c(\xv - \xv_i),
\end{equation}
where $\gamma_i$ is a random weight sampled from a gamma distribution with unit mean and variance $1/k$ as before, and $\rho_c$ is a Gaussian constituent density
\begin{equation}
  \rho_c(\xv) = \frac{1}{(2 \pi v^2)^{3/2}} \exp\left(- \frac{|\xv|^2}{2 v^2}\right)
\end{equation}
of variable width $v$.
The constituent positions $\xv_i$ were sampled independently (ignoring correlations) from a Gaussian radial distribution
\begin{equation}
  \label{eq:radial_dist}
  \rho_\text{pos}(\xv_i; \xv_n) = \frac{1}{(2\pi r^2)^{3/2}} \exp\left(-\frac{|\xv_i - \xv_n|^2}{2 r^2}\right),
\end{equation}
where $\xv_n$ is the position of each nucleon, and $r$ is a parameter which varies the sampling radius of the constituent positions about each nucleon position.
This allows the model to generate a diverse range of initial nucleon shapes as shown in figure \ref{fig:substructure}.

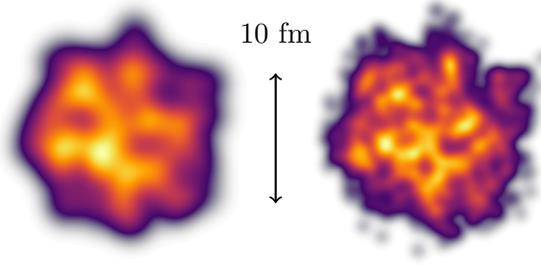
\begin{figure}[t]
  \centering
  \begin{tikzpicture}[scale=1]
    \node {\includegraphics[scale=1]{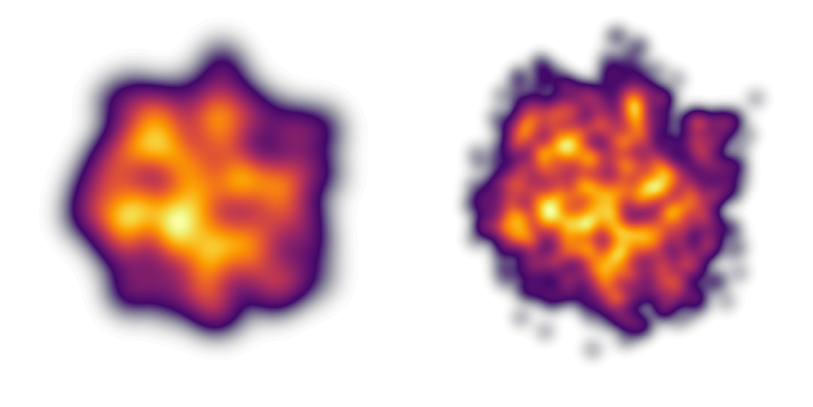}};
    \node[label=above:{10 fm}] (a) at (0, 1) {};
    \node (b) at (0, -1) {};
    \draw [<->, thick] (a) to (b);
  \end{tikzpicture}
  \caption{
    \label{fig:thickness}
    Effect of nucleon substructure on the nuclear thickness function $T(\xv_\perp) = \int dz\, \rho(\xv_\perp, z)$ of a $^{208}\mathrm{Pb}$ nucleus.
    The nucleus on the left has Gaussian nucleons of width $0.8$~fm, while the nucleus on the right has composite nucleons, each containing six constituents of width $0.4$~fm.
  }
\end{figure}

As a matter of convenience, we sampled the nucleon positions \emph{before} determining the constituent positions.
This creates small discrepancies between the designated nucleon positions $\xv_n$ and the actual position of each nucleon's center-of-mass, owing to fluctuations in the constituent positions $\xv_i$.
The sampling radius $r$ should thus be interpreted with care.
It is \emph{not} the Gaussian width of the sampled nucleons in their center-of-mass frame.

Once each lumpy nucleon is sampled, the nucleon density is treated as a singular field $\rho_n$.
The constituents are merely used as a mechanism to add deformity and fluctuations to this field.
This distinguishes the \trento\ model from the wounded quark model which calculates individual quark participants, depositing for each wounded quark a fixed amount of energy or entropy \cite{Broniowski:2016pvx, ANISOVICH1978477, PhysRevC.67.064905}.
Protons and neutrons are bound by the strong force, so I'd argue that there is no sensible concept of spectator quarks like there is for spectator nucleons.
If one constituent collides inelastically, the entire nucleon is ripped apart and contributes inelastically.

The \trento\ model essentially treats nucleus-nucleus collisions as a superposition of individual proton-proton collisions so nucleon substructure modifies the overall thickness function of a macroscopic nucleus as well.
Figure~\ref{fig:thickness} shows the effect of nucleon substructure on the thickness function of a generic lead nucleus.
Additional fluctuations emerge over the length scale of a nucleon, but the macroscopic geometry of the nucleus is largely unchanged.
We therefore expect nucleon substructure to have a small effect on heavy-ion collisions, but it is impossible to know for certain without running the calculations directly.

\subsection{Parameter design and observables}
\label{sec:observables}

This study estimated the joint posterior distribution for fifteen model parameters.
Seven parameters varied the initial conditions modeled by \trento:
\begin{enumerate}[leftmargin=2\parindent]
  \item
    the overall normalization factor for initial energy deposition (same for $p$-Pb and Pb-Pb),
  \item
    the generalized mean parameter $p$ for the scaling of initial energy deposition as a function of nuclear thickness,
  \item
    the number of nucleon constituents $n_c$,
  \item
    the sampling radius for the nucleon constituent positions $r$,
  \item
    the nucleon constituents' width $v$,
  \item
    the nucleons' effective fluctuation standard deviation $\sigf = (n_c k)^{-1/2}$, where $k$ is the of the inverse variance of the gamma random variable used to fluctuate each constituent density, and
  \item
    the cube of the minimum inter-nucleon distance $\dmin^3$.
\end{enumerate}
Meanwhile, another eight parameters varied the properties of the hybrid model simulation:
\begin{enumerate}[leftmargin=2\parindent]
  \setcounter{enumi}{7}
  \item
    the pre-equilibrium free streaming time $\tfs$,
  \item[9--11.]
    three parameters ($\etas$ min, slope, and curvature) for the temperature dependence of the QGP shear viscosity,
  \item[12--14.]
    another three parameters ($\zetas$ max, width, and location) for the temperature dependence of the QGP bulk viscosity, and
  \item[15.]
    a particlization temperature \Tsw\ that defined the isotherm for Cooper-Frye particle emission.
\end{enumerate}
We assigned each parameter the conservative range of prior values listed in table \ref{tab:design_v3} and used Latin hypercube sampling to distribute 500 design points uniformly throughout the fifteen dimensional parameter space.
The selected parameter ranges were chosen to be intentionally wide to avoid clipping the calibrated posterior.
For example, even though references \cite{Bernhard:2016tnd, Bernhard:2019ntr} found $p \sim 0$, we decided to use a prior range $p \in [-1, 1]$ to account for differences in the present model, e.g.\ nucleon substructure, which could modify its posterior.

One exception is the constituent number $n_c$ which we limited for practical considerations.
Recall that each constituent fluctuates independently, weighted by a gamma random variable.
Hence for constituent numbers $n_c \gg 1$, the fluctuations average out, and the resulting nucleon fluctuations vanish.
To counteract this effect, the constituent fluctuation variance must increase as $n_c$ increases.
Eventually, these fluctuations become unreasonably large.
We found that for $n_c < 10$, the energy density fluctuations are reasonable, and hence we limited $n_c$ to this prior range.

\begin{table}[t]
  \centering
  \caption{
    Input parameter ranges for the nuclear collision model.
    \label{tab:design_v3}
  }
  \small
  \begin{tabular}{lll}
    \toprule
    Parameter         & Description                          & Range           \\
    \midrule
    Norm              & Normalization factor                 & 9--28 GeV       \\
    $p$               & Energy deposition parameter          & $-1$ to $+1$    \\
    $\sigmaf$         & Nucleon fluctuation std.\ dev.\      & 0--2            \\
    $r$               & Constituent sampling radius          & 0--1.2 fm       \\
    $n_c$             & Number of nucleon constituents       & 1--9            \\
    $v$               & Constituent width                    & 0.2--1.2 fm     \\
    $\dmin^3$         & Minimum inter-nucleon distance       & 0--4.9 fm$^3$   \\
    $\tfs$            & Free streaming time                  & 0.1--1.5 \fmc   \\
    $\etasmin$        & Minimum value of $\eta/s$ (at $T_c$) & 0--0.2          \\
    $\etasslope$      & Slope of $\eta/s$ above $T_c$        & 0--8 GeV$^{-1}$ \\
    $\etascrv$        & Curvature of $\eta/s$ above $T_c$    & $-1$ to $+1$    \\
    $\zetasmax$       & Maximum value of $\zeta/s$           & 0--0.1          \\
    $\zetaswidth$     & Width of $\zeta/s$ peak              & 0--0.1 GeV      \\
    $\zetasloc$       & Temperature of $\zeta/s$ maximum     & 150--200 MeV    \\
    $\Tsw$            & Switching/particlization temp.       & 135--165 MeV    \\
    \bottomrule
  \end{tabular}
\end{table}

We executed \order{10^4} minimum-bias and multiplicity-triggered $p$-Pb and Pb-Pb events at $\sqrts=5.02$~TeV at each design point, and calculated the model predictions for a number of experimental observables at midrapidity:
\begin{itemize}[itemsep=0pt]
  \item
    the charged-particle density $d\nch/d\eta$ \cite{Adam:2015ptt, Adam:2014qja} and identified-particle densities $dN/dy$ of pions, kaons, and protons,
  \item
    the transverse energy production $dE_T/d\eta$,
  \item
    the mean transverse momentum $\langle p_T \rangle$ for charged-particles, pions, kaons, and protons \cite{Abelev:2013bla},
  \item
    the mean transverse momentum fluctuations $\delta p_T / \langle p_T \rangle$ (defined shortly),
  \item
    the two-particle flow cumulants $\vnk{n}{2}$ for $n=2,3,4$ and the four-particle flow cumulant $\vnk{2}{4}$ for charged-particles \cite{Adam:2016izf, Chatrchyan:2013nka}, and
  \item
    the symmetric cumulants $\SC(4, 2)$ and $\SC(3,2)$.
\end{itemize}
We matched the kinematic cuts of each observable to experiment with two exceptions: we used a larger rapidity interval $|\eta| < 0.8$ for some boost-invariant observables to improve our finite particle statistics, and we did not apply a rapidity gap, e.g.\ $|\Delta \eta| > 1$, between pairs of particles when calculating the two-particle cumulant $\vnk{n}{2}$ since we already oversample particles from each hydrodynamic event, and this oversampling suppresses non-flow correlations.

\begin{table}[b]
  \centering
  \caption{
    \label{tab:observables_v3}
    Experimental data used to calibrate the model parameters.
  }
  \small
  \makebox[\textwidth]{
    \begin{tabular}{ccccc}
      \toprule
      Collision system & Observable & Rapidity cut & Momentum cut & Ref. \\
      \midrule
      \multirow{3}{*}{Pb-Pb, 5.02 TeV}
      & Yield $d\nch/d\eta$ & $|\eta| < 0.5$ & --- & \cite{Adam:2015ptt} \\
      & Flow cumulants $\vnk{n}{2}$, & \multirow{2}{*}{$|\eta| < 0.8$,\, $|\Delta\eta| > 1$} & \multirow{2}{*}{$0.2 < p_T < 5.0$~GeV} & \multirow{2}{*}{\cite{Adam:2016izf}} \\
      & $n = 2$, 3, 4 & & & \\[1ex]
      \multirow{3}{*}{$p$-Pb, 5.02 TeV}
      & Yield $d\nch/d\eta$ & $|\eta| < 1.4$ & --- & \cite{Adam:2014qja} \\
      & Mean transverse momentum $p_T$ & $|\eta| < 0.3$ & $0.15 < p_T < 10$~GeV & \cite{Abelev:2013bla} \\
      & Flow cumulants $\vnk n 2$, & \multirow{2}{*}{$|\eta| < 2.4$,\, $|\Delta\eta| > 2$} & \multirow{2}{*}{$0.3 < p_T < 3.0$~GeV} & \multirow{2}{*}{\cite{Chatrchyan:2013nka}} \\
      & $n = 2$, 3 & & & \\
      \bottomrule
    \end{tabular}
  }
\end{table}

At the time of this dissertation writing, many of the aforementioned experimental observables are not yet available for $p$-Pb and Pb-Pb collisions at $\sqrts=5.02$~TeV.
Therefore, we restricted our calibration to the subset of measured and published observables listed in table~\ref{tab:observables_v3}.
Notably absent from this list are the four-particle cumulants $\vnk{n}{4}$ at $\sqrts=5.02$~TeV despite being measured and published.
Unfortunately, the four-particle cumulants require minimum-bias event statistics an order of magnitude larger than those used in this work.
Therefore we refrained from \emph{calibrating} on the four-particle cumulants, although I'll show calculations of the four-particle cumulant $\vnk{2}{4}$ later in this text, using a single set of calibration parameters.

Most of the calibration observables listed in table~\ref{tab:observables_v3} are presented as a function of collision centrality, where centrality is defined using some measure of the underlying event activity, e.g.\ the charged-particle yield in a given rapidity window.
To calculate these observables, we generated \order{10^4} minimum-bias events at each design point and divided the events into centrality bins using the charged-particle yield at midrapidity similar to the procedure used by experiment.

However, for some observables such as $p$-Pb mean $p_T$ \cite{Abelev:2013bla} and flow cumulants $\vnk{n}{k}$ \cite{Chatrchyan:2013nka}, the experiments used a special high-multiplicity trigger to select rare, ultra-central events according to the number of charged-particles produced $\nch$ or detector tracks offline $\ntrk$.
These high-multiplicity bins are too selective for our modest minimum-bias event sample, and so a different procedure is required.
We exploited, for this purpose, the approximate monotonic relation between each event's initial transverse energy density
\begin{equation}
  \frac{dE_T}{d\eta_s} \bigg\vert_{\eta_s=0} = \tau_0 \int d^2x_\perp e(\xv_\perp, \eta_s=0, \tau_0),
\end{equation}
and its final charged-particle density $(d\nch/d\eta) \vert_{\eta=0}$ at midrapidity.

Consider, for example, a single multiplicity bin $[\nch^\text{low}, \nch^\text{high}]$ which selects events from a minimum-bias event sample with $p_T^\text{min} < p_T < p_T^\text{max}$ and $|\eta| < \eta^\text{max}$.
Let $\langle\nch\rangle$ denote the average charged-particle multiplicity of these events.
We first rescaled the experimental multiplicity bin edges
\begin{equation}
  [\nch^\text{low}, \nch^\text{high}] \rightarrow \left[\frac{\nch^\text{low}}{\langle \nch \rangle},\, \frac{\nch^\text{high}}{\langle \nch \rangle}\right ],
\end{equation}
in order to reexpress each bin edge as a unitless variable.
These bin edges were then associated with a pair of transverse energy bin edges
\begin{equation}
  \label{eq:mult_trigger}
  \left [\frac{E_T^\text{min}}{\langle E_T \rangle}, \frac{E_T^\text{max}}{\langle E_T \rangle} \right] \leftrightarrow \left [\frac{\nch^\text{low}}{\langle \nch \rangle}, \frac{\nch^\text{high}}{\langle \nch \rangle} \right ],
\end{equation}
where $E_T \propto (dE_T/d\eta_s) \vert_{\eta_s=0}$ is the midrapidity transverse energy of a single event in the desired kinematic range, and $\langle E_T \rangle$ is the corresponding average transverse energy over the full minimum-bias event sample.

Finally, we mimicked the method used by experiment and applied equation~\eqref{eq:mult_trigger} to select rare high-multiplicity events from a continuous stream of minimum-bias \trento\ events satisfying the correct bin edges.
This of course means that, in addition to running a large sample of minimum-bias events for centrality binned observables, we also had to generate (much like experiment) a separate sample of multiplicity-triggered events.
In practice, we used a few hundred to a few thousand events per multiplicity bin, depending on the type of observable.

\subsubsection{Treatment of uncertainties}

We also took stock of the statistical and systematic errors reported by each experiment and incorporated their uncertainty into the likelihood covariance matrix
\begin{equation}
  \Sigma = \Sigma_m + \Sigma_e
\end{equation}
described in subsection \ref{subsec:likelihood}, which includes uncertainty contributions from both the model emulator $\Sigma_m$ and the experimental data $\Sigma_e$.
The experimental contribution to the covariance $\Sigma_e$ can be further broken down into its statistical and systematic components
\begin{equation}
  \Sigma_e = \Sigma_e^\text{stat} + \Sigma_e^\text{sys}.
\end{equation}
The statistical errors in $\Sigma_e^\text{stat}$ are uncorrelated, so its covariance matrix is diagonal:
\begin{equation}
  \Sigma_e^\text{stat} = \diag[(\sigma^\text{stat}_1)^2, (\sigma^\text{stat}_2)^2, \dots (\sigma^\text{stat}_m)^2 ],
\end{equation}
where $\sigma^\text{stat}_i$ is the statistical uncertainty of observable $y_i$ in the experimental observable vector $\yv_e = (y_1, \dots, y_m)$.
The systematic errors, meanwhile, are typically correlated, but the correlation structure is not reported by the experiments so we asserted a reasonable form.
We expanded the systematic covariance matrix as
\begin{equation}
  (\Sigma_e^\text{sys})_{ij} =  \rho_{ij} \sigma_i \sigma_j,
\end{equation}
where $\sigma_i$ and $\sigma_j$ are the systematic errors of observables $y_i$ and $y_j$ respectively, and $\rho_{ij}$ is the Pearson correlation coefficient between observable $y_i$ and $y_j$:
\begin{equation}
  \rho_{ij} = \frac{\cov(y_i, y_j)}{\sigma_i \sigma_j},
\end{equation}
which satisfies $\rho_{ij}=1$ for $i=j$ and $|\rho_{ij}| \le 1$ for $i \ne j$.
We assumed that each observable is correlated across different centrality/multiplicity bins, and uncorrelated with observables of a different type, e.g.\ correlations between yields and flows.
This is a crude simplifying assumption but it is better than neglecting the correlation structure of the experimental data entirely.

\begin{figure}[t]
  \centering
  \makebox[\textwidth]{
    \includegraphics{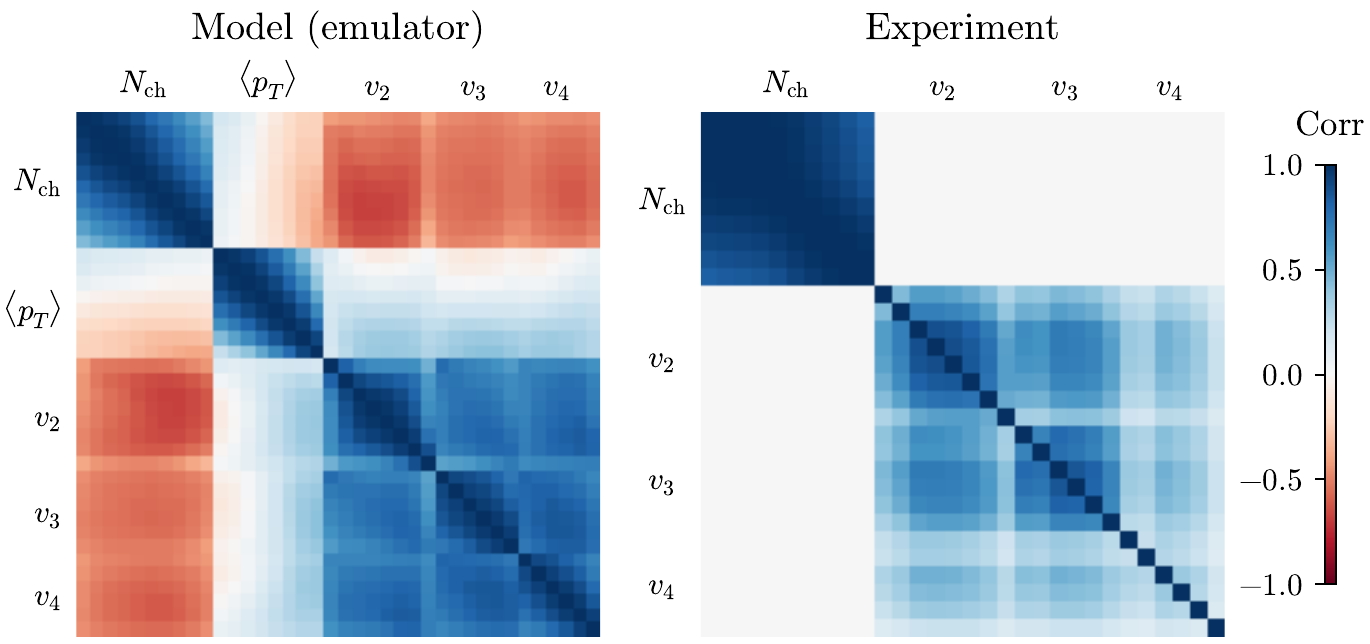}
  }
  \caption{
    \label{fig:correlation}
    Visualization of the Pb-Pb correlation matrix $\corr(y_i, y_j) = \cov(y_i,y_j)/(\sigma_i \sigma_j)$ for the model (emulator) at a random point in parameter space (left-side) and for the experimental data (right-side).
    Each cell represents an observable in a single centrality bin. Experimental statistical and systematic errors are from ALICE \cite{Adam:2015ptt, Adam:2016izf}.
    The experimental correlation structure is modeled using equation~\eqref{eq:corr}.
  }
\end{figure}

For the correlation structure between different observable bins, we asserted a simple Gaussian form
\begin{equation}
  \label{eq:corr}
  \rho_{ij}^\text{sys} = \exp \left[ -\frac{1}{2} \left(\frac{b_i - b_j}{l} \right)^2 \right],
\end{equation}
where $b_i$ and $b_j$ are the midpoints of two observable bins of a single type (centrality or relative multiplicity), and $l$ is a correlation length which describes how quickly the observable bins decorrelate as the distance between the bins increases.
We used centrality correlation lengths $l=100$ for all of the centrality binned Pb-Pb observables and $l=30$ for the centrality binned $p$-Pb charged-particle yield $d\nch/d\eta$.
The \mbox{$p$-Pb} mean $p_T$ and flow observables, meanwhile, use relative multiplicity bins $\nch / \langle \nch \rangle$ and $\ntrk / \langle \ntrk \rangle$ which necessitate a smaller correlation length $l=5$.
The right-side of figure~\ref{fig:correlation} shows an example correlation matrix
\begin{equation}
  \mathrm{corr}(y_i, y_j) = \cov(y_i, y_j)/(\sigma_i \sigma_j),
\end{equation}
for the Pb-Pb experimental data constructed using equation~\eqref{eq:corr}.
Here $y_i$ denotes an element of the experimental data $\yv_e$ and $\sigma_i$ its corresponding uncertainty.
The correlation matrix is block diagonal, with each block representing the correlations within a single class of observable.

For the model covariance matrix $\Sigma_m$, we used the procedure described in subsection \ref{subsec:emu_uncertainty} to calculate the emulator covariance in principal component space and transform it to the physical observable space.
This covariance matrix accounts for the statistical uncertainty caused by our limited event statistics and the interpolation uncertainty caused by our limited number of design points.
Figure~\ref{fig:correlation} shows the resulting Pb-Pb correlation matrix $\mathrm{corr}(y_i,y_j)$ for the model (emulator) at a random parameter point $\xv$ in the design space (left-side) along side the same correlation matrix for the experimental data (right-side) discussed previously.

\subsection{Bayesian parameter estimation}

We calibrated the model on two different collision systems by expanding the likelihood, equation \eqref{eq:likelihood}, into a joint likelihood
\begin{equation}
  \label{eq:joint_likelihood}
  P(E | \xv) = P(E_\text{Pb-Pb} | \xv) \cdot P(E_\text{$p$-Pb} | \xv),
\end{equation}
where $E$ subsumes all evidence from the $p$-Pb and Pb-Pb collision systems and $\xv$ is shorthand for the hypothesis that $\xv=\xv_\text{true}$.
We then used \textsc{emcee} \cite{emcee} to perform Markov-chain Monte Carlo (MCMC) importance sampling on the posterior distribution $P(\xv | E)$ defined by equation~\eqref{eq:bayes}, assuming a flat prior on each parameter that was constant within the design range \ref{tab:design_v3} and zero outside it.
The MCMC chain was allowed to ``burn-in'' before generating \order{7} posterior samples.

\begin{figure}[t]
  \centering
  \includegraphics{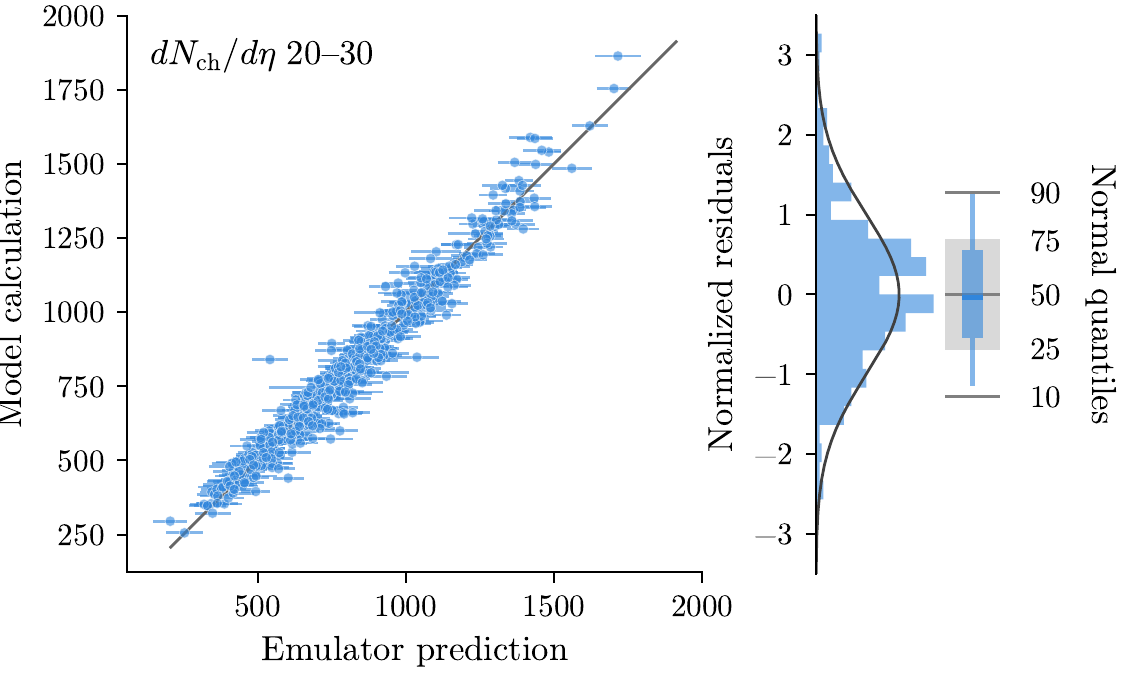}
  \caption{
    \label{fig:validation_example}
    Example emulator validation for one observable, the Pb-Pb charged-particle yield $d\nch/d\eta$ in the 20-30\% centrality class.
    We use the k-fold cross validation method (explained in the text) to partition the model inputs $X$ and outputs $Y$ into training and validation data.
    The scatter plot on the left shows the emulator predictions and one sigma error bars (x-axis) against explicit model calculations (y-axis).
    Perfect emulator/model agreement is indicated by the black like $y_\text{pred}=y_\text{obs}$.
    The histogram on the right shows that the errors are properly accounted for, i.e.\ the normalized residuals follow a normal distribution with unit variance and zero mean.
  }
\end{figure}

We validated the emulator predictions using a method known as k-fold cross validation.
The $d=500$ training points were randomly partitioned into $k=20$ equal sized subsamples or ``folds''.
For each partition, we used one subsample to validate the emulator and the remaining $k-1$ subsamples to train it.
The process was then repeated for each of the subsamples to validate all of the training data.
Figure~\ref{fig:validation_example} shows a scatter plot of the emulator predictions with one-sigma error bars (x-axis) against explicit model calculations (y-axis).
Perfect emulator and model agreement is indicated by the black line $y_\text{pred} = y_\text{obs}$.
If the emulator errors are properly accounted for, then the normalized residuals ${z=(y_\text{pred} - y_\text{obs})/\sigma_\text{pred}}$ sample a unit normal distribution as discussed in subsection \ref{subsec:emu_validation}, written as
\begin{equation}
  \label{eq:frac_error}
  P(z) \sim \mathcal{N}(\mu=0,\sigma=1).
\end{equation}

\begin{figure}[p]
  \centering
  \makebox[\textwidth]{
    \includegraphics{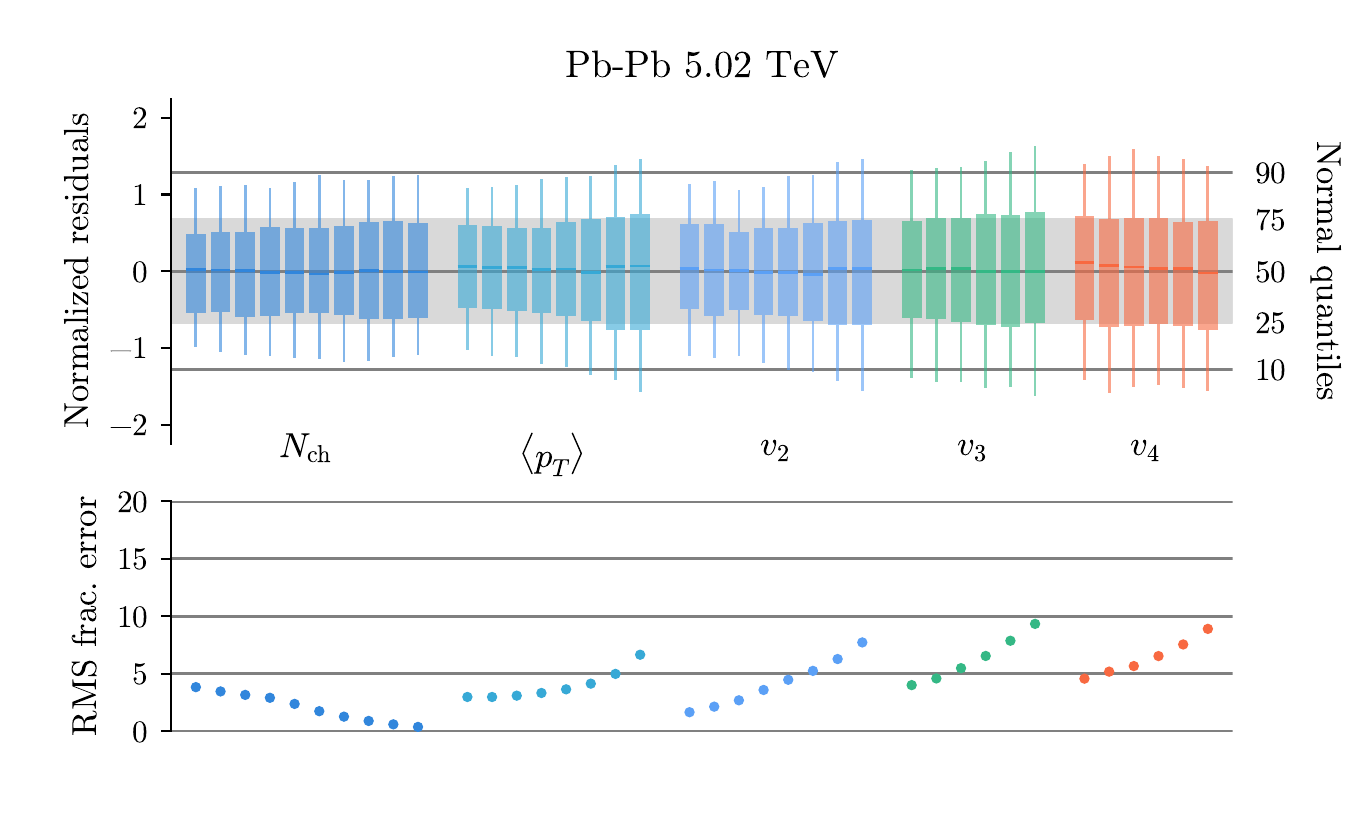}
  } \\
  \makebox[\textwidth]{
    \includegraphics{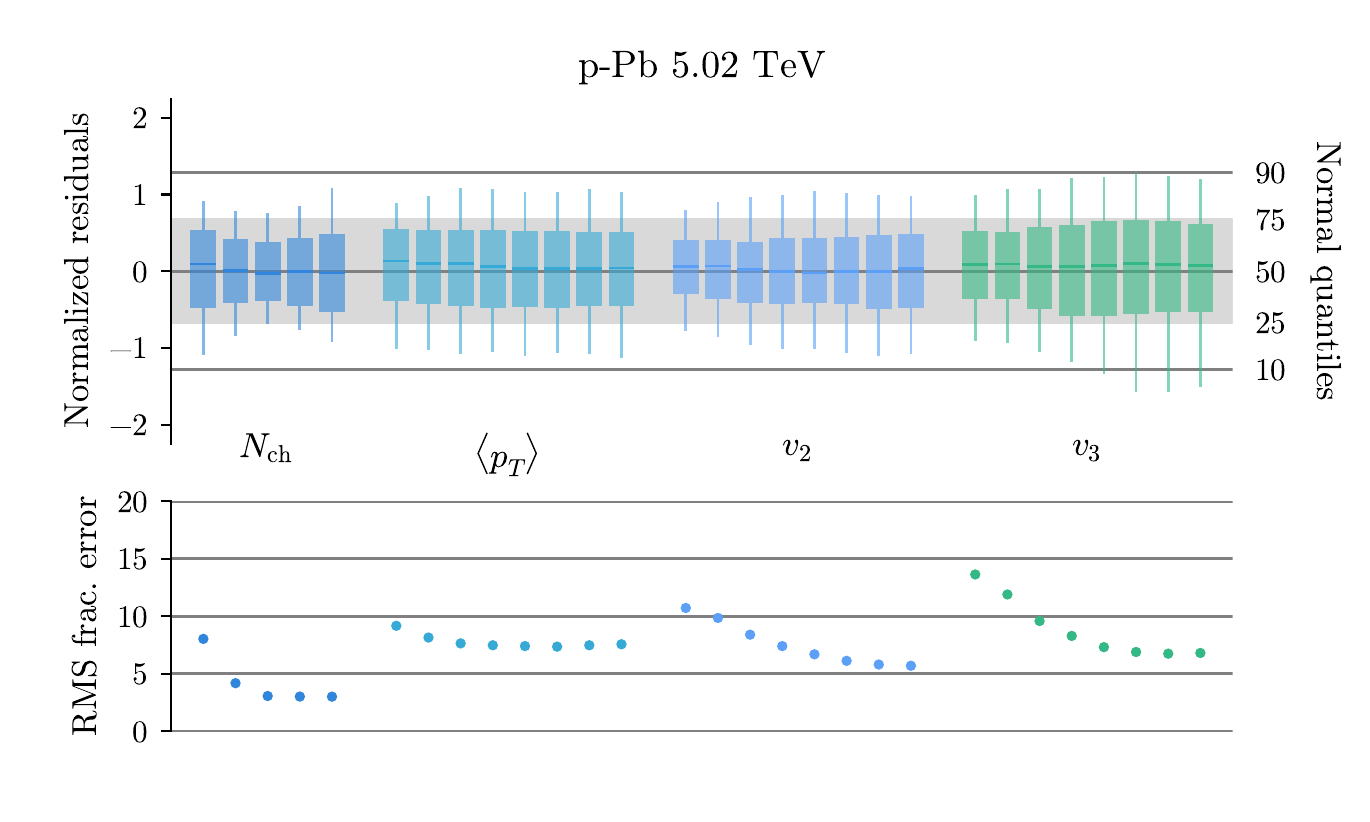}
  }
  \caption{
    \label{fig:validation_all}
    Emulator validation for the Pb-Pb collision system (top) and $p$-Pb collision system (bottom) at $\sqrts=5.02$~TeV.
    The ``piano keys'' in the top row of each figure are horizontally stacked box plots for the normalized residuals of each model observable.
    The boxes are 50\% interquartile ranges and whiskers are the 90\% interquantiles.
    The bottom row of each figure is the RMS fractional error defined by equation~\eqref{eq:frac_error}.
  }
\end{figure}

This comparison is shown by the histogram and box plot on the right side of figure~\ref{fig:validation_example}.
The emulator error is clearly significant, but it is also properly modeled, as indicated by the agreement between the normalized residuals and the unit normal distribution on the right (black curve).
Moreover, since we included this uncertainty in the likelihood covariance matrix \eqref{eq:likelihood}, we expect our results to be robust to the emulator limitations.
This is an important point that bears repeating.
The emulator uncertainty does not erode the veracity of the posterior distribution if it is correctly modeled and accounted for.

Figure \ref{fig:validation_all} applies this $z$-score validation test to \emph{every} observable $y \in \yv$ for the $p$-Pb and Pb-Pb collision systems.
The top row of each figure shows a box-plot for the normalized residuals of each observable compared to the quantiles of a unit normal distribution.
The thin horizontal black lines correspond to the 10th and 90th percentiles of a unit normal distribution, and the gray band its interquartile range.
These visual references should be compared to the whiskers and interquartile range respectively of each box plot, analogous to the comparison test of figure~\ref{fig:validation_example}.
The emulators generally behave as expected, although the $p$-Pb uncertainties are noticeably over estimated.
This is not necessarily a problem, but it is suboptimal.
Evidently our $p$-Pb emulator is somewhat \emph{more} accurate than it purports to be.
Therefore our posterior uncertainty estimates are conservative.

We also show in figure~\ref{fig:validation_all} an estimate of the emulator error magnitude.
This error is expressed as a fraction
\begin{equation}
  f_\text{err} = \frac{y_\text{pred} - y_\text{obs}}{(\Delta y)_{99\%}},
\end{equation}
where $(\Delta y)_{99\%}$ is 99\% of the full variability of $y$ across the design.
Thus $f_\text{err}$ can be thought of as a fractional emulator error relative to the full design variability.
The bottom row of each figure shows the root-mean-square (RMS) value of $f_\text{err}$.
It ranges from a few percent for most observables to a maximum value of 15\% for the $p$-Pb triangular flow $v_3\{2\}$ in the lowest multiplicity bin.
There are at least two ways which we could reduce this error: we could run more $p$-Pb events to reduce our finite statistical error, or we could add more design points to suppress our systematic emulator interpolation error.

\subsection{Posterior parameter estimates}
\label{subsec:results}

\begin{figure}[p]
  \centering
  \makebox[\textwidth]{
    \includegraphics{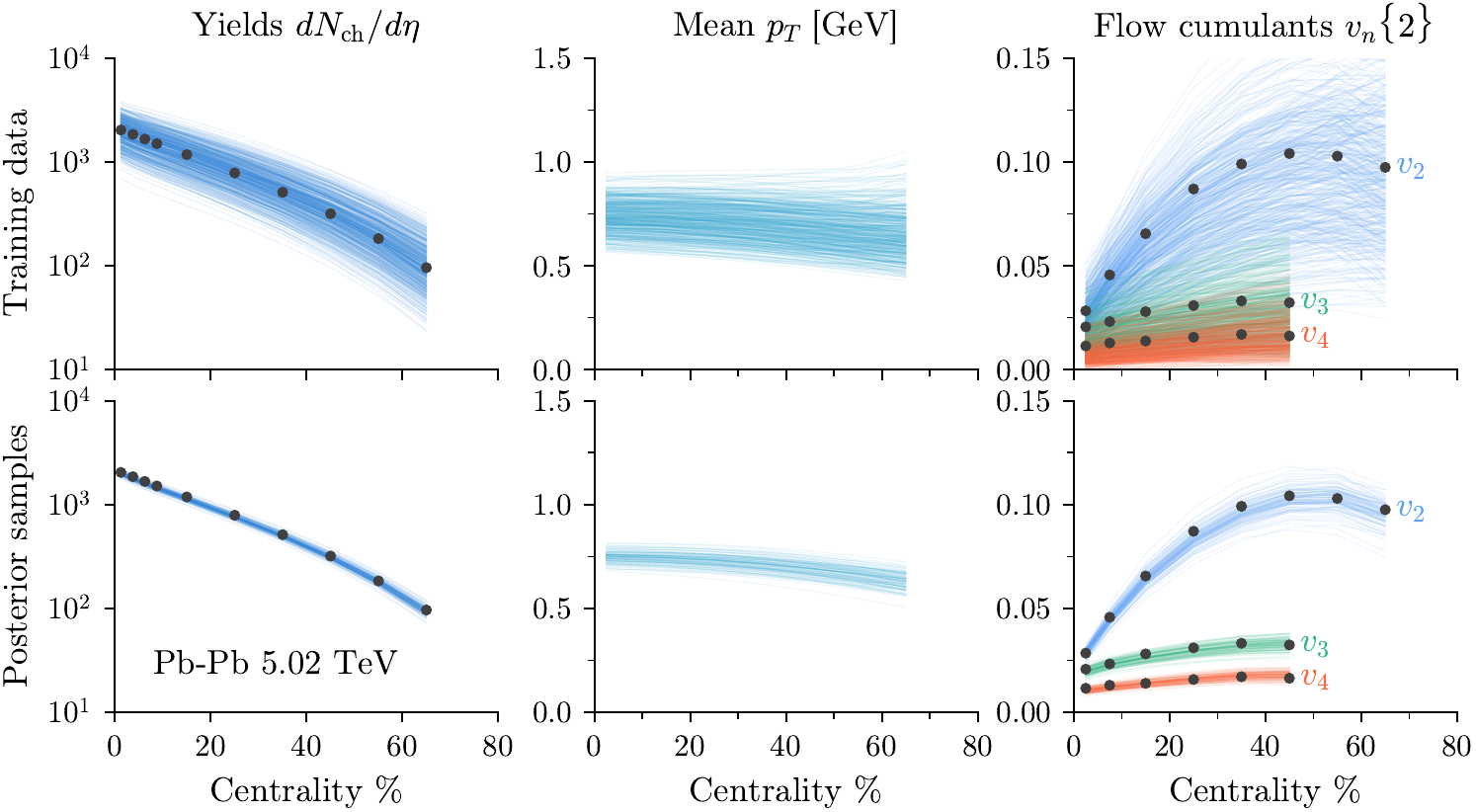}
  }
  \caption{
    \label{fig:obs_pbpb}
    Simulated observables compared to experimental data for Pb-Pb collisions at $\sqrts=5.02$~TeV.
    Top row: Explicit model calculations (no emulator) for each of the $d=500$ design points.
    Bottom row: Emulator predictions for $n=100$ random samples drawn from the posterior.
    Black symbols are experimental data from ALICE with statistical and systematic errors added in quadrature \cite{Adam:2015ptt, Adam:2016izf}.
  }
  \bigskip
  \makebox[\textwidth]{
    \includegraphics{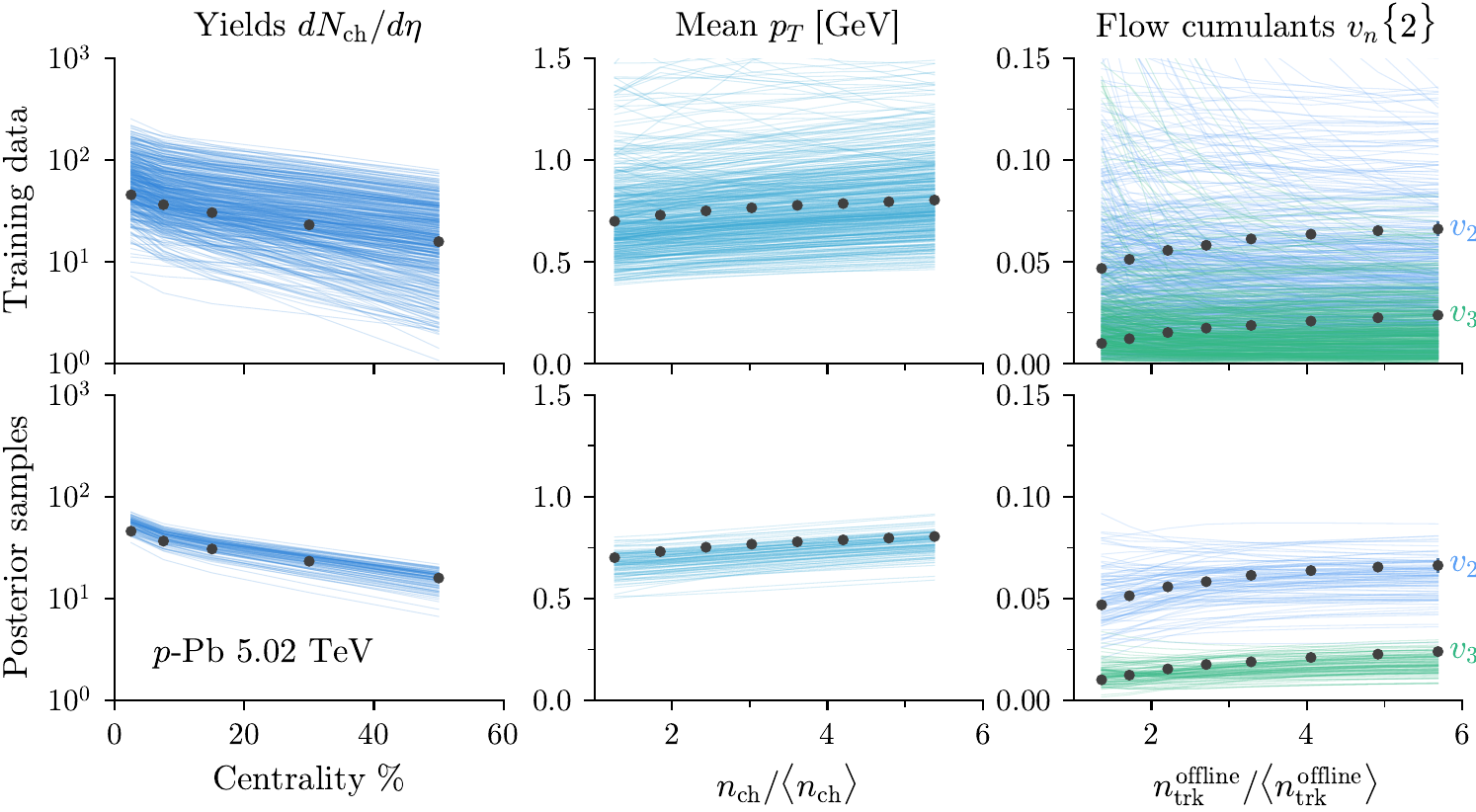}
  }
  \captionsetup{width=.7\paperwidth}
  \caption{
    \label{fig:obs_ppb}
    Same as Fig.~\ref{fig:obs_pbpb} but for $p$-Pb collisions at $\sqrts=5.02$~TeV.
    Multiplicity bins are used for mean $p_T$ and flow cumulant observables to match the bins used by experiment.
    Experimental data are from ALICE \cite{Adam:2014qja, Abelev:2013bla} and CMS \cite{Chatrchyan:2013nka}.
  }
\end{figure}

Figures \ref{fig:obs_pbpb} and \ref{fig:obs_ppb} show model calculations for the Pb-Pb and $p$-Pb systems respectively at $\sqrts=5.02$~TeV (thin colored lines) compared to experimental data from the CMS \cite{Chatrchyan:2013nka} and ALICE collaborations \cite{Adam:2015ptt, Adam:2016izf, Adam:2014qja, Abelev:2013bla}.
The top row of each figure shows explicit model calculations at each of the $d=500$ design points (training data), while the bottom row shows emulator predictions for $n=100$ random parameter samples drawn from the Bayesian posterior (sampled from the MCMC chain).
Each column shows a different class of observable.
The charged-particle yield $d\nch/d\eta$ is shown on the left, mean $p_T$ is in the middle, and two-particle flow cumulants $\vnk{n}{2}$ for $n=2,3,4$ are on the right.
The Pb-Pb mean $p_T$ and $p$-Pb $\vnk{4}{2}$ datasets are missing and hence are omitted from the present calibration.

Notice the large spread of the observables calculated at the training points (top row of each figure).
The design is constructed to vary each parameter across a wide range of values, specified in table~\ref{tab:design_v3}, and hence the corresponding model calculations are equally uncertain.
We also point out that there is considerably more variance in the $p$-Pb training data than the Pb-Pb training data.
The $p$-Pb yields, mean $p_T$, and flow cumulants all vary wildly within the chosen parameter ranges.
For instance, we can turn the $p$-Pb flows completely \emph{off} with suitably chosen parameters which is not possible in the Pb-Pb system.
Evidently the $p$-Pb model predictions are far more sensitive to modeling uncertainties.

\begin{table}[t]
  \centering
  \caption{
    \label{tab:post_param}
    Posterior parameter estimates corresponding to Fig.~\ref{fig:posterior_substruct}.
    The reported values are for the distribution median and 90\% HPD credible interval.
  }
  \small
  \newlength{\cellwidth}
  \settowidth{\cellwidth}{$-0.00$}
  \newcommand{\est}[3]{\parbox{\cellwidth}{\hfill$#1$}$_{-#2}^{+#3}$}
  \begin{tabular}{llll}
    \toprule
    \multicolumn{2}{c}{Initial condition / Pre-eq}     & \multicolumn{2}{c}{QGP medium}              \\
    \cmidrule(r){1-2}                                    \cmidrule(l){3-4}
    \addlinespace[.4ex]
    Norm       & \est{20.0}{2.5}{2.6} GeV      & $\etasmin$     & \est{0.08}{0.07}{0.07}            \\[1.1ex]
    $p$        & \est{0.002}{0.180}{0.157}     & $\etasslope$   & \est{1.24}{1.24}{1.46} GeV$^{-1}$ \\[1.1ex]
    $\sigmaf$  & \est{0.91}{0.33}{0.32}        & $\etascrv$     & \est{-0.09}{0.91}{0.80}           \\[1.1ex]
    $r$        & \est{0.88}{0.23}{0.26} fm     & $\zetasmax$    & \est{0.026}{0.026}{0.032}         \\[1.1ex]
    $n_c$      & \est{6.0}{3.4}{3.0}           & $\zetaswidth$  & \est{0.035}{0.035}{0.043} GeV     \\[1.1ex]
    $v$        & \est{0.52}{0.20}{0.28} fm     & $\zetasloc$    & \est{0.174}{0.024}{0.020} GeV     \\[1.1ex]
    $\dmin$    & \est{1.12}{0.49}{0.58} fm     & $\Tsw$         & \est{0.149}{0.014}{0.013} GeV     \\[1.1ex]
    $\tfs$     & \est{0.47}{0.37}{0.55} \fmc   & \\
    \addlinespace[.4ex]
    \bottomrule
  \end{tabular}
\end{table}

Conversely, the calibrated (posterior sampled) emulator predictions (bottom row of each figure) are far better constrained and nicely track the experimental data points.
We emphasize here that the posterior parameter values are obtained from a \emph{simultaneous} calibration to $p$-Pb and Pb-Pb data, and thus they are self-consistent between the two systems.
The spread in the posterior samples reflects different sources of model and experimental uncertainty as well as tension in the optimal fit parameters which describe each observable.
I'll demonstrate later in the text that a single set of model parameters well describes all of the calibration data.
Therefore it appears that much of the spread in the posterior samples is uncertainty contributed by our emulator.
We also note that although the $p$-Pb posterior samples have a somewhat larger spread than the Pb-Pb samples, the percentage uncertainty of the $p$-Pb emulator is similar to that of the Pb-Pb emulator, and thus the difference is likely due to the larger variance of the $p$-Pb training data.
The uncertainty in the posterior distribution could thus be improved by running the calibration with more design points or with a narrower range of parameter values to increase the density of the training points and reduce interpolation uncertainty.

\begin{figure}[p]
  \makebox[\textwidth]{
    \includegraphics[width=.95\paperwidth]{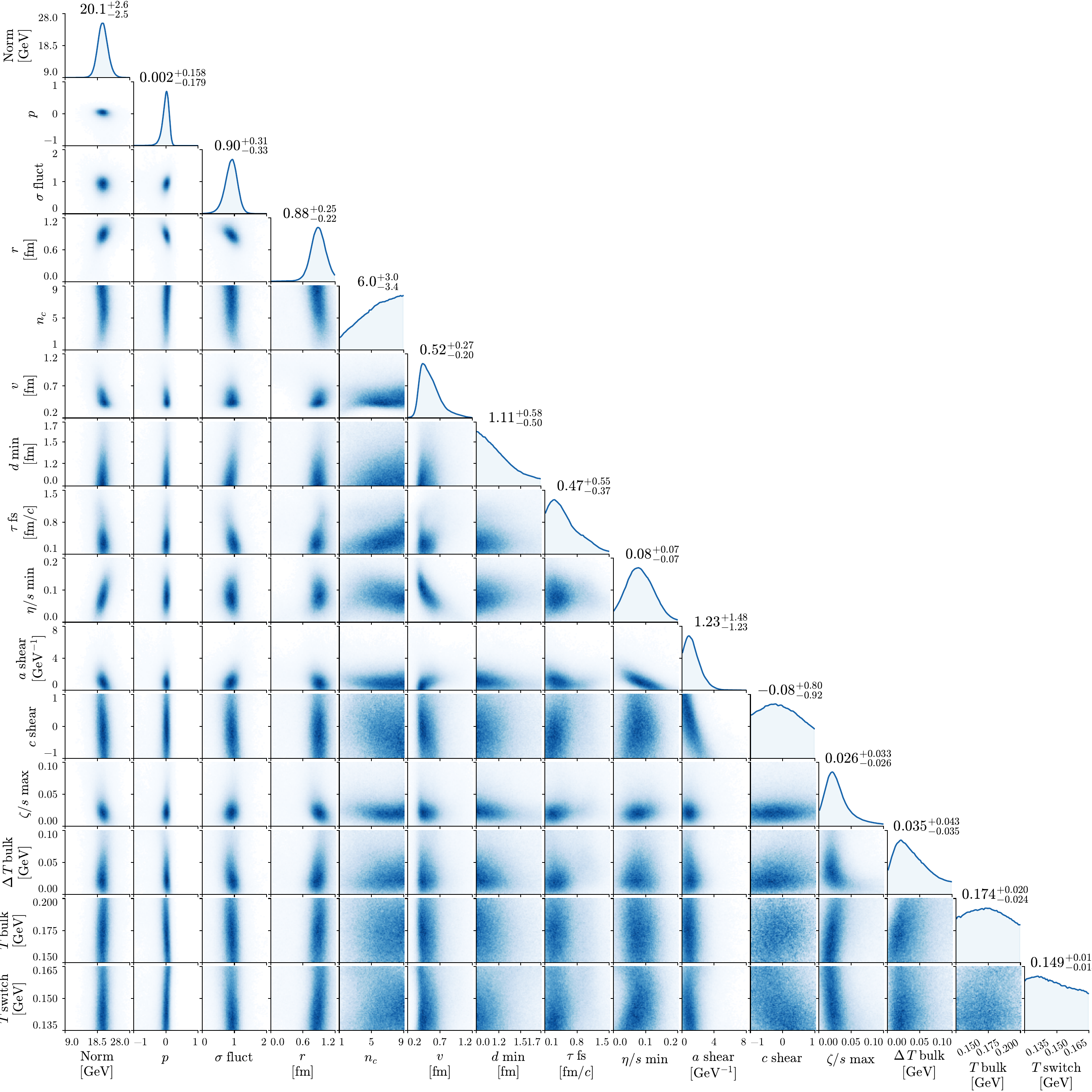}
  }
  \captionsetup{width=.85\paperwidth}
  \caption{
    \label{fig:posterior_substruct}
    Bayesian posterior distribution of the model input parameters.
    The diagonal panels show the marginalized distributions of individual model parameters, while off-diagonal panels show the joint distributions for pairs of model parameters, visualizing their correlations.
    The marginalized distribution medians and 90\% credible intervals are annotated along the diagonal.
  }
\end{figure}

Figure~\ref{fig:posterior_substruct} shows the main result of this work, the fifteen dimensional posterior distribution for the model input parameters.
Table \ref{tab:post_param} also lists each parameter's median and 90\% HPD credible interval.
I'll proceed as before by describing the constraints on each model parameter one-by-one, starting with the initial condition parameters located in the top-left corner of the posterior diagonal and ending with the QGP medium parameters located in the bottom-right.

\subsubsection{Initial condition properties}

The \trento\ normalization factor ${\text{Norm} = 20.0^{+2.6}_{-2.5}}$ and generalized mean energy deposition parameter $p=0.002^{+0.157}_{-0.180}$ are well constrained by the present analysis.
Moreover, figures~\ref{fig:obs_pbpb} and \ref{fig:obs_ppb} show that the model predictions using these values nicely describe both the $p$-Pb and Pb-Pb calibration observables.
While it would not be surprising to fit one or two of these observables using a narrow range of parameter values, the quality of the combined fit (more on this later) and the number of observables described is highly non-trivial.
For example, consider the ratio of the $p$-Pb charged-particle yield to the Pb-Pb charged-particle yield.
As the generalized mean parameter $p$ trends toward positive (negative) infinity, particle production scales like the maximum (minimum) of the two nuclear thickness functions.
This has a much stronger effect on the highly asymmetric $p$-Pb system than it does on the Pb-Pb system; hence the parameter $p$ strongly affects the ratio of the two average yields.

It is therefore compelling that $p \sim 0$ correctly describes the charged-particle yield $d\nch/d\eta$ of both systems while simultaneously describing the centrality dependence of $\vnk{n}{k}$, an observable which is also known to strongly depend on $p$ \cite{Bernhard:2018hnz}.
Specifically, this value corresponds to an energy deposition mapping proportional to the geometric mean of participant nuclear thickness
\begin{equation}
  \label{eq:geometric_mean}
  e(\xv_\perp, \eta_s=0, \tau_0) \propto \sqrt{\T_A\, \T_B}.
\end{equation}
We caution, however, that this specific analytic form should not be interpreted too literally.
For instance, a generalized mean described by $p=0.05$ is well within our 90\% credible interval, but it does not equal the geometric mean in equation~\eqref{eq:geometric_mean}.
We also note that this scaling is somewhat different than the scaling obtained by reference \cite{Bernhard:2016tnd}, which parametrized the \emph{entropy} density using a framework which assumed instant thermalization and zero pre-equilibrium flow.
Evidently, both prescriptions prefer geometric mean scaling, but each prescription leads to a somewhat different interpretation of the initially produced quantity.

\begin{figure}[t]
  \centering
  \makebox[\textwidth]{
    \includegraphics{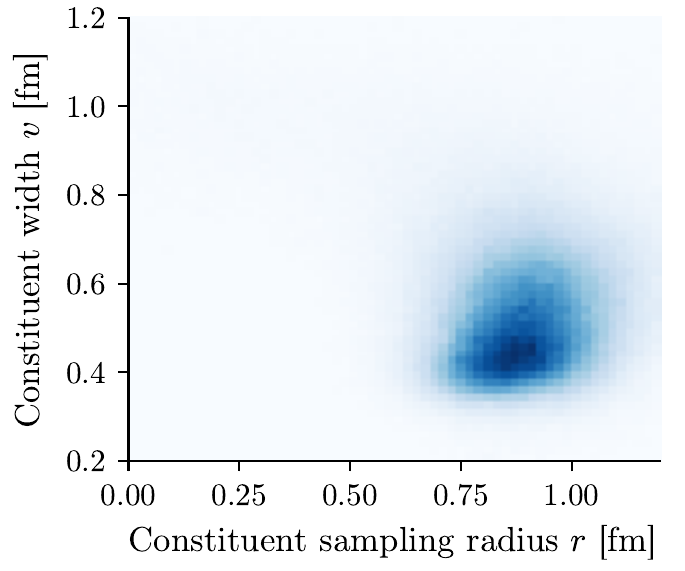}
    \hspace{3ex}
    \includegraphics{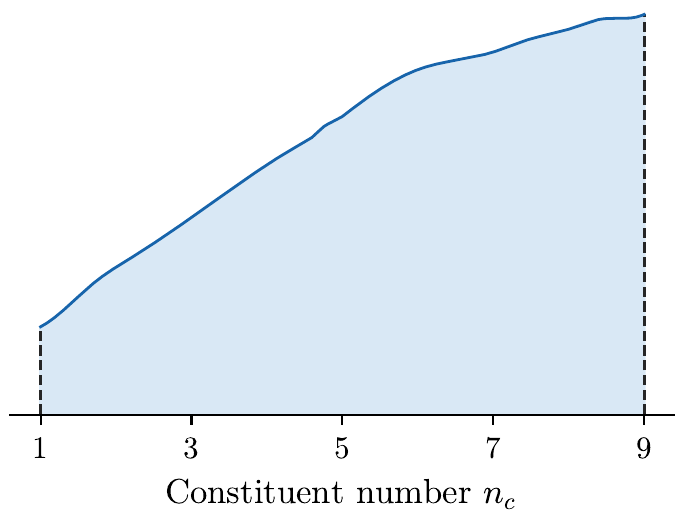}
  }
  \caption{
    \label{fig:posterior_proton}
    Left: Joint posterior distribution for the constituent sampling radius $r$ and constituent width $v$.
    The prior range for $r$ and $v$ spans the full plot range.
    The posterior distribution, shown in blue, indicates the preferred values for $r$ and $v$ determined by the analysis.
    Right: Marginal posterior distribution on the constituent number $n_c$ which was varied in the initial range $n_c \in [1, 9]$.
  }
\end{figure}

Continuing down the diagonal in figure~\ref{fig:posterior_substruct}, we see that the constituent sampling radius $r$ and the constituent width $v$ are both tightly constrained.
Figure~\ref{fig:posterior_proton} shows the joint posterior distribution of both parameters (left side), plotted for the region scanned by the Bayesian prior.
This figure suggests that we can infer the nucleon's fluctuating size and shape from the collective properties of bulk particle production, a feat largely unimaginable a decade ago.
While the sampling radius $r$ varies the size of the nucleons, we caution that its specific meaning should be interpreted with care; it specifies a computational \emph{sampling} radius, not a physical nucleon width.
Consider, for instance, a single nucleon with $n_c = 2$ constituents.
If the two constituent positions land on the same side of the nucleon, the effective nucleon size will be smaller than the Gaussian sampling radius $r$.
Despite this idiosyncrasy, one can easily define a physical nucleon width in the nucleon center-of-mass frame \emph{ex post facto}, given specific values for the sampling radius $r$, constituent width $v$, and constituent number $n_c$.

For example, using the posterior distribution's median values, $r=0.88$~fm, $n_c=6$, and $v=0.52$~fm, we can generate a large ensemble of random nucleon configurations and average their density in each nucleon's center-of-mass frame.
The resulting ensemble-averaged nucleon density
\begin{equation}
  \langle \rho_n(\xv) \rangle = \frac{1}{(2 \pi w^2)^{3/2}} \exp \left( -\frac{|\xv|^2}{2 w^2} \right),
\end{equation}
is described by a single Gaussian of width $w = 0.96$~fm.
This nucleon width is consistent with a previous estimate, $w = 0.96_{-0.05}^{+0.04}$~fm, obtained by a similar Bayesian analysis of Pb-Pb collisions at $\sqrts=2.76$ and 5.02~TeV using a physics model without nucleon substructure \cite{Bernhard:2018hnz}.

This is perhaps the single largest difference between our work and the conclusions of recent saturation-based calculations which constrained the event-by-event fluctuations of the proton using a color-dipole picture of vector meson production \cite{Mantysaari:2016ykx, Mantysaari:2016jaz}.
Those studies find that the measured coherent and incoherent $J/\Psi$ spectra at HERA prefer a compact gluon distribution inside each nucleon, with a Gaussian width $w_g \approx 0.4$~fm which is roughly \emph{half} the Gaussian width preferred by our analysis.
Evidently, it may be necessary to place an informative prior on our nucleon substructure parameters in order to resolve the apparent tension between our parameter values and those needed to describe DIS measurements at HERA.
Alternatively, it is also possible that the fluctuations probed by coherent and incoherent $J/\Psi$ production are different than those probed by minimum-bias particle production.

Moving on, we redirect our attention to the posterior on the constituent number $n_c$ shown enlarged in figure \ref{fig:posterior_proton} (right side).
The distribution is not sharply peaked, and hence we refrain from quoting a distribution median and 90\% credible interval.
Note, however, that the posterior clearly favors $n_c > 1$ constituents.
This is not surprising.
Sans nucleon substructure, saturation based models produce proton-sized fireballs in $p$-Pb collisions characterized by small eccentricities \cite{Bzdak:2013zma}.
Saturation-based models are therefore unable to describe the significant flow measured in high-multiplicity $p$-Pb collisions without nucleon substructure, or alternatively, some other source of additional correlations \cite{Schenke:2017bog}.
It's also worth noting that we see no special preference for $n_c=3$ constituents as commonly used in the literature.
In fact, larger constituent numbers $n_c$ generally improve the description of the data.

\subsubsection{QGP medium properties}

\begin{figure}[t]
  \centering
  \makebox[\textwidth]{
    \includegraphics{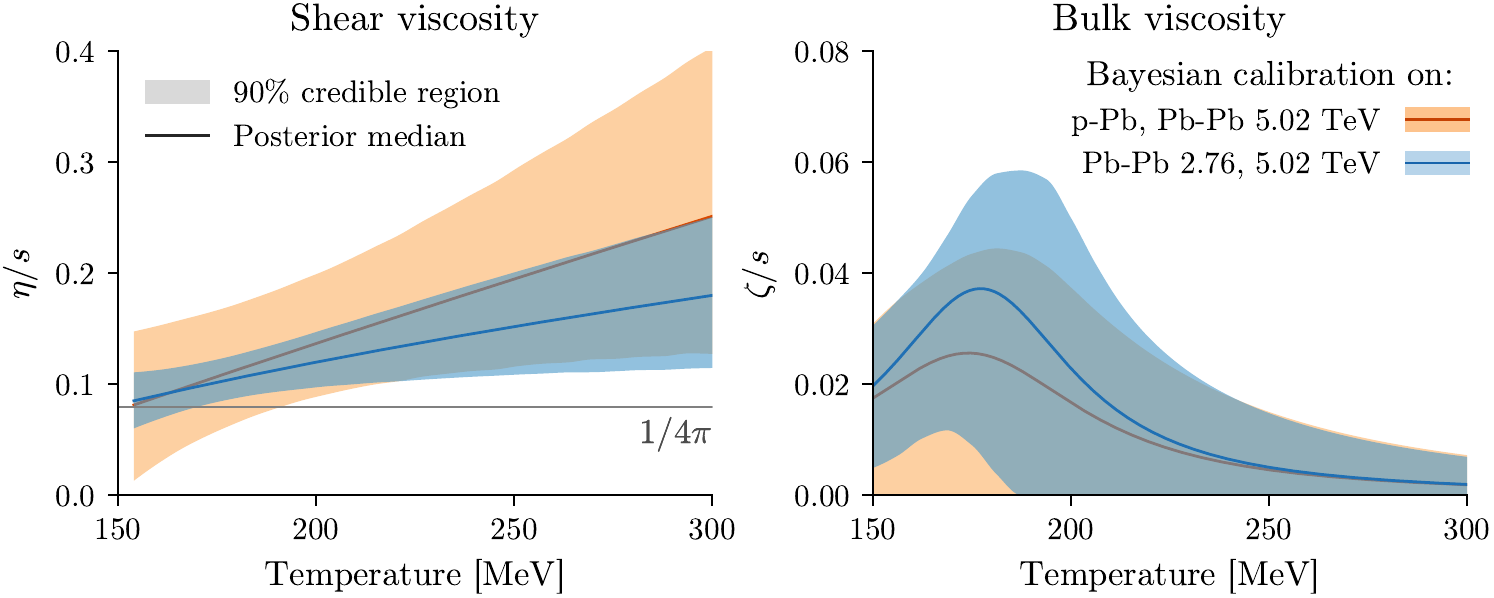}
  }
  \caption{
    \label{fig:region_shear_bulk}
    Left figure: Estimated temperature dependence of the QGP specific shear viscosity $(\eta/s)(T)$ determined by the present Bayesian analysis of $p$-Pb and Pb-Pb collisions at $\sqrts=5.02$~TeV (orange line/band) compared to a previous Bayesian analysis of Pb-Pb collisions at $\sqrts=2.76$ and 5.02~TeV (blue line/band) \cite{Bernhard:2019ntr}.
    The lines are the medians of each posterior distribution, and the bands are their 90\% credible regions.
    Right figure: Same as before, but for the temperature dependence of the QGP specific bulk viscosity $(\zeta/s)(T)$.
  }
\end{figure}

It's interesting to compare the posterior estimates for the shear and bulk viscosities obtained by this study to those of reference \cite{Bernhard:2019ntr} which used an (almost) identical version of the present physics model.
The only modeling difference is the inclusion of nucleon substructure in the present study which was absent from reference~\cite{Bernhard:2019ntr}.
Several calibration details, however, are different between the two analyses.
This work used a modest number of $p$-Pb and Pb-Pb observables at $\sqrts=5.02$~TeV (limited by availability), whereas reference~\cite{Bernhard:2019ntr} calibrated on a much larger number of Pb-Pb observables at $\sqrts=2.76$ and 5.02~TeV.

The posterior free streaming time $\tfs=0.47_{-0.37}^{+0.55}~\fmc$ obtained in this work is significantly smaller than the estimate $\tfs=1.16_{-0.25}^{+0.29}\ \fmc$ obtained by reference~\cite{Bernhard:2019ntr}.
We point out that the present study is missing several important observables which could affect the estimated free streaming time, e.g.\ the Pb-Pb mean $p_T$ and mean $p_T$ fluctuations at $\sqrts=5.02$~TeV.
Nevertheless, it appears that the inclusion of nucleon substructure significantly reduces the maximum allowed free streaming time, although more work is needed to establish if this is indeed the case.

We also compare in figure~\ref{fig:region_shear_bulk} our estimates for the temperature dependence of the QGP specific shear viscosity $(\eta/s)(T)$ and bulk viscosity $(\zeta/s)(T)$ with those of reference~\cite{Bernhard:2019ntr}.
The lines are the distribution medians, and the bands are their 90\% credible regions.
The results of this work are shown in orange, and the results of reference~\cite{Bernhard:2019ntr} are shown in blue.
In general, our estimates are broader and less certain than reference \cite{Bernhard:2019ntr} but otherwise self-consistent.
Evidently, the combined analysis of Pb-Pb data at $\sqrts=2.76$ and 5.02~TeV in reference~\cite{Bernhard:2019ntr} provides a better constraint on the QGP viscosities which is not surprising given the additional observables and multiple beam energies studied.
The $p$-Pb data used in this study, meanwhile, does not appear to provide any unique viscous constraints.

\begin{figure}[p]
  \centering
  \makebox[\textwidth]{
    \includegraphics{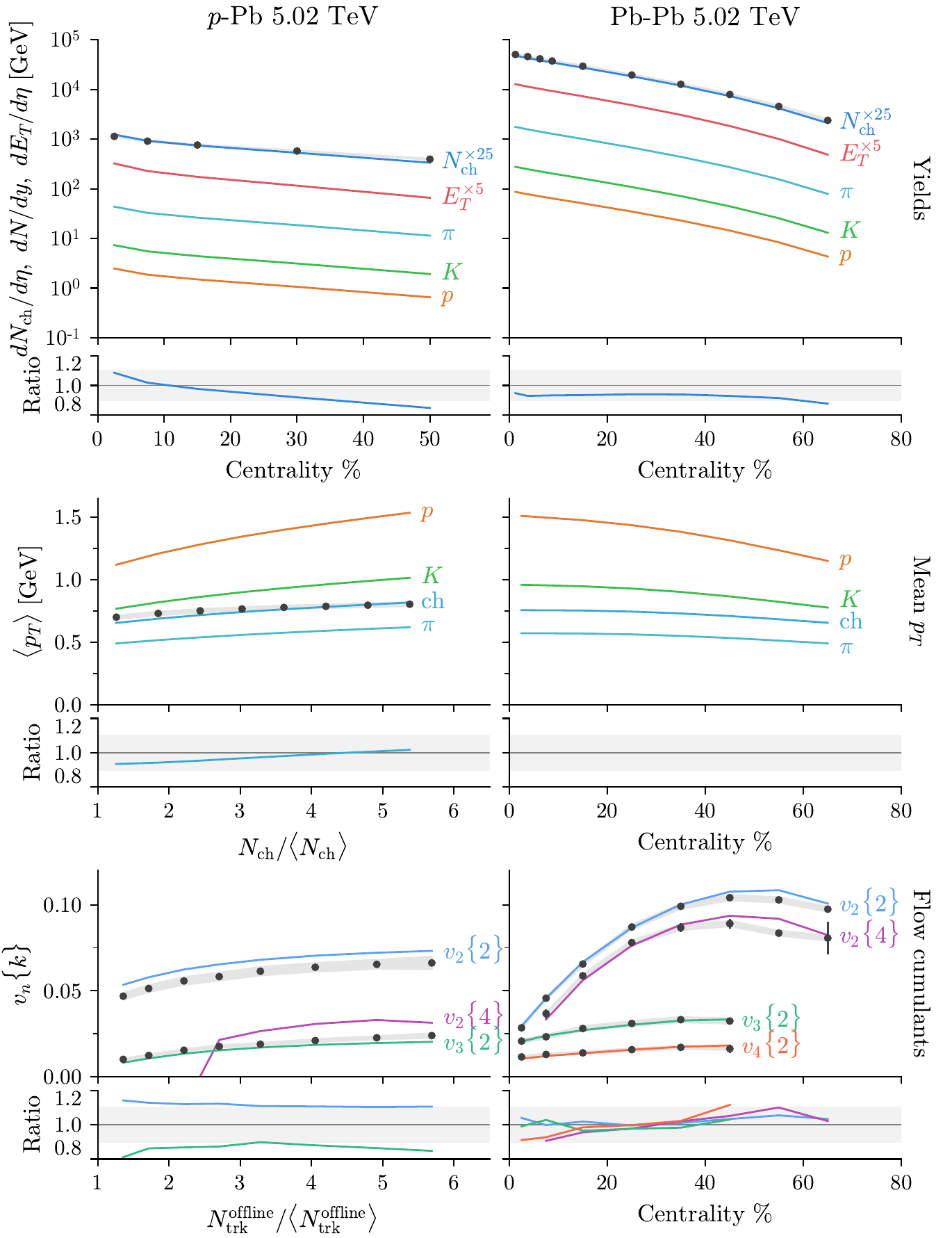}
  }
  \captionsetup{width=.62\paperwidth}
  \caption{
    \label{fig:obs_map}
    Model calculations using the \emph{maximum a posterior} (MAP) parameters compared to experiment.
    Colored lines are model calculations for $p$-Pb collisions (left) and Pb-Pb collisions (right) at $\sqrts=5.02$~TeV.
    Points with error bars are the experimental data with statistical uncertainties, and gray bands their corresponding systematic uncertainties, from CMS \cite{Chatrchyan:2013nka} and ALICE \cite{Adam:2015ptt, Adam:2016izf, Adam:2014qja, Abelev:2013bla}.
    The sub-axes show the ratio of model over data where available with gray bands indicating $\pm 10\%$.
  }
\end{figure}

\subsubsection{Verification of high-probability parameters}

Finally, we verified the emulator and tested the accuracy of our physics model framework using parameters selected from the mode of the Bayesian posterior distribution, listed in table \ref{tab:mode_params}.
Recall that these parameters characterize the approximate ``best fit'' values of the calibrated model, and are commonly referred to as the \emph{maximum a posteriori} (MAP) estimate.
We then ran \order{10^6} minimum-bias and multiplicity-triggered events using the MAP estimate and computed all of the model observables listed in section~\ref{sec:observables}.
The resulting model calculations are shown in figure~\ref{fig:obs_map} alongside experimental data from CMS \cite{Chatrchyan:2013nka} and ALICE \cite{Adam:2015ptt, Adam:2016izf, Adam:2014qja, Abelev:2013bla}.
The left and right columns show the results for the $p$-Pb and Pb-Pb collision systems respectively, and each row shows a different group of related observables.

\begin{table}[h]
  \centering
  \caption{
    \label{tab:mode_params}
    High-probability parameters selected from the posterior distribution and used to generate Fig.~\ref{fig:obs_map}.
    The posterior distribution on the particlization temperature $\Tsw$ is flat (agnostic), so we fix it's value using Ref.~\cite{Bernhard:2019ntr}.
  }
  \small
  \begin{tabular}{ll@{\hspace{2em}}ll}
    \toprule
    \multicolumn{2}{c}{Initial condition / Pre-eq} & \multicolumn{2}{c}{QGP medium} \\
    \midrule
    Norm     & 20. GeV        & $\etasmin$      & 0.11           \\
    $p$      & 0.0            & $\etasslope$    & 1.6 GeV$^{-1}$ \\
    $k$      & 0.19           & $\etascrv$      & -0.29          \\
    $n_c$    & 6              & $\zetasmax$     & 0.032          \\
    $r$      & 0.81 fm        & $\zetaswidth$   & 0.024 GeV      \\
    $v$      & 0.43 fm        & $\zetasloc$     & 175 MeV        \\
    $\dmin$  & 0.81 fm        & $\Tsw$          & 151 MeV        \\
    $\tfs$   & 0.37 \fmc \\
    \bottomrule
  \end{tabular}
\end{table}

The global agreement of the MAP model calculations with the experimental data is very good.
The largest tension is observed in the two-particle cumulants $\vnk{2}{2}$ and $\vnk{3}{2}$ of the $p$-Pb system, although even that tension is only about 10--15\%.
Quite remarkably, the model perfectly describes the shape of the $p$-Pb and Pb-Pb two-particle correlations which is strong evidence that these correlations are hydrodynamic in origin.
Moreover, we obtain an excellent description of the \mbox{$p$-Pb} mean $p_T$, although this fit is somewhat less meaningful since we are unable to calibrate on the Pb-Pb mean $p_T$ simultaneously (data is not yet available).
Additionally, the model provides a simultaneous description of the \mbox{$p$-Pb} and Pb-Pb charged-particle yields using a single energy deposition parameter $p=0$.
This is the \emph{exact same} generalized mean $p$-value supported by multiple previous studies \cite{Moreland:2014oya, Bernhard:2016tnd, Ke:2016jrd, Bernhard:2018hnz}.
Evidently, this scaling continues to hold for initial conditions with sizable nucleon substructure.

We also present calculations for several observables which were omitted from the calibration due to missing experimental data and the statistical limitations of our training data.
Here our MAP event sample is several orders of magnitude larger so the statistics are no issue.
The bottom-right panel of figure~\ref{fig:obs_map} shows our model calculation for the four-particle elliptic flow cumulant $\vnk{2}{4}$ along with the measured data points from \mbox{ALICE} \cite{Adam:2016izf}.
We see that the MAP estimate nicely describes the measured $\vnk{2}{4}$ data which is encouraging since this particular observable was never used to calibrate the model.

\begin{figure}[t]
  \centering
  \makebox[\textwidth]{
    \includegraphics{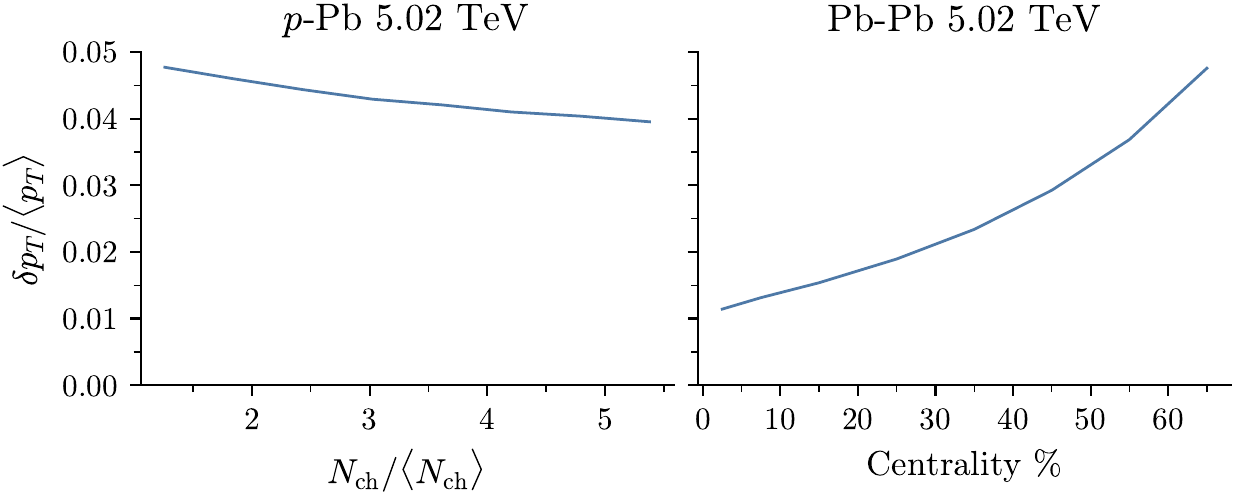}
  }
  \caption{
    \label{fig:pT_fluct}
    Relative mean transverse momentum fluctuations $\delta p_T / \langle p_T \rangle$ plotted for high-multiplicity $p$-Pb collisions (left) and Pb-Pb collisions (right) at $\sqrts=5.02$~TeV.
  }
\end{figure}

The relative mean $p_T$ fluctuation $\delta p_T / \langle p_T \rangle$ is another important bulk observable to test the predictions of the calibrated model.
It measures the \emph{dynamical} component of event-by-event mean $p_T$ fluctuations, quantified by the two-particle correlator
\begin{equation}
  \label{eq:mean_pT_corr}
  (\delta p_T)^2 = \langle \langle (p_{T,i} - \langle p_T \rangle) (p_{T,j} - \langle p_T \rangle) \rangle \rangle.
\end{equation}
The inner-average in equation~\eqref{eq:mean_pT_corr} runs over all pairs of particles $i,j$ in the same event, the outer average runs over all events in a given bin (centrality or multiplicity), and the symbol $\langle p_T \rangle$ denotes the usual mean transverse momentum of particles in the bin.
The observable is typically presented in terms of the dimensionless ratio $\delta p_T / \langle p_T \rangle$ which quantifies the strength of dynamical fluctuations in units of the average transverse momentum $\langle p_T \rangle$.

Figure \ref{fig:pT_fluct} shows the MAP estimate predictions for the $p$-Pb and Pb-Pb relative mean $p_T$ fluctuations $\delta p_T / \langle p_T \rangle$ at $\sqrts=5.02$~TeV.
For the Pb-Pb system, we used centrality bins, and for the $p$-Pb system we used the same relative multiplicity bins used for the $p$-Pb charged-particle mean $p_T$.
The relative mean $p_T$ fluctuations have been shown to be particularly sensitive to the existence of nucleon substructure \cite{Bozek:2017elk}, and thus it would be interesting to ultimately include this observable in the calibration when the data becomes available.

\begin{figure}[t]
  \centering
  \makebox[\textwidth]{
    \includegraphics{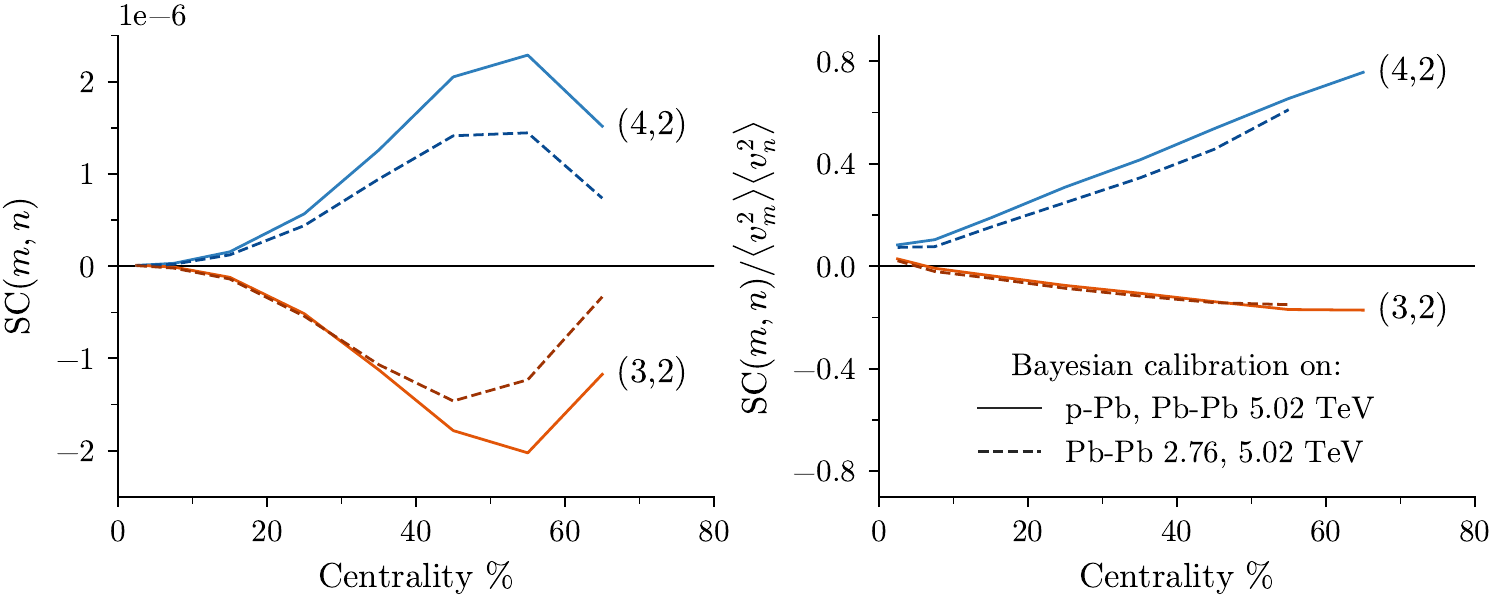}
  }
  \caption{
    \label{fig:flow_corr}
    Model calculations of the symmetric cumulants (top) and normalized symmetric cumulants (bottom) for Pb-Pb collisions at $\sqrts=5.02$~TeV using the \emph{maximum a posteriori} (MAP) parameters.
    The solid lines are the MAP estimate of the present analysis (with nucleon substructure), and the dashed lines are the MAP estimate of Ref.~\cite{Bernhard:2018hnz} (without nucleon substructure) which was calibrated on Pb-Pb observables at $\sqrts=2.76$ and 5.02~TeV.
    In general, most model parameters are somewhat different between the two studies.
  }
\end{figure}

Finally, we present in figure \ref{fig:flow_corr} our calculations for the symmetric cumulants $\SC(m, n)$ (left) and normalized symmetric cumulants $\NSC(m,n)$ (right) for $(m, n) = (4, 2)$ (blue lines) and $(3, 2)$ (orange lines), which were previously defined in subsection \ref{subsec:verification_trento3d}.
Recall that these observables quantify correlations between event-by-event fluctuations of the flow harmonics of different order \cite{Bilandzic:2013kga, ALICE:2016kpq}.
The solid lines are calculations using the current model and MAP estimate, while the dashed lines are calculations using the model and MAP estimate of reference \cite{Bernhard:2019ntr}, calibrated to Pb-Pb collisions at $\sqrts=2.76$ and 5.02 TeV \emph{without} nucleon substructure.
We observe that the gap between $\SC(3,2)$ and $\SC(4,2)$ is generally wider in the present analysis than in reference~\cite{Bernhard:2018hnz}, as is the gap between the normalized symmetric cumulants $\NSC(3,2)$ and $\NSC(4,2)$.

Multiple aspects of the two analyses are different such as the collision systems and beam energies considered, the observables which were included in each calibration, and the existence of nucleon substructure in the model.
Therefore, we can only speculate what might have caused the large difference in the symmetric flow cumulants between the two analyses.
Two reasonable culprits would be the inclusion of nucleon substructure and the large difference in the preferred pre-equilibrium free streaming time determined by the two studies.
Future studies of the symmetric cumulants therefore promise to provide additional constraints on the initial condition and QGP medium parameters.

\section{Topics for future study}

This dissertation shed new light on the QGP initial condition problem, but the problem itself is hardly solved.
As I conclude this results chapter, I would like to identify some areas of concern and outline new ideas for future improvements to the present work.
The following topics are those which I consider to be the highest priority.

\subsection{\trento\ model}

\paragraph{Nuclear structure}

The starting point of every initial condition model is a Monte Carlo generator for the nuclear density inside the nucleus.
If this nuclear density is miscalculated, then it can severely hamstring the model predictions.
Initial condition models typically use low-energy nuclear structure measurements to model the nucleus at high-energy.
This is a somewhat worrisome approximation that should be investigated further.
Presumably, if the inelastic cross section grows strongly with collision energy, the size of the nucleus could as well.

\paragraph{Multiplicity fluctuations}

Minimum-bias proton-proton collisions produce large multiplicity fluctuations which cannot be described by geometric effects alone.
To account for these fluctuations, I multiplied each nucleon (or constituent) by an \emph{ad hoc} random weight following previous work \cite{Shen:2014gfa, Dumitru:2012yr, Bozek:2013uha}.
Personally, I find this prescription somewhat unsettling because it associates the fluctuations with the participant matter instead of with the dynamical process itself.
What is the physical origin of these multiplicity fluctuations?

\paragraph{Generalized mean ansatz}

The generalized mean ansatz is flexible, but it is still a one parameter function.
It is possible that we've missed a degree of freedom that is important to describe the data.
The merits of the generalized mean ansatz rest heavily on its assumption of scale invariance, equation \eqref{eq:scale_invariant}, so it is important to test the validity of this assertion.
If future experiments show that scale invariance breaks, then the generalized mean ansatz should be modified or discarded.
Alternatively, one could try expanding the energy density $e(\T_A, \T_B)$ as a sum of symmetric orthogonal functions.

\subsection{Calibration methodology}

\paragraph{New collision systems}

I have dedicated the vast majority of this dissertation to studying $p$-Pb, Xe-Xe, and Pb-Pb collisions.
There are, of course, many other collision systems which have been studied by experiments at RHIC and the LHC, including $p$-$p$, $d$-Au, $^3$He-Au, Cu-Cu, Cu-Au, Au-Au, and U-U to name a few.
I've presented limited results for a few of these systems, but I have not yet studied them at great length.
Once the \trento\ model is calibrated at a specific beam energy, it should be able to describe all collision systems without additional fine tuning.
Therefore, this additional data should be used to validate (or invalidate) the model.

\paragraph{Parameter energy dependence}

In principle, the \trento\ parameters could all vary as functions of the beam energy $\sqrts$.
It would be interesting, for example, to see if we obtain different posterior estimates for the Gaussian nucleon width at different beam energies, or if we observe a beam energy dependence for the \trento\ generalized mean parameter $p$.
This could signal the emergence of new physics at high-energies.

\paragraph{Sizing the nucleon}

We obtained relatively consistent posterior estimates for the \trento\ model parameters with one notable exception.
The Gaussian nucleon width $w$ varied considerably from study to study.
To complicate matters further, our most recent posterior estimate \cite{Bernhard:2019ntr} is incompatible with independent proton size measurements based on coherent and incoherent $J/\Psi$ production \cite{Mantysaari:2016ykx, Mantysaari:2016jaz, Aaron:2009aa, Abramowicz:2015mha}.
It would be helpful to add new observables to the model calibration which are directly sensitive to the fluctuating size and shape of the proton.

\paragraph{Expanded nucleon substructure study}

I presented an exploratory study of $p$-Pb and Pb-Pb collisions at $\sqrts=5.02$~TeV using an updated version of the \trento\ model with nucleon substructure \cite{Moreland:2018gsh}.
Future work should expand this substructure analysis to include more collision systems and beam energies.
Nucleon substructure could resolve several existing puzzles, such as the centrality dependence of mean transverse momentum fluctuations $\delta p_T / \langle p_T \rangle$ and symmetric flow cumulants $\SC(m, n)$.

\chapter{Conclusion}
\label{ch:conclusion}

\lettrine{H}{ydrodynamics} is an indispensable tool to study the bulk properties of hot and dense nuclear matter produced by ultrarelativistic nuclear collisions.
Notably, the framework has been used to extract dynamical properties of the strongly-coupled QGP liquid produced in the first ${\sim}10^{-23}$~seconds of the collision, such as its specific shear viscosity $\etas$ and bulk viscosity $\zetas$.
For many years, decades even, theoretical uncertainty in the simulation's initial conditions has complicated efforts to rigorously constrain the QGP transport coefficients using heavy-ion collision data.
This \emph{initial condition problem} has correspondingly been identified as one of the primary factors impeding our understanding of QGP matter.

Over the last decade, theoretical models of the QGP initial conditions have improved greatly.
\emph{Ab initio} calculations based on general concepts of gluon saturation are now able to reproduce a wide variety of experimental bulk observables---including the collision's final particle yields, mean $p_T$, and multiparticle correlations---to impressive precision.
There remains, however, differences between these calculations, and there is no uniform consensus within the theoretical community regarding the validity of each approach.
Moreover, it is generally fair to say that no single initial condition model describes the experimental data perfectly.
Therefore, it is natural to wonder how this residual error affects extracted values of the QGP transport coefficients.

In this dissertation, I motivated and developed---in collaboration with Jonah Bernhard and Steffen Bass---a simple parametric model for the QGP initial conditions called \trento\ \cite{Moreland:2014oya}, based on a family of functions known as the generalized means.
This functional form parametrizes a semi-exhaustive subspace of all physically reasonable initial condition models, allowing the model to not only mimic certain calculations in the literature but also interpolate between them.
Hence, it effectively represents a meta-model for the landscape of mutually incompatible theory calculations.

This parametric initial condition model was then embedded in a realistic hydrodynamic simulation and constrained using a Bayesian parameter estimation framework developed by Jonah Bernhard \cite{Bernhard:2018hnz}.
Leveraging the statistical machinery provided by his model-to-data comparison framework, we were able to show that the initial condition and QGP medium properties are simultaneously constrained by the experimental data \cite{Bernhard:2019ntr}.
This analysis resulted in the first quantitative estimates for $(\etas)(T)$ and $(\zetas)(T)$ obtained from hydrodynamic simulations with rigorously defined modeling uncertainties (see figure \ref{fig:region_shear_bulk_v2}).
Determining these quantities has been a primary goal of the RHIC and LHC heavy-ion programs and has been the subject of numerous published papers.
Notably, we found that the QGP specific shear viscosity obtains a minimum value $\etasmin = 0.085_{-0.025}^{+0.026}$ at $T_c = 154$~MeV which is conspicuously close to the conjectured KSS bound $\etas \ge 1/4\pi$ \cite{Danielewicz:1984ww, Policastro:2001yc, Kovtun:2004de}.

Studies of the \trento\ initial condition model at midrapidity also suggest that the collision's initial entropy density \cite{Bernhard:2016tnd} or energy density \cite{Bernhard:2019ntr, Moreland:2018gsh} scales approximately as
\begin{equation}
  \left \{
  \begin{aligned}
    e_0\\
    s_0
  \end{aligned}
  \right \}
  \propto \sqrt{\T_A\, \T_B},
\end{equation}
where $\T_A, \T_B$ are the participant thickness functions of each nucleus defined by equation \eqref{eq:participant_thickness}.

The specific meaning of this expression, i.e.\ whether the quantity on the left-side is an energy or entropy density, depends on the assumed hydrodynamic matching procedure.
Although one should be careful interpreting this analytic form too literally---the analyses constrain an approximate region of function space, not a specific analytic form---the result is so simple that it merits further investigation.
Furthermore, we find that this scaling persists for different beam energies and nuclear collision systems.

This dissertation research further expands on the previous studies by performing additional exploratory analyses of the QGP initial conditions far from midrapidity and in small collision systems.
In both cases, the present hydrodynamic model framework provides a compelling simultaneous description of bulk particle properties, corroborating the assumptions of the \trento\ initial condition framework and the broad success of hydrodynamic descriptions.

Perhaps most intriguingly, the small-system study revealed that a simultaneous quantitative description of $p$-Pb and Pb-Pb bulk observables at $\sqrts=5.02$~TeV is obtainable if one simply replaces spherically symmetric round nucleons with deformed ``lumpy'' nucleons, parametrized by two additional degrees of freedom.

This observation evidences the hydrodynamic nature of small-system multi-particle correlations and suggests that a unified hydrodynamic description of small and large collision systems is emerging.
While additional work is needed to establish if this is indeed the case, it is compelling that the \trento\ model with nucleon structure simultaneously describes the $p$-Pb and Pb-Pb collision systems using the \emph{exact} same model parameters and model-to-data comparison methods.

\bigskip

I want to conclude by pausing to appreciate the remarkable success of the hydrodynamic framework pioneered by Landau, Bjorken, and the many other scientists which have contributed to its formulation over the years.
Extracting the QGP specific shear viscosity to $\pm 30\%$ accuracy from hydrodynamic simulations would have been unthinkable fifty years ago.
Moreover, no one could have imagined that hydrodynamic simulations might evidence the fluctuating shape of constituent sources inside the proton.
These developments indicate that hydrodynamic simulations are entering a new precision age.

\backmatter

\chapter{Acknowledgments}

\noindent
This work was supported by the Department of Energy National Nuclear Security Administration Stewardship Science Graduate Fellowship (DOE NNSA SSGF) under grant number DE-FC52-08NA28752, and by the U.S.\ Department of Energy (DOE) under grant number DE-FG02-05ER41367.

\vspace{1.25em}
\noindent
Computing resources were provided by the Open Science Grid (OSG), funded by the DOE and National Science Foundation (NSF), and by the National Energy Research Scientific Computing Center (NERSC), funded by the DOE.

{\raggedright\printbibliography[heading=bibintoc, title={References}]}

\end{document}